\documentclass[twocolumn]{aastex63}
\usepackage{graphicx}
\usepackage{color}
\usepackage{xcolor}
\usepackage{amsmath,amssymb}
\usepackage{ulem}
\usepackage{bm}

\makeatletter
\renewcommand{\paragraph}{\@startsection{paragraph}{4}{0ex}%
   {-3.25ex plus -1ex minus -0.2ex}%
   {1.5ex plus 0.2ex}%
   {\normalfont\small\it\center}}
\makeatother

\stepcounter{secnumdepth}
\stepcounter{tocdepth}

\DeclareRobustCommand{\dl}{\bgroup\markoverwith{\textcolor{red}{\rule[.5ex]{2pt}{0.4pt}}}\ULon}

\def\vs#1{\vspace*{#1cm}}
\def\hs#1{\hspace*{#1cm}}

\def\vsp#1{\vspace{ 0.#1cm}}
\def\vsm#1{\vspace{-0.#1cm}}

\def\tnm#1{\tablenotemark{\,\scriptsize \rm #1}}
\def\tnt#1#2{\tablenotetext{\scriptsize \rm #1}{\,#2}}

\def\lra{\longrightarrow}

\accepted{December 28, 2023}
\submitjournal{ApJS}
\shorttitle{The impact of mixing on molecule formation in SN~1987A}
\shortauthors{Ono et al.}

\begin{document}

\title{The impact of effective matter mixing based on three-dimensional hydrodynamical models on the molecule formation in the ejecta of SN~1987A}

\correspondingauthor{Masaomi Ono}
\email{masaomi@asiaa.sinica.edu.tw}

\author[0000-0002-0603-918X]{Masaomi Ono}
\affil{Institute of Astronomy and Astrophysics, Academia Sinica, Taipei 10617, Taiwan}
\affil{Astrophysical Big Bang Laboratory (ABBL), RIKEN Cluster for Pioneering Research, 2-1 Hirosawa, Wako, Saitama 351-0198, Japan}
\affil{RIKEN Interdisciplinary Theoretical and Mathematical Sciences Program (iTHEMS), 2-1 Hirosawa, Wako, Saitama 351-0198, Japan}
\author{Takaya Nozawa}
\affil{Division of Theoretical Astronomy, National Astronomical Observatory of Japan, Mitaka, Tokyo 181-8588, Japan}
\author{Shigehiro Nagataki}
\affil{Astrophysical Big Bang Laboratory (ABBL), RIKEN Cluster for Pioneering Research, 2-1 Hirosawa, Wako, Saitama 351-0198, Japan}
\affil{RIKEN Interdisciplinary Theoretical and Mathematical Sciences Program (iTHEMS), 2-1 Hirosawa, Wako, Saitama 351-0198, Japan}
\affil{Astrophysical Big Bang Group (ABBG), Okinawa Institute of Science and Technology Graduate University (OIST), 1919-1 Tancha, Onna-son, Kunigami-gun, Okinawa 904-0495, Japan}
\author{Alexandra Kozyreva}
\affil{Heidelberger Institut f\"ur Theoretische Studien, Schloss-Wolfsbrunnenweg 35, 69118 Heidelberg, Germany}
%
\author{Salvatore Orlando}
\affil{INAF-Osservatorio Astronomico di Palermo, Piazza del Parlamento 1, I-90134 Palermo, Italy}
\author{Marco Miceli}
\affil{Universit\`a degli Studi di Palermo, Dipartimento di Fisica e Chimica, Piazza del Parlamento 1, I-90134 Palermo, Italy}
\affil{INAF-Osservatorio Astronomico di Palermo, Piazza del Parlamento 1, I-90134 Palermo, Italy}
\author{Ke-Jung Chen}
\affil{Institute of Astronomy and Astrophysics, Academia Sinica, Taipei 10617, Taiwan}

\begin{abstract}
To investigate the impact of matter mixing on the formation of molecules in the ejecta of SN~1987A, time-dependent rate equations for chemical reactions are solved for one-zone and one-dimensional ejecta models of SN~1987A. %
The latter models are based on the one-dimensional profiles obtained by angle-averaging of the three-dimensional hydrodynamical models \citep{2020ApJ...888..111O}, which effectively reflect the 3D matter mixing; the impact is demonstrated, for the first time, based on three-dimensional hydrodynamical models. %
The distributions of initial seed atoms and radioactive $^{56}$Ni influenced by the mixing could affect the formation of molecules. 
By comparing the calculations for spherical cases and for several specified directions in the bipolar-like explosions in the three-dimensional hydrodynamical models, the impact is discussed. %
The decay of $^{56}$Ni, practically $^{56}$Co at later phases, could heat the gas and delay the molecule formation. %
Additionally, Compton electrons produced by the decay could ionize atoms and molecules and could destruct molecules. %
Several chemical reactions involved with ions such as H$^+$ and He$^+$ could also destruct molecules. %
The mixing of $^{56}$Ni plays a non-negligible role in both the formation and destruction of molecules through the processes above. %
The destructive processes of carbon monoxide and silicon monoxide due to the decay of $^{56}$Ni generally reduce the amounts. %
However, if the molecule formation is sufficiently delayed under a certain condition, the decay of $^{56}$Ni could locally increase the amounts through a sequence of reactions. %

\end{abstract}

\section{Introduction} \label{sec:intro} 

More than 30 years have passed since the discovery of supernova~1987A (SN 1987A) in LMC on February 23, 1987; it has entered a phase of young supernova remnant (SNR). %
Therefore, SN~1987A provides a unique opportunity to study the evolution of a core-collapse supernova (CCSN) from the explosion to an early phase of the SNR. %
Inside the ejecta, as the temperature goes down mainly due to the expansion, molecules and dust could form during this transition phase to a young SNR. %
Since SN~1987A is the last nearby supernova (SN) and the inner ejecta are spatially resolved (as mentioned below, spatially resolved distribution of the emission from molecules and dust has already been observed), SN~1987A is also a unique target for the study of molecule and dust formation in CCSNe. %

Actually, in infrared spectra of early observations of SN~1987A, carbon monoxide (CO) \citep{1987IAUC.4457....1C,1987IAUC.4468....2M,1987Msngr..50...18O,1988Natur.334..327S,1989MNRAS.238..193M,1993ApJS...88..477W} and silicon monoxide (SiO) \citep{1988MNRAS.235P..19A,1989Natur.337..533R,1991MNRAS.252P..39R,1993MNRAS.261..522R,1993A&A...273..451B,1993ApJS...88..477W} have been detected as the emission of rotational-vibrational (hereafter, ro-vibrational) transitions. %
The light curves of the CO fundamental ($\sim$ 4.6 $\mu$m) \citep{1993A&A...273..451B} and the first overtone ($\sim$ 2.3 $\mu$m) \citep{1993A&A...273..451B,1993MNRAS.261..535M} bands and the SiO fundamental ($\sim$ 8 $\mu$m) band \citep{1993A&A...273..451B} have also been delineated. %
From the observations above the masses of CO and SiO have been derived approximately as $10^{-5}$--$10^{-4}$ $M_{\odot}$ depending on the age and the model \citep{1987Msngr..50...18O,1988Natur.334..327S,1989MNRAS.238..193M} and $4 \pm 2 \times 10^{-6}$ $M_{\odot}$ \citep{1991MNRAS.252P..39R,1993MNRAS.261..522R}, respectively. %
The mass of CO has been re-interpreted as 10$^{-3}$ $M_{\odot}$ with a non-LTE and an optically thick assumption \citep{1992ApJ...396..679L}, and later approximately as (0.3--4) $\times 10^{-2}$ $M_{\odot}$ \citep{1995ApJ...454..472L}. %
Similarly, the mass of SiO was also re-interpreted as 10$^{-4}$ $M_{\odot}$ \citep{1994ApJ...428..769L}. %
CO has also been detected in other SNe such as Type~II~SN~1995ad \citep{1996MNRAS.283L..89S}, Type~IIn~SN~1998S \citep{2000AJ....119.2968G,2001MNRAS.325..907F}, Type~Ic~SN~2000ew \citep{2002PASJ...54..905G}, Type~II-P~SN~2004et \citep[][SiO was also detected]{2009ApJ...704..306K}, and Type~Ib/c~SN~2013eg \citep{2016ApJ...821...57D}. %
CO has been found even in the young SNR Cassiopeia A (Cas A) with the age of $>$ 300 yr, in which CO may have survived in the ejecta without significant destruction \citep{2009ApJ...693L..39R,2012ApJ...747L...6R}. %
For a more complete list of the detections of molecules in SNe and SNRs earlier than 2017 including the ones by the emission from rotational transitions, see Table~1 in \cite{2017hsn..book.2125M}. %


There have been theoretical studies of chemistry in the ejecta of CCSNe. %
Motivated by the early detections of CO and SiO in SN~1987A, the formation of molecules in the core of SN~1987A has been studied with a chemical reaction model (network) mainly focusing on CO \citep{1989ApJ...342..406P,1990ApJ...358..262L,1990MNRAS.246..208R}. %
Later, \cite{1995ApJ...454..472L} modeled CO vibrational transitions based on thermal-chemical calculations with the optically thick regime (already mentioned above), in which time-dependent chemical equations were solved by simultaneously solving the thermal balance with the effects of heating due to the energy deposition from the decay of $^{56}$Co and cooling via CO vibrational transitions; %
the observed light curves of the fundamental (${\it \Delta} v = 1$, here $v$ denotes the vibrational level) and the first overtone (${\it \Delta} v = 2$) bands were successfully reproduced \citep{1995ApJ...454..472L}, although an arbitrary time-dependent factor to control the energy deposition efficiency is necessary; %
even now there has been no other study that reproduces the CO light curves. There have been similar studies on the chemistry targeted at SiO \citep{1994ApJ...435..909L,1996ApJ...471..480L}. %
Apart from SN~1987A, the chemistry, i.e. the formation and/or destruction of molecules, clusters, and dust, in the ejecta of primordial (Population III) SNe has been investigated with comprehensive (large) chemical reaction networks in a series of works \citep{2008ApJ...683L.123C,2009ApJ...703..642C,2010ApJ...713....1C}. %
The chemistry of molecules and dust in the ejecta of Cas A (Type IIb SN) \citep{2014A&A...564A..25B,2016A&A...589A.132B} and Type II-P SNe \citep{2013ApJ...776..107S,2015A&A...575A..95S} has also been explored with similar approaches based on chemical reaction networks. %
Recently, \cite{2018MNRAS.480.5580S} revisited the chemistry in the ejecta of SN~1987A with a large chemical reaction network including the nucleation of clusters, and comprehensively applied a dust formation theory. %
\cite{2020A&A...642A.135L} studied the formation of CO with the cooling via ro-vibrational transitions of CO. %

It should be noted that in all the studies mentioned in the previous paragraph, one-zone or multi-zone ejecta models have been used; %
some models include the effects of matter mixing (see the next paragraph) with an ad-hoc assumption, i.e., a uniform composition throughout the core/a specific zone. %
There has been no study on molecule formation in CCSN ejecta that takes into account the effect of matter mixing based on 3D hydrodynamical models. %
Additionally, as pointed out in several studies, chemical reactions involved with He$^+$, Ne$^+$, Ar$^+$, could play a role in the destruction of molecules \citep{1990ApJ...358..262L,1996ApJ...471..480L,2009ApJ...703..642C}, which means that how He, Ne, and Ar are mixed in the core and ionized by high-energy electrons produced via the decay of $^{56}$Co could potentially be crucial for the survival of molecules. %

Matter mixing has been evoked by several early observations of SN~1987A, i.e., the early detections of hard X-ray emission \citep{1987Natur.330..230D,1987Natur.330..227S}, direct gamma-ray lines from the decay of $^{56}$Co \citep{1988Natur.331..416M,1990MNRAS.245..570V}, the fine structure in H$_{\alpha}$ line \citep{1988MNRAS.234P..41H}, and [Fe~II] lines with high-velocity tails ($\sim 4000$ km s$^{-1}$) \citep{1990ApJ...360..257H}; %
some mixing of radioactive $^{56}$Ni to be conveyed into high-velocity outer layers which consist of helium and hydrogen is necessary. Motivated by the failures of early numerical studies based on spherical CCSN explosions \citep{1989ApJ...341L..63A,1990ApJ...358L..57H, 1992ApJ...390..230H,1991ApJ...367..619F,1991A&A...251..505M, 1991ApJ...370L..81H,1992ApJ...387..294H}, which have mainly focused on the impact of fluid instabilities such as Rayleigh-Taylor instability, the impact of non-spherical CCSN explosions \citep{1991ApJ...382..594Y,1998ApJ...495..413N,2000ApJS..127..141N,2000ApJ...531L.123K,2003ApJ...594..390H,2003A&A...408..621K,2005ApJ...635..487H,2006A&A...453..661K,2009ApJ...693.1780J,2010ApJ...709...11J,2010ApJ...723..353J,2010ApJ...714.1371H,2010A&A...521A..38G,2012ApJ...755..160E,2013ApJ...773..161O,2015ApJ...808..164M,2015A&A...577A..48W,2017ApJ...842...13W,2021MNRAS.502.3264G} on the matter mixing has been studied based on multi-dimensional hydrodynamical simulations. %
It was shown that a bipolar-like explosion works to produce a large amount of $^{44}$Ti, which has the advantage to explain the bright light curve of SN 1987A in the late phase \citep{1997ApJ...486.1026N,2000ApJS..127..141N}. %
From the point of view of the mechanisms of CCSN explosions, multi-dimensionality in the explosion due to neutrino-driven convection, Standing Accretion Shock Instability (SASI) \citep{2003ApJ...584..971B}, and magnetorotational effects are essential for a successful explosion \citep[for reviews of the mechanisms of CCSN explosions, see][]{2012PTEP.2012aA301K,2012AdAst2012E..39K,2012ARNPS..62..407J,2012PTEP.2012aA309J,2013RvMP...85..245B,2016PASA...33...48M}. %
\cite{2021MNRAS.502.3264G} investigated the so-called ``Ni bubble" effect \citep[e.g.,][]{1994ApJ...425..264B}, i.e., the acceleration and inflation of the ejecta by the heating due to the radioactive decay of $^{56}$Ni ($^{56}$Ni $\lra$ $^{56}$Co $\lra$ $^{56}$Fe), with 3D hydrodynamical models with an approximated model for the energy deposition. %
3D hydrodynamical models with 3D matter mixing have also been applied to other CCSN objects, Cas A \citep{2017ApJ...842...13W}, Crab \citep{2020MNRAS.496.2039S}, and SN~2007Y \citep{2023MNRAS.523..954V}. %
Among studies above, \cite{2015A&A...577A..48W} pointed out the importance of the density structure of the progenitor stars on the matter mixing. %
Actually, there have been debates \citep{1989ARA&A..27..629A} on the progenitor star of SN~1987A, Sanduleak $-$69$^{\circ}$ 202 
(Sk $-$69$^{\circ}$ 202), which is known as a blue supergiant (BSG) \citep{1987A&A...177L...1W,1987ApJ...321L..41W}, since it had been difficult to explain observational features of Sk $-$69$^{\circ}$ 202, e.g., the effective temperature, luminosity, anomalous abundance in the nebula, and how to form the well-known triple-ring structure, by a single-star evolution model without fine-tuning parameters. %
Recently, progenitor star models, which satisfy all the observational features of Sk $-$69$^{\circ}$ 202, based on binary mergers have been developed \citep{2017MNRAS.469.4649M,2018MNRAS.473L.101U}, following the binary merger scenarios \citep{1990A&A...227L...9P,1992ApJ...391..246P,2007Sci...315.1103M,2009MNRAS.399..515M}. %
A series of theoretical studies on the optical light curves of SN~1987A  \citep{2015A&A...581A..40U,2019MNRAS.482..438M,2019A&A...624A.116U,2021ApJ...914....4U} have shown that binary merger progenitor models better explain the observed light curves than single-star progenitor models. %
On the basis of part of the studies above, we performed \citep{2020ApJ...888..111O} three-dimensional (3D) hydrodynamical simulations of globally asymmetric (bipolar-like) SN explosions with progenitor models including a binary merger progenitor model \citep{2018MNRAS.473L.101U}. %
As a result, an explosion model with the binary merger model succeeds to explain the observed features of [Fe II] lines \citep{1990ApJ...360..257H} compared with other explosion models with single-star progenitor models at least in the paper. %
Then, to link such aspherical explosions to the consequential observables in early SNR phases, further evolutions were followed by 3D magnetohydrodynamical (MHD) simulations \citep{2020A&A...636A..22O}; %
it was found that the explosion model with the binary merger progenitor model also well explains the observed X-ray spectra, morphology, and light curves \citep[e.g.,][]{2016ApJ...829...40F,2019NatAs...3..236M}. %
Additionally, as discussed in \cite{2020A&A...636A..22O}, thanks to the modeled bipolar-like explosion, the model is also able to reproduce the Doppler shift and broadening of the observed $^{44}$Ti lines with NuSTAR \citep{2015Sci...348..670B} and the distributions of CO and SiO inferred from the observations by ALMA \citep[][see the next paragraph]{2017ApJ...842L..24A}. %

Recent breakthrough observations of SN~1987A by ALMA \citep{2017ApJ...842L..24A} have revealed the 3D distribution of emission lines from rotational transitions of CO and SiO molecules in the ejecta. The distribution is not spherically symmetric and is rather complex; %
in particular, the distribution of CO has a torus-like structure perpendicular to the observed equatorial ring in the nebula of SN~1987A. %
It is noted that the approximated distributions of CO and SiO with the 3D MHD model above \citep[see, Fig.~10 in][]{2020A&A...636A..22O} are consistent with the sizes of the observed molecular-rich structures and the torus-like distribution of CO and its orientation to the equatorial ring; %
in the 3D model, the bipolar-like explosion axis is perpendicular to the CO torus. %
The consistency with the ALMA observations further supports the progenitor model, asymmetric explosion, and matter mixing demonstrated in the 3D MHD model \citep{2020A&A...636A..22O}. %
Further observations by ALMA have also delineated the distributions of the emission from both dust and the two molecules above \citep{2019ApJ...886...51C}. %
Interestingly, from the latter ALMA observations, a blob (hotspot) of dust emission has been found and it has been interpreted that the required extra heating to maintain the blob may be attributed to the emission from the putative neutron star of SN 1987A (NS 1987A), which has not directly been detected yet regardless of more than 30 years of searches \citep[e.g.,][]{2018ApJ...864..174A}. Recent spectral fittings of X-ray observations of SN 1987A by Chandra, NuSTAR, and XMM-Newton \citep{2021ApJ...908L..45G,2022ApJ...931..132G} have suggested a non-thermal component which probably stems from a pulsar wind nebula (PWN) activity, which has further supported the existence of NS 1987A \citep[for an alternative interpretation of the X-ray data, see][]{2021ApJ...916...76A}. %
There have also been attempts to constrain the properties of NS~1987A and to predict its evolution \citep{2020ApJ...898..125P,2023ApJ...949...97D}. %
Observations mentioned above have further motivated us to investigate the impacts of aspherical explosions and matter mixing on the formation of molecules in the ejecta of SN~1987A. %
Moreover, SN~1987A is a target of the newly launched James Webb Space Telescope (JWST) to figure out features of shocked dust grains \citep[approved JWST proposal for Cycle 1:][]{2021jwst.prop.1726M}. %
Recently, the first observational results of SN~1987A by JWST NIRSpec have been reported \citep{2023ApJ...949L..27L}. 
The ejecta distribution traced by [Fe I] line looks like a broken dipole. %
It is remarkable that the morphology and orientation are roughly consistent with the modeled bipolar-like explosions in our previous studies \citep{2000ApJS..127..141N,2020ApJ...888..111O,2020A&A...636A..22O}. %

In this study, we investigate the impact of effective matter mixing (before the molecule formation in the ejecta) on the formation of diatomic molecules approximately up to ten thousand days after the explosion, for the first time, based on 3D CCSN explosion models for SN 1987A \citep{2020ApJ...888..111O}. %
We construct a chemical reaction network to follow the formation of molecules in the ejecta. %
The thermal evolution of the ejecta is crucial for molecule formation. %
Then, heating by the radioactive decay of $^{56}$Ni\footnote{Hereafter, ``the decay of $^{56}$Ni" means ``the decay of $^{56}$Ni and/or $^{56}$Co", if not specifically mentioned. This is because $^{56}$Co is involved with the same decay sequence of $^{56}$Ni $\lra$ $^{56}$Co $\lra$ $^{56}$Fe.} (and/or $^{56}$Co) and cooling by CO ro-vibrational transitions are somehow phenomenologically introduced (in the calculations adiabatic cooling is also taken into account). %
By comparing with early observations of the light curves of CO vibrational bands, the parameters involved with the phenomenological approaches for the heating and cooling are calibrated to some extent. %
A perfect reproduction of the CO light curves is out of the scope of this paper because of the lack of realistic treatments for radiative transfer, excitation, ionization, and energy depositions by the decay of $^{56}$Ni. %
Instead, we focus on the qualitative impact of the matter mixing during the SN shock propagation inside the progenitor star, which determines the initial composition of the seed atoms. %
Before fully applying our methodology to the 3D hydrodynamical models \citep{2020ApJ...888..111O,2020A&A...636A..22O}, we utilize one-zone models and one-dimensional (1D) radial profiles derived from 3D SN explosion models \citep{2020ApJ...888..111O} as the initial conditions in this paper. %
By comparing with model results of spherical explosion cases and ones with angle-specified 1D profiles, the qualitative impact of effective matter mixing on the molecule formation in the ejecta is investigated. %

This paper is organized as follows. In Section~\ref{sec:method}, the methodology is described. %
In Section~\ref{sec:strategy}, the strategy and models for investigating the impact of matter mixing on the molecule formation in the ejecta of SN~1987A are delineated. %
Section~\ref{sec:results} is dedicated to the results. %
In Section~\ref{sec:discussion}, several issues are discussed on the basis of the results. %
In Section~\ref{sec:summary}, the study in this paper is summarized. %

\section{Method} \label{sec:method} 

In this study, the dimensionalities of the models are restricted up to 1D. To derive the initial conditions of the ejecta, i.e., the density, temperature, and composition of the seed atoms, 3D CCSN explosion models in one of our previous studies \citep{2020ApJ...888..111O} are utilized. %
By averaging the last snapshots ($\sim$ 1 day after the explosion) of the 3D models, the inner regions of the ejecta which consist of seed atoms are approximated, first, as one-zones for understanding overall trends and, then, as 1D radial profiles to investigate the impact of matter mixing. For the former case, the temperature and density evolutions are basically obtained by power-laws. %
For the latter case, the evolutions are obtained by performing 1D (spherical) hydrodynamical simulations with the 1D profiles as the initial conditions. %
For both cases, the effects of extra heating and cooling on the thermal evolution of the ejecta are taken into account. %
Then, rate equations for chemical reactions are solved for the ejecta models. %
Here, the method is described in detail. %

\subsection{Chemical reaction network} \label{subsec:chemical} 

\begin{deluxetable*}{ll}
\tabletypesize{\footnotesize}
\tablewidth{0pt}
\tablenum{1}
\tablecolumns{2}
\tablecaption{Included species.\label{table:species}} %
\startdata
\vs{-0.1}\\
Atoms 
& \hs{0.5} H, He, C, N, O, Ne, Mg, Si, S, Ar, Fe \vs{0.2}\\ \hline
Diatomic molecules 
& \hs{0.5} H$_2$, CH, C$_2$, CN, CO, CS, NH, N$_2$, NO, \\
& \hs{0.5} OH, O$_2$, MgO, MgS, SiH, SiC, SiN, SiO, \\
& \hs{0.5} Si$_2$, SiS, SO, S$_2$, FeO, FeS, Fe$_2$ \vs{0.2}\\ \hline
Ions 
& \hs{0.5} e$^-$, H$^-$, C$^-$, O$^-$, H$^+$, He$^+$, C$^+$, N$^+$, O$^+$, \\
& \hs{0.5} Ne$^+$, Mg$^+$, Si$^+$, S$^+$, Ar$^+$, Fe$^+$, H$_2^+$, HeH$^+$, \\
& \hs{0.5} CH$^+$, C$_2^+$, CN$^+$, CO$^+$, CS$^+$, NH$^+$, N$_2^+$, NO$^+$,  \\ 
& \hs{0.5} OH$^+$, O$_2^+$, MgO$^+$, MgS$^+$, SiH$^+$, SiC$^+$, SiN$^+$ \\
& \hs{0.5} SiO$^+$, Si$_2^+$, SiS$^+$, SO$^+$, S$_2^+$, FeO$^+$, FeS$^+$, Fe$_2^+$ \\
\enddata
\end{deluxetable*}
%

\begin{deluxetable}{ll}
\tabletypesize{\footnotesize}
\tablewidth{0pt}
\tablenum{2}
\tablecolumns{2}
\tablecaption{Types of chemical reactions.\label{table:types}} %
\tablehead{Code & \hs{0.5} Reaction type} %
\startdata
\texttt{3B} & \hs{0.5} Three-body reaction \\
\texttt{CE}  & \hs{0.5} Charge exchange reaction \\
\texttt{IN}  & \hs{0.5} Ion-neutral reaction \\
\texttt{MN}  & \hs{0.5} Mutual neutralization reaction \\
\texttt{NN}  & \hs{0.5} Neutral-neutral reaction \\
\texttt{AD}  & \hs{0.5} Associative electron detachment \\
\texttt{DR}  & \hs{0.5} Dissociative recombination \\
\texttt{RA}  & \hs{0.5} Radiative association \\
\texttt{REA} & \hs{0.5} Radiative electron attachment \\
\texttt{RR}  & \hs{0.5} Radiative recombination \\
\texttt{CD}  & \hs{0.5} Collisional dissociation \\
\hline
\texttt{TF}\tnm{a}  & \hs{0.5} Thermal fragmentation reaction \\
\texttt{CM}\tnm{b}  & \hs{0.5} Ionization/destruction by Compton electrons \\
\texttt{UV}\tnm{b}  & \hs{0.5} Dissociation by UV photons
\enddata
\tnt{a}{Reactions described in Section~\ref{subsubsec:thermal_frag}}
\tnt{b}{Reactions described in Section~\ref{subsubsec:compton}}
\end{deluxetable}

Here, the chemical reaction network adopted in this paper is described. To follow the formation and destruction of (diatomic) molecules, rate equations are solved. Species taken into account in this paper are listed in Table~\ref{table:species}; %
11 atoms, 24 diatomic molecules, electrons, and 39 singly-ionized ions, i.e., in total, 75 species, are included. 
In this paper, we limit our discussion to the formation of molecules; the nucleation of clusters of dust, and grain growth are beyond the scope of this study. %
Since more than triatomic species are more related to the nucleation of clusters, we take into account diatomic molecules as molecular species\footnote{MgO, MgS, SiC, FeO, and FeS can be regarded as monomers of corresponding $n$-mers. C$_2$, Si$_2$, Mg$_2$, and Fe$_2$ can be regarded as clusters of corresponding single-element grains.} for simplicity. %
All diatomic molecules considered in the previous theoretical studies on molecule formation in CCSNe \citep{2009ApJ...703..642C,2013ApJ...776..107S,2015A&A...575A..95S,2018MNRAS.480.5580S} are included except for aluminum oxide (AlO) that can not form from the seed atoms corresponding to the 19 nuclei taken into account in the 3D hydrodynamical models \citep[][in which aluminum was not included]{2020ApJ...888..111O}. %
As for ion species, considering the low ionization fraction in the environment in this paper (see, Section~\ref{subsubsec:init_chemi}), singly-ionized species\footnote{In \cite{2020A&A...642A.135L}, doubly ionized carbon (C$^{2+}$) and oxygen (O$^{2+}$) were taken into account in the molecule formation calculation focusing on carbon monoxide; the fractions of the doubly-ionized ions are roughly three orders of magnitude smaller than those of corresponding singly-ionized ions in the environment, and the doubly-ionized ions seem to play only a minor role at least in the formation of carbon monoxide.} corresponding to the adopted atoms and diatomic molecules are taken into account if the data for the rates of the reactions involved with those species are available. %

The rate equations to be solved are as follows. %
\begin{equation}
\frac{dc_i}{dt} = \sum F_i - \sum D_i, \label{eq:rate_eq}
\end{equation}
where $c_i$ is the number density of species $i$. $F_i$ and $D_i$ are formation and destruction rates of species $i$, respectively. %
Here, the former (later) summation is taken for the reactions which include species $i$ in the products (reactants). %
In the case of two-body reactions, $F_i$ for $j + k \lra i + \cdots$ reaction and $D_i$ for $i + j \lra k + \cdots$ reaction are given by 
\begin{equation}
F_i = k_i(T) \, c_j \,c_k, \ \ \ \ D_i = k_i(T) \, c_i \,c_j, \label{eq:form_dest}
\end{equation}
respectively, where $k_i (T)$ is the temperature ($T$) dependent rate coefficient. %
$c_i$, $c_j$, and $c_k$ are the number density of the reactants, $i$, $j$ and $k$, respectively. %
For the case of three-body reactions, $F_i$ for $j + k + l \lra i + \cdots$ reaction and $D_i$ for $i + j + k \lra l + \cdots$ reaction are given by 
\begin{equation}
F_i = k_i(T) \, c_j \,c_k \, c_l, \ \ \ \ D_i = k_i(T) \, c_i \,c_j \,c_k,
\end{equation}
respectively. %
The rate coefficient, $k_i (T)$, is conventionally expressed by the so-called Arrhenius form, 
\begin{equation}
k_i(T) = A_i \left( \frac{T}{300 \ {\rm K}}\right)^{\alpha_i} \exp \,(-\beta_i/T), \label{eq:arrhenius}
\end{equation}
where $A_i$, $\alpha_i$, and $\beta_i$ are specific coefficients for each reaction. Given the species listed in Table~\ref{table:species}, all the reactions for the types listed in Table~\ref{table:types} (other than \texttt{TF}, \texttt{CM}, and \texttt{UV}) in which the coefficient data is available in the UMIST database\footnote{\texttt{http://udfa.ajmarkwick.net/}} for astrochemistry \citep{2013...550A..36M} or in the literature \citep{2007...466.1197W,2009ApJ...703..642C,2010ApJ...713....1C,2018MNRAS.480.5580S} are included in the network. %
The adopted coefficient values for each reaction and the corresponding reference are listed in Table~\ref{table:reaction} in Appendix~\ref{app:reac}. %
Other than the reactions above, thermal fragmentation reactions of diatomic molecules via thermal collisions with arbitrary particles (\texttt{TF} reactions), ionization, or destruction reactions by Compton electrons that stem from the decay of radioactive $^{56}$Ni and $^{56}$Co (\texttt{CM} reactions), and dissociation by UV photons (\texttt{UV} reactions) are also taken into account (see, Sections~\ref{subsubsec:thermal_frag} and \ref{subsubsec:compton}). %

The rate equations to be solved are essentially the same as the nucleosynthesis calculations performed in our previous studies 
\citep[e.g.,][]{2009PThPh.122..755O,2012PThPh.128..741O}; in this paper, the equations are iteratively solved by the same full-implicit method in the time integration but with Gauss elimination with LU decomposition as a matrix solver, which is different from the one (a sparse matrix solver) used in our previous studies above, for solving the matrix robustly. %

\subsubsection{Thermal fragmentation reactions} \label{subsubsec:thermal_frag} 

\begin{deluxetable}{lc}
\tabletypesize{\footnotesize}
\tablewidth{0pt}
\tablenum{3}
\tablecolumns{12}
\tablecaption{Atomic radii. %
The values are taken from \cite{1967JChPh..47.1300C}.\label{table:atoms}} %
\tablehead
{
\vsm{4} \\
Atom ($i$) & \ \ \ \ \ $a_{0,i}$ \ \ \ \ \ \\
           &  (\AA) 
}
\startdata
\vsm{4} \\
H    & 0.53 \\
He   & 0.31 \\
C    & 0.67 \\
N    & 0.56 \\
O    & 0.48 \\
Ne   & 0.38 \\
Mg   & 1.45 \\
Si   & 1.11 \\
S    & 0.88 \\
Ar   & 0.71 \\
Fe   & 1.56 \\

\vsm{4} \\
\enddata
\end{deluxetable}
%

\begin{deluxetable*}{lrrrrrcc}
\tabletypesize{\footnotesize}
\tablewidth{0pt}
\tablenum{4}
\tablecolumns{8}
\tablecaption{Radii and values related to the bond-dissociation energies of the diatomic molecules.\label{table:molecules}} %
\tablehead{\vsm{3} \\
Molecule ($i$) & \ \ $a_{0,i}$\tnm{a} & \ \ \ \ \ \ \ $\epsilon_{i,1}$\tnm{b} & $\epsilon_{i,2}$\tnm{b} & $\epsilon_{i,3}$\tnm{b} & $\epsilon_{i,4}$\tnm{b} & $E_{b,i}$\tnm{c} ($T = 2000$ K) & $E_{b,i}$\tnm{c} ($T = 4000$ K)\tnm{c} \\
 & (\AA) & & & & & (eV) & (eV)} %
\startdata
\\
${\rm H}_2 $ & 0.668 &  4.4737 &  4.6595 (-1) & -2.0450 (-1) &  2.8627 (-2) &  4.7019 &  4.7992 \\
${\rm CH}  $ & 0.766 &  3.4906 &  4.4935 (-1) & -3.3945 (-1) &  8.7186 (-2) &  3.6651 &  3.6929 \\
${\rm C}_2 $ & 0.844 &  6.1532 &  1.8501 (-1) & -6.7833 (-2) &  8.7819 (-3) &  6.2490 &  6.3001 \\
${\rm CN}  $ & 0.781 &  7.7735 &  4.7013 (-1) & -3.5102 (-1) &  8.4160 (-2) &  7.9558 &  7.9758 \\
${\rm CO}  $ & 0.744 & 11.1200 &  3.9010 (-1) & -2.0736 (-1) &  6.1460 (-2) & 11.3070 & 11.4180 \\
${\rm CS}  $ & 0.994 &  7.3673 &  2.9418 (-1) & -1.1130 (-1) &  3.3346 (-2) &  7.5238 &  7.6407 \\
${\rm NH}  $ & 0.687 &  3.2134 &  4.5679 (-1) & -2.4998 (-1) &  5.5345 (-2) &  3.4232 &  3.5092 \\
${\rm N}_2 $ & 0.706 &  9.7592 &  4.3238 (-1) & -2.8255 (-1) &  9.4590 (-2) &  9.9499 & 10.0580 \\
${\rm NO}  $ & 0.659 &  6.5114 &  3.8560 (-1) & -2.4091 (-1) &  7.6626 (-2) &  6.6841 &  6.7789 \\
${\rm OH}  $ & 0.638 &  4.3952 &  4.5843 (-1) & -2.2279 (-1) &  4.8458 (-2) &  4.6162 &  4.7253 \\
${\rm O}_2 $ & 0.605 &  5.1285 &  3.9141 (-1) & -2.5362 (-1) &  5.8071 (-2) &  5.2939 &  5.3372 \\
${\rm MgO} $ & 1.467 &  3.5241 & -1.9805 (-1) &  2.2559 (-2) &  4.1024 (-2) &  3.4143 &  3.3974 \\
${\rm MgS} $ & 1.551 &  2.9102 & -2.4538 (-1) &  1.2113 (-1) &  1.2405 (-2) &  2.8041 &  2.8277 \\
${\rm SiH} $ & 1.149 &  2.9886 &  3.3233 (-1) & -1.2256 (-1) &  2.0183 (-2) &  3.1617 &  3.2617 \\
${\rm SiC} $ & 1.186 &  4.6197 &  1.3931 (-2) &  6.0514 (-2) & -8.0209 (-3) &  4.6535 &  4.7268 \\
${\rm SiN} $ & 1.156 &  5.6667 &  3.9114 (-1) & -3.6496 (-1) &  1.0959 (-1) &  5.7977 &  5.7992 \\
${\rm SiO} $ & 1.139 &  8.2581 &  2.8743 (-1) & -1.0153 (-1) &  2.9247 (-2) &  8.4132 &  8.5301 \\
${\rm Si}_2$ & 1.399 &  3.2074 &  2.1689 (-2) &  2.2195 (-2) &  2.6906 (-3) &  3.2325 &  3.2822 \\
${\rm SiS} $ & 1.270 &  6.4142 &  2.1404 (-1) & -7.8795 (-3) & -1.3752 (-3) &  6.5530 &  6.6823 \\
${\rm SO}  $ & 0.925 &  5.3678 &  3.6931 (-1) & -2.5105 (-1) &  6.5632 (-2) &  5.5219 &  5.5695 \\
${\rm S}_2 $ & 1.109 &  4.3814 &  3.1299 (-1) & -2.0727 (-1) &  5.5606 (-2) &  4.5145 &  4.5621 \\
${\rm FeO} $ & 1.575 &  4.2540 &  3.5731 (-1) & -2.0600 (-1) &  8.8265 (-2) &  4.4268 &  4.5734 \\
${\rm FeS} $ & 1.648 &  3.3050 &  3.2418 (-1) & -1.6523 (-1) &  7.8517 (-2) &  3.4710 &  3.6296 \\
${\rm Fe}_2$\tnm{d} & 1.965 &  4.3814 &  3.1299 (-1) & -2.0727 (-1) &  5.5606 (-2) &  4.5145 &  4.5621 \\

\enddata
\rm
{
\tnt{a}{Approximated radius of the molecule, $i$, calculated as $a_{0,i} = (a_{0,j}^3 + a_{0,k}^3)^{1/3}$ with atomic radii in Table~\ref{table:atoms}, where $a_{0,j}$ and $a_{0,k}$ are the atomic radii of the constituent atoms.} %
\tnt{b}{Coefficients of the fitting formula for the bond-dissociation energies in Equation~(\ref{eq:be}). The bond-dissociation energies are estimated from the changes in the enthalpies of formation (${\it \Delta}_{\rm f} H^{\circ}$) in the reaction systems (\texttt{https://janaf.nist.gov/}).} %
\tnt{c}{Bond-dissociation energies at $T =$ 2000 K and 4000 K obtained by Equation~(\ref{eq:be}).} %
\tnt{d}{The values for the bond-dissociation energy of Fe$_2$ are assumed to be same as S$_2$.} %
}
\end{deluxetable*}

For thermal fragmentation reactions (\texttt{TF} reactions in Table~\ref{table:types}),\footnote{It is noted that several \texttt{CD} reactions adopted in this study (see, the last part of Table~\ref{table:reaction}), e.g., H $+$ H$_2$ $\lra$ H $+$ H $+$ H, have the same reactants and products as some of thermal fragmentation reactions. We adopt both of such corresponding reactions regarding the two as independent. The response to the gas temperature seems to be different between the two, and the corresponding two reactions seem not to be comparable with each other at the same time.} 
we adopt a kinetic approach, i.e., it is assumed that particles whose kinetic energy is greater than the bond-dissociation energy of the molecule $i$ can destruct the molecule by thermal collisions. Therefore, $D_i$ is given by 
\begin{equation}
D_i = c_i \sum_j \frac{1}{1 + \delta_{ij}} c_j \, S_{ij} \,G_{ij} (T), \label{eq:di}
\end{equation}
where $S_{ij}$ is the cross section for the collision between the molecule $i$ and a particle $j$; $S_{ij}$ is assumed to be 
\begin{equation}
S_{ij} = \pi (a_{0,i} + a_{0,j})^2,
\end{equation}
where $a_{0,i}$ is the radius of the species $i$. %
The values of the radii of species are listed in Table~\ref{table:atoms} and Table~\ref{table:molecules}. %
$G_{ij} (T)$ corresponds to the mean relative velocity  between the particles $i$ and $j$ in which it is assumed that the relative velocities follow the Maxwell-Boltzmann distribution. %
Then, $G_{ij} (T)$ is given by
\begin{equation}
\begin{aligned}
G_{ij} (T) = \frac{4}{(2 \pi \mu_{ij})^{1/2} \,(k_B T)^{3/2}} \\
\times \int_{E_{{\rm b},i}}^{\infty} E \exp\left(-\frac{E}{k_B T}\right) dE, 
%
\end{aligned}
\end{equation}
where $\mu_{ij}$ is the reduced mass of the two body system of particles $i$ and $j$, $k_{\rm B}$ is the Boltzmann constant, $E$ denotes the kinetic energy of the relative motion and $E_{{\rm b},i}$ is the bond-dissociation energy of the molecule $i$. %
$E_{{\rm b},i}$ is given by a fitting form as below. %
%
\begin{equation}
E_{{\rm b},i} (T) = \sum_{l=1}^4 \epsilon_{i,l} \left( \frac{T}{3000\,{\rm K}} \right)^{l-1}, \label{eq:be}
\end{equation}
where $\epsilon_{i,l}$ are the fitting coefficients listed in Table~\ref{table:molecules}. In Table~\ref{table:molecules}, the bond-dissociation energies at representative temperatures, $T=2000$ K and 4000 K, are also listed. It is noted that the factor, $1/(1 + \delta_{ij})$, in Equation~(\ref{eq:di}) is introduced to avoid double counting of the rate in the case of $i=j$. %
Here, $\delta_{ij}$ is the Kronecker delta. %

\subsubsection{Collisional ionization and destruction by Compton electrons; dissociation by UV photons} \label{subsubsec:compton} 

\begin{deluxetable}{lr}
\tabletypesize{\footnotesize}
\tablewidth{0pt}
\tablenum{5}
\tablecolumns{2}
\tablecaption{Reactions involved with Compton electrons e$_{\rm C}^-$ and the adopted mean energy per ion-pair $W$. %
Here, A and B denote constituent atoms other than oxygen. %
X denotes an arbitrary atom.\label{table:mean_engy}} %
\tablehead{\vsm{3} \\Reaction (type) & \hs{1} $W$ (eV)} %
\startdata
CO + e$_{\rm C}^-$ $\lra$ C + O$^+$ + e$^-$ + e$_{\rm C}^-$ & 768 \\
CO + e$_{\rm C}^-$ $\lra$ C$^+$ + O + e$^-$ + e$_{\rm C}^-$ & 274 \\
CO + e$_{\rm C}^-$ $\lra$ C + O + e$_{\rm C}^-$ & 125 \\
CO + e$_{\rm C}^-$ $\lra$ CO$^+$ + e$^-$ + e$_{\rm C}^-$ & 34 \vsp{2}\\ \hline
SiO + e$_{\rm C}^-$ $\lra$ Si + O$^+$ + e$^-$ + e$_{\rm C}^-$ & 678 \\
SiO + e$_{\rm C}^-$ $\lra$ Si$^+$ + O + e$^-$ + e$_{\rm C}^-$ & 218 \\
SiO + e$_{\rm C}^-$ $\lra$ Si + O + e$_{\rm C}^-$ & 110 \\
SiO + e$_{\rm C}^-$ $\lra$ SiO$^+$ + e$^-$ + e$_{\rm C}^-$ & 30 \vsp{2}\\ \hline
H$_2$ + e$_{\rm C}^-$ $\lra$ H$^+$ + H + e$^-$ + e$_{\rm C}^-$ & 829 \\
H$_2$ + e$_{\rm C}^-$ $\lra$ H + H + e$_{\rm C}^-$ & 77 \\
H$_2$ + e$_{\rm C}^-$ $\lra$ H$^+_2$ + e$^-$ + e$_{\rm C}^-$ & 37.7 \vsp{2}\\ \hline
N$_2$ + e$_{\rm C}^-$ $\lra$ N$^+$ + N + e$^-$ + e$_{\rm C}^-$ & 264 \\
N$_2$ + e$_{\rm C}^-$ $\lra$ N + N + e$_{\rm C}^-$ & 133.5 \\
N$_2$ + e$_{\rm C}^-$ $\lra$ N$^+_2$ + e$^-$ + e$_{\rm C}^-$ & 36.3 \vsp{2}\\ \hline
O$_2$ + e$_{\rm C}^-$ $\lra$ O$^+$ + O + e$^-$ + e$_{\rm C}^-$ & 768 \\
O$_2$ + e$_{\rm C}^-$ $\lra$ O + O + e$_{\rm C}^-$ & 125 \\
O$_2$ + e$_{\rm C}^-$ $\lra$ O$^+_2$ + e$^-$ + e$_{\rm C}^-$ & 34 \vsp{2}\\ \hline
For oxygen involving molecules other than above \\
AO + e$_{\rm C}^-$ $\lra$ A + O$^+$ + e$^-$ + e$_{\rm C}^-$ & 768 \\
AO + e$_{\rm C}^-$ $\lra$ A$^+$ + O + e$^-$ + e$_{\rm C}^-$ & 274 \\
AO + e$_{\rm C}^-$ $\lra$ A + O + e$_{\rm C}^-$ & 125 \\
AO + e$_{\rm C}^-$ $\lra$ AO$^+$ + e$^-$ + e$_{\rm C}^-$ & 34 \vsp{2}\\ \hline
For molecules other than above \\
AB + e$_{\rm C}^-$ $\lra$ A + B$^+$ + e$^-$ + e$_{\rm C}^-$ & 274 \\
AB + e$_{\rm C}^-$ $\lra$ A$^+$ + B + e$^-$ + e$_{\rm C}^-$ & 274 \\
AB + e$_{\rm C}^-$ $\lra$ A + B + e$_{\rm C}^-$ & 125 \\
AB + e$_{\rm C}^-$ $\lra$ AB$^+$ + e$^-$ + e$_{\rm C}^-$ & 34 \vsp{2}\\ \hline
Ionization of atoms \\
X  + e$_{\rm C}^-$ $\lra$ X$^+$ + e$^-$ + e$_{\rm C}^-$ & 47 \vsp{2}\\
\enddata
\end{deluxetable}

Compton electrons produced by the decay of $^{56}$Ni and $^{56}$Co could collisionally ionize atoms and molecules and could destruct (dissociate) molecules (\texttt{CM} reactions in Table~\ref{table:types}). %
Subsequent UV photons caused by the deposited energy could also destruct molecules (\texttt{UV} reactions in Table~\ref{table:types}). Such ionization and destruction are taken into account in a similar way to ones in the literature \citep[e.g.,][]{2009ApJ...703..642C,2020A&A...642A.135L}. %

First, the local energy generation rates by the decay of $^{56}$Ni and $^{56}$Co, $\epsilon_{\rm Ni}$ and $\epsilon_{\rm Co}$, respectively, are considered as described in \cite{2009ApJ...693.1780J}. $\epsilon_{\rm Ni}$ is given by
\begin{equation}
\epsilon_{\rm Ni} = \lambda_{\rm Ni} \,e^{-\lambda_{\rm Ni} t} \,q_{\rm Ni} \,X_{\rm Ni} \ \ \ {\rm erg \,g}^{-1} {\rm s}^{-1}, 
\label{eq:epsilon_ni}
\end{equation}
where $\lambda_{\rm Ni}$, $t$, $q_{\rm Ni}$, and $X_{\rm Ni}$ are the decay rate of $^{56}$Ni, the time after the explosion, the $q$-value of the decay from $^{56}$Ni to $^{56}$Co, and the local mass fraction of $^{56}$Ni, respectively. %
The values of $\lambda_{\rm Ni}$, $q_{\rm Ni}$ are $1.315 \times 10^{-6}$ s$^{-1}$ and $2.96 \times 10^{16}$ erg~g$^{-1}$, respectively. %
$\epsilon_{\rm Co}$ is given by
\begin{equation}
\begin{aligned}
\epsilon_{\rm Co} = \frac{\lambda_{\rm Ni}}{\lambda_{\rm Co}-\lambda_{\rm Ni}} \lambda_{\rm Co} 
\left(e^{-\lambda_{\rm Ni} t} - e^{-\lambda_{\rm Co} t} \right) q_{\rm Co} \,X_{\rm Ni} \\ 
\ \ \ {\rm erg \,\,g}^{-1} {\rm s}^{-1},
\label{eq:epsilon_co}
\end{aligned}
\end{equation}
where $\lambda_{\rm Co}$ and $q_{\rm Co}$ are the decay rate of $^{56}$Co and the $q$-value of the decay from $^{56}$Co to $^{56}$Fe, 
respectively. %
The values of $\lambda_{\rm Co}$, $q_{\rm Co}$ are $1.042 \times 10^{-7}$ s$^{-1}$ and $6.4 \times 10^{16}$ erg~g$^{-1}$, respectively. %
Then, the local energy deposition rate $\epsilon$ is given as follows. %
\begin{equation}
\epsilon = \left(\epsilon_{\rm Ni} + \epsilon_{\rm Co} \right) D_{\gamma} \ \ \ {\rm erg \,\,g}^{-1} {\rm s}^{-1}, \label{eq:epsilon}
\end{equation}
where $D_{\gamma}$ is the fraction of the energy deposition, which depends on the effective optical depth, $\tau_{\gamma}$, to gamma rays that emerged from the decay as follows. %
\begin{equation}
D_{\gamma} = 1 - \exp(- \tau_{\gamma}). \label{eq:d_gamma}
\end{equation}
The optical depth is obtained differently for one-zone and 1D cases. For the one-zone case, the optical depth is obtained as in \cite{1995ApJ...454..472L}, i.e. 
\begin{equation}
\tau_{\gamma} = \tau_0 \,(t/t_0)^{-2},
\end{equation}
where $\tau_0$ is the optical depth at a reference time $t_0$. $\tau_0$ and $t_0$ are 31.1 and 100 days, respectively. %
For the 1D case, the optical depth depends on the radius, $r$, and it is derived as 
\begin{equation}
\tau_{\gamma} (r) = \int^{\infty}_{r} \frac{1}{\,\,l_{\gamma}\,}\,dr,
\label{eq:tau_gamma}
\end{equation}
where $l_{\gamma}$ is the effective mean free path of the gamma rays. The upper limit of the integration is practically taken to be the radius of the upper computational domain. $l_{\gamma}$ is taken from \cite{1990ApJ...360..242S} as 
\begin{equation}
l_{\gamma} = 1.67 \times 10^{13} \ Y_{\rm e}^{-1} \left( \frac{\rho}{10^{-12} \ {\rm g} \,{\rm cm}^{-3}} \right)^{-1} \ \ \ {\rm cm},
\label{eq:mean_free_gamma}
\end{equation}
where $Y_{\rm e}$ and $\rho$ are the electron fraction (the total number of free and bound electrons per nucleon) and the density, respectively. %
$l_{\gamma}$ corresponds to the effective distance that emerged 1.28 MeV gamma-rays travel before being degraded to 60 keV and absorbed \citep{1990ApJ...360..242S}. %

It is noted that as described above, non-local energy deposition is not taken into account. %
In reality, emitted energy can non-locally be deposited over a region approximately corresponding to the length scale of the effective mean free path in Equation~(\ref{eq:mean_free_gamma}) (practically over several tracer particles; see, Section~\ref{subsec:1d_calc}), which may cause smoother gas temperature profiles across the optically thick-to-thin regions and may allow energy deposition outside the regions where $^{56}$Ni exists. %
Therefore, the potential impact of non-local energy deposition is discussed with a simple model in Appendix~\ref{app:non_local_edep}. %
The methodology to count the effects of the energy deposition (without non-local energy deposition) on thermal quantities of the gas is described in Sections~\ref{subsec:one_zone_calc} and \ref{subsec:1d_calc}. %

Additionally, as described in Equation~(\ref{eq:tau_gamma}), the optical depth only in the radial direction, i.e., the escape of the emitted energies only in the radial direction, is considered. %
Optical depths in non-radial directions \citep[e.g.,][]{2021MNRAS.502.3264G} for counting non-radial energy escape, which may be effective if the distribution of $^{56}$Ni is globally asymmetric, are not also taken into account since the hydrodynamical models presented in this study are assumed to be spherically symmetric, and the effect cannot be treated appropriately. %

Given the energy deposition rate $\epsilon$ in Equation~(\ref{eq:epsilon}), the rate coefficients, $k_{\rm C}$, of the ionization and destruction by Compton electrons are given as follows. %
\begin{equation}
k_{\,\rm C} = \frac{\epsilon \, \rho}{W \, n_{\rm tot}} \ \ \ {\rm s}^{-1}, \label{eq:kc}
\end{equation}
where $W$ is the mean energy per ion pair and the values depend on the target and the reaction type. $n_{\rm tot}$ is the total number density of the particles. The values of $W$ could also depend on the ionization fraction \citep[see, Figure~2 in][]{1995ApJ...454..472L}. However, we take the values as constant for simplicity. %
The constant values of $W$ are taken from \cite{2009ApJ...703..642C} and \cite{2018MNRAS.480.5580S}. %
For the ionization and destruction reactions by Compton electrons, reaction types of the forms described below are taken into account. %
Ionization of an atom X:
\begin{equation}
{\rm X} + {\rm e}^-_{\rm C} \lra {\rm X}^+ + {\rm e}^- + {\rm e}^-_{\rm C}, \label{eq:ion_x}
\end{equation}
where e$^-_{\rm C}$ denotes the Compton electron. Ionization of a diatomic molecule AB (A and B denote the constituent atoms):
\begin{equation}
{\rm AB} + {\rm e}^-_{\rm C} \lra {\rm AB}^+ + {\rm e}^- + {\rm e}^-_{\rm C}. \label{eq:ion_ab}
\end{equation}
Following dissociative reactions:
\begin{equation}
\begin{aligned}
&{\rm AB} + {\rm e}^-_{\rm C} \lra {\rm A}   + {\rm B}   + {\rm e}^-_{\rm C}, \\
&{\rm AB} + {\rm e}^-_{\rm C} \lra {\rm A}^+ + {\rm B}   + {\rm e}^- + {\rm e}^-_{\rm C}, \\
&{\rm AB} + {\rm e}^-_{\rm C} \lra {\rm A}   + {\rm B}^+ + {\rm e}^- + {\rm e}^-_{\rm C}.
\end{aligned}
\label{eq:destruction}
\end{equation}
The adopted values of the mean energy per ion pair $W$ for the ionization and destruction reactions described above are listed in Table~\ref{table:mean_engy}. %
Similarly, secondary UV photons destruct a diatomic molecule AB to produce A and B. %
The rate coefficient $k_{\rm UV}$ is given by
\begin{equation}
k_{\,\rm UV} = \frac{\alpha \,\epsilon \, \rho}{E_{\rm UV} \, n_{\rm tot}} \ \ \ {\rm s}^{-1}, \label{eq:kuv}
\end{equation}
where, $E_{\rm UV}$ is the representative energy of UV photons. The value of $E_{\rm UV}$ is set to be the energy of a fiducial photon with the wavelength of 1302 ${\rm \AA}$ as in \cite{2009ApJ...703..642C}. %
$\alpha$ is the fraction of the deposited energy to be converted to UV photons. %
For simplicity, the values of $\alpha$ at 200 days and 1000 days are set to be 0.25 and 0.4, respectively, according to a discussion in \cite{2009ApJ...703..642C}. %
For an epoch between 200 days and 1000 days, the value is linearly interpolated and on the outside, the values are extrapolated but arbitrarily the maximum and minimum values are set to be 0.5 and 0.1, respectively. %
Both the Compton electrons and UV photons are not included as one of the explicit species in the chemical reaction network. %
For the reactions mentioned above, the destruction rate $D_i$ in Equation~(\ref{eq:rate_eq}) is obtained by regarding the reaction as a transition (one-body reaction); for example, the destruction rate $D_i$ in Equation~(\ref{eq:rate_eq}) for the case of the ionization of an atom X by Compton electrons is given by 
\begin{align}
&D_{\rm X} = k_{\rm C} \,c_{\rm X}. 
\end{align}
As described above, the derived ionization and destruction rates are simply proportional to the number of available Compton electrons per particle estimated with the constant mean energy $W$; there may be large uncertainties in the derived rate coefficients. %
Therefore, to see the impact of such uncertainties, a constant factor $f_{\rm d}$ is introduced as a model parameter to control the efficiency of the rates as in \cite{2018MNRAS.480.5580S}. %
Then, the ionization and destruction rates in Equations~(\ref{eq:kc}) and~(\ref{eq:kuv}) are practically introduced as $f_{\rm d}\,k_{\rm C}$ and $f_{\rm d}\,k_{\rm UV}$ in the chemical reaction calculations. %

\subsubsection{Initialization of the chemical reaction calculations} \label{subsubsec:init_chemi} 

Generally, if the gas temperature is higher than 10$^{4}$ K, the destruction by thermal fragmentation reactions dominates the formation reactions of molecules. %
The transition temperature when molecules start to form, however, is not so clear. %
Then, in this study, the chemical reaction calculations are started if the gas temperature drops to a bit higher temperature than above, i.e., 1.5~$\times$~10$^4$ K, to be safe. %
The initial abundances of the species for the chemical reaction calculations are determined as follows. %
As the initial conditions, only the 11 atoms, singly ionized atoms, and electrons have non-zero abundances initially. %
The abundances of the atoms are determined to be consistent with the compositions derived from the 3D hydrodynamical calculation results \citep{2020ApJ...888..111O}. %
In \cite{2020ApJ...888..111O}, in total 19 nuclei were taken into account. %
Then, the abundances of the nuclei are assigned to the corresponding atoms depending on the isotopes. %
For example, the abundances of $^{52}$Fe, $^{54}$Fe, and $^{56}$Ni ($^{56}$Ni $\lra$ $^{56}$Co $\lra$ $^{56}$Fe) are assigned as the iron atom Fe. %
Stable isotopes are simply summed up to count the corresponding atomic abundances. %
It is noted that the electron fraction $Y_{\rm e}$ in Equation~(\ref{eq:mean_free_gamma}) does not depend on ionization and chemical reactions. Hence, the electron fraction is obtained from the original abundances of the 19 nuclei. %
The change in the electron fraction due to the radioactive decay is neglected since the change does not affect much the optical depth in Equation~(\ref{eq:tau_gamma}). %

Before starting the molecule formation, some fraction of atoms could be ionized. %
Therefore, the ionization before starting the chemical reaction calculation is determined manually by introducing the ionization fraction\footnote{Sometime ``ionization fraction" may be called as ``electron fraction" but the definition is different from $Y_{\rm e}$ in Equation~(\ref{eq:mean_free_gamma}).} $X_{\rm e}$, i.e., the number of free electrons per all the nuclei. %
In the ejecta core of SN~1987A, $X_{\rm e}$ could range from $\sim$ 10$^{-1}$--10$^{-2}$ to 10$^{-3}$--10$^{-4}$ corresponding to from 200 days to 2000 days after the explosion depending on the position in the core layers \citep[see, Fig.~8 in][]{1998ApJ...496..946K}. %
Therefore, two parameters, the initial ($\sim$ 1 day after the explosion) ionization fraction $X_{{\rm e},i}$ and the ionization fraction at 2000 days $X_{{\rm e},f}$ are introduced for simplicity. %
Then, $X_{\rm e}$ before 2000 days is obtained by linear interpolation of the powers of ten with the values of $X_{{\rm e},i}$ and $X_{{\rm e},f}$. %
$X_{\rm e}$ after 2000 days is set to be the same value at 2000 days. %
As the reference values of $X_{{\rm e},i}$ and $X_{{\rm e},f}$, 10$^{-1}$ and 10$^{-3}$ are adopted, respectively. %

With the $X_{\rm e}$ just before the chemical reaction calculation, the initial abundances of the species are determined to be consistent with it; a fraction of $X_{\rm e}$ of the abundance of each atom is assigned to the abundance of the corresponding singly ionized atom. Then, the same fraction of the abundance is added to the electron abundance. %

\subsection{Cooling by emission from ro-vibrational transitions of CO} \label{subsec:cooling} 

Emission from CO ro-vibrational transitions could cool the ejecta gas and eventually affect the chemistry in the ejecta. %
To estimate emission lines of CO ro-vibrational transitions, rate equations for ``vibrational" levels are solved. In this study, rotational levels are assumed to be thermally populated. %
Here, we basically adopt the method described in \cite{1995ApJ...454..472L}. %
The rate equations for vibrational levels (vibrational quantum number, $v = 0, 1, \ldots, v_{\rm max}$) are as follows. %
\begin{equation}
\frac{dX_i}{dt} = \sum_{j} \left( R_{ji} X_{j} - R_{ij} X_i \right), \label{eq:rate_eq_co}
\end{equation}
where $X_{i}$ and $R_{ij}$ is the population of the vibrationally excited state $i$ and the rate coefficient for the transition from the vibrational level $i$ to $j$, respectively. %
Here, the rate coefficient is expressed as
\begin{equation}
\begin{aligned}
R_{ij} &= \overline{A_{ij}} + \overline{B_{ij} \,J_{ij}}+ n_{\rm e} \,q_{ij}, &i > j, \\
&= \overline{B_{ij} \,J_{ji}} + n_{\rm e}\,q_{ij}, &i < j,
\label{eq:rij}
\end{aligned}
\end{equation}
where $\overline{A_{ij}}$, $\overline{B_{ij} \,J_{ij}}$ are derived (explained later below) from the Einstein's $A$-coefficient 
$A_{\kappa \lambda}$ for spontaneous emission of the ro-vibrational transition, similarly $B$-coefficient $B_{\kappa \lambda}$ for stimulated emission, and the mean radiation field $J_{\kappa \lambda}$ corresponding to the ro-vibrational transition, respectively. %
Here, Greek characters in subscripts denote ro-vibrational levels. %
Hence, a ro-vibrational level is specified by the vibrational level $v$ and the rotational level, $J = 0, 1, \ldots, J_{\rm max}$; %
the state $\kappa$ can be expressed more specifically as $\kappa$ = ($v$,$J$). %
The term, $n_{\rm e}\,q_{ij}$, is the contribution from electron impact excitation/de-excitation. %
$n_{\rm e}$ is the number density of free (thermal) electrons. %
$q_{ij}$ is the rate coefficient of the excitation ($i < j$) or de-excitation ($i > j$) of the vibrational level $i$ to $j$ (for the description, see, the last paragraph in this section and Appendix~\ref{app:elec_impact}). %
Einstein's coefficients are followed by the two relations below. %
\begin{equation}
A_{\kappa \lambda} = \frac{2 h \nu_{\kappa \lambda}^3}{c^2} B_{\kappa \lambda}, \ \ \ E_{\kappa} > E_{\lambda},
\end{equation}
%
\begin{equation}
g_{\kappa} B_{\kappa \lambda} = g_{\lambda} B_{\lambda \kappa},
\end{equation}
where $\nu_{\kappa \lambda}$ is the frequency of the corresponding photon. %
$h$ and $c$ are the Planck constant and the speed of light, respectively. %
$g_{\kappa}$ is the statistical weight of the state $\kappa$. %
$E_{\kappa}$ is the energy of the ro-vibrational state $\kappa$. %

As mentioned above, rotational levels are assumed to be thermally populated; the population probability of a rotational level $J$ among the same vibrational level $v$ is obtained as 
\begin{equation}
P_{\,(v,J)} = \frac{g_{\,(v,J)} \exp(- E_{\,(v,J)}/k_{\rm B} T)}{\displaystyle \sum_{J'=0}^{J_{\rm max}} g_{\,(v,J')} 
\exp(- E_{\,(v,J')}/k_{\rm B} T)},
\end{equation}
where $g_{(v,J)}$ and $E_{(v,J)}$ are the statistical weight and the energy of the ro-vibrational state ($v$,$J$), respectively. %
$T$ and $k_{\rm B}$ are the temperature and the Boltzmann constant, respectively. %
Then, $\overline{A_{ij}}$ are derived from the summation of the ro-vibrational Einstein's $A$ coefficients over the transitions of the vibrational level $i$ to $j$ with the weight of the initial state's population probability as 
\begin{equation}
\overline{A_{ij}} = \sum_{J=0}^{J_{\rm max}} \sum_{J'=0}^{J_{\rm max}} P_{\,(i,J)} 
\,A_{\,(i,J)(j,J')}. 
\label{eq:abar}
\end{equation}
%

\begin{figure*}[htbp]
\begin{tabular}{cc}
\begin{minipage}{0.5\hsize}
\begin{center}
\hs{-2}
\includegraphics[width=7.5cm,keepaspectratio,clip]{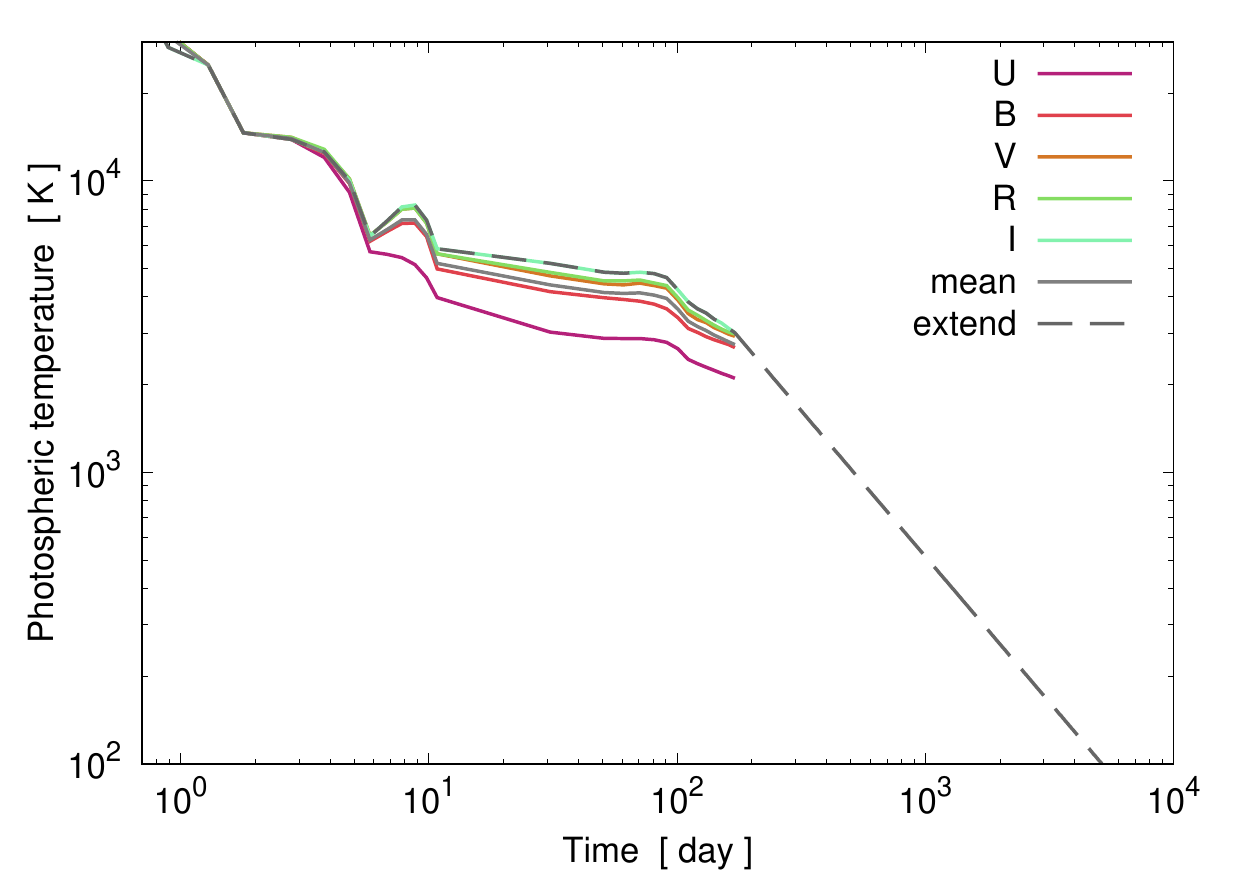}
\end{center}
\end{minipage}
\begin{minipage}{0.5\hsize}
\begin{center}
\hs{-2}
\includegraphics[width=7.5cm,keepaspectratio,clip]{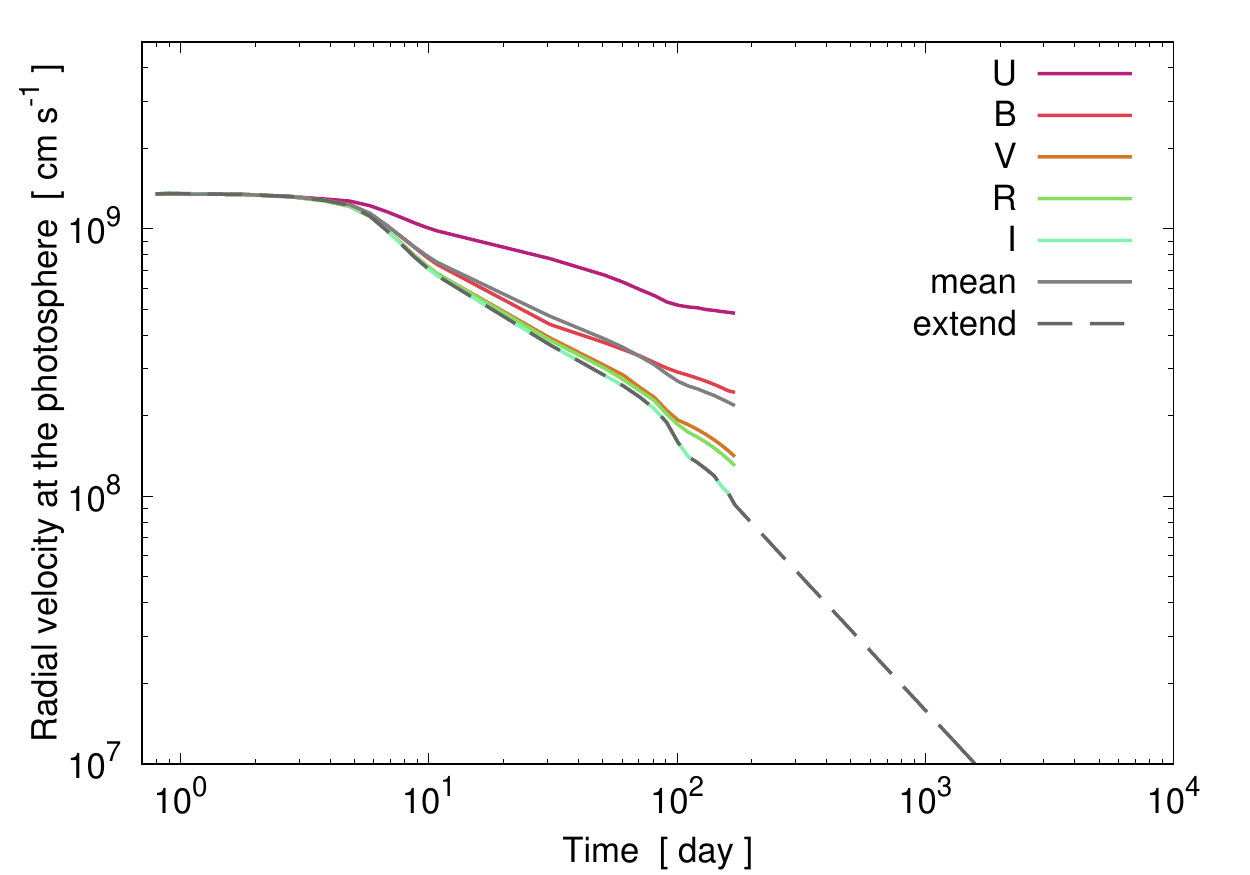}
\end{center}
\end{minipage}
\end{tabular}
\caption{The photospheric temperatures (Left) and the radial velocities at the photospheres (Right) as a function of time after the explosion for the U, B, V, R, and I bands calculated by a 1D radiation hydrodynamical code with the angle-averaged 1D profiles based 
on the 3D hydrodynamical model \citep[the model b18.3-high:][]{2020ApJ...888..111O} as the initial condition. %
Arithmetic averages (mean) and arbitrary extrapolations (extend) (see the text) are also plotted.} %
\label{fig:photo}
\end{figure*}
The mean radiation field $J_{\kappa \lambda}$ is obtained with the so-called Sobolev approximation \citep{1960mes..book.....S}, in which as shown below the line optical depth is derived with only local variables under the condition that the velocity gradient is steep enough, as follows. %
\begin{equation}
\begin{aligned}
J_{\kappa \lambda} = (1-\beta_{\kappa \lambda}) \,\frac{2 h \nu_{\kappa \lambda}^3}{c^2} 
\left(\frac{X_{\lambda}/g_{\lambda}}{X_{\kappa}/g_{\kappa}}-1 \right)^{-1} \\
+ \ \beta_{\kappa \lambda} \,B_{\rm d} (\nu_{\kappa \lambda},T_{\rm ph}), \ \ \  E_{\kappa} > E_{\lambda},
\label{eq:J}
\end{aligned}
\end{equation}
where $\beta_{\kappa \lambda}$ is the escape probability. %
$X_{\kappa}$ ($X_{\lambda}$) and $g_{\kappa}$ ($g_{\lambda}$) are the population of the ro-vibrational state $\kappa$ ($\lambda$) and the statistical weight of the state $\kappa$ ($\lambda$), respectively. %
The escape probability $\beta_{\kappa \lambda}$ is obtained by 
\begin{equation}
\beta_{\kappa \lambda} = \frac{1-\exp(-\tau_{\kappa \lambda})}{\tau_{\kappa \lambda}}, 
\label{eq:beta}
\end{equation}
where $\tau_{\kappa \lambda}$ is the line optical depth expressed as
\begin{equation}
\tau_{\kappa \lambda} = \frac{A_{\kappa \lambda} \, c^3 n_{\kappa}}{8 \pi \nu_{\kappa \lambda}^3} 
\left(\frac{X_{\lambda}/g_{\lambda}}{X_{\kappa}/g_{\kappa}}-1 \right) t,
\end{equation}
where $n_{\kappa}$ is the number density of the state $\kappa$ and $t$ is the time after the explosion. %
The second term of Equation~(\ref{eq:J}) is the contribution from the background photons from the photosphere and $B_{\rm d}$ is the Planck function diluted by a factor $\alpha_{\rm d}$ (dilution factor); %
\begin{equation}
B_{\rm d}(\nu_{\kappa \lambda},T_{\rm ph}) = \frac{2 h \nu_{\kappa \lambda}^3}{c^2} 
\frac{\alpha_{\rm d}}{\exp(h \nu_{\kappa \lambda}/k_{\rm B} T_{\rm ph})-1},
\end{equation}
where $T_{\rm ph}$ is the temperature at the photosphere. %
The dilution factor $\alpha_{\rm d}$ is calculated as
\begin{equation}
\alpha_{\rm d} = \frac{1}{2} \left(1 - \sqrt{1- \left(v_{r,{\rm ph}}/v_{r}\right)^2} \right), 
\end{equation}
where $v_{r,{\rm ph}}$ is the radial velocity at the photosphere and $v_{r}$ is the local radial velocity. %
If $v_r < v_{r,{\rm ph}}$, i.e., inside the photosphere, the contribution from the background photons is turned off, i.e., $B_{\rm d}(\nu_{\kappa \lambda},T_{\rm ph}) = 0$, by practically setting the dilution factor to zero. %
The properties of the photosphere are unknown. %
Then, throughout this paper, the results of one of 1D radiative transfer calculations are adopted. %
With the angle-averaged 1D profiles based on the 3D hydrodynamical calculation for SN~1987A \citep[the model b18.3-high:][]{2020ApJ...888..111O} (see Section~\ref{subsec:1d_calc}) as the initial condition, the radiative transfer is calculated\footnote{Some of the calculation results, bolometric and broad-band light curves, are presented in the url below. \texttt{https://wwwmpa.mpa-garching.mpg.de/ccsnarchive/data/}
\texttt{Kozyreva87A/}.} (Kozyreva et al.,~in prep.) by using a 1D radiation hydrodynamical code \citep[STELLA, see e.g.,][]{1998ApJ...496..454B, 2006A&A...453..229B, 2020MNRAS.499.4312K}. %
Figure~\ref{fig:photo} shows the photospheric temperatures, $T_{\rm ph}$, and the radial velocities at the photospheres, $v_{r,{\rm ph}}$, for several bands (U, B, V, R, I). %
Since the emission from ro-vibrational transitions (${\it \Delta} v=1$ and ${\it \Delta} v=2$ transitions correspond to $\sim$ 4.6 $\mu$m and $\sim$ 2.3 $\mu$m, respectively) are observed outside the I band, we adopt the values of the I band as the reference. %
The properties of the photospheres are obtained only up to $\sim$ 170 days after the explosion. %
Therefore, the values after 170 days are crudely extrapolated with a power-law ($\sim$ $t^{-1}$, see, the dashed lines in Figure~\ref{fig:photo}). %

Then, $\overline{B_{ij} \,J_{ij}}$ in the first line in Equation~(\ref{eq:rij}) is obtained similarly as in Equation~(\ref{eq:abar}), i.e. 
\begin{equation}
\overline{B_{ij} \,J_{ij}} = \sum_{J=0}^{J_{\rm max}} \sum_{J'=0}^{J_{\rm max}} P_{\,(i,J)} 
\,\,B_{\,(i,J)(j,J')} \,J_{\,(i,J)(j,J')}.
\end{equation}
Once the number density of the state $\kappa$, $n_{\kappa}$, is obtained by solving Equation~(\ref{eq:rate_eq_co}), the line emissivity for the ro-vibrational transition from the state $\kappa$ to $\lambda$ is eventually obtained by 
\begin{equation}
\begin{aligned}
j_{\kappa \lambda} = \frac{h \nu_{\kappa \lambda}}{4 \pi} n_{\kappa} \,A_{\kappa \lambda} \,\beta_{\kappa \lambda}, 
\ \ \  {\rm erg} \,{\rm cm}^{-3} {\rm s}^{-1} {\rm sr}^{-1}, \\ 
\ \ \ E_{\kappa} > E_{\lambda}.
\label{eq:line_emiss}
\end{aligned}
\end{equation}
The rate equations described in Equation~(\ref{eq:rate_eq_co}) are solved with the same methodology for the rate equations in Equation~(\ref{eq:rate_eq}). %

As for the data necessary for the ro-vibrational transitions, i.e., Einstein's coefficients, energy levels (transition frequencies), and statistical weights, are taken from \cite{2015ApJS..216...15L}\footnote{By assuming that $^{12}$C and $^{16}$O are dominated among the isotopes, $^{12,13,14}$C and $^{16,17,18}$O, we adopt the values for $^{12}$C$^{16}$O in \cite{2015ApJS..216...15L}.}. %
The rate coefficients of the electron impact excitation/de-excitation of vibrational levels, i.e. $q_{ij}$, are derived from the values of the cross sections in \cite{2008JPCA..11212296P} by taking Maxwellian averaging (see, Apendix~\ref{app:elec_impact}). %
As for the maximum quantum numbers (levels) of vibrational and rotational states to be considered in this study, we set $v_{\rm max} = 6$ and $J_{\rm max} = 128$, respectively. %
It is noted that in a recent theoretical study on the formation of CO in a CCSN ejecta \citep{2020A&A...642A.135L}, vibrational levels up to $v=6$ were taken into account. We confirm that taking into account higher levels does not affect the results much. %
As for the initial population, it is assumed that only the grand state, i.e. $v=0$, is initially populated; %
it is confirmed that changing the initial population does not affect the main results of this study. %
Once CO forms, Equation~(\ref{eq:rate_eq_co}) is solved to obtain the population of (ro)-vibrational levels and the emissivity as in Equation~(\ref{eq:line_emiss}). %
The calculation for the CO (ro)-vibrational levels is performed up to 1000 days after the explosion. After this epoch there have been no observations of CO vibrational bands in SN~1987A and the CO cooling would play only a minor role. %

\subsection{Calculations with the one-zone approximation} \label{subsec:one_zone_calc} 

\begin{deluxetable}{llllc}
\tabletypesize{\footnotesize}
\tablewidth{0pt}
\tablenum{6}
\tablecolumns{4}
\tablecaption{Initial conditions for one-zone calculations obtained based on the 3D hydrodynamical model b18.3-high \citep{2020ApJ...888..111O}.\label{table:init_one_zone}} %
\tablehead{\vsm{3} \\Description & Parameter \hs{1} & Value \hs{0.2} & Unit} %
%
\startdata
Hydrogen mass & $M_{\rm H}$ & 1.28 &$M_{\odot}$\\
Helium mass & $M_{\rm He}$ & 1.66 &$M_{\odot}$\\
Carbon mass & $M_{\rm C}$ & 1.17 (-1)\tnm{a} &$M_{\odot}$\\
Nitrogen mass & $M_{\rm N}$ & 5.59 (-3) &$M_{\odot}$\\
Oxygen mass & $M_{\rm O}$ & 3.21 (-1) &$M_{\odot}$\\
Neon mass & $M_{\rm Ne}$ & 7.66 (-3) &$M_{\odot}$\\
Magnesium mass & $M_{\rm Mg}$ & 1.30 (-2) &$M_{\odot}$\\
Silicon mass & $M_{\rm Si}$ & 1.57 (-1) &$M_{\odot}$\\
Sulfur mass & $M_{\rm S}$ & 4.36 (-2) &$M_{\odot}$\\
Argon mass & $M_{\rm Ar}$ & 6.30 (-3) &$M_{\odot}$\\
Iron mass & $M_{\rm Fe}$ & 8.49 (-2) &$M_{\odot}$\\ \hline
%
Total mass & $M_{\rm tot}$ & 3.70 &$M_{\odot}$\\ \hline
%
Density & $\rho$ & 7.86 (-7) &g cm$^{-3}$ \\
Temperature & $T$ & 5.76 (5) &K \\
Radial velocity & $v_r$ & 1.43 (8) &km s$^{-1}$ \\
%
\enddata
\tnt{a}{The numbers in the parentheses denote the powers of ten.} %
\end{deluxetable}

\begin{figure}
\begin{center}
\includegraphics[width=8cm,keepaspectratio,clip]{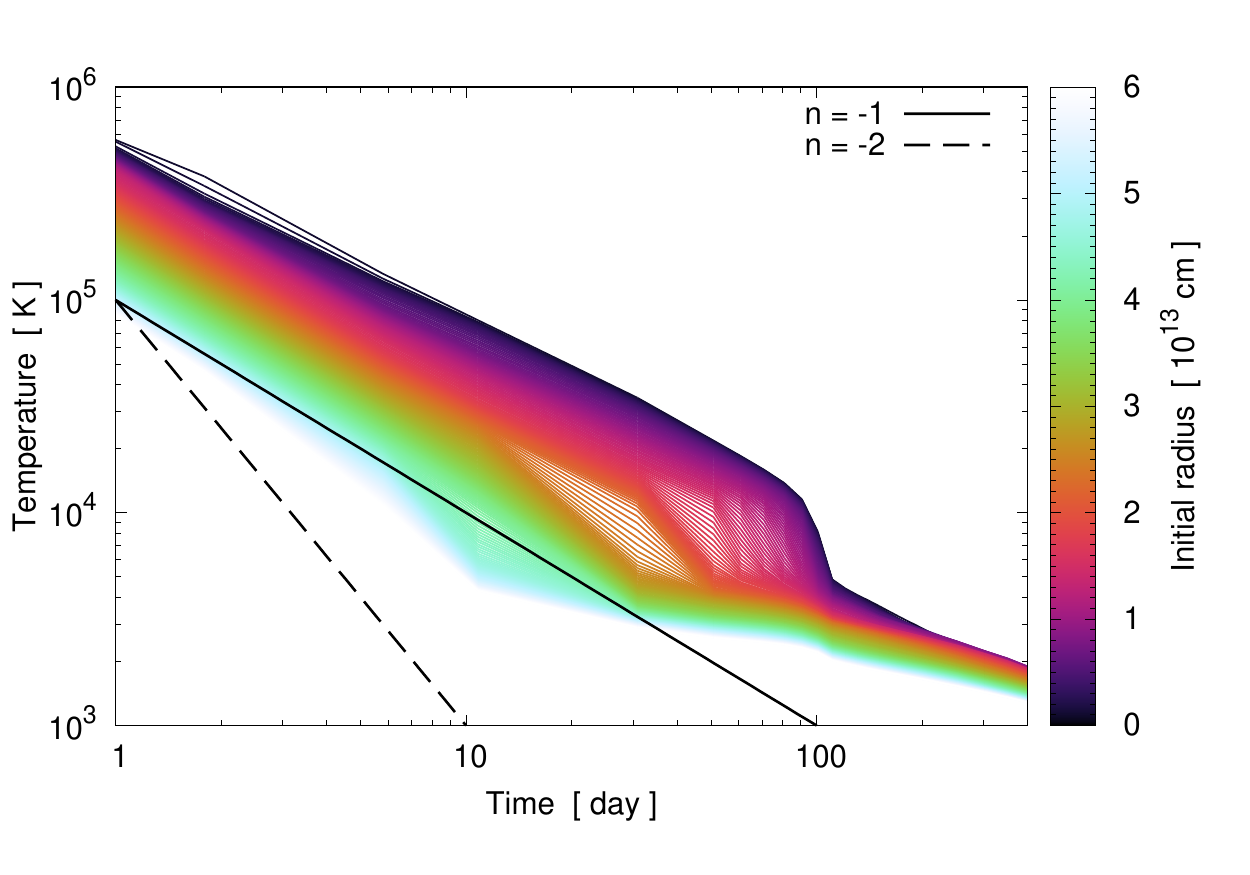}
\vs{-0.5}
\end{center}
\caption{The gas temperatures of Lagrange meshes as a function of time after the explosion. %
The values are obtained from the 1D radiation hydrodynamical calculation results with the angle-averaged 1D profiles based on the 3D hydrodynamical model \citep[the model b18.3-high:][]{2020ApJ...888..111O}. %
Each line shows the time evolution for each Lagrange mesh. The colors denote the corresponding initial positions (radii). %
The two straight lines represent the power-law evolutions with the powers of $-1$ and $-2$ corresponding to $\gamma_{\rm ad} = 4/3$ and $\gamma_{\rm ad} = 5/3$, respectively.} %
\label{fig:temp_1d_rad}
\end{figure}

In order to see the chemical evolution in the ejecta, first, for simplicity we apply our method to one-zone models which regard the inner core of the ejecta as one-zone. %
The initial condition of the ejecta is obtained based on the 3D hydrodynamical simulation results for SN~1987A \citep{2020ApJ...888..111O}. %
Among the models, we adopt the best model b18.3-high, i.e., an asymmetric bipolar-like explosion with a BSG progenitor star formed through a binary merger. %
The initial condition is obtained by averaging the physical quantities in a core region of the 3D model. %
First, an angle-averaged 1D profile is derived based on the last snapshot ($\sim$ 1 day after the explosion) of the 3D model. %
Then, the physical quantities from the inner boundary to the mass coordinate of about 6 $M_{\odot}$ are averaged with the density as a weight for quantities other than the density. %
The obtained initial physical values, i.e. the masses of the atoms, the total mass, the density, temperature, and radial velocity, are listed\,\footnote{Here, the values of the masses are derived without the effects of ionization. For the inputs of the rate equations of the chemical reactions, some fractions of these atoms are converted to the corresponding ions depending on the ionization fraction $X_{\rm e}$ as mentioned in Section~\ref{subsubsec:init_chemi}.} in Table~\ref{table:init_one_zone}. %
Time evolutions of the density $\rho$ and the specific internal energy $e$ are followed by power-laws with the powers of $-3$ and $-3\,(\gamma_{\rm ad}-1)$ as below, respectively, assuming an adiabatic expansion, 
\begin{align}
\label{eq:rho_adiabatic}
\rho \,(t) &= \rho_0 \left(t/t_0\right)^{-3}, \\
\label{eq:e_adiabatic}
e\,(t) &= e_0 \left( t/t_0 \right)^{-3 \,(\gamma_{\rm ad} -1)},
\end{align}
where $\gamma_{\rm ad}$ is the adiabatic index. $t_0$ is the time of the previous time step and $\rho_0$ and $e_0$ are the density and specific internal energy at $t = t_0$, respectively. The temperature is derived through the ideal equation of state (EoS), 
\begin{equation}
P = \frac{R}{\mu} \,\rho \,T = e \,(\gamma_{\rm ad} - 1) \,\rho, \label{eq:eos}
\end{equation}
where $P$ is the pressure, $R$ is the gas constant, and $\mu$ is the mean molecular weight. %
If the radiation is dominant in the system, $\gamma_{\rm ad} = 4/3$. %
On the other hand, if the monoatomic gas (diatomic gas) is dominant, $\gamma_{\rm ad} = 5/3$ $(7/5)$. %
At the age of the initial conditions ($\sim$ 1 day after the explosion), the density of the inner core (see, Table~\ref{table:init_one_zone}) is still high and the radiation is dominant. %
As the expansion proceeds, the ejecta gas becomes optically thin, and a transition in the adiabatic index $\gamma_{\rm ad}$ occurs at some point as shown in the 1D radiation hydrodynamical calculation results mentioned in Section~\ref{subsec:cooling} (see, Figure~\ref{fig:temp_1d_rad}). %

Therefore, when the density drops to a certain critical value ($\rho_{\,\rm break}$) for the first time, the adiabatic index is changed manually (in the calculations it is automatically switched according to the density) from 4/3 to 5/3 if the timing is before the start of the molecule formation calculation. %
If the timing of the change is after the start of the molecule formation calculation, the adiabatic index is changed from 4/3 to an effective adiabatic index derived from the averaging of ones for diatomic species and the others with the mass fractions as a weight. %
Such sudden change in the adiabatic index results in a jump in the gas temperature before and after the change with the same internal energy. %
In reality, after the system becomes optically thin, the energy of radiation escapes from the gas. %
In this study, some fraction of the internal energy of the gas is subtracted at the transition point in the adiabatic index to keep the same gas temperature before and after it. %
As seen in Figure~\ref{fig:temp_1d_rad}, after the break gas temperatures seem to remain at temperatures $\gtrsim$ 10$^3$ K; %
the trend may be attributed to not a physical reason but a computational one in the 1D radiation hydrodynamical calculations (Kozyreva et al., in prep.). %
Therefore, we adopt the single break in the specific internal energy (temperature) as a fiducial thermal evolution. Based on the power-law evolution in Equations~(\ref{eq:rho_adiabatic}) and (\ref{eq:e_adiabatic}), additional heating and cooling effects are taken into account as follows. %

As mentioned in Section~\ref{subsubsec:compton}, the decay of radioactive $^{56}$Ni and $^{56}$Co could deposit some energies into the gas depending on the optical depth $\tau_{\gamma}$. %
In reality, the deposition processes, i.e., how the emerged gamma-rays produce Compton electrons and how the energies of Compton electrons are degraded by the interaction with gas particles are complicated; %
the deposited energies could be consumed by the excitation of electrons, ionization, and heating of gas \citep{1994ApJ...435..909L}. %
Due to the lack of accurate treatments for the time evolution of the electron energy distribution, the ionization, and the radiative transfer in this paper, we adopt a simple approach to the heating of gas via the decay of $^{56}$Ni, i.e., some fraction of the ``deposited energy" expressed in Equation~(\ref{eq:epsilon}) is ``locally" used for the heating of the gas. %
The potential impact of non-local energy deposition is discussed in Appendix~\ref{app:non_local_edep} as mentioned in Section~\ref{subsubsec:compton}. %
To count the effective efficiency of the heating of the gas, a constant factor, $f_{\rm h}$, which denotes the fraction of the deposited energy to be used for the heating of the gas, is introduced as a parameter. %

It is noted that in this paper, radiative cooling by line emission from ionized atomic species \citep{1998ApJ...496..946K} is not taken into account. %
In particular in the metal-rich ejecta core, such radiative cooling could be faster than adiabatic cooling \citep[see, e.g., Fig.~5 in][]{1998ApJ...496..946K}. %
Actually, as can be seen in Figure~\ref{fig:temp_1d_rad}, some of the inner particles have steeper temperature slopes than that of the case of adiabatic cooling for the ideal gas ($n = -2$: dashed line). %
Such inner particles tend to contain $^{56}$Ni inside. %
Therefore, in such inner particles, the heating of gas due to the decay of $^{56}$Ni may compete with the radiative cooling by ionized atomic species; neglecting the radiative cooling may partly be compensated by adopting lower $f_{\rm h}$ values. %
We would regard that the parameter, $f_{\rm h}$, effectively covers the uncertainties due to neglecting the radiative cooling by atomic species, although the effects may not be represented by such a constant factor. %
More realistic treatments of gas heating and cooling are beyond the scope of this study. %

Given the effective energy deposition rates, the change in the specific internal energy of gas per unit time is expressed as below. %
\begin{equation}
\frac{de}{dt} = f_{\rm h} \,\epsilon = f_{\rm h} \, (\epsilon_{\rm Ni} + \epsilon_{\rm Co}) \,D_{\gamma}.
\label{eq:heating_rate}
\end{equation}

With the temperature evolution, the abundance of CO is obtained through the chemical reaction calculations described in Section~\ref{subsec:chemical}. %
Once CO forms in the ejecta, it could be an important coolant through the emission via ro-vibrational transitions. %
As the emissivity is given in Equation~(\ref{eq:line_emiss}), the cooling rate of gas can be calculated as follows. %
\begin{equation}
\begin{aligned}
\frac{de}{dt}
= - \frac{h}{\rho} \sum_{E_{\kappa} > E_{\lambda}} \, \nu_{\kappa \lambda} \,n_{\kappa} \,A_{\kappa \lambda} 
\,\beta_{\kappa \lambda} \\
\times  \left( \frac{S_{\kappa \lambda} - B_{\rm d} (\nu_{\kappa \lambda}, 
T_{\rm ph})}{S_{\kappa \lambda}} \right) \\ 
\ \ \ {\rm erg} \,{\rm g}^{-1} {\rm s}^{-1},
%
\end{aligned}
\label{eq:cooling_rate}
\end{equation}
where $S_{\kappa \lambda}$ is the so-called source function, which can also be seen in the first term of Equation~(\ref{eq:J}), and is expressed below. %
\begin{equation}
S_{\kappa \lambda} = \frac{2 h \nu_{\kappa \lambda}^3}{c^2} 
\left(\frac{X_{\lambda}/g_{\lambda}}{X_{\kappa}/g_{\kappa}}-1 \right)^{-1}.
\end{equation}
Here, if the excitations are mainly maintained by the radiation from the background photons, the radiated energies should not be subtracted from the internal energy of the gas. %
Then, a controlling factor, i.e., inside the parentheses in Equation~(\ref{eq:cooling_rate}), is introduced as in \cite{1975ApJ...199...69D}. %
The source function, $S_{\kappa \lambda}$, and the diluted Planck function, $B_{\rm d} (\nu_{\kappa \lambda}, T_{\rm ph})$, represent the contributions from the local and background photons, respectively. %
If $S_{\kappa \lambda} < B_{\rm d} (\nu_{\kappa \lambda}, T_{\rm ph})$ (the contribution of the background photons dominates the local one), the controlling factor is set to zero, i.e., no energy subtraction from the internal energy of the gas. %
In the case of $S_{\kappa \lambda} \geq B_{\rm d} (\nu_{\kappa \lambda}, T_{\rm ph})$, the energy subtraction from the internal energy of the gas is controlled by the factor of less than unity depending on the magnitudes of the source function and the diluted Planck function. %
In the case of $B_{\rm d} (\nu_{\kappa \lambda}, T_{\rm ph}) = 0$, i.e., no contribution from the background photons, the controlling factor becomes unity. %

If the specific internal energy is updated by the heating and/or cooling in an operator-splitting manner as in Equations~(\ref{eq:heating_rate}) and (\ref{eq:cooling_rate}), through the equation of state in Equation~(\ref{eq:eos}), the temperature of the gas is also updated. %
Additionally, if the extra heating (cooling) is dominant, the local gas could be inflated (deflated). %
To mimic such an effect, by assuming a pressure balance between the local gas and the surroundings, the density of the gas with such extra heating and/or cooling, $\rho$, is obtained by equating the two cases of the right-hand side of Equation~(\ref{eq:eos}), i.e., with and without net extra heating/cooling as follows. %
\begin{equation}
\rho = \rho_{\rm ad} \,(e_{\rm ad}/e),
\label{eq:thermal_i}
\end{equation}
where $\rho_{\rm ad}$ and $e_{\rm ad}$ denote the density and the specific internal energy, respectively, obtained from the evolution with no extra heating by the decay of $^{56}$Ni and cooling by CO ro-vibrational transitions, i.e., only with the adiabatic cooling described in Equations~(\ref{eq:rho_adiabatic}) and (\ref{eq:e_adiabatic}) throughout the calculation. %
$e$ is the specific internal energy obtained from the evolution with net extra heating and/or cooling. %
Once the extra cooling becomes dominant, the local gas could be deflated and the density increases. %
Such an enhancement in density may cause additional emissions, which may lead to thermal instability. %
In this manner, such a potential impact is effectively taken into account in the calculations. %
The methodology described in Equation~(\ref{eq:thermal_i}) is, however, post-processing; the feedback of the inflation and deflation to the hydrodynamics is not taken into account. %
The feedback to the hydrodynamics may result in expanded bubble-like structures around $^{56}$Ni-rich regions \citep[as seen in][]{2021MNRAS.502.3264G} and/or denser clumps around CO-rich regions. %

Because of the adiabatic cooling and/or cooling by CO ro-vibrational transitions, with the methodology above, the gas temperature drops to $\sim$ 10$^2$ K at some point. %
In this study, practically, we set a minimum gas temperature to be 10$^2$ K assuming that the ejecta keeps the minimum temperature thanks to some heating e.g., by the radioactive decay of $^{44}$Ti.\footnote{$^{44}$Ti decays as $^{44}$Ti $\lra$ $^{44}$Sc $\lra$ $^{44}$Ca. The long half-life of the former decay is 58.9$\pm$0.3 yr \citep{2006PhRvC..74f5803A}. The latter decay is quick with a timescale of hours.} %
It is noted that the ALMA observations of molecular lines of CO, SiO, and HCO$^+$ in the ejecta of SN~1987A have shown the temperature of the ejecta gas ranges over 13-132 K \citep{2013ApJ...773L..34K,2017MNRAS.469.3347M,2019ApJ...886...51C}. %
Because of this minimum temperature, the minimum specific internal energy is also implicitly set through the EoS. %
Practically, by turning off the adiabatic cooling described in Equation~(\ref{eq:e_adiabatic}) after both $e$ and $e_{\rm ad}$ become the minimum value, $e$ can be equal to $e_{\rm ad}$ at some point without effective heating due to the decay of $^{56}$Ni, i.e., the density evolution becomes again consistent with the original power-law evolution after that. %
In this way, we avoid introducing the effect described in Equation~(\ref{eq:thermal_i}) at such low temperature environment.

As a summary, the chemical reaction calculation, the calculation of CO (ro)-vibrational levels for CO line cooling, and the updating of the specific internal energy and gas temperature are recurrently performed to obtain the time evolution. %
The time step for the iteration is empirically determined not to cause a change of a large fraction of the internal energy per step. %
Additionally, for the chemical reaction calculation and the rate equation calculation for the CO (ro)-vibrational levels, sub-time steps are independently introduced. %

As will be discussed in Section~\ref{subsec:one_zone_results} and Appendix~\ref{app:fred}, the calculations with the method described above tend to result in higher CO line emissions than that corresponding to the peak fluxes of the observed CO light curves \citep{1993A&A...273..451B,1993MNRAS.261..535M} at an early phase ($\lesssim$ 200 days). %
With out of consideration for the lack of realistic treatments for the radiative transfer, ionization, and energy deposition through the decay of $^{56}$Ni in this study, to avoid a significant cooling in the ejecta gas, a time-dependent reduction factor (function) $f_{\rm red}$ for the escape probability $\beta_{\kappa \lambda}$ found in Equations~(\ref{eq:J}), (\ref{eq:line_emiss}), and (\ref{eq:cooling_rate}) is introduced. %
The function is arbitrarily given by
\begin{equation}
f_{\rm red}= \exp\left[- (t_{\rm c} - t)/t_{\rm s} \right] \ \ \ t < t_{\rm c}, \label{eq:fred}
\end{equation}
where $t_{\rm c}$ is an arbitrary critical time up to when the factor is effective. %
$t_{\rm s}$ is a timescale to control the magnitude of the factor. %
Then, the escape probability $\beta_{\kappa \lambda}$ is practically multiplied by the reduction factor $f_{\rm red}$ to reduce 
the escape probability (to increase the line optical depth effectively). %
As seen in Equation~(\ref{eq:beta}), the $\beta_{\kappa \lambda}$ is dependent on the line optical depth $\tau_{\kappa \lambda}$ in which the age $t$ is introduced as a rough approximation for the reciprocal of the velocity gradient of the system \citep[see e.g.,][]{1975ApJ...199...69D}. %
Then, we consider that there is some uncertainty at least in the approximation. %
The reduction factor is not dependent on the transition levels but the resultant escape probability, $f_{\rm red}\, \beta_{\kappa \lambda}$, is varied depending on the transitions. %
The values of $t_{\rm c}$ and $t_{\rm s}$ are empirically determined to avoid a significant cooling incompatible with the observed light curves. %
The impact of the reduction factor $f_{\rm red}$ through different values of the parameter $t_{\rm s}$ on the CO ro-vibrational emissions is discussed in Section~\ref{subsec:one_zone_results} and Appendix~\ref{app:fred}. %

The model parameters related to the calculations with the one-zone approximation are listed in Table~\ref{table:param}. %

\begin{deluxetable*}{llcc}
\tabletypesize{\footnotesize}
\tablewidth{0pt}
\tablenum{7}
\tablecolumns{4}
\tablecaption{Model parameters for one-zone and 1D calculations.\label{table:param}} %
\tablehead{\vsm{3} \\Parameter & Explanation & \hs{0.5} & Value/range} %
\startdata
%
$f_{\rm h}$ & Efficiency of gas heating by the decay of $^{56}$Ni & & 10$^{-4}$ -- 10$^{-2}$\\
$f_{\rm d}$ & Efficiency of the destruction/ionization via the decay of $^{56}$Ni & &10$^{-2}$ -- 1.0\\
$t_{\rm c}$ & Critical time in the reduction factor $f_{\rm red}$ in Equation~(\ref{eq:fred}) for the escape probability 
$\beta_{\kappa \lambda}$ & & 1000 days (fixed)\\
$t_{\rm s}$ & Timescale in the reduction factor $f_{\rm red}$ in Equation~(\ref{eq:fred}) for the escape probability 
$\beta_{\kappa \lambda}$ & & 200 -- $\infty$ days\\
$X_{{\rm e},i}$ & Initial ionization fraction & & 10$^{-1}$ (fixed) \\
$X_{{\rm e},f}$ & Ionization fraction at 2000 days & & 10$^{-3}$ (fixed)\\
$\rho_{\rm break}$\tnm{a} & Critical density for the transition of the adiabatic index $\gamma_{\rm ad}$ & & 10$^{-9}$ g cm$^{-3}$ (fixed)\\
\enddata
\tnt{a}{This parameter is only for the one-zone calculations.} %
\end{deluxetable*}

\subsection{Calculations with 1D radial profiles} \label{subsec:1d_calc} 

To figure out the chemical evolution in the ejecta in a more realistic way, next, 1D (radial) profiles derived from 3D hydrodynamical models for SN~1987A \citep{2020ApJ...888..111O} are utilized as the initial conditions. %
By implementing angle-averaging with the density as a weight for quantities other than the density on the results of the last snapshots at $\sim$ 1 day after the explosion, 1D profiles are obtained (as partly mentioned in Section~\ref{subsec:one_zone_calc}). %
Among the 3D hydrodynamical models in \cite{2020ApJ...888..111O}, the models b18.3-high and n16.3-high are selected. %
The former model is the one with an aspherical (bipolar-like) explosion with the binary merger progenitor model \citep{2018MNRAS.473L.101U}, which explains the observational features of the progenitor star of SN~1987A, Sk $-$69$^{\circ}$ 202; %
the explosion model also best explains the observed [Fe II] line profiles \citep{1990ApJ...360..257H} among the models in \cite{2020ApJ...888..111O} and its further evolution \citep{2020A&A...636A..22O} also well reproduces the observed X-ray morphology and light curves \citep[e.g.][]{2016ApJ...829...40F} and the spatial distribution of CO and SiO \citep{2017ApJ...842L..24A} in SN~1987A. %
The latter model is the model with the explosion with the same asymmetry and explosion energy as the model b18.3-high but the progenitor model \citep{1988PhR...163...13N,1990ApJ...360..242S} is based on a single star evolution and it was modeled somehow artificially to fit the observations of Sk $-$69$^{\circ}$ 202. %
As mentioned in Section~\ref{sec:intro}, the density structure of the progenitor star could play an important role in the matter mixing \citep{2015A&A...577A..48W}. %
Actually, as seen in \cite{2020ApJ...888..111O}, because of the differences between the two progenitor models, the mixing of the radioactive $^{56}$Ni into high-velocity outer laters changes and affects the velocity distribution of $^{56}$Ni corresponding to the observed [Fe II] line profiles. %
Therefore, we investigate the dependence of chemical evolution on the two progenitor models. %

To further see the impact of the matter mixing, we also investigate the molecule formation for the two cases where the supernova explosion itself is spherical but the evolution is 3D, in which mixing via hydrodynamical instabilities such as Rayleigh-Taylor instability could be involved, and the whole evolution is purely spherical. %
To obtain the 1D profiles as the initial conditions for the two cases, additional hydrodynamical calculations are performed as follows. %
For the former case, two 3D hydrodynamical simulations are newly performed up to $\sim$ 1 day after the explosion as the counterparts of the models b18.3-high and n16.3-high. %
With the same method in \cite{2020ApJ...888..111O}, i.e., the explosions are artificially initiated by injecting thermal and kinetic energies asymmetrically, 3D hydrodynamical simulations are performed with the same progenitor models and injection energies but the energy injections are spherical. %
Then, angle-averaged 1D profiles are obtained based on the last snapshots of the 3D hydrodynamical simulation results. %
In the latter case, two 1D hydrodynamical simulations are performed with the same method in \cite{2020ApJ...888..111O} and the same progenitor models and injection energies but with a 1D version of the same open-source hydrodynamical code, FLASH \citep{2000ApJS..131..273F}, with the spherical coordinate (for the details of the 1D code, see later in this section). %

\begin{figure*}
\begin{minipage}{0.5\hsize}
\begin{center}
\includegraphics[width=8cm,keepaspectratio,clip]{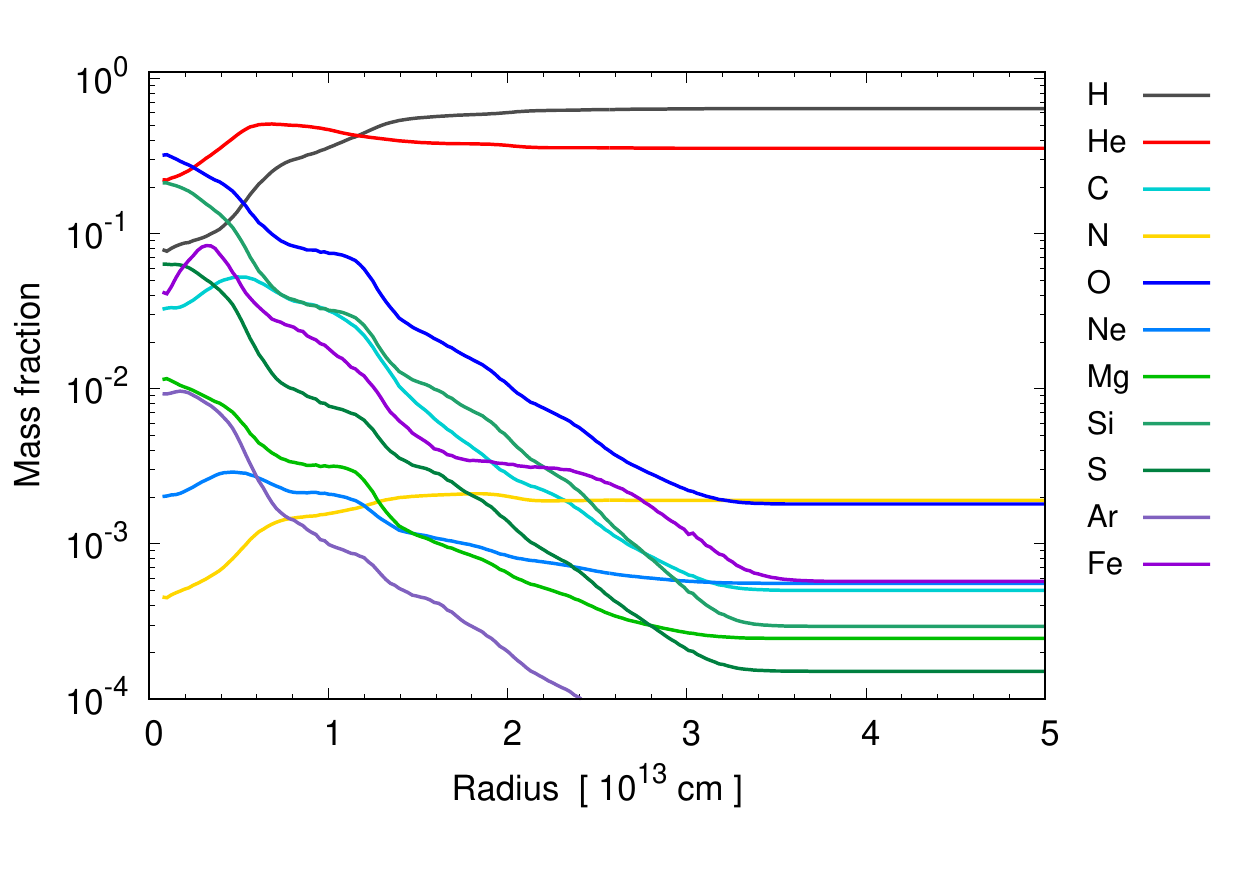}
\end{center}
\end{minipage}
\begin{minipage}{0.5\hsize}
\begin{center}
\includegraphics[width=8.5cm,keepaspectratio,clip]{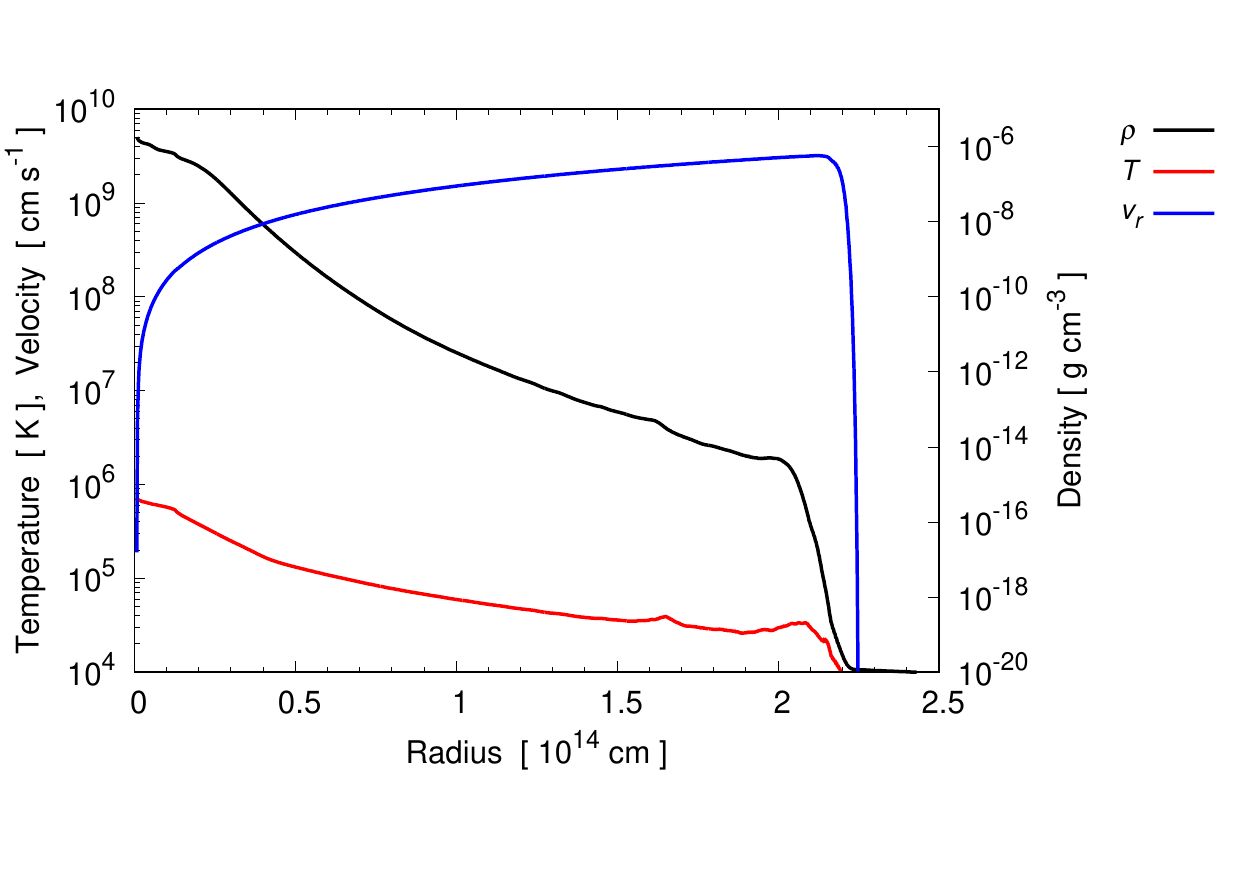}
\end{center}
\end{minipage}
\\
\begin{minipage}{0.5\hsize}
\vs{-1.}
\begin{center}
\includegraphics[width=8cm,keepaspectratio,clip]{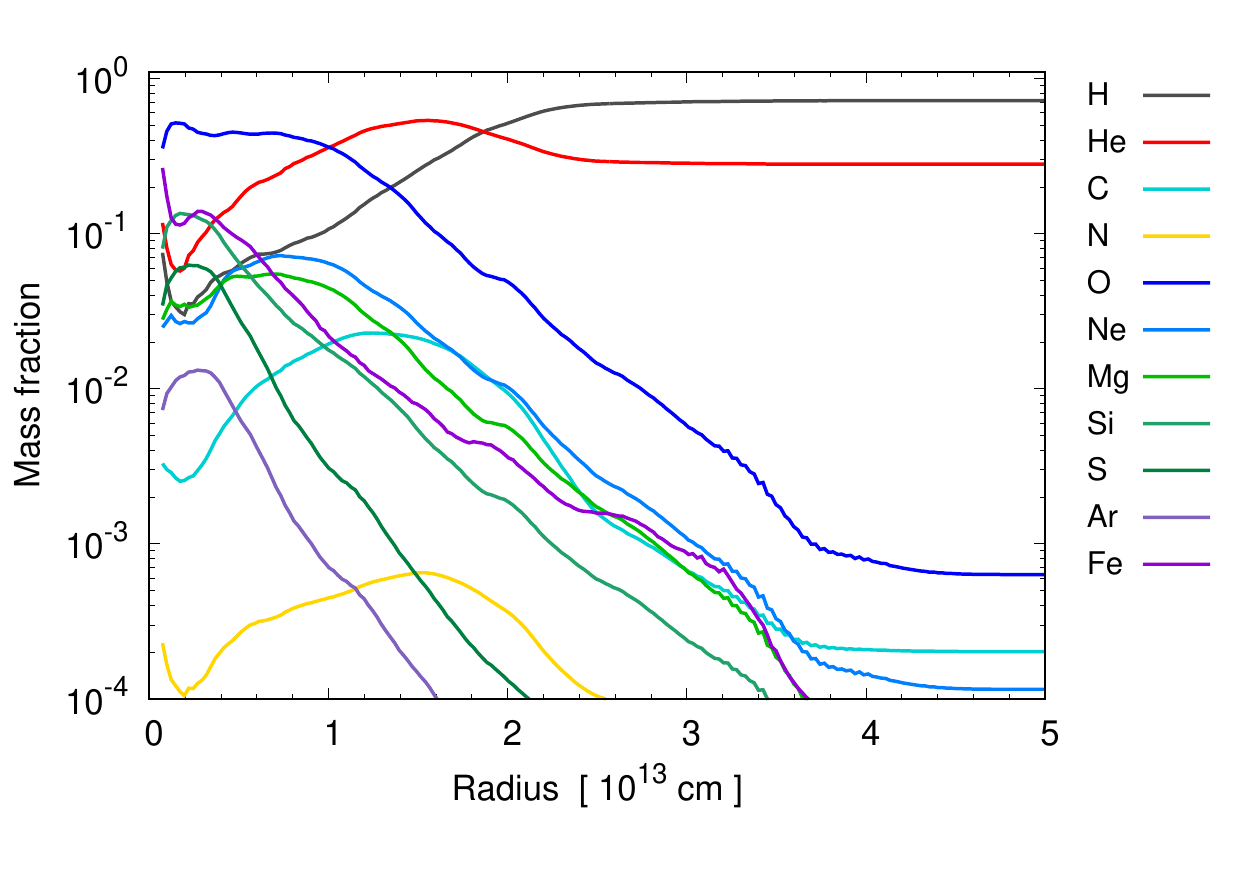}
\end{center}
\vs{-1.}
\end{minipage}
\begin{minipage}{0.5\hsize}
\vs{-1.}
\begin{center}
\includegraphics[width=8.5cm,keepaspectratio,clip]{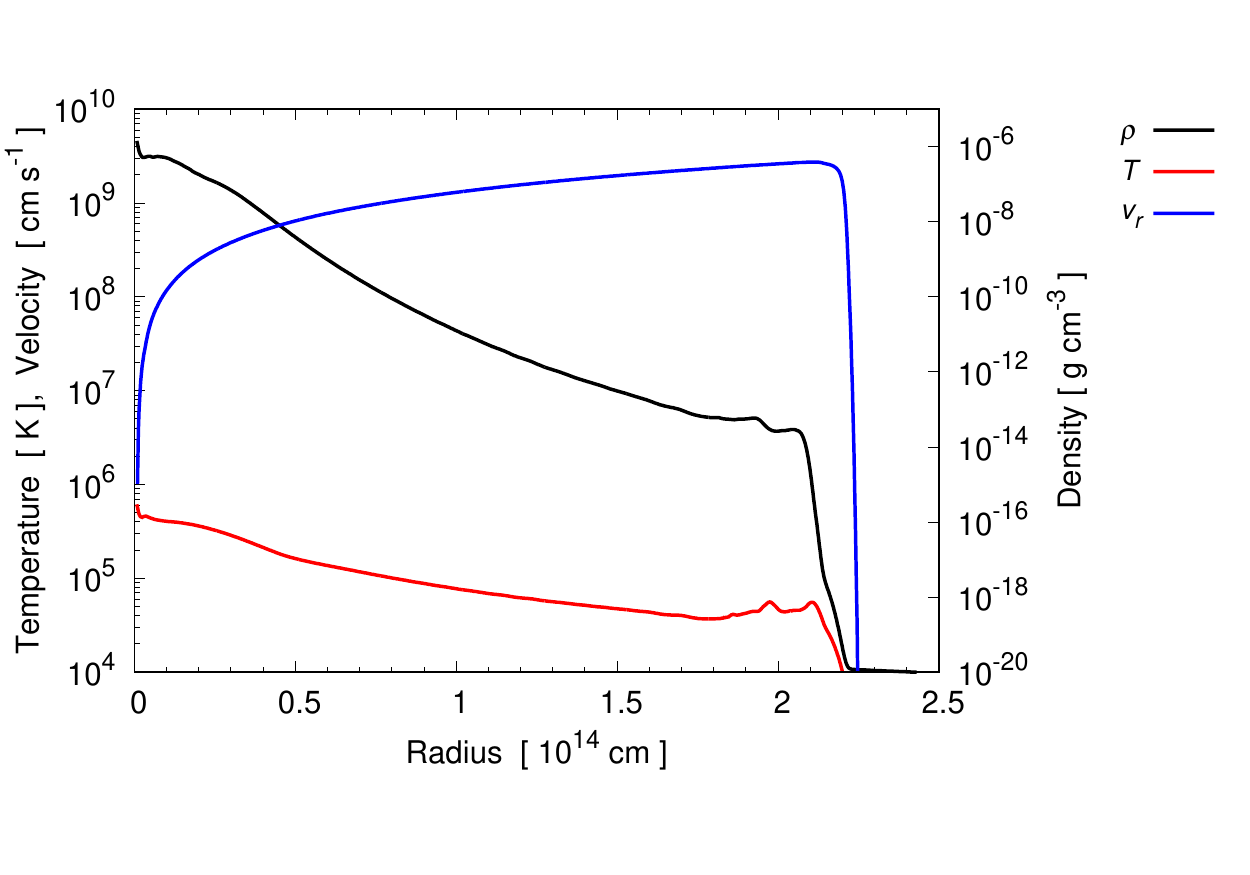}
\end{center}
\vs{-1.}
\end{minipage}
\caption{Angle-averaged 1D profiles (initial conditions) for the models b18.3-high (top) and n16.3-high (bottom). 
Mass fractions of species as a function of the radius (Left) and the density, temperature, and radial velocity as a 
function of radius (Right). 
The ages are 19.0 hours and 22.3 hours 
after the explosion for the former and latter models, respectively.}
\label{fig:prof_mean}
\end{figure*}

In Figure~\ref{fig:prof_mean}, the angle-averaged 1D profiles of the mass fractions of atoms\footnote{Here, the mass fraction of each atom is derived from the summation of ones of the corresponding stable and radioactive isotopes as described in Section~\ref{subsubsec:init_chemi}. For example, $^{56}$Ni is counted as iron.} and hydrodynamical quantities, the density, temperature, and radial velocity, based on the models b18.3-high and n16.3-high \citep{2020ApJ...888..111O}, are shown. %
Overall elements heavier than carbon (left panels) are smoothly extended to the radius of $\sim$ 3 $\times$ 10$^{13}$ cm ($\sim$ 3.5 $\times$ 10$^{13}$ cm) for the model b18.3-high (n16.3-high). %
Between the two models, there are significant differences in the composition. %
For example, the mass fraction of oxygen (O) in the model n16.3-high is higher than that of b18.3-high inside the radius of $\sim$ 10$^{13}$ cm. %
On the other hand, the mass fraction of carbon (C) in the model b18.3-high is higher than that of n16.3-high at the inner region. %
Additionally, parts of hydrogen (H) and helium (He) originally in the outer layers are mixed into the inner regions. %

As mentioned above, matter mixing sensitively depends on the density structure of the progenitor star \citep{2015A&A...577A..48W,2020ApJ...888..111O}; %
the progenitor model of b18.3-high \citep[BSG through a binary merger;][]{2018MNRAS.473L.101U} results in an efficient matter mixing thanks to the compact helium core. %
Then, in the model b18.3-high, the mass fractions of hydrogen and helium are higher than that of n16.3-high at the inner region. %
For the same reason, the mass fraction of iron (Fe) (mostly coming from $^{56}$Ni at inner regions) is affected to be mixed into outer layers. The mass fraction of iron in the model b18.3-high has a plateau between the radii of (1.6--2.6) $\times$ 10$^{13}$ cm. %
It is noted that in the model b18.3-high, the abundance of nitrogen (N) in the hydrogen envelope is apparently higher than that of the model n16.3-high. %
This is because the progenitor model of the b18.3-high is the binary merger model; during the merger process of the companion, additional hydrogen burning including CNO cycle is triggered. %
Then, CNO cycle-processed materials (nitrogen, more specifically $^{14}$N, is one of the main products of the CNO cycle) are mixed into the envelope. As for the hydrodynamical values, i.e., the density, temperature, and radial velocity, there are no significant differences between the two models. %

%
\begin{figure*}
\begin{minipage}{0.5\hsize}
\begin{center}
\includegraphics[width=8cm,keepaspectratio,clip]{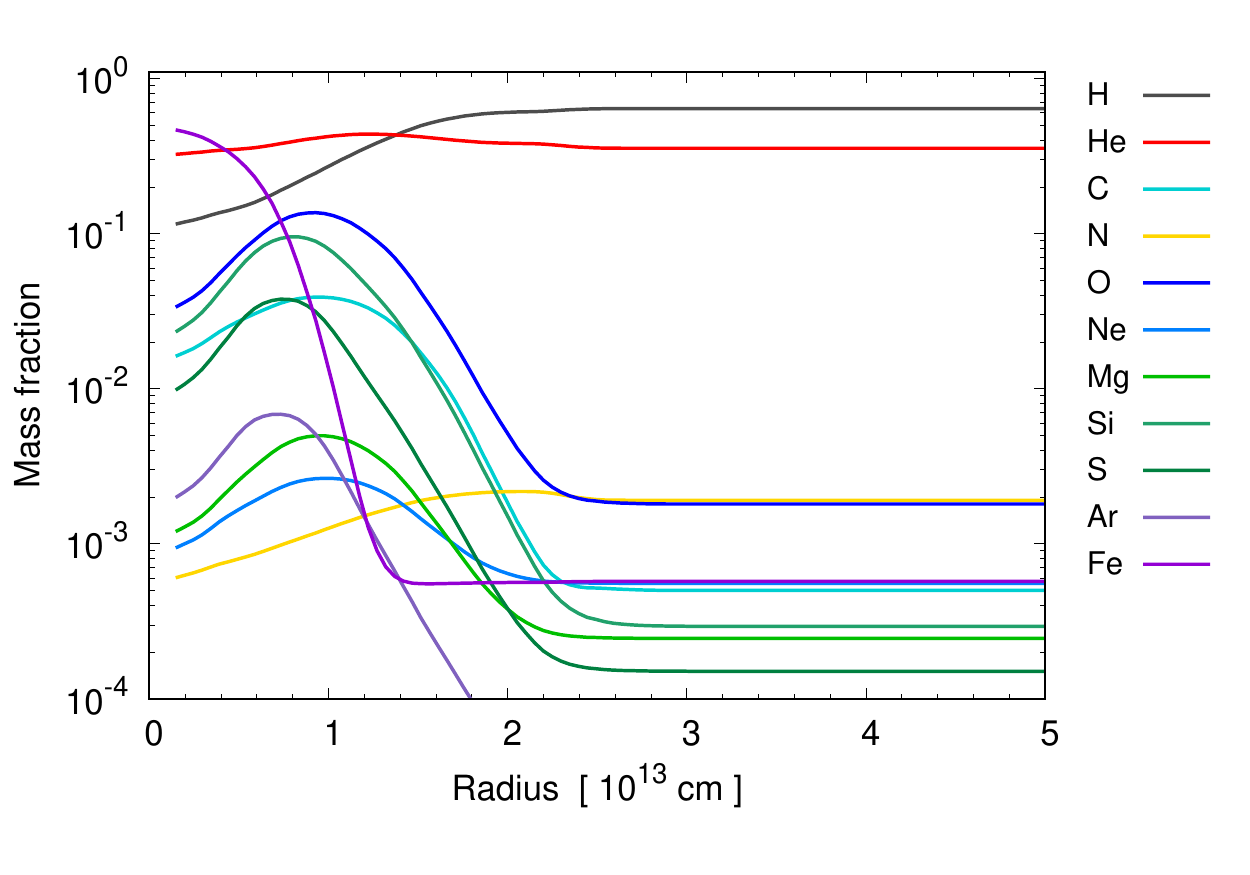}
\end{center}
\end{minipage}
\begin{minipage}{0.5\hsize}
\begin{center}
\includegraphics[width=8.5cm,keepaspectratio,clip]{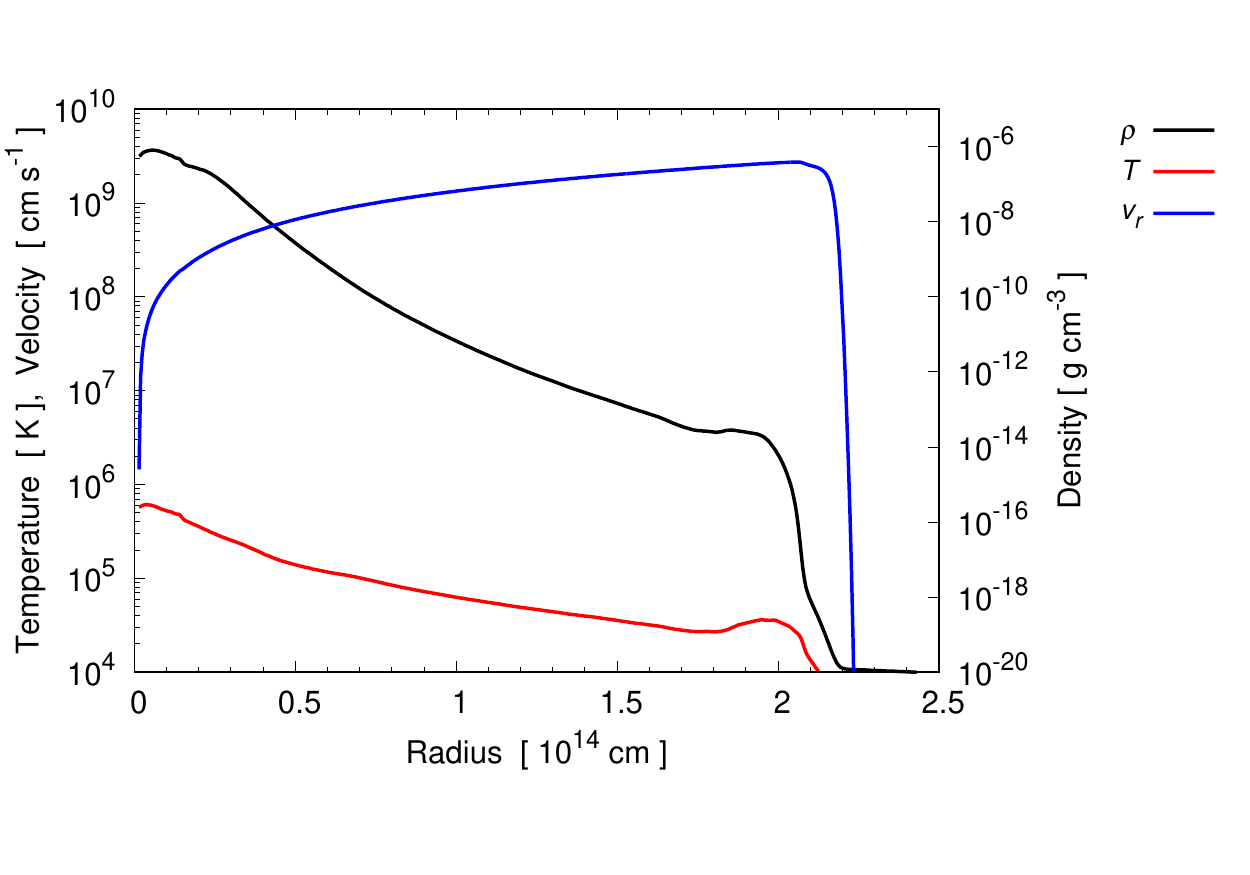}
\end{center}
\end{minipage}
\\
\begin{minipage}{0.5\hsize}
\vs{-1.}
\begin{center}
\includegraphics[width=8cm,keepaspectratio,clip]{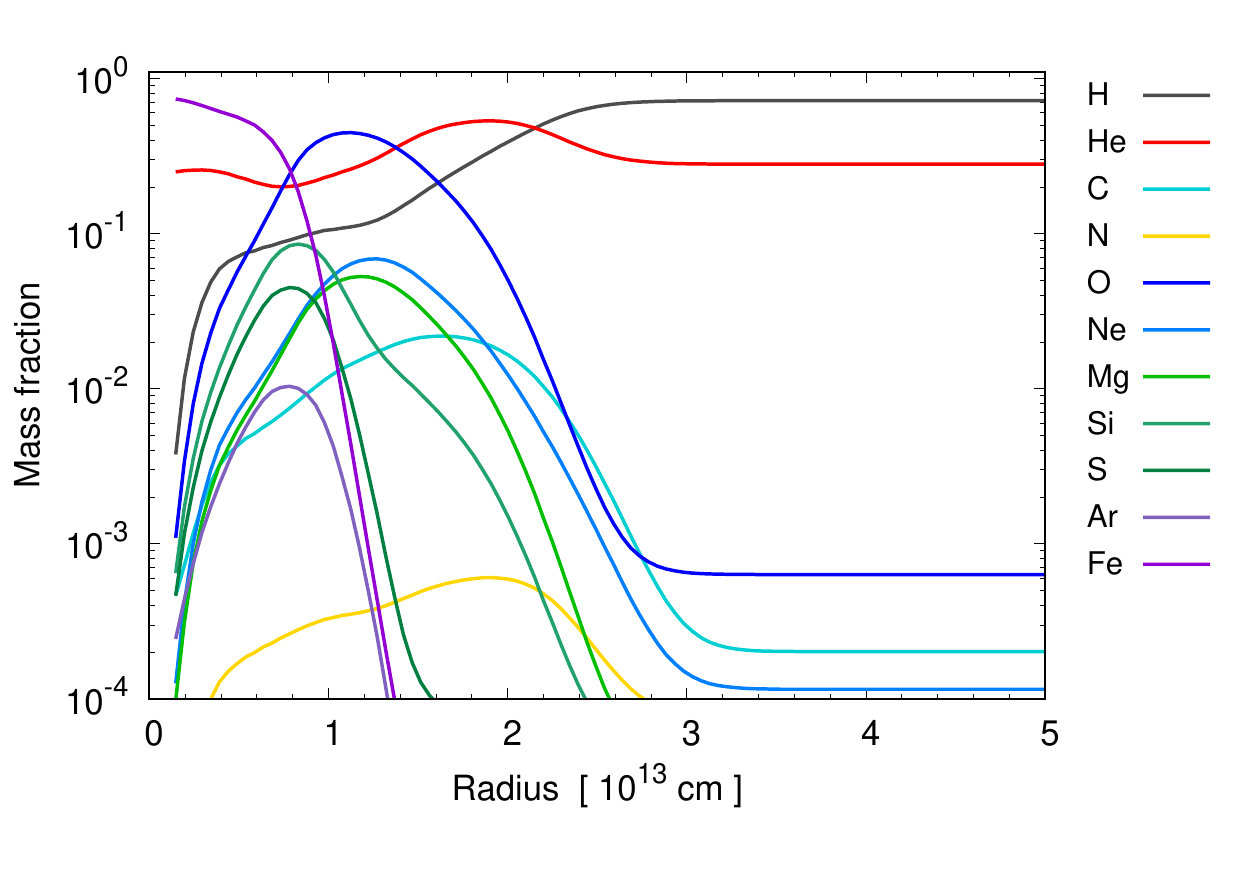}
\end{center}
\vs{-1.}
\end{minipage}
\begin{minipage}{0.5\hsize}
\vs{-1.}
\begin{center}
\includegraphics[width=8.5cm,keepaspectratio,clip]{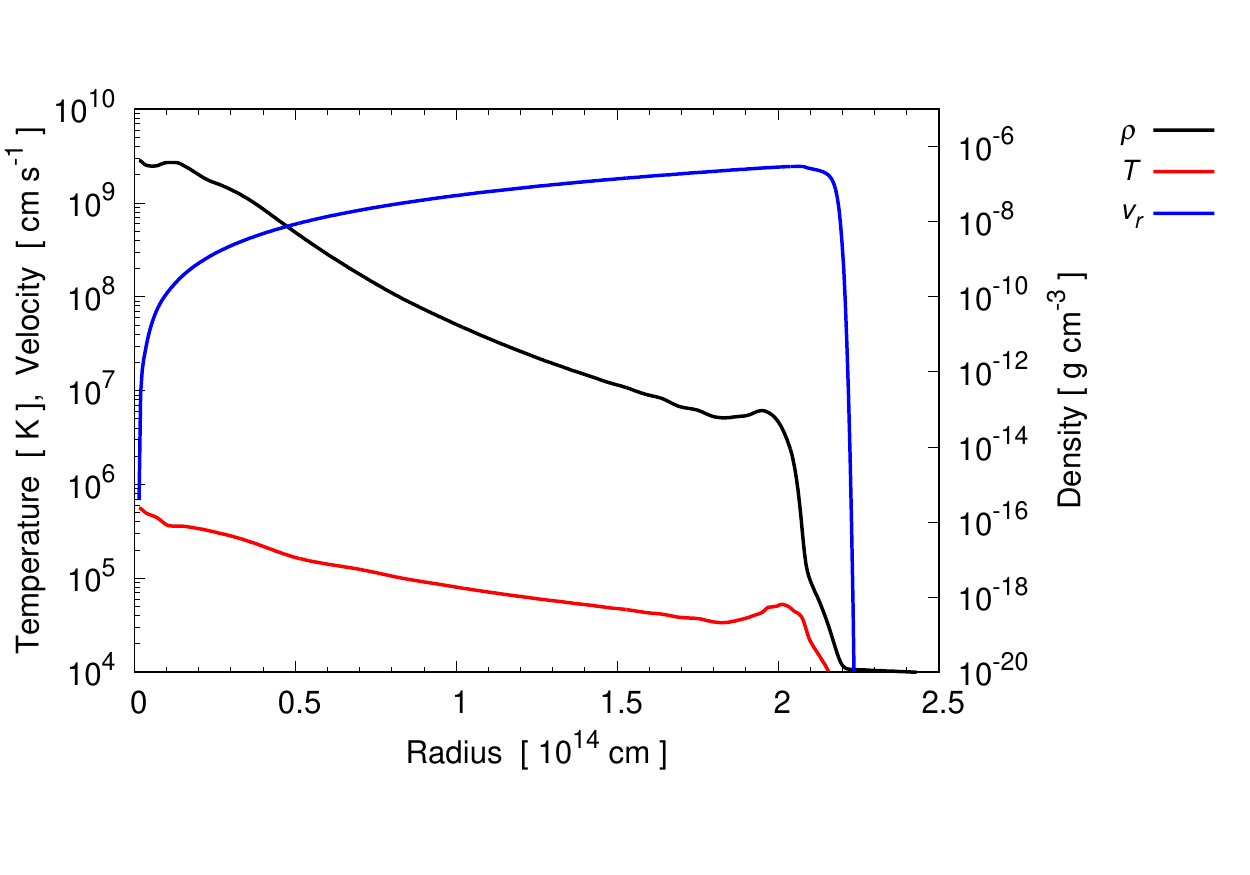}
\end{center}
\vs{-1.}
\end{minipage}
\caption{Same as Figure~\ref{fig:prof_mean} but for the spherical explosion case for the models b18.3-high (top) and 
n16.3-high (bottom) and 
the ages are 21.4 hours and 24.0 hours 
after the explosion for the former and latter models, respectively.}
\label{fig:prof_sphel}
\end{figure*}
Figure~\ref{fig:prof_sphel} shows the 1D profiles of the spherical explosion case. %
The profiles of the composition are apparently smoother (with lesser radial bumps/irregularities) than that of the previous case. %
However, the mass fractions of elements heavier than carbon other than iron are roughly peaked around the radius of $\sim$ 10$^{13}$ cm and the one of iron is concentrated in the innermost region for both models. %
Since iron (practically $^{56}$Ni) is limited to the radius of $\sim$ 10$^{13}$ cm, the destruction and ionization of molecules at outer regions due to the decay of $^{56}$Ni may be restricted compared with the previous case. %
It is noted that the amount of $^{56}$Ni obtained by the 3D hydrodynamical simulations in the spherical explosion case is higher than that of the original 3D models, i.e., b18.3-high and n16.3-high. The amount of $^{56}$Ni in the model b18.3-high (n16.3-high) is 8.64 $\times$ 10$^{-2}$ $M_{\odot}$ (9.67 $\times$ 10$^{-2}$ $M_{\odot}$). %
On the other hand, the amount of $^{56}$Ni of the spherical explosion case corresponding to the model b18.3-high (n16.3-high) is 1.92 $\times$ 10$^{-1}$ $M_{\odot}$ (1.89 $\times$ 10$^{-1}$ $M_{\odot}$). %
Therefore, the amounts of $^{56}$Ni in the spherical explosion case are higher roughly by a factor of two than those in the 3D models. The high amounts of $^{56}$Ni in the spherical case are attributed to the fact that in the spherical explosion models, larger regions are heated to enough high temperature \citep[$\sim$ 5 $\times$ 10$^9$ K; e.g.,][]{1990ApJ...349..222T} for the synthesis of $^{56}$Ni during the explosive nucleosynthesis. In the 3D (bipolar-like explosion) models, only limited regions (steradians) have enough high temperatures. On the other hand, in the spherical explosion models, all directions are heated evenly to have enough high temperatures. %
The differences in oxygen, hydrogen, and helium between the two models due to the different adopted progenitor models seen in the previous case, i.e., angle-averaged 1D profiles (Figure~\ref{fig:prof_mean}), are roughly kept in the spherical explosion case. %

%
\begin{figure*}
\begin{minipage}{0.5\hsize}
\begin{center}
\includegraphics[width=8cm,keepaspectratio,clip]{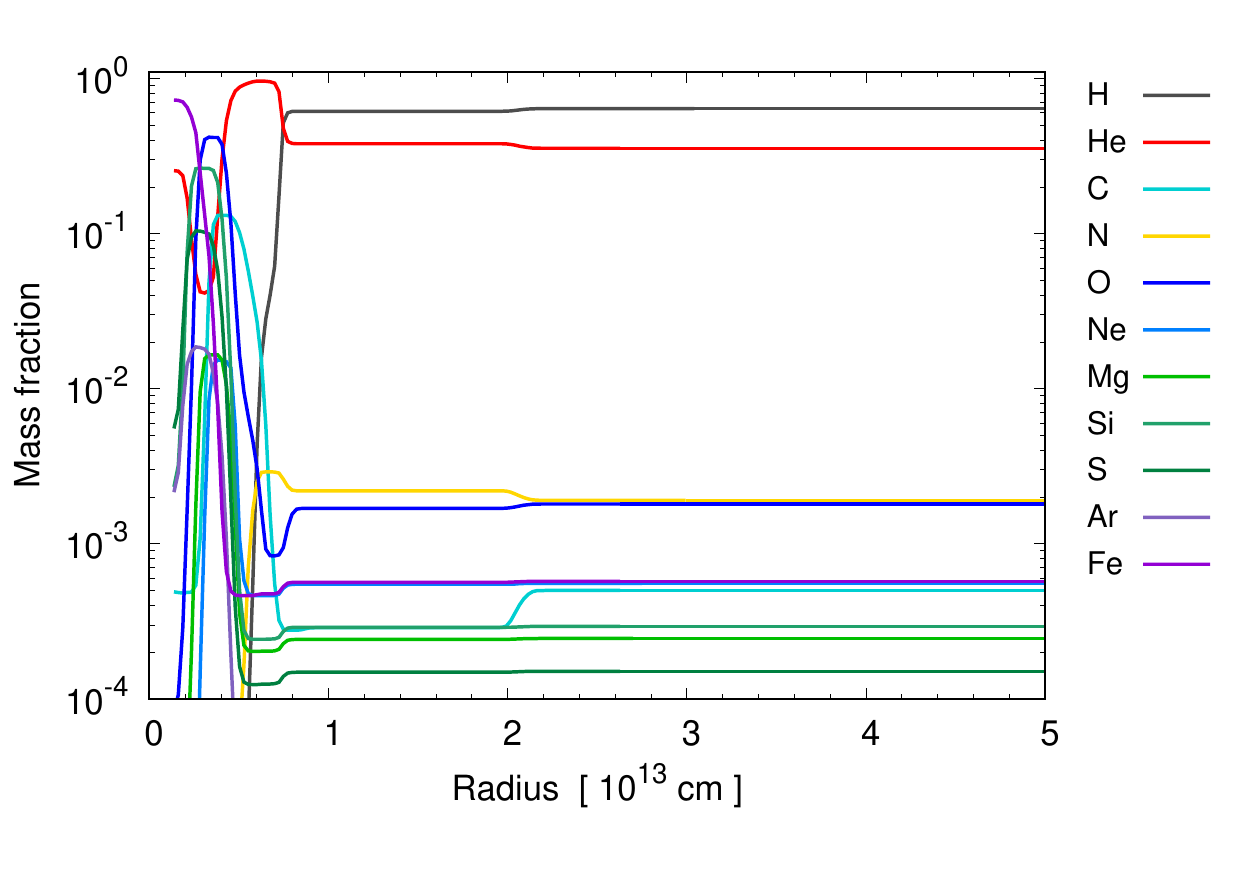}
\end{center}
\end{minipage}
\begin{minipage}{0.5\hsize}
\includegraphics[width=8.5cm,keepaspectratio,clip]{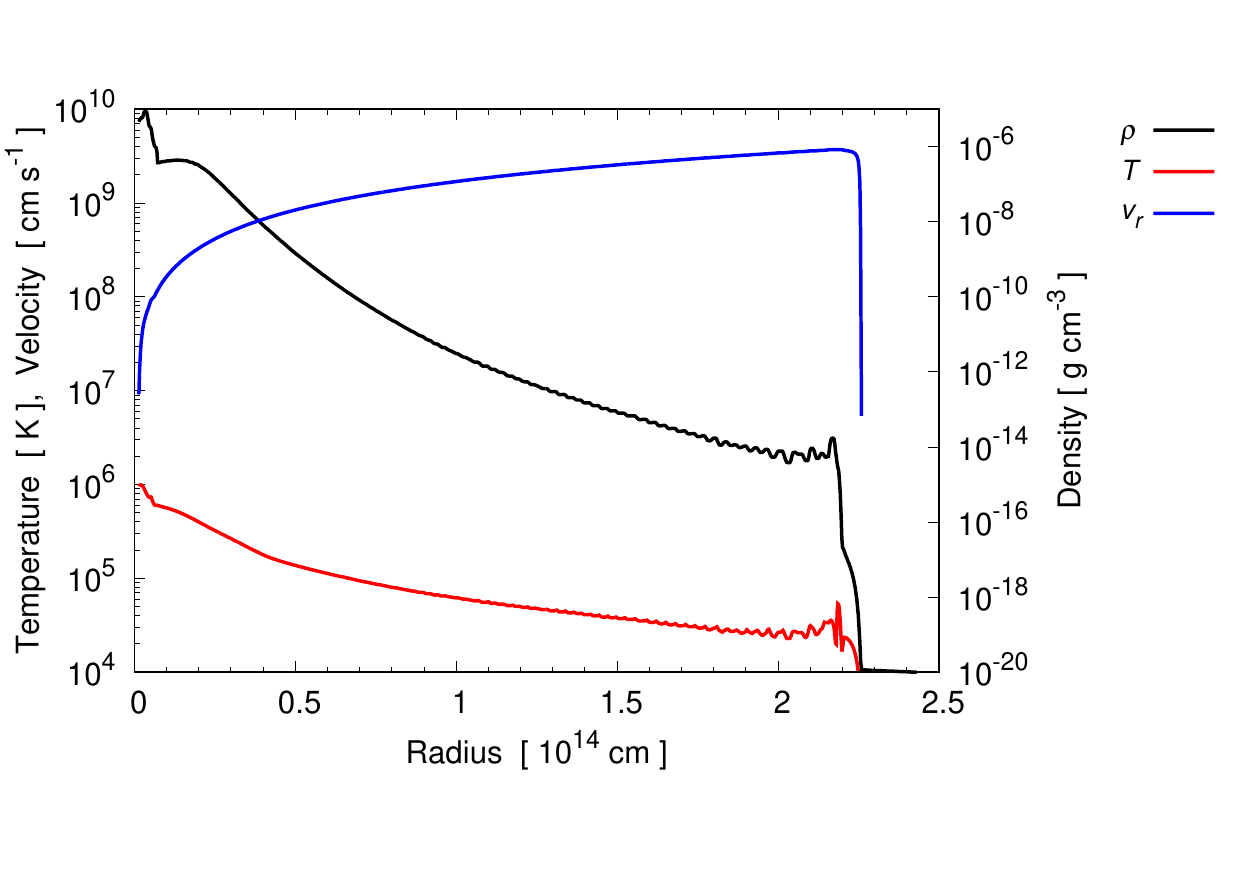}
\end{minipage}
\\
\begin{minipage}{0.5\hsize}
\vs{-1.}
\begin{center}
\includegraphics[width=8cm,keepaspectratio,clip]{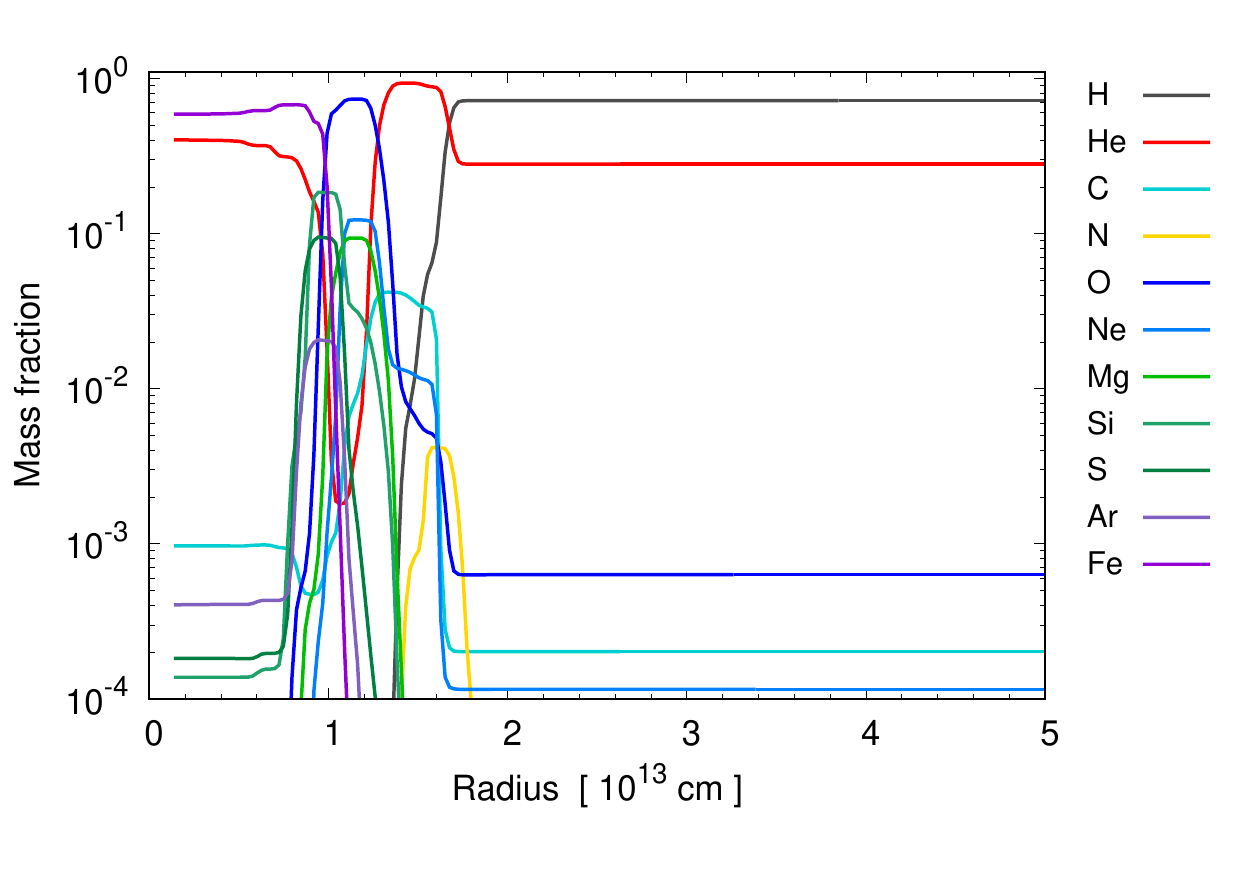}
\end{center}
\vs{-1.}
\end{minipage}
\begin{minipage}{0.5\hsize}
\vs{-1.}
\begin{center}
\includegraphics[width=8.5cm,keepaspectratio,clip]{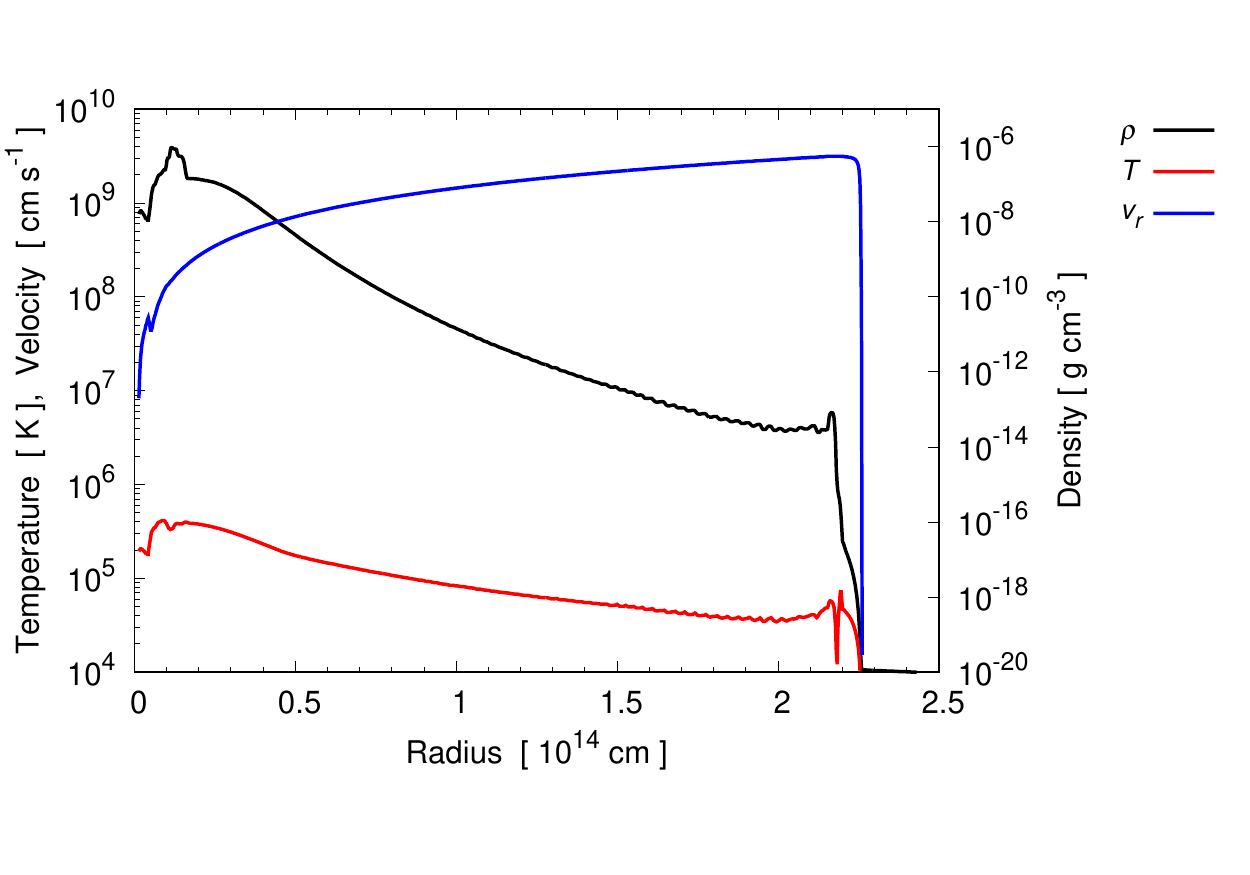}
\end{center}
\vs{-1.}
\end{minipage}
\caption{Same as Figure~\ref{fig:prof_mean} but for the purely spherical case for the models b18.3-high (top) 
and n16.3-high (bottom) and 
the ages are 17.0 hours and 20.0 hours 
after the explosion for the former and latter models, respectively.}
\label{fig:prof_sphel_pure}
\end{figure*}
In Figure~\ref{fig:prof_sphel_pure}, the 1D profiles of the purely spherical case are shown. %
The composition profiles are apparently more stratified compared with the previous two cases; %
from the outside, hydrogen, helium, carbon plus oxygen, and iron-rich layers are distinct. %
In particular, $^{56}$Ni is confined to the innermost regions. %
The positions of the distinct layers are different between the two models depending on the size of the cores of the progenitor star and its density structure, which affects the acceleration and deceleration of the supernova shock. %
Since the size and mass of the helium core of the progenitor model for the model n16.3-high are both larger than those for the model b18.3-high, the inner cores (layers) are pushed to the helium layers due to the stronger deceleration than that for the model b18.3-high. %
The peak of the density profile for the model n16.3-high (bottom right panel) is positioned outward compared with that for the model b18.3-high (top right panel). %
The amount of $^{56}$Ni of the purely spherical case corresponding to the model b18.3-high (n16.3-high) is 1.82 $\times$ 10$^{-1}$ $M_{\odot}$ (1.95 $\times$ 10$^{-1}$ $M_{\odot}$). %
The amounts of $^{56}$Ni in the purely spherical case are also higher than the 3D models because of the same reason mentioned above. %
The purely spherical case is useful to see the impact of matter mixing as the representative case of no mixing as mentioned. %

It is noted that the amounts of $^{56}$Ni presented above may have some uncertainties. %
The explosions in the models in this study, i.e., the 3D models \citep{2020ApJ...888..111O} and the spherical cases, are initiated by a kind of thermal bomb (thermal and kinetic energies are artificially injected around the surface of the original iron core). %
Conventionally, in the spherical CCSN explosion models, the so-called ``mass cut" \citep[e.g.,][]{1995ApJS..101..181W}, i.e., the mass coordinate from where the matter inside should fall back to the compact object, has been introduced to compensate for the limitations in the theories and to obtain an appropriate $^{56}$Ni mass consistent with the light curve observations \citep[$\sim$ 0.07 $M_{\odot}$: e.g.,][]{1988A&A...196..141S,1988ApJ...330..218W}. %
In the 3D explosion models \citep{2020ApJ...888..111O}, mass cut is not introduced to adjust the $^{56}$Ni mass since the obtained $^{56}$Ni mass is not far from the value obtained from the observations, and how to determine the boundary between the ejecta and compact object is not simple for the case of globally asymmetric explosions \citep[see,][]{2000ApJS..127..141N}. %
In the spherical (spherical explosion and purely spherical) models in this paper, we also do not consider mass cut to have consistent hydrodynamical conditions, e.g., the explosion energy, other than the explosion asymmetry for simple comparison. %
Considering the adopted method in the explosion models in this study, more realistic explosion models may result in different $^{56}$Ni masses. %
Additionally, the nucleosynthesis calculations may also introduce uncertainties in the amounts of $^{56}$Ni as shown in our previous study \citep{2015ApJ...808..164M}.\footnote{By comparing the results of a spherical explosion model with a small nuclear reaction network (19 nuclei; the same as one in this study) to that with a larger nuclear reaction network (464 nuclei), it was found that the small nuclear reaction network could overestimate the mass of $^{56}$Ni approximately 50\% larger than that with the larger nuclear reaction network.} %
We expect that the uncertainties in the amounts of $^{56}$Ni due to the points above may be at most a factor of two as seen in the results in this paper unless the explosion energy and the progenitor model are significantly different from those in this study. %
As shown in the later sections, the distribution of $^{56}$Ni (how $^{56}$Ni is extensively distributed in outer layers) is more important for the formation of molecules. %
Uncertainty within a factor of two in the amount of $^{56}$Ni is acceptable for the purpose of this study (to see the qualitative impact of the matter mixing on the formation of molecules).

In order to obtain the hydrodynamical evolution from $\sim$ 1 day to $\sim$ 10$^4$ days, 1D hydrodynamical simulations are further performed with the angle-averaged 1D profiles as the initial conditions. %
For the calculations, the 1D version of the hydrodynamical code FLASH \citep{2000ApJS..131..273F} mentioned above is used again. %
In the code, the directionally-split Eulerian hydrodynamical unit with the piecewise parabolic method \citep{1984JCoPh..54..174C} is utilized. %
As for the Riemann solver, a hybrid solver with an HLLE solver inside shocks is adopted. %
The coordinate system is the spherical coordinates. In the code, the so-called adaptive mesh refinement (AMR) is implemented but a uniform grid with 1024 meshes is used in the calculations. %
As in our previous studies on the matter mixing \citep{2013ApJ...773..161O,2015ApJ...808..164M,2020ApJ...888..111O}, as the SN shock approaches the outer computational boundary, the computational domains are expanded with a scaling factor of 1.2 to resolve the structure inside the SN shock and follow the long-term evolution. %
In the pre-shocked regions outside the progenitor star, a spherically symmetric blue supergiant wind with a constant mass-loss rate of $\dot{M_{\rm w}} = 10^{-7} M_{\odot}$ yr$^{-1}$ and wind velocity of $v_{\rm w} = 500$ km s$^{-1}$ \citep{2007Sci...315.1103M} is assumed as the initial condition as in \cite{2020ApJ...888..111O}, i.e., the density profile follows the power-law of $\rho (r) \propto r^{-2}$, where $r$ is the radius from the center of the explosion. %
The following points are different from the 1D hydrodynamical simulation mentioned in the previous paragraph. %
During the simulations, the unit for the nucleosynthesis is turned off, since at epochs $\sim$ 1 day after the explosion, explosive nucleosynthesis no longer occurs. %
In the equation of state, the contribution from the radiation is omitted, i.e., the EoS is the ideal gas EoS with the adiabatic index of $\gamma_{\rm ad} = 5/3$. %
The energy deposition from the decay of $^{56}$Ni is also turned off. %
Neglecting the energy deposition from the decay does not affect the overall dynamics but the effect of the decay of $^{56}$Ni on the thermal evolution of the ejecta and the chemistry is taken into account similar to the one-zone calculations as mentioned later. %

The chemical evolution in the ejecta is calculated in post-processing. %
For the sake of this, first, based on the 1D hydrodynamical simulation results, the time evolution of tracer particles (imaginary Lagrange particles) is obtained with the procedures described below. %
In total, one hundred tracer particles are radially distributed in the initial computational domain of the hydrodynamical simulation to cover the inner metal-rich ejecta which includes seed atoms. %
Initially for each particle, the mass and composition are assigned depending on its initial position (the values are obtained with the same interpolation method as described later in this paragraph); %
the mass is assigned to be a constant value so that the particle covers the mass corresponding to a ``shell" in which the particle initially is. %
As for the composition, the initial one is respected, i.e., without a change via chemical reactions, each particle's composition is assumed to be constant. %
Then, the tracer particles at $t = t^n$, here integer $n$ denotes the time step, are moved along the velocity field obtained from the hydrodynamical simulation to estimate the positions at $t^{n+1} = t^n + {\it \Delta t}$. %
To obtain the time evolution accurately, as a time integration method, the so-called predictor-corrector method is used, i.e., generally, the position of a particle at $t=t^{n+1}$, $\bm{x}^{n+1}$, is obtained with the velocity field $\bm{v} (\bm{x})$ as follows. %
\begin{equation}
\begin{aligned}
&\bm{x}' = \bm{x}^n + \bm{v}^n (\bm{x}^n) \,{\it \Delta} t, \\
&\bm{x}^{n+1} = \bm{x}^n + \frac{\bm{v}^{n+1} (\bm{x}') + \bm{v}^n (\bm{x}^n)}{2} \,{\it \Delta} t,
\end{aligned}
\end{equation}
where $\bm{x}'$ is the temporal value (predictor) to obtain the correct value $\bm{x}^{n+1}$ (corrector). Shortly, we plan to apply our method to 3D Eulerian hydrodynamical simulation results. %
In such a 3D case, practically it is difficult to save the data for all the time snapshots. %
Therefore, we check among several time integration methods, e.g., the 4-th order Runge-Kutta method and an implicit midpoint method, which 
method best reproduces the full-time step results with limited time slices (outputs every ten or hundred time steps). %
Then, we conclude that the predictor-corrector method is the most appropriate for this purpose. %
Even though there is no such a time snapshot problem in the calculations limited in 1D in this paper, the predictor-corrector method is adopted. %
The physical quantities, e.g., the density, temperature, and velocity, for each particle are derived by time and spacial interpolations with the data for nearby time slices and grid points, respectively. %
A monotonic cubic interpolation method \citep{1990A&A...239..443S} is utilized for the interpolations of physical quantities. %

The thermal and chemical evolutions are calculated for each tracer particle. %
The basic methodology is the same as the case of the one-zone approximation as described in Section~\ref{subsec:one_zone_calc}. %
From here, the points different from the one-zone case are mainly delineated. %
Although the time evolution of gas temperatures (internal energies) can be obtained from the interpolations with the 1D hydrodynamical simulation results, the thermal evolution of each particle is obtained by the basic power-laws plus contributions from the additional heating and/or cooling as in the one-zone case. %
This is because in the 1D hydrodynamical simulations, the contribution from the radiation (in the ideal EoS, the constant adiabatic index of $\gamma_{\rm ad} = 5/3$ is used), the possible energy depositions via the decay of $^{56}$Ni, and the unknown CO line cooling are neglected. %
As for the time evolution of the density and radial velocity, values obtained by the interpolations, i.e., the hydrodynamical simulation results, are respected except for the effect described in Equation~(\ref{eq:thermal_i}) in the density. %
In the case of the 1D calculations, $\rho_{\rm ad}$ in Equation~(\ref{eq:thermal_i}) is replaced with the density obtained by the interpolation. %
As in the one-zone case, it is assumed that the gas temperature (specific internal energy) has a break at some time point when the gas becomes optically thin. %
At such point the adiabatic index $\gamma_{\rm ad}$ is manually changed. %
However, the transition point is differently determined from the one-zone case as described below. %
To determine the transition point, as a reference, the optical depth to Thomson scattering, $\tau_{\rm Th}$, is derived for each particle as below. %
\begin{equation}
\tau_{\rm Th} (r) = \int_r^{\infty} \sigma_{\rm Th} \,n_{\rm e} \,dr,
\end{equation}
where $\sigma_{\rm Th}$ is the Thomson scattering cross section. %
Here, the number density of electrons $n_{\rm e}$ reflects the gas density and the ionization and recombination through the chemical reaction calculation. %
In deriving $n_{\rm e}$, the gas density obtained from Equation~(\ref{eq:thermal_i}) is used. %
As described in Section~\ref{subsubsec:init_chemi}, before starting the chemical reaction calculation, the degree of the ionization is controlled by the ionization fraction $X_{\rm e}$. %
After starting the chemical reaction calculation, $n_{\rm e}$, is derived from the resultant abundance of (thermal) electrons, although most of the particles may have a break before the chemical reaction calculation starts. %
For each particle, once the $\tau_{\rm Th}$ becomes less than 2/3, regarding that the gas becomes optically thin, the adiabatic index $\gamma_{\rm ad}$ is automatically switched from 4/3 to 5/3 (before the start of the molecule formation calculation) or the effective adiabatic index derived from the averaging of ones for the diatomic and monoatomic species (after that). %
Other parts of the calculations are the same as the one-zone case; %
the chemical reaction calculations, the calculations of CO (ro)-vibrational levels for CO line cooling, and the updating of the internal energies and gas temperatures are recurrently carried out for each tracer particle. %
The model parameters related to the calculations with the 1D profiles are listed in Table~\ref{table:param}. %

\section{Strategy and models} \label{sec:strategy} 

\begin{figure*}
\begin{minipage}{0.5\hsize}
\begin{center}
\includegraphics[width=8cm,keepaspectratio,clip]{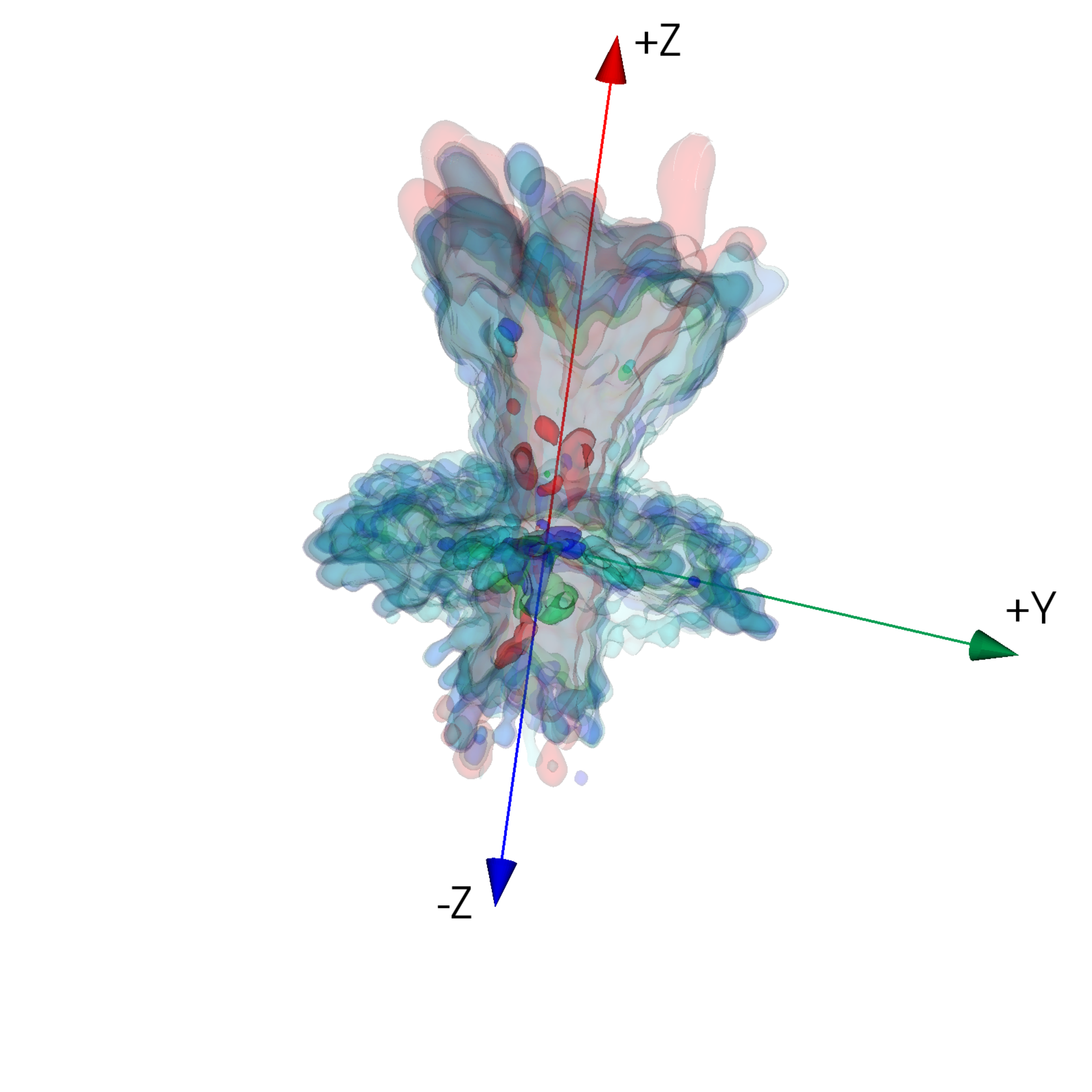}
\end{center}
\vs{-1.5}
\end{minipage}
\begin{minipage}{0.5\hsize}
\begin{center}
\includegraphics[width=8cm,keepaspectratio,clip]{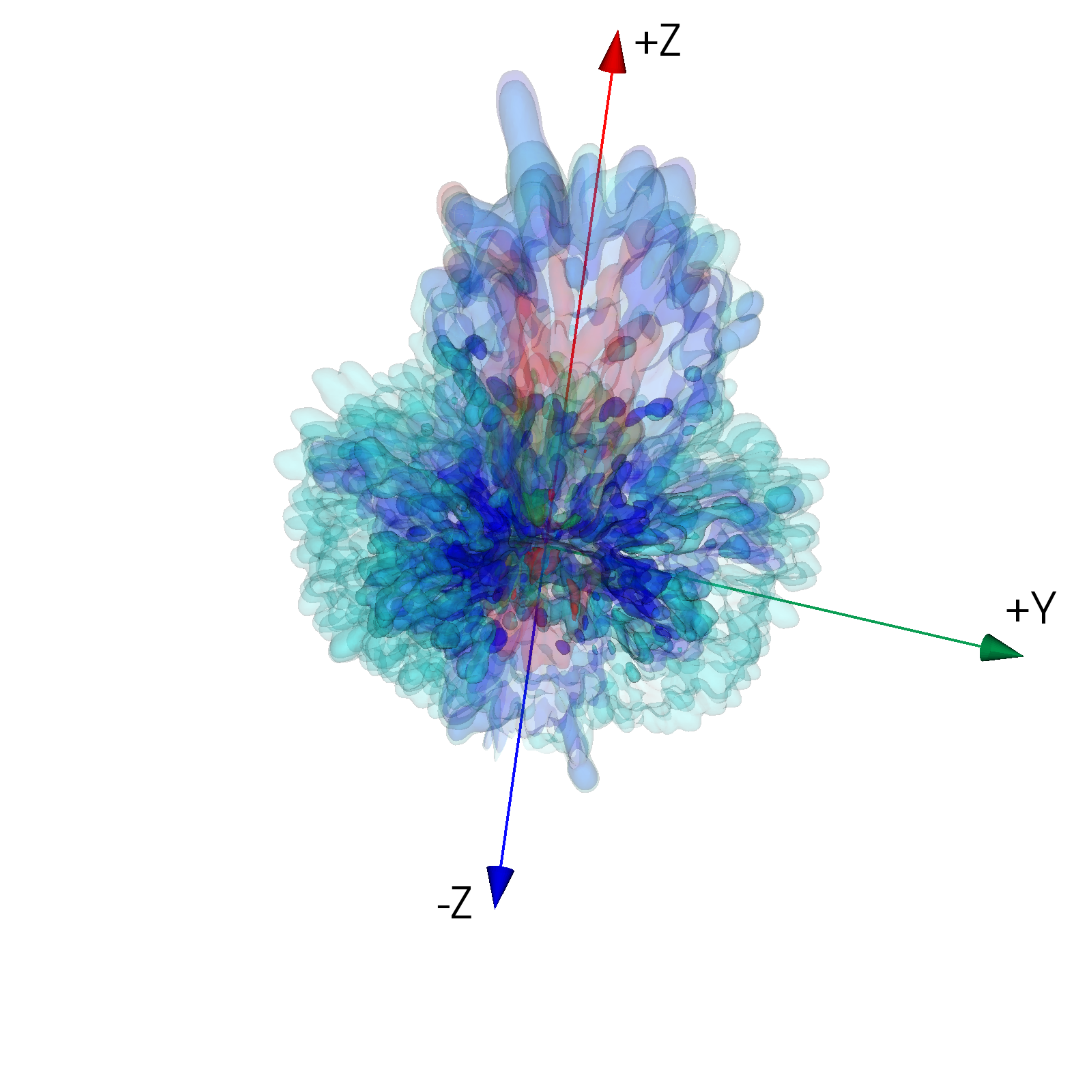}
\end{center}
\vs{-1.5}
\end{minipage}
\caption{Isosurfaces of the mass fractions of elements, $^{56}$Ni (red), $^{28}$Si (green), $^{16}$O (blue), and $^{12}$C (cyan), in the 3D hydrodynamical models \citep{2020ApJ...888..111O} b18.3-high (left) and n16.3-high (right) with arrows denoting the $+Z$, $-Z$, and $+Y$ directions. %
The snapshots for the models b18.3-high and n16.3-high are at 19.0 hours and 22.3 hours after the explosion, respectively. %
Isosurfaces corresponding to 10\% (lighter color) and 70\% (darker color) of the maximum value for each element are shown.\label{fig:3d_direc}} %
\end{figure*}
In this section, the strategy to investigate the impact of the matter mixing on molecule formation in the ejecta with the method delineated 
in the previous section and the related models are described. %

First, with one-zone calculations, the impacts of the parameters and overall trends are briefly explained. By comparing the results of the time evolutions of the masses of CO and SiO estimated in previous studies \citep[e.g.,][]{1995ApJ...454..472L} and observed light curves of CO vibrational bands in SN~1987A \citep{1989MNRAS.238..193M,1993MNRAS.261..535M,1993A&A...273..451B,1993ApJS...88..477W}, better parameter values for the one-zone approximation are derived. %
Additionally, important chemical reactions in particular for CO and SiO molecules are extracted according to the one-zone calculation results for the parameter values derived above. %
The results are presented in Section~\ref{subsec:one_zone_results}. %

Second, with the angle-averaged 1D profiles for the model b18.3-high \citep{2020ApJ...888..111O}, similarly to the one-zone calculations, an acceptable set (acceptable sets) of parameter values for 1D calculations are derived by comparing the results with the previous studies and observations mentioned above. %
Then, the parameter values are fixed to be the acceptable values to investigate the impact of the matter mixing on the molecule formation for the latter discussion. %
The impact of the matter mixing in the ejecta is effectively discussed by comparing the model results for angle-averaged 1D profiles (full mixing) and profiles corresponding to the spherical explosion (but the further evolution is in 3D) case (partial mixing), and purely spherical case (no mixing). %
The results are shown in Section~\ref{subsec:1d_results}. %

A caveat on the discussion with the initial profiles above is that the angle-averaged profiles may overestimate the mixing and the molecule 
formation results could be different between the molecule formation calculation with a single angle-averaged profile and ones for all radial 
profiles in 3D models. %
Therefore, similar molecule formation calculations for a few radial profiles along several specified directions (angle-specified profiles) are performed to obtain an insight into the impact of more realistic matter mixing based on full 3D hydrodynamical models. %
To do this, a 1D hydrodynamical simulation is performed with each angle-specified profile as the initial conditions to obtain the evolution from $\sim$ 1 day to $\sim$ 10$^4$ days after the explosion with the same method described in Section~\ref{subsec:1d_calc}. %
Therefore, implicitly, it is assumed that the profile is spherically distributed. %
Here, selected representative directions are described. As already mentioned in Section~\ref{subsec:one_zone_calc}, among the models investigated in \cite{2020ApJ...888..111O}, the asymmetric bipolar-like explosion model, i.e. the model b18.3-high, with the BSG progenitor model through a binary merger (b18.3) \citep{2018MNRAS.473L.101U}, best reproduces the observed [Fe II] line profiles \citep{1990ApJ...360..257H} among the models. %
Additionally, with the same explosion energy and the explosion asymmetry, a different matter mixing was demonstrated based on the explosion model (n16.3-high) with another BSG progenitor model of a single star evolution (n16.3) \citep{1988PhR...163...13N,1990ApJ...360..242S}. %
Then, radial profiles are extracted from the two explosion models, b18.3-high and n16.3-high. %
As for the directions specified, two directions along the bipolar-like explosion axis, i.e., $+Z$ and $-Z$ directions, are selected, where the strongest explosion is directed to the $+Z$ direction and the explosion is weaker in the $-Z$ direction \citep[see, Figure~23 in][]{2020ApJ...888..111O}. %
As another representative direction, the $+Y$ direction on the equatorial plane, against which the bipolar explosion is asymmetric, is selected. %

%
\begin{figure*}
\begin{minipage}{0.5\hsize}
\begin{center}
\includegraphics[width=8cm,keepaspectratio,clip]{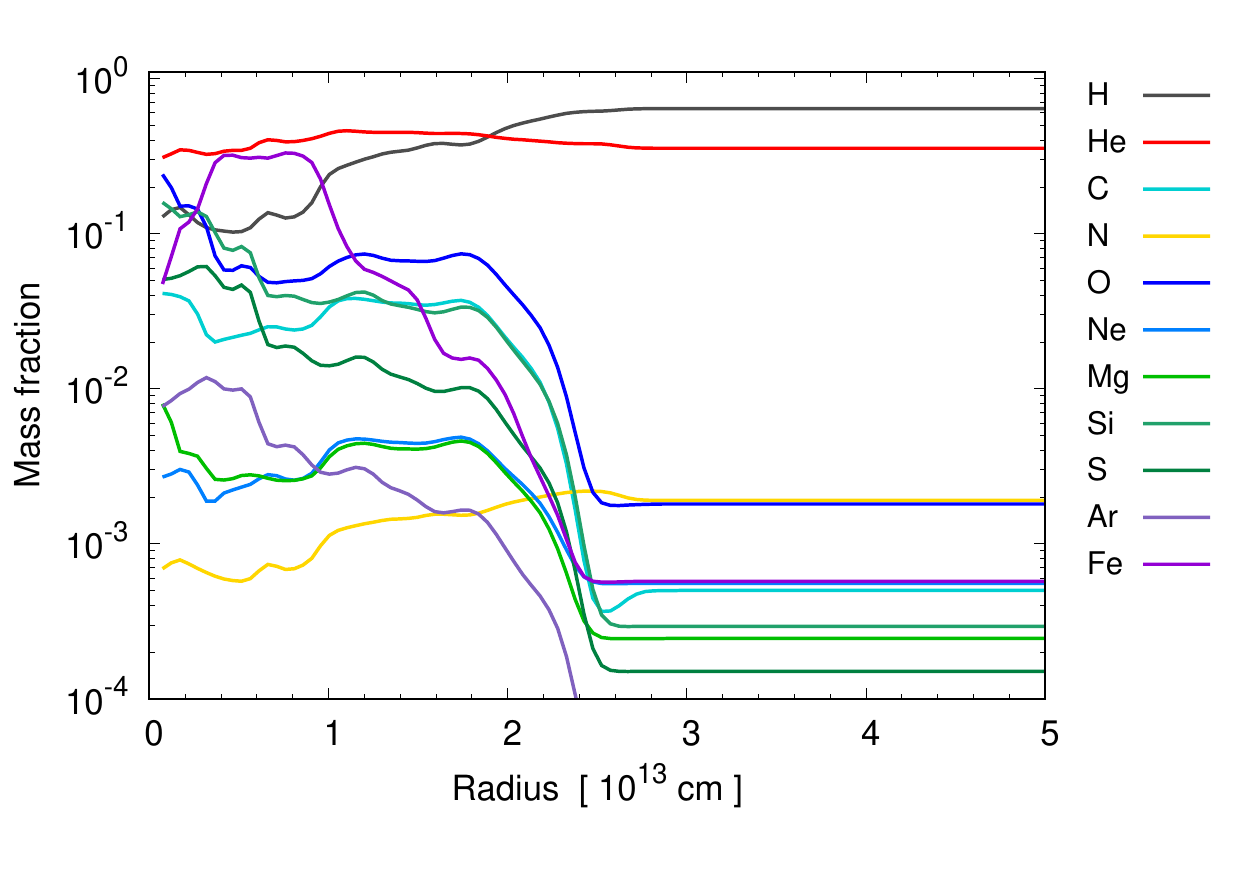}
\end{center}
\end{minipage}
\begin{minipage}{0.5\hsize}
\begin{center}
\includegraphics[width=8.5cm,keepaspectratio,clip]{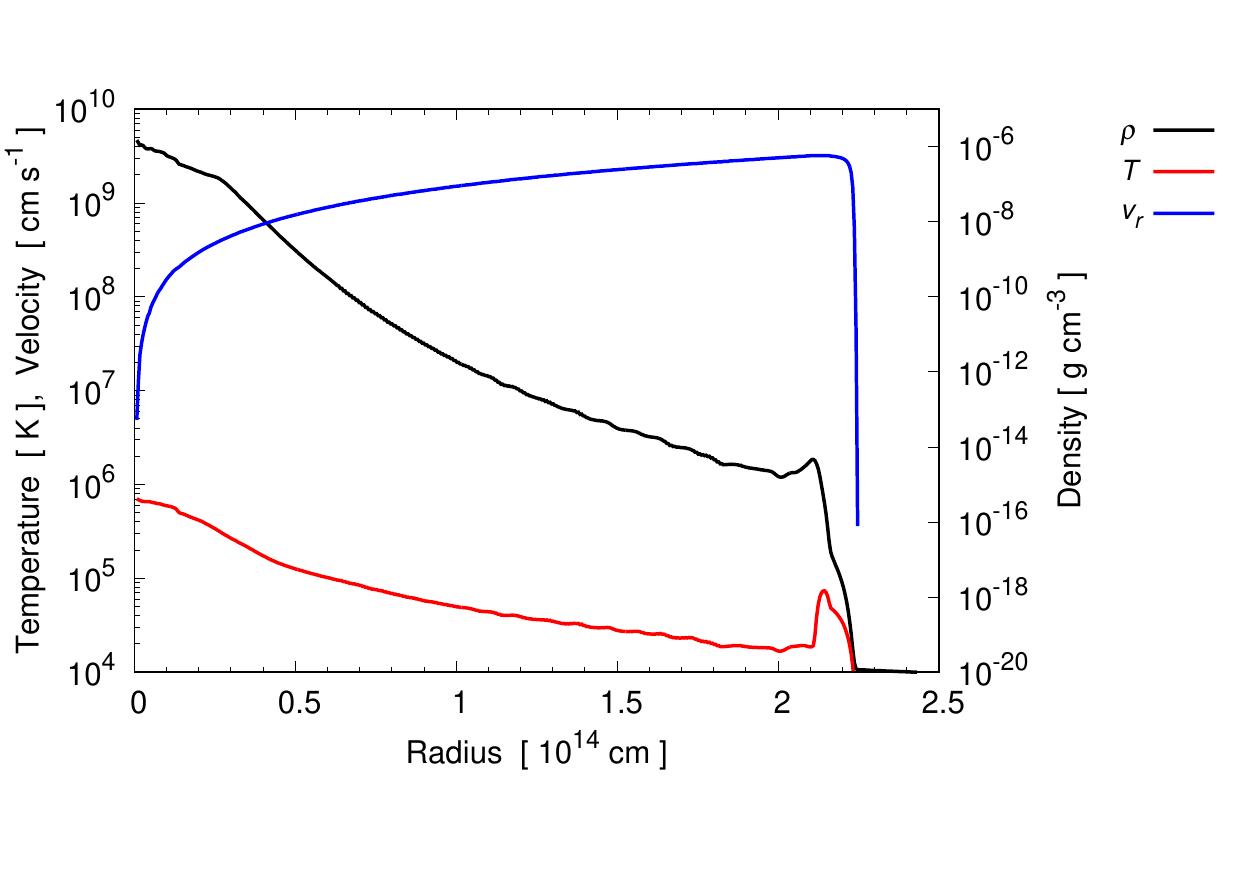}
\end{center}
\end{minipage}
\\
\begin{minipage}{0.5\hsize}
\vs{-1}
\begin{center}
\includegraphics[width=8cm,keepaspectratio,clip]{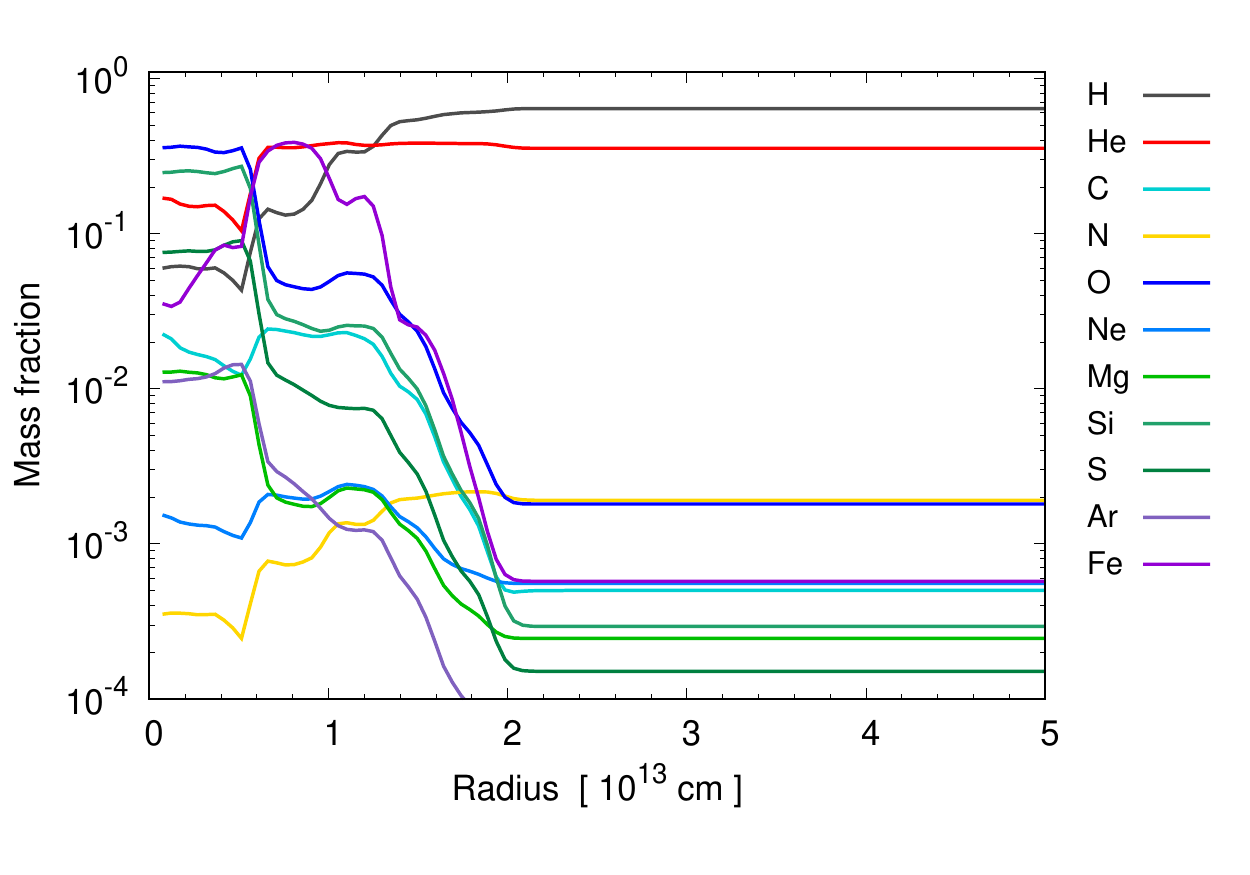}
\end{center}
\end{minipage}
\begin{minipage}{0.5\hsize}
\vs{-1}
\begin{center}
\includegraphics[width=8.5cm,keepaspectratio,clip]{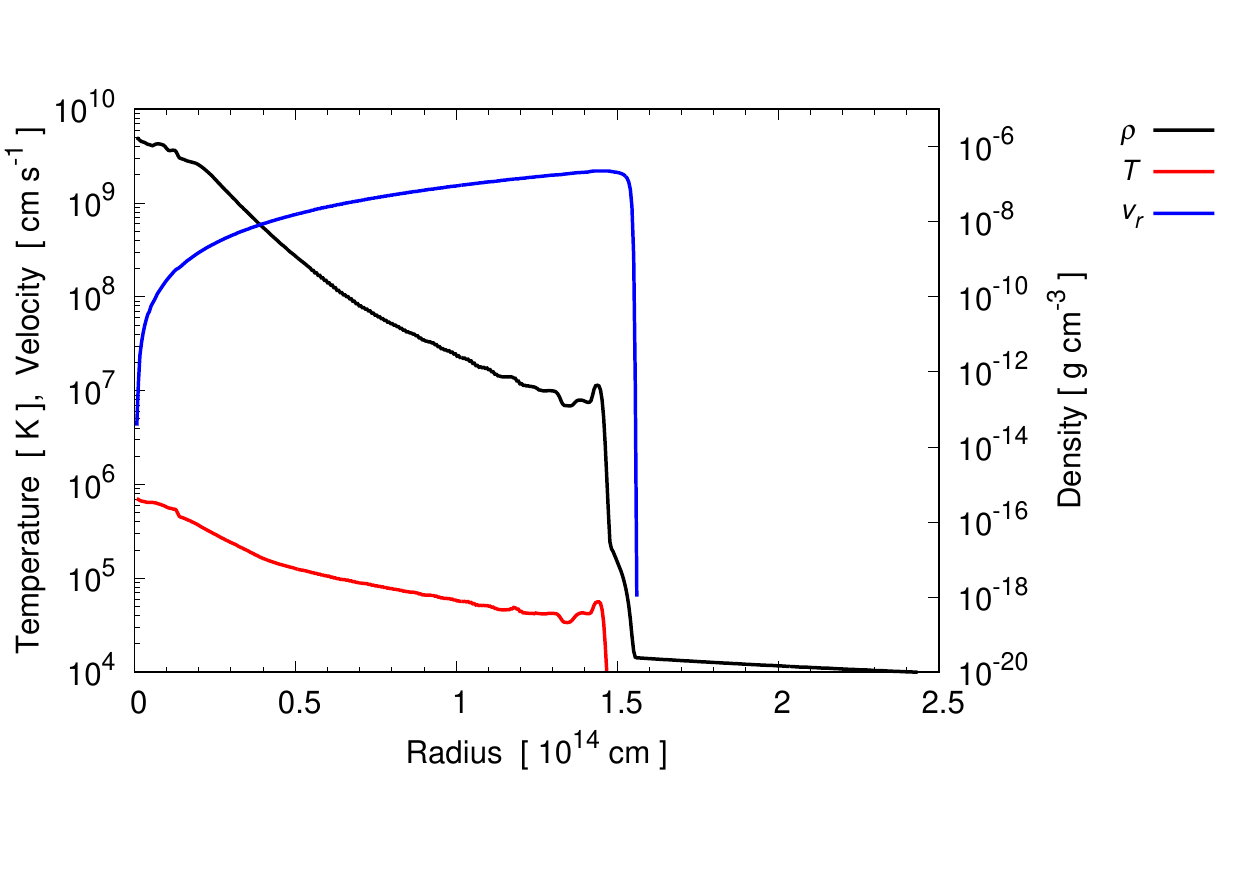}
\end{center}
\end{minipage}
\\
\begin{minipage}{0.5\hsize}
\vs{-1}
\begin{center}
\includegraphics[width=8cm,keepaspectratio,clip]{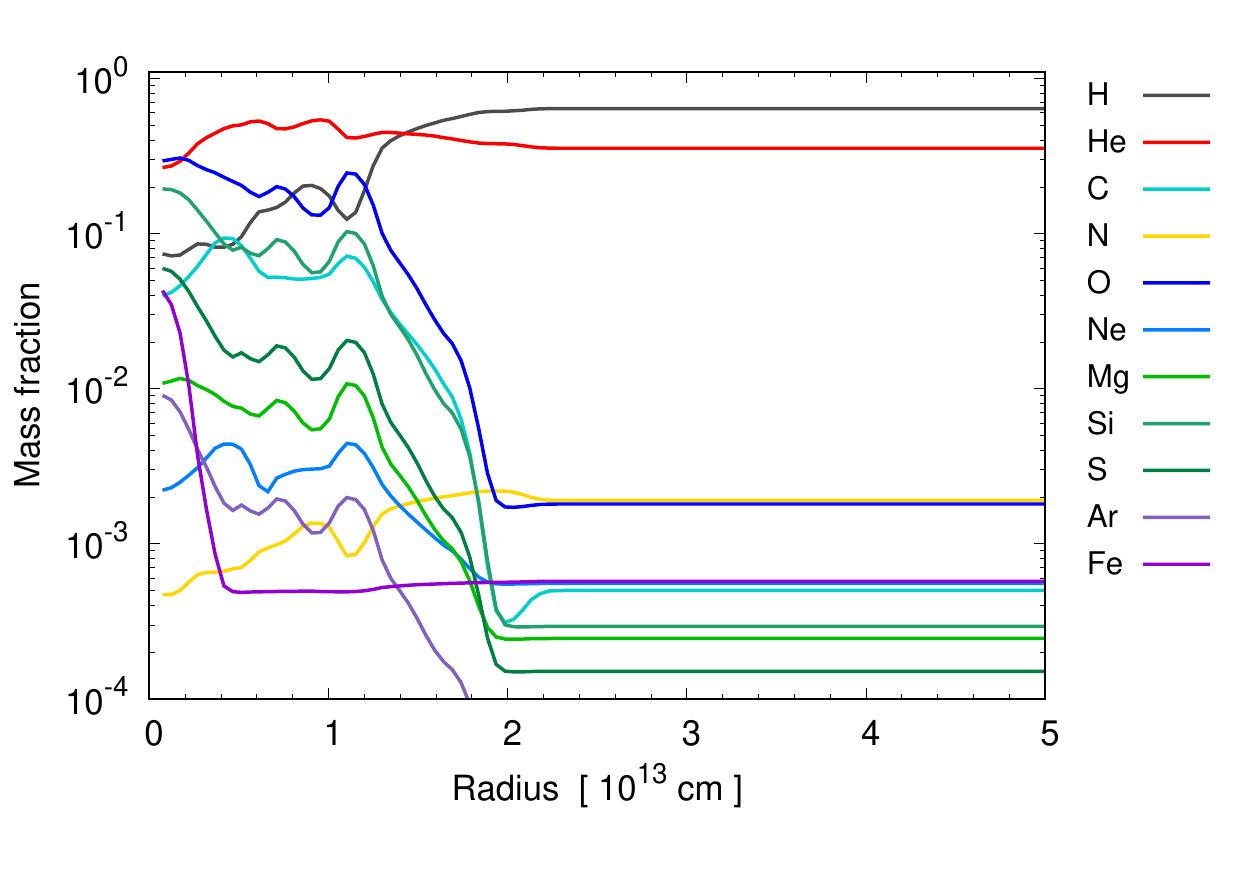}
\end{center}
\vs{-1}
\end{minipage}
\begin{minipage}{0.5\hsize}
\vs{-1}
\begin{center}
\includegraphics[width=8.5cm,keepaspectratio,clip]{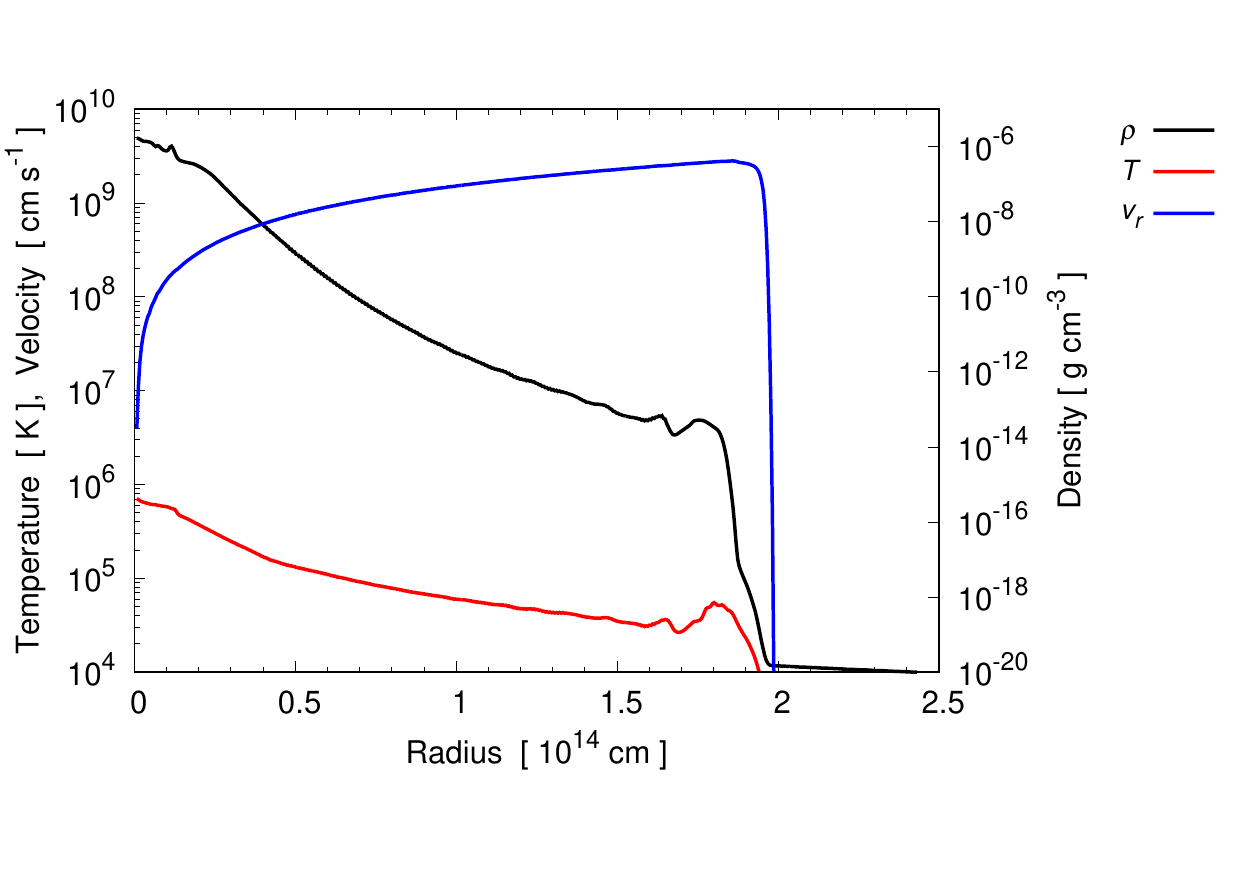}
\end{center}
\vs{-1}
\end{minipage}
\caption{Initial radial profiles along the specific directions, $+Z$ (Top), $-Z$ (Middle), and $+Y$ (Bottom), for the model b18.3-high at 
the age of 19.0 hours after the explosion. 
Mass fractions of species as a function of the radius (Left) and the density, temperature, and radial velocity as a function of radius (Right).\label{fig:b18.3_angle}} %
\end{figure*}

Figure~\ref{fig:3d_direc}, for a reference, shows the distributions of representative elements, $^{56}$Ni, $^{28}$Si, $^{16}$O, and $^{12}$C, at the end of the simulation in the 3D models b18.3-high and n16.3-high \citep{2020ApJ...888..111O} are shown with the arrows denoting the $+Z$, $-Z$, and $+Y$ directions. %
The distributions reflect the asymmetric bipolar-like explosions and those are different between the two models with the different progenitor models. %
It is noted that compared with the distributions just before the shock breakout \citep[see, Figure 23 in][]{2020ApJ...888..111O}, smaller-scale structures are lost after the shock breakout. %
This is because the forward shock (FS) is accelerated due to the steep pressure gradient across the stellar surface after the shock breakout, and the inner ejecta becomes left far behind the FS. %
Then, to keep the resolution inside the FS with the limited number of meshes, the resolution of the inner ejecta relatively becomes low after the shock breakout. %
Even though the smaller-scale structures are lost, the asymmetry caused by the bipolar-like explosions and the overall distributions of the seed atoms remain; such a numerical artifact should not change the main findings presented in later sections. 

%
\begin{figure*}
\begin{minipage}{0.5\hsize}
\begin{center}
\includegraphics[width=8cm,keepaspectratio,clip]{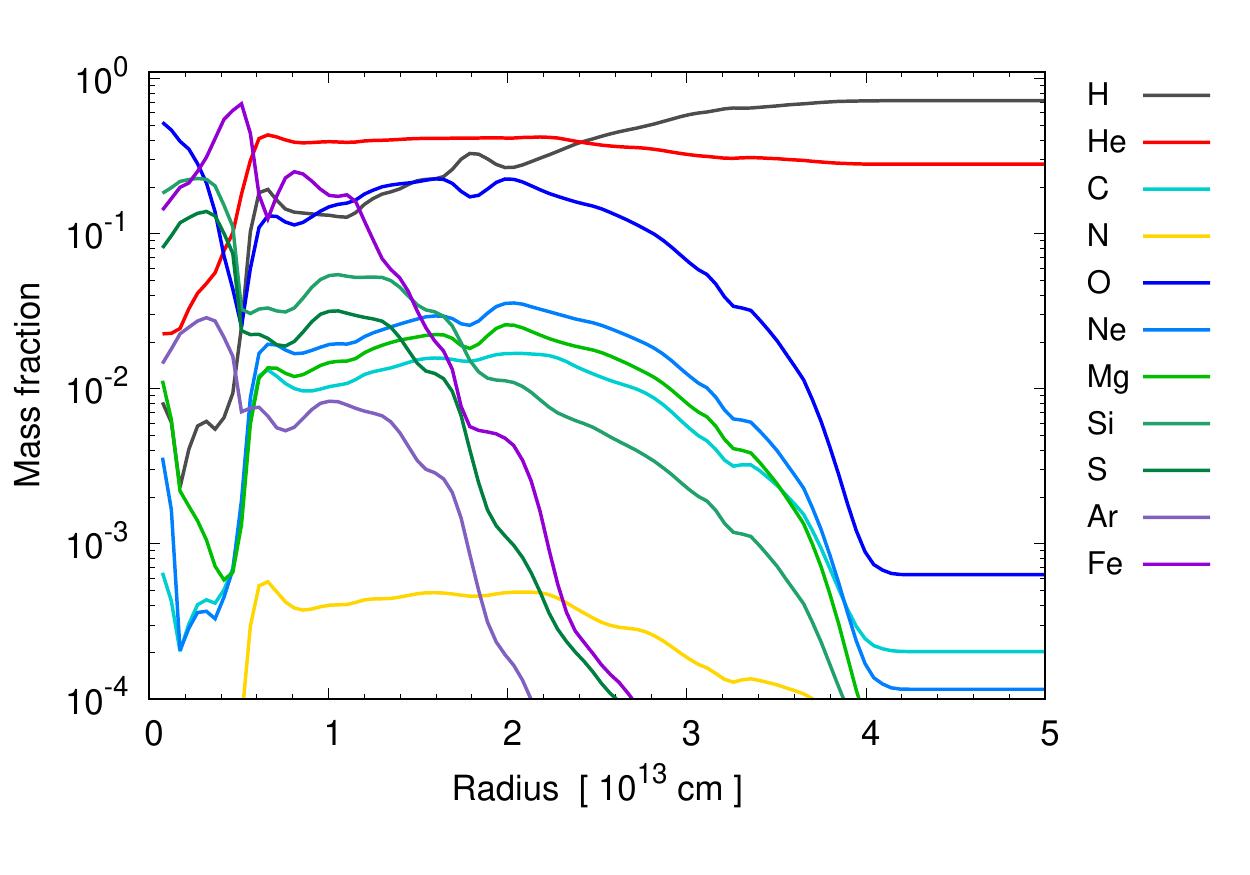}
\end{center}
\end{minipage}
\begin{minipage}{0.5\hsize}
\begin{center}
\includegraphics[width=8.5cm,keepaspectratio,clip]{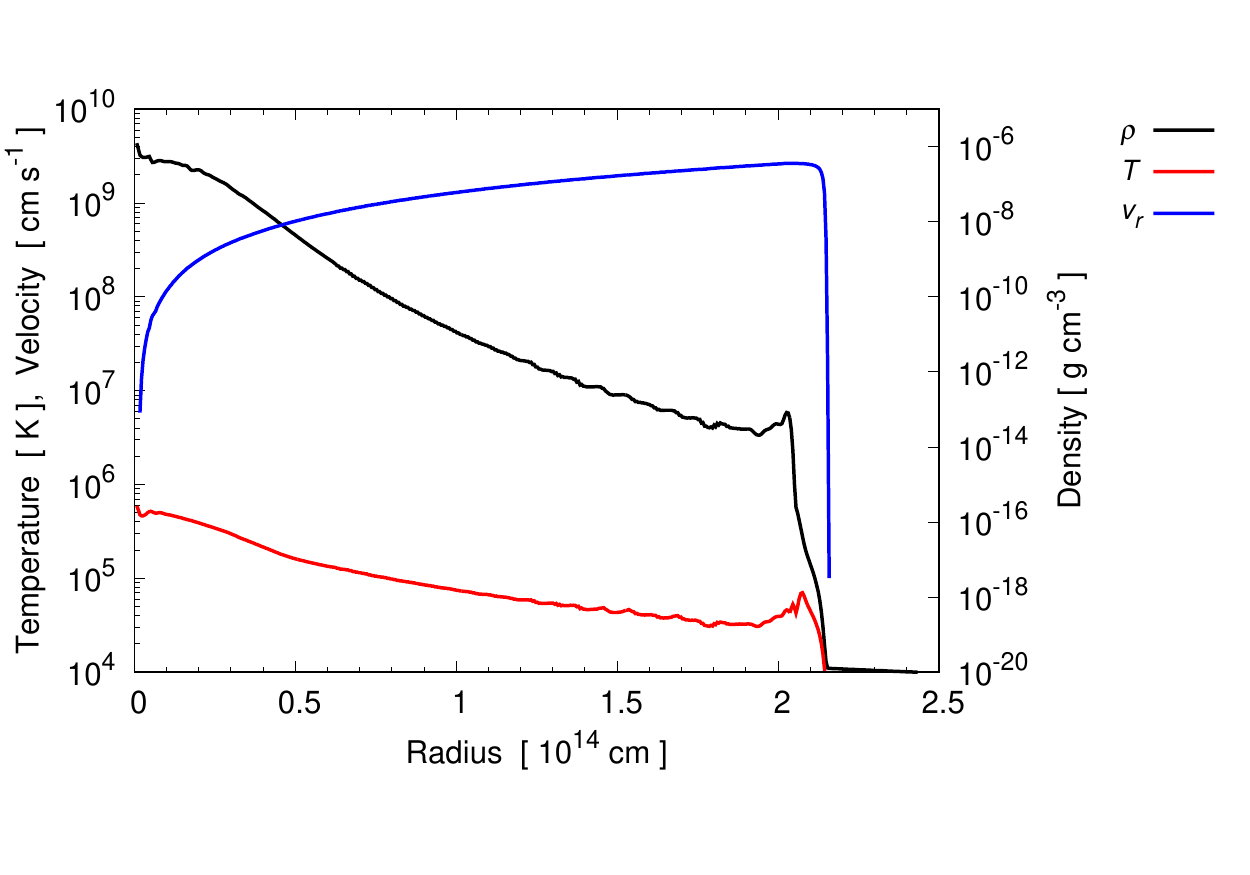}
\end{center}
\end{minipage}
\\
\begin{minipage}{0.5\hsize}
\vs{-1}
\begin{center}
\includegraphics[width=8cm,keepaspectratio,clip]{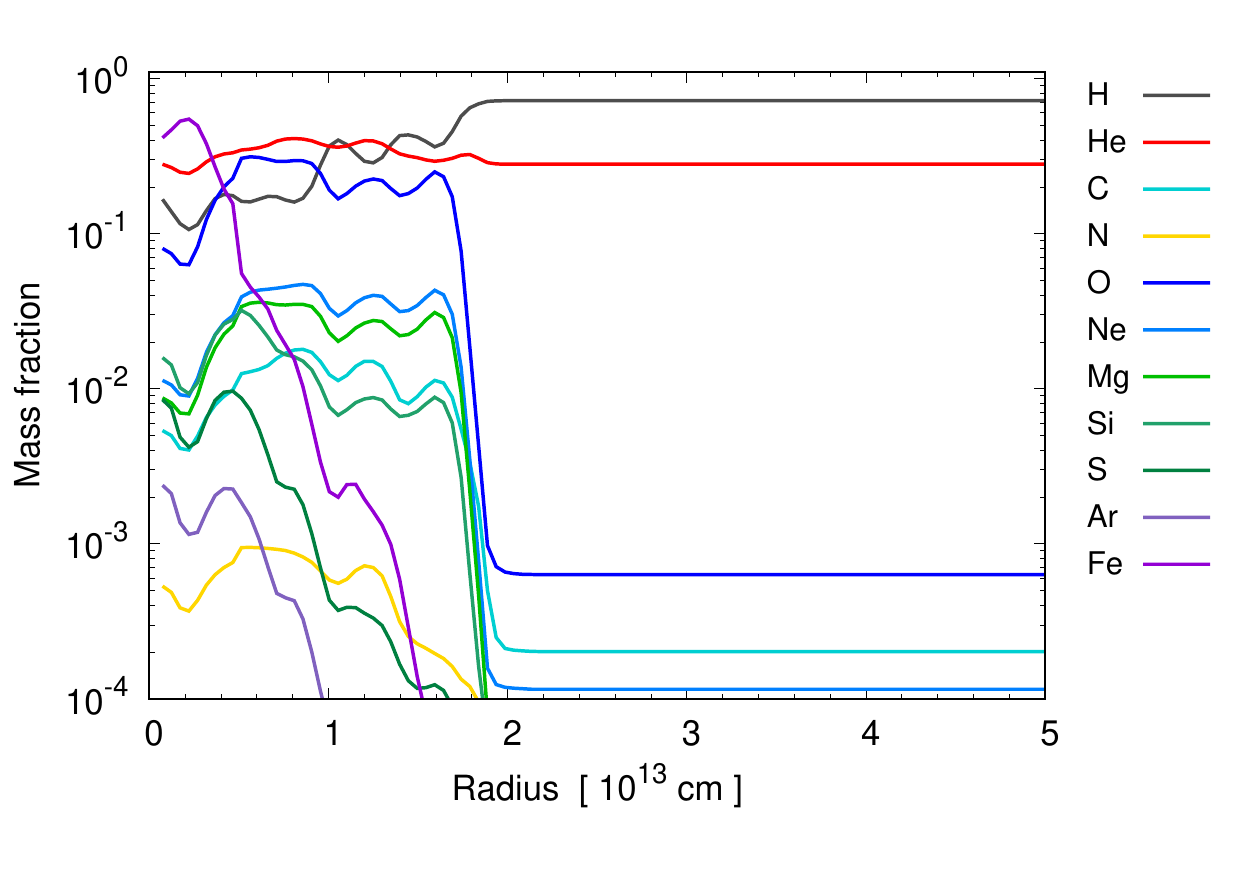}
\end{center}
\end{minipage}
\begin{minipage}{0.5\hsize}
\vs{-1}
\begin{center}
\includegraphics[width=8.5cm,keepaspectratio,clip]{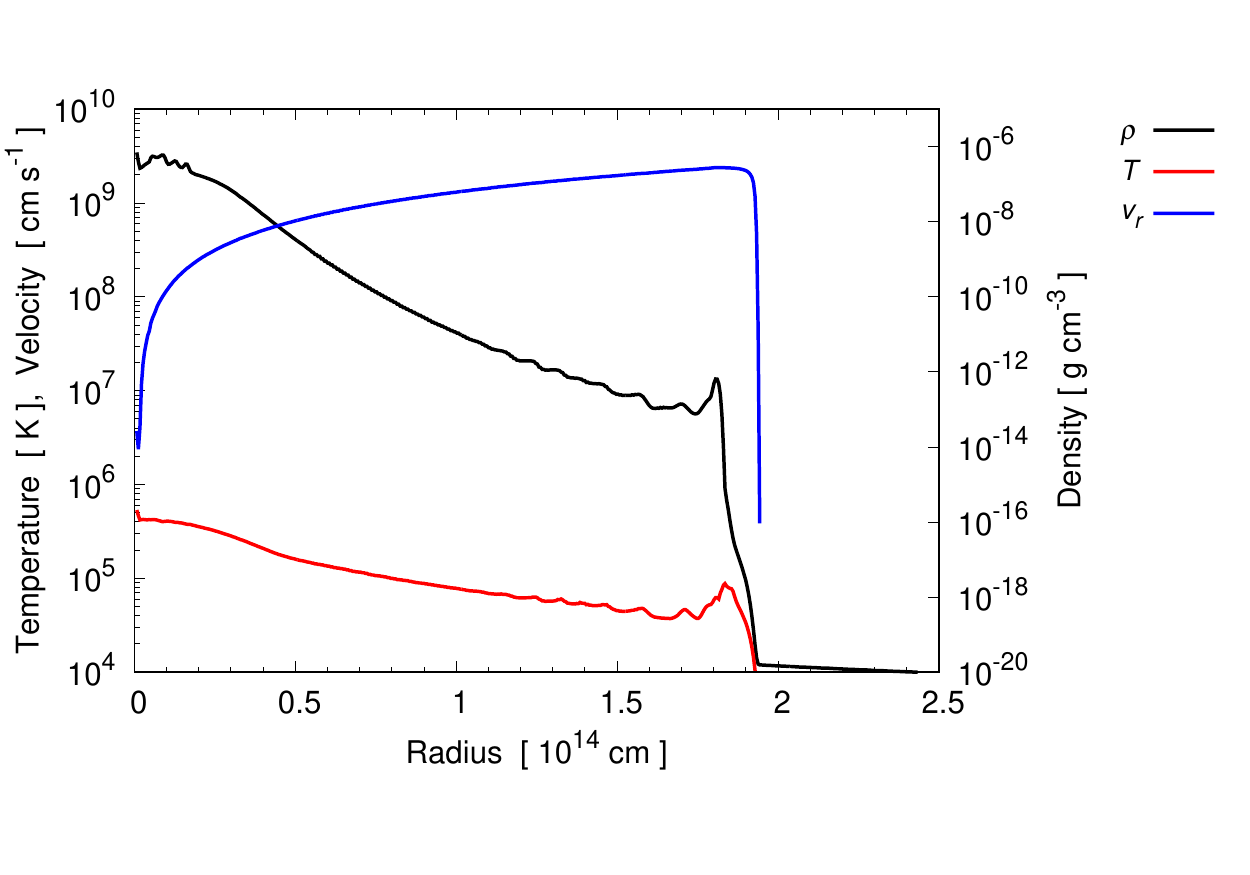}
\end{center}
\end{minipage}
\\
\begin{minipage}{0.5\hsize}
\vs{-1}
\begin{center}
\includegraphics[width=8cm,keepaspectratio,clip]{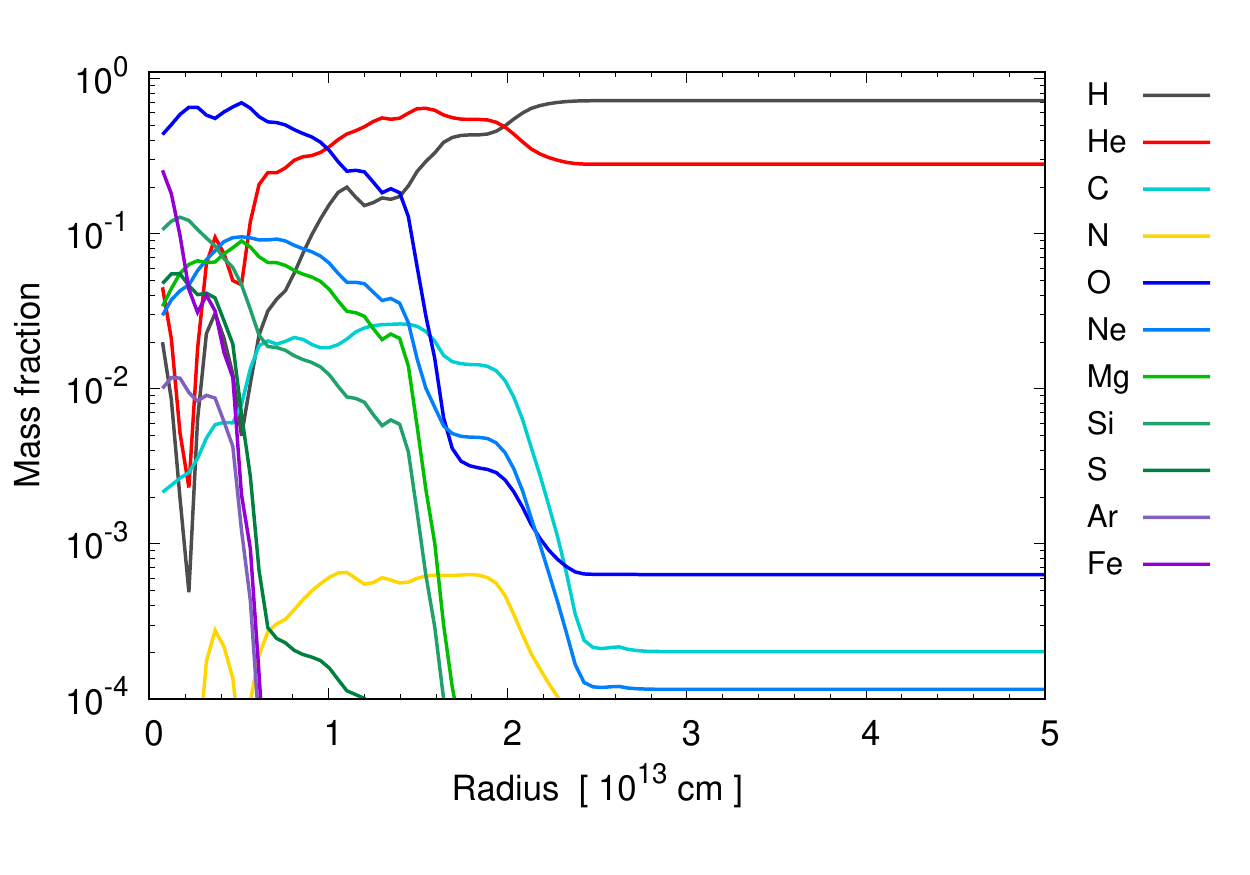}
\end{center}
\vs{-1}
\end{minipage}
\begin{minipage}{0.5\hsize}
\vs{-1}
\begin{center}
\includegraphics[width=8.5cm,keepaspectratio,clip]{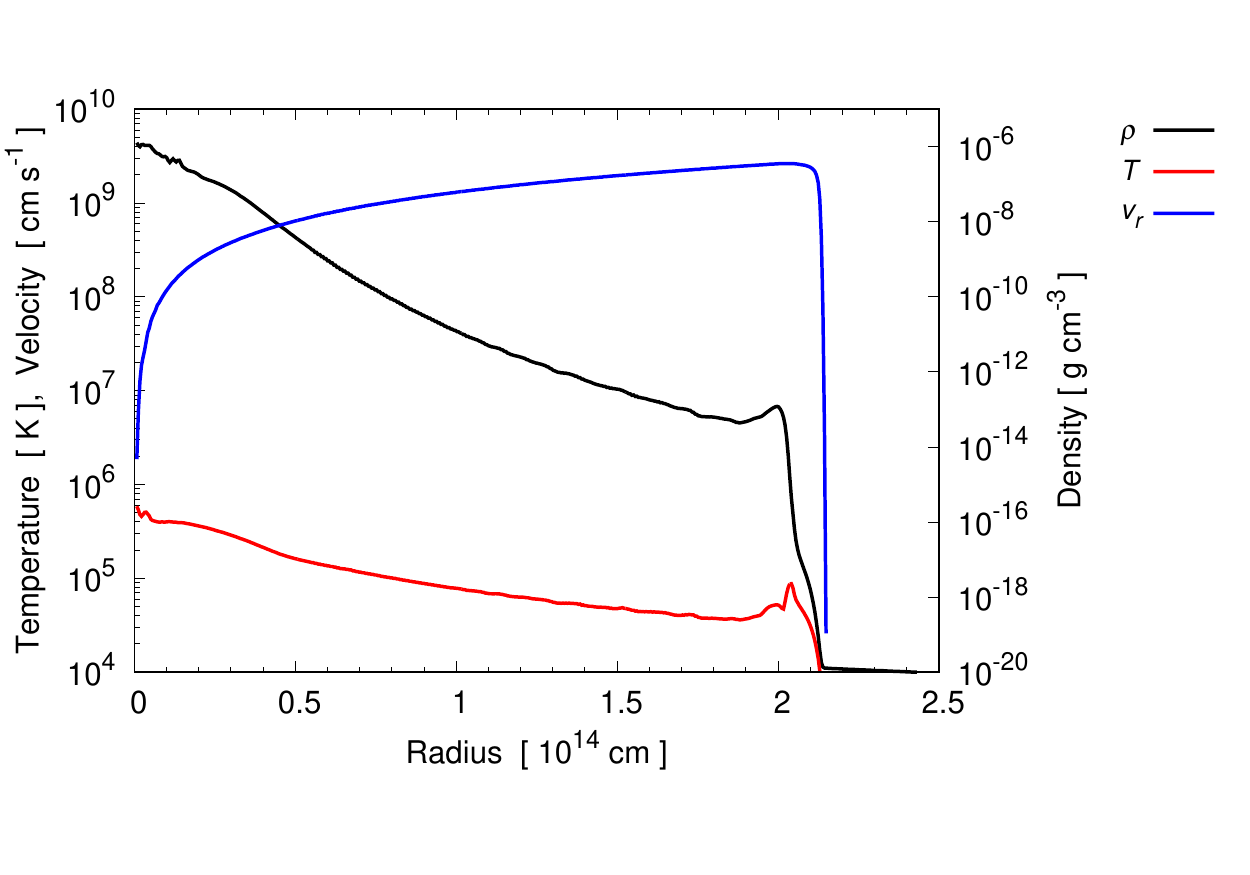}
\end{center}
\vs{-1}
\end{minipage}
\caption{Same as Figure~\ref{fig:b18.3_angle} but for the model n16.3-high and at the age of 22.3 hours after the explosion.\label{fig:n16.3_angle}} 
\end{figure*}

In Figures~\ref{fig:b18.3_angle} and~\ref{fig:n16.3_angle}, angle-specified radial profiles of the mass fractions of atoms and hydrodynamical quantities, the density, temperature, and radial velocity, at the end of the simulation (corresponding to initial conditions in this paper) are plotted for the models b18.3-high and n16.3-high, respectively. 
The positions (radii) of the forward shock depend on the directions (see the left panels). %
In the case of the model b18.3-high (Figure~\ref{fig:b18.3_angle}), the radii of the forward shock along the three directions are $\sim$ 2.2 $\times$ 10$^{14}$ cm, $\sim$ 2.0 $\times$ 10$^{14}$ cm, and $\sim$ 1.5 $\times$ 10$^{14}$ cm in descending order corresponding to the directions, $+Z$, $+Y$, and $-Z$, respectively, depending on the initial explosion asymmetry. %
The fact that the forward shock radius in the $+Y$ direction is larger than that in the $-Z$ direction sounds a bit strange since at the shock-revival phase, the explosion (energy injection) is weakest in the $+Y$ direction. %
It is difficult to specify the reason clearly but it may be interpreted as follows. %
Because of the initial weakest explosion in the $+Y$ direction (due to the artificially imposed bipolar-like explosions), the density of the expanding ejecta in the $+Y$ direction tends to be higher than those along the bipolar-like explosion axis during the propagation of the supernova shock inside the star \citep[see, the left panels in Figure 19 in][]{2020ApJ...888..111O}\footnote{The distinct density contrast between the $+Y$ and the other two directions gradually diminishes from the outer part to the inner part during the shock wave propagation inside the spherical progenitor structure. This is especially evident after the strong reverse shock (which develops during the shock propagation in the hydrogen envelope) passes through the ejecta (see, the snapshot after the shock breakout in the right panels in Figure~\ref{fig:b18.3_angle}).}; %
the dense material in the $+Y$ direction maintains its outward velocity for a longer time than that in other directions, thanks to its high inertia. Consequently, the forward shock in the $+Y$ direction is influenced and deformed by the dense material. %
Additionally, Rayleigh-Taylor (RT) instability develops in the $+Y$ direction after the shock wave enters the hydrogen envelope, due to the steep density gradient between the dense material and the surroundings.\footnote{The growth rate of RT instability is proportional to the square root of the absolute density gradient \citep[e.g.,][]{1991A&A...251..505M}.} This instability may partly play a role in exaggerating the expansion of the dens material and the deformation of the forward shock. %
Actually, extended finger-like structures can be recognized along the equatorial plane \citep[see, the left panel in Figure~23 in][]{2020ApJ...888..111O}. %
On the other hand, in the model n16.3-high, the positions of the forward shock are more identical to each other. %
The composition profiles are radially fluctuated compared with the 1D profiles described above (Figures~\ref{fig:prof_mean},~\ref{fig:prof_sphel},~and~\ref{fig:prof_sphel_pure}). %
The positions of the outer boundaries of the metal-rich cores do not necessarily reflect the forward shock radii; %
for example, in the model b18.3-high (Figure~\ref{fig:b18.3_angle}), the outer boundaries along $-Z$ and $+Y$ directions are similar to each other in contrast to the forward shock radii; %
in the model n16.3-high, the outer boundary along the $+Z$ direction is more extended than those along the $-Z$ and $+Y$ directions approximately by a factor of two. %
It is noted that $^{56}$Ni in the $+Y$ direction is very concentrated in the innermost regions for both b18.3-high and n16.3-high models. %
This is because $^{56}$Ni is not abundantly synthesized during the explosive nucleosynthesis due to the less energetic explosion in the equatorial plane. %
Again, although the amounts of $^{56}$Ni could be uncertain at most by a factor of two, the qualitative differences among the directions (how $^{56}$Ni is extensively distributed) due to the bipolar-like explosions may not be affected by bipolar-like explosions obtained by more realistic models, e.g., with a better fallback model and/or a larger nuclear reaction network, with the similar explosion energies and the progenitor models. %

By the way, in reality, depending on the explosion mechanism, e.g., the delayed neutrino-driven explosion aided by convection/SASI or the magnetorotationally-driven explosion as mentioned in Section~\ref{sec:intro}, the distribution of $^{56}$Ni may be different from the idealized bipolar-like explosions in this paper, although some neutrino-driven explosion models seem to result in an asymmetric dipole-like morphology \citep[e.g.,][]{2019MNRAS.482..351V}. %
As shown in the later sections, the distribution of $^{56}$Ni could affect the formation of molecules through several processes. %
The comparisons of theoretical models like those presented in this study with observations may provide insights to pin down the explosion mechanism in the future. 

Here, in order to easily distinguish calculation models, we name the models for 1D calculations as follows. %
Each model is distinguished by specifying the adopted progenitor model (b18.3 or n16.3) and the type of the initial 1D profiles (angle-averaged/spherical explosion/purely spherical/angle-specified) after fixing the values of the parameters, $f_{\rm h}$, $f_{\rm d}$, and $t_{\rm s}$. %
Then, a model name is expressed as ``Progenitor model"-``Type of 1D profile", for example, b18.3-mean. %
Here, ``b18.3" and ``mean" denote the binary merger progenitor model (b18.3) and the angle-averaged profiles, respectively. %
Similarly, for the first key, ``n16.3" denotes the single-star progenitor model (n16.3). %
For the second key, ``sphel", ``sphel-pure", ``zp", ``zn", and ``yp" denote the 1D profiles for the spherical explosion case, the purely spherical case, along the $+Z$, $-Z$, and $+Y$ axes, respectively. %

\section{Results} \label{sec:results} 

\subsection{One zone calculation results} \label{subsec:one_zone_results} 

To find a set (sets) of values of the parameters which provides (provide) reasonable results compared with previous studies and observed 
CO light curves, the impact of those parameters is investigated. %
Showing all the calculated results is rather lengthy and the purpose of the paper is not the investigation of the impact of those parameters but the impact of matter mixing. %
Instead, only several calculation results are presented to describe the overall evolution and features. %
Among several parameters listed in Table~\ref{table:param}, critical parameters which affect the molecule formation results significantly are $f_{\rm h}$, $f_{\rm d}$, and $t_{\rm s}$. %
Other parameters, i.e., $t_{\rm c}$, $X_{{\rm e},i}$, $X_{{\rm e},f}$, and $\rho_{\rm break}$, are practically fixed to be a reasonable value. %
For simplicity, the value of $t_{\rm c}$ is set to be 1000 days up to when the calculation of CO (ro)-vibrational transitions is performed. %
The values of $X_{{\rm e},i}$ and $X_{{\rm e},f}$, are set to be 0.1 and 10$^{-3}$, respectively by reference to the values shown in Fig.~8 in \cite{1998ApJ...496..946K}, in which the thermal and ionization evolutions in the ejecta of SN~1987A were modeled. %
We confirmed that changing the values of $X_{{\rm e},i}$ and $X_{{\rm e},f}$ affects negligibly the molecule formation results. %
The value of $\rho_{\rm break}$ changes the timing of the break in the gas temperature and a higher (lower) $\rho_{\rm break}$ value causes an early (late) break and an early (late) molecule formation, since the gas temperature drops to $\sim$ 10$^{4}$ K, at which molecules start to form, at an early (late) phase. %
The value of $\rho_{\,\rm break}$ is set to be 10$^{-9}$ g cm$^{-3}$, which is consistent with the treatment in the EoS of the hydrodynamical simulations \citep{2020ApJ...888..111O}. %
Then, by changing the values of the parameters, $f_{\rm h}$, $f_{\rm d}$, and $t_{\rm s}$, in total 100 cases are calculated: %
the cases investigated are the combinations of $f_{\rm h} = 10^{-4}$, 5~$\times$~10$^{-4}$, 10$^{-3}$, 5~$\times$~10$^{-3}$, 10$^{-2}$, $f_{\rm d} = 10^{-2}$, 5~$\times$~10$^{-2}$, 10$^{-1}$, 5~$\times$~10$^{-1}$, 1.0, $t_{\rm s} =$ 200, 300, 500, and $\infty$ (practically $f_{\rm red} = 1.0$) days. %

In Section~\ref{subsubsec:one_zone_trend}, the evolution of physical quantities is described with several representative cases and the most reasonable parameter set for one-zone calculations is selected. %
Section~\ref{subsubsec:chemi_reac} is devoted to explaining important chemical reactions with the best parameter set as a representative. %

\subsubsection{Overall trends of physical quantities and the selection of reasonable parameter sets} \label{subsubsec:one_zone_trend} 

In the following subsections (Sections~\ref{para:one_zone_param1}, \ref{para:one_zone_param2}, and \ref{para:one_zone_param3}), the results of the representative three cases are presented as candidates of the reasonable parameter set, respectively. %
Among all the calculated cases, the three cases are arbitrarily selected according to the three criteria described in Section~\ref{para:one_zone_selec}; %
then, among the three cases, the most reasonable one is selected. %

\paragraph{The case with $f_{\rm h} = 10^{-4}$, $f_{\rm d} = 5 \times 10^{-2}$, and 
$t_{\rm s} = 200$ days} \label{para:one_zone_param1}

\begin{figure*}
\begin{minipage}{0.5\hsize}
\begin{center}
\includegraphics[width=8.5cm,keepaspectratio,clip]{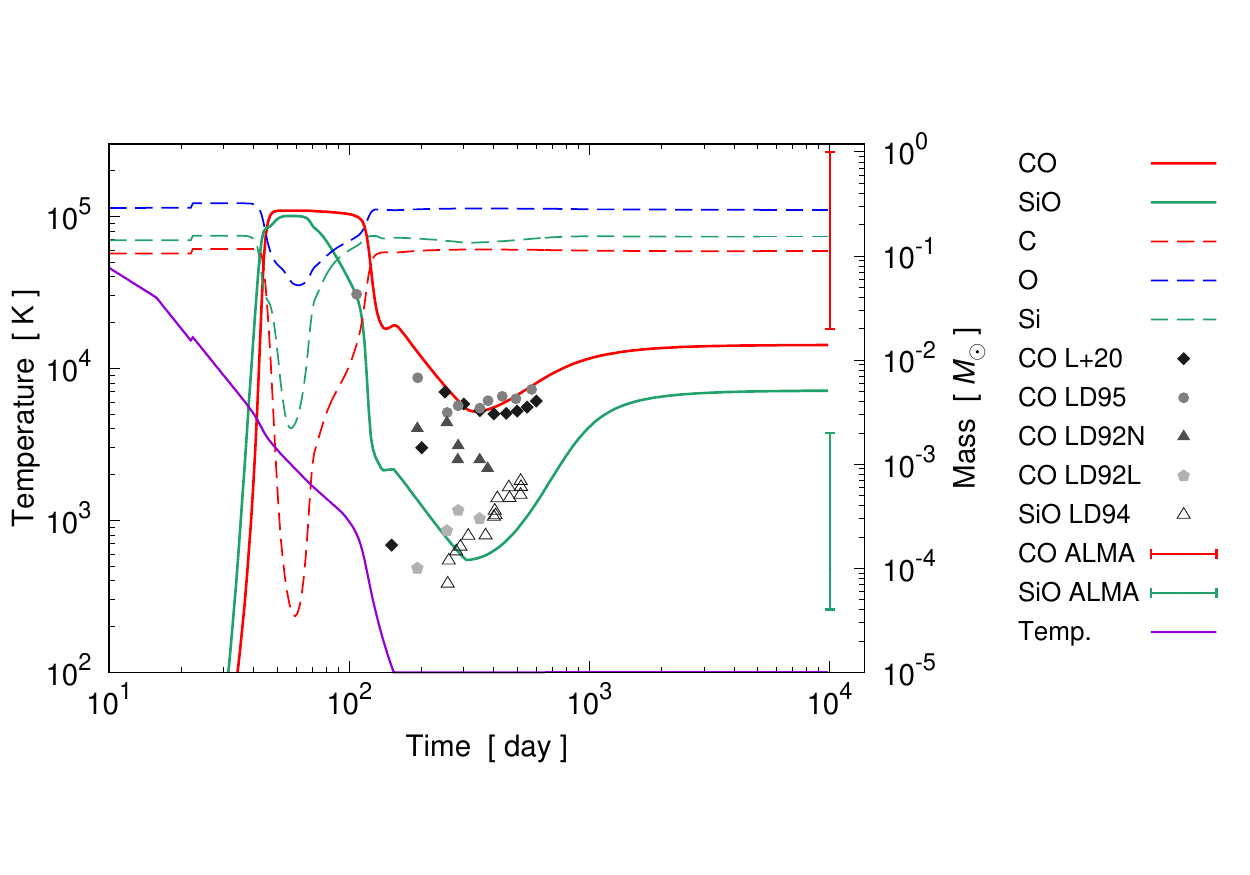}
\end{center}
\end{minipage}
\begin{minipage}{0.5\hsize}
\begin{center}
\includegraphics[width=8.cm,keepaspectratio,clip]{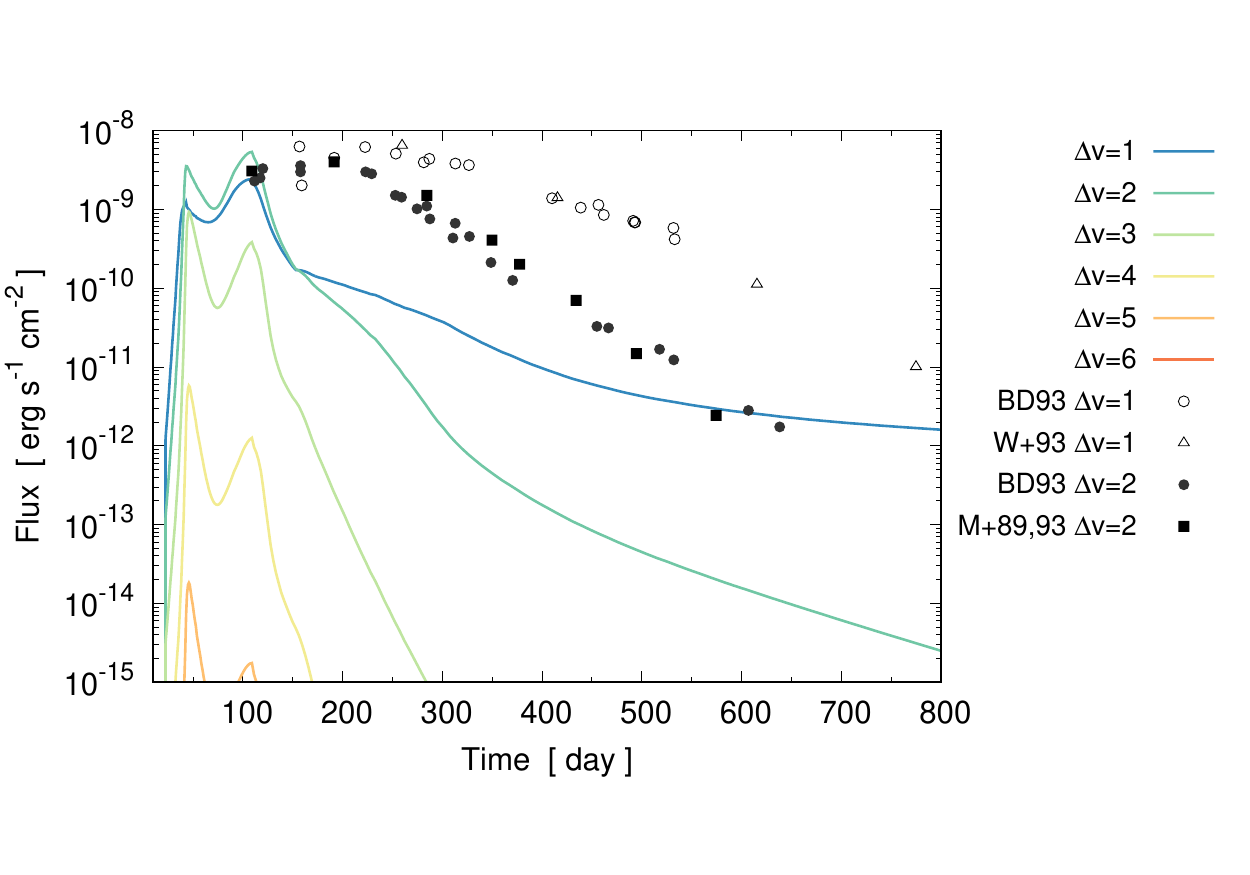}
\end{center}
\end{minipage}
\\
%
\begin{minipage}{0.5\hsize}
\vs{-1.5}
\begin{center}
\includegraphics[width=8.5cm,keepaspectratio,clip]{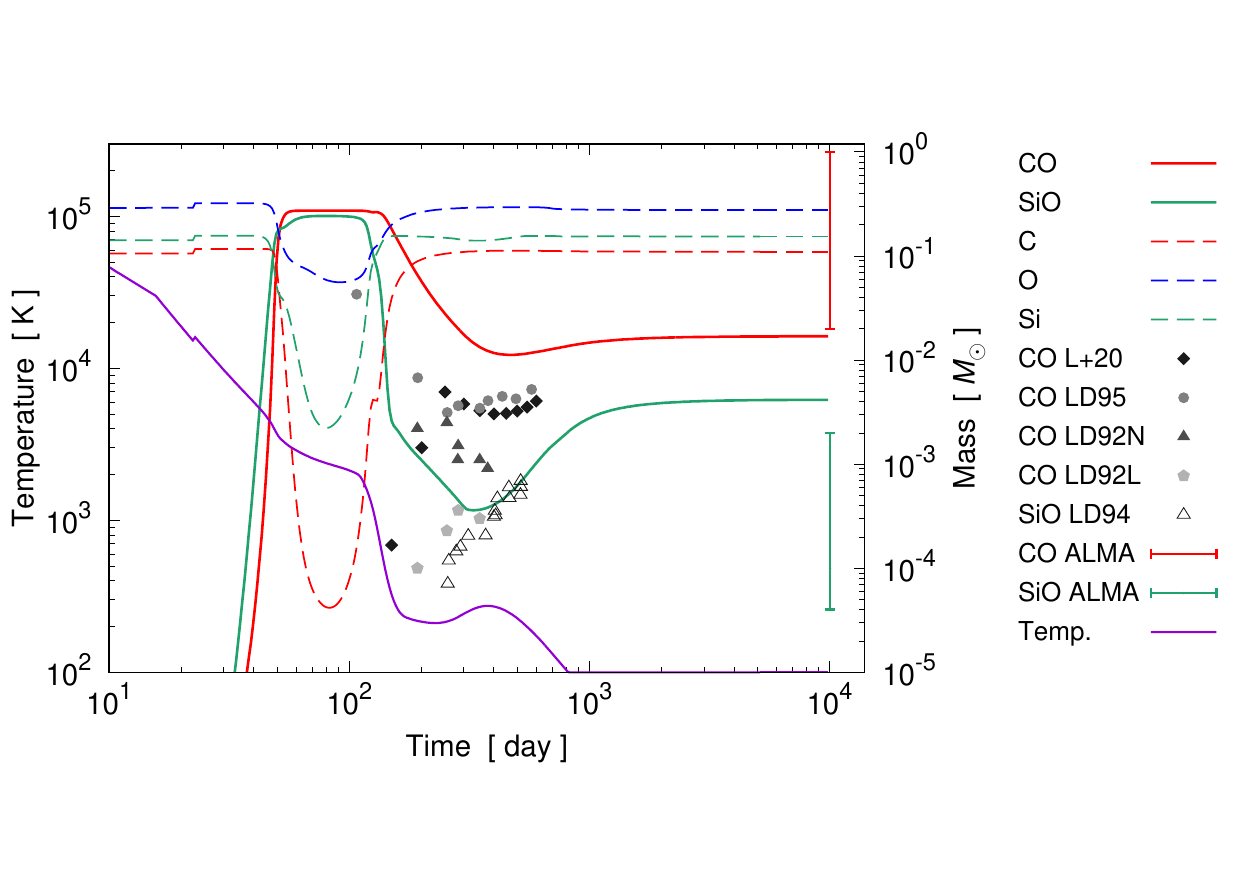}
\end{center}
\end{minipage}
\begin{minipage}{0.5\hsize}
\vs{-1.5}
\begin{center}
\includegraphics[width=8.cm,keepaspectratio,clip]{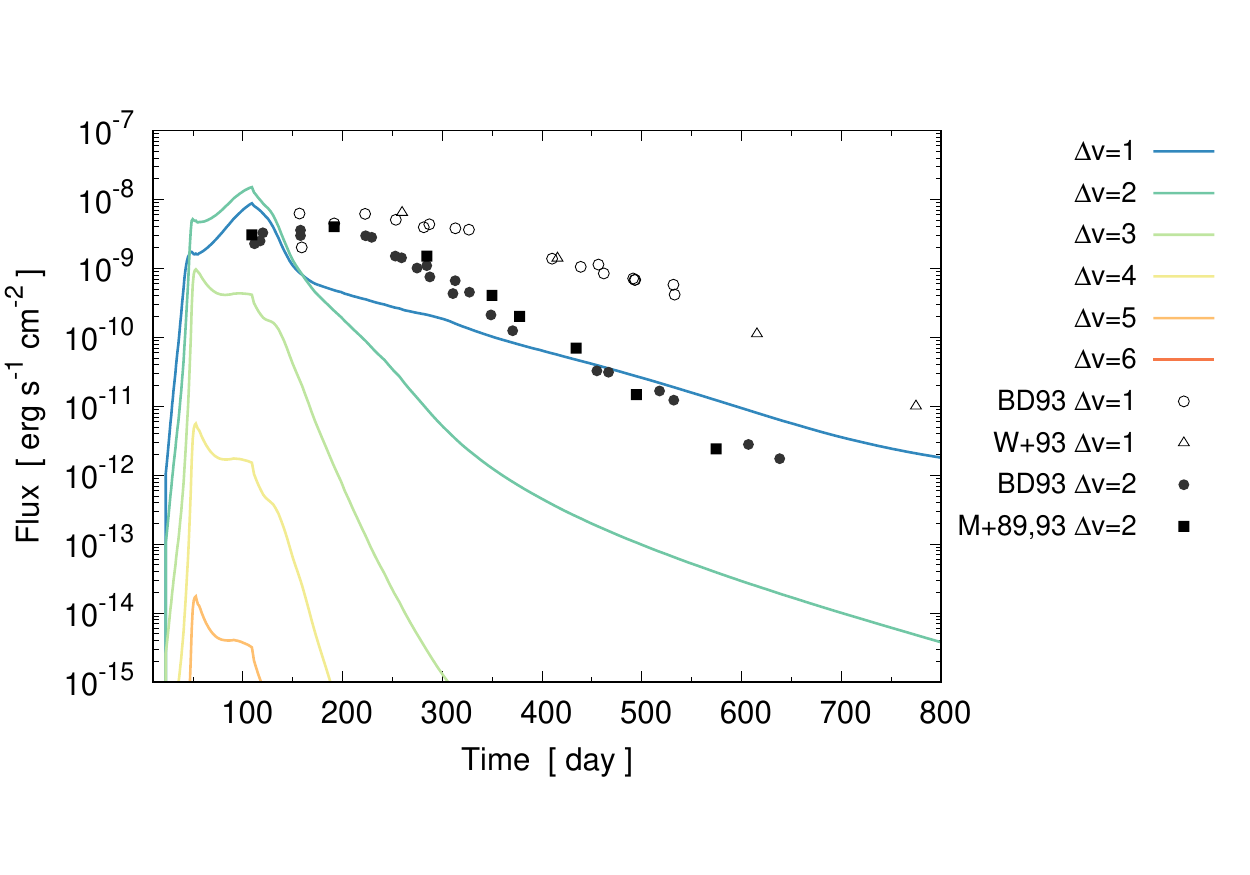}
\end{center}
\end{minipage}
\\
%
\begin{minipage}{0.5\hsize}
\vs{-1.5}
\begin{center}
\includegraphics[width=8.5cm,keepaspectratio,clip]{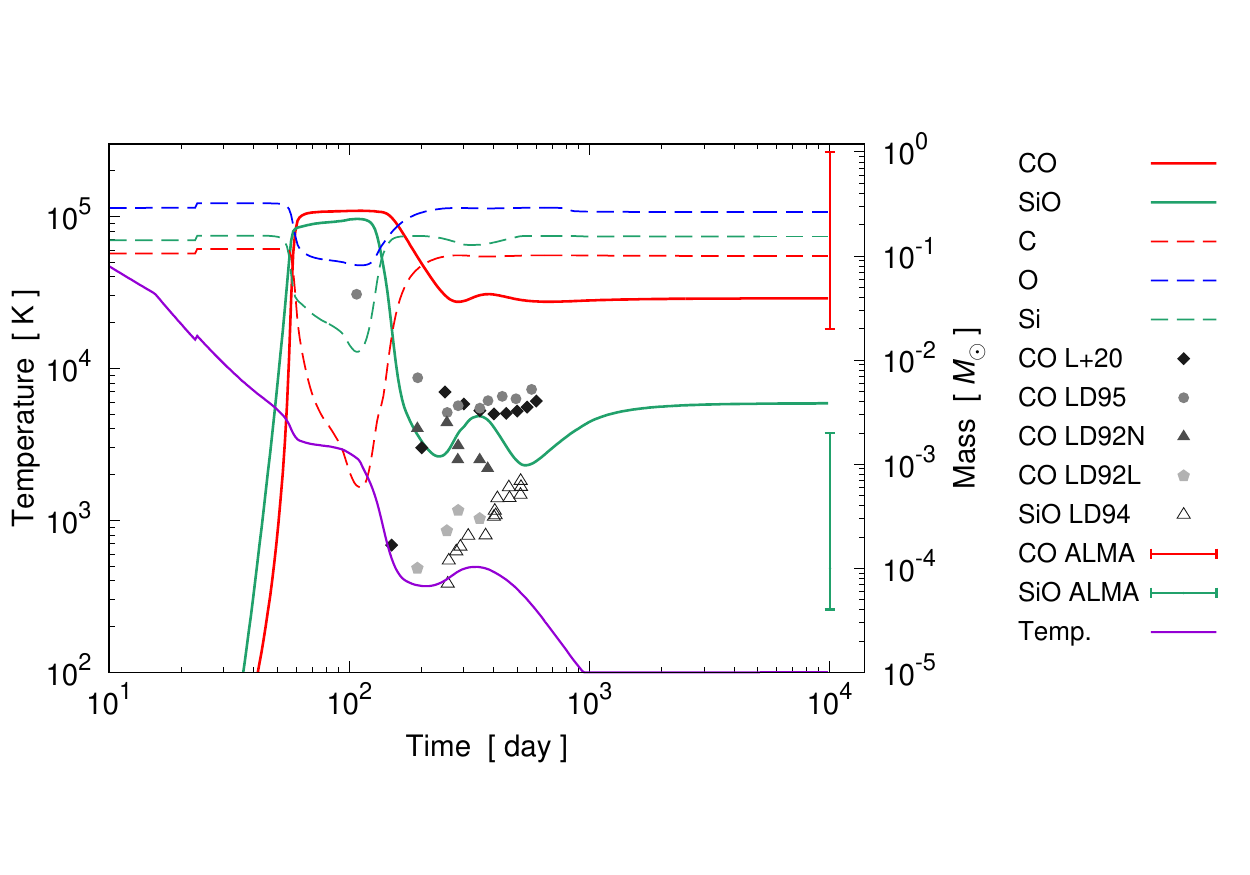}
\end{center}
\vs{-1.}
\end{minipage}
\begin{minipage}{0.5\hsize}
\vs{-1.5}
\begin{center}
\includegraphics[width=8.cm,keepaspectratio,clip]{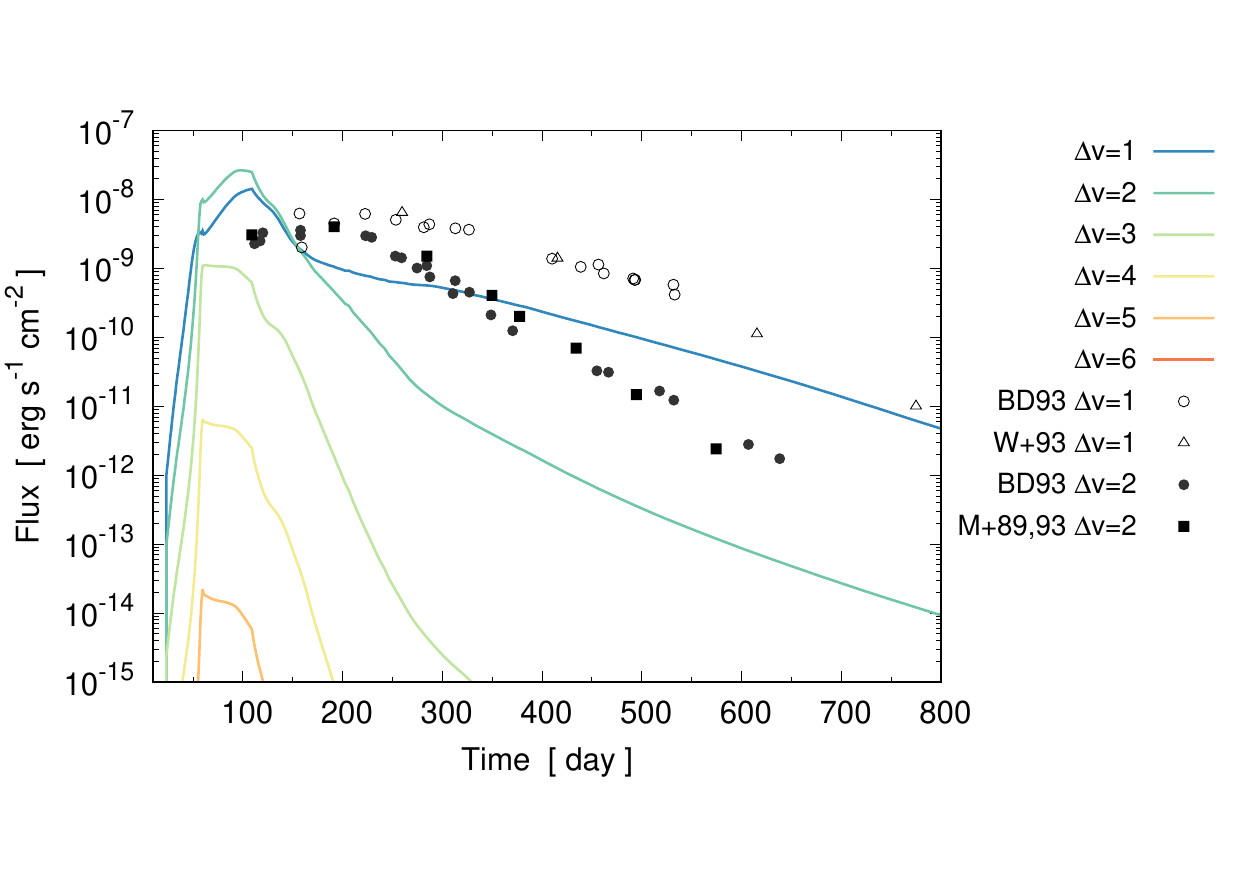}
\end{center}
\vs{-1.}
\end{minipage}
\caption{One-zone calculation results for three sets of the parameters, 
$f_{\rm h} = 10^{-4}$, $f_{\rm d} = 5 \times 10^{-2}$, and $t_{\rm s} = 200$ days (top panels); 
$f_{\rm h} = 5 \times 10^{-4}$, $f_{\rm d} = 10^{-2}$, and $t_{\rm s} = 200$ days (middle panels); 
$f_{\rm h} = 10^{-3}$, $f_{\rm d} = 10^{-2}$, and $t_{\rm s} = 200$ days (bottom panels). %
Left panels: time evolution of the amounts of CO and SiO and the seed atoms, carbon, oxygen, and silicon, compared with the estimations (including theoretical calculations) for CO and SiO in previous studies: LD92 \citep{1992ApJ...396..679L}, LD94 \citep{1994ApJ...428..769L}, LD95 \citep{1995ApJ...454..472L}, ALMA \citep{2017MNRAS.469.3347M}, and L+20 \citep{2020A&A...642A.135L}. The time evolution of gas temperature is also plotted. %
Right panels: time evolution of the fluxes for CO vibrational bands, ${\it \Delta}v=1$ (fundamental), 
${\it \Delta}v=2$ (first overtone), \ldots, ${\it \Delta}v=6$, compared with the observed light curves 
for ${\it \Delta}v=1$ 
\citep[BD93; W$+$93:][respectively]{1993A&A...273..451B,1993ApJS...88..477W} and ${\it \Delta}v=2$ 
\citep[M+89, 93; BD93:][respectively]{1989MNRAS.238..193M,1993MNRAS.261..535M,1993A&A...273..451B}.} %
\label{fig:one_zone_param1}
\end{figure*}

\begin{figure}
\begin{minipage}{0.5\hsize}
\vs{-0.5}
\begin{center}
\hs{0.1}
\includegraphics[width=8.cm,keepaspectratio,clip]{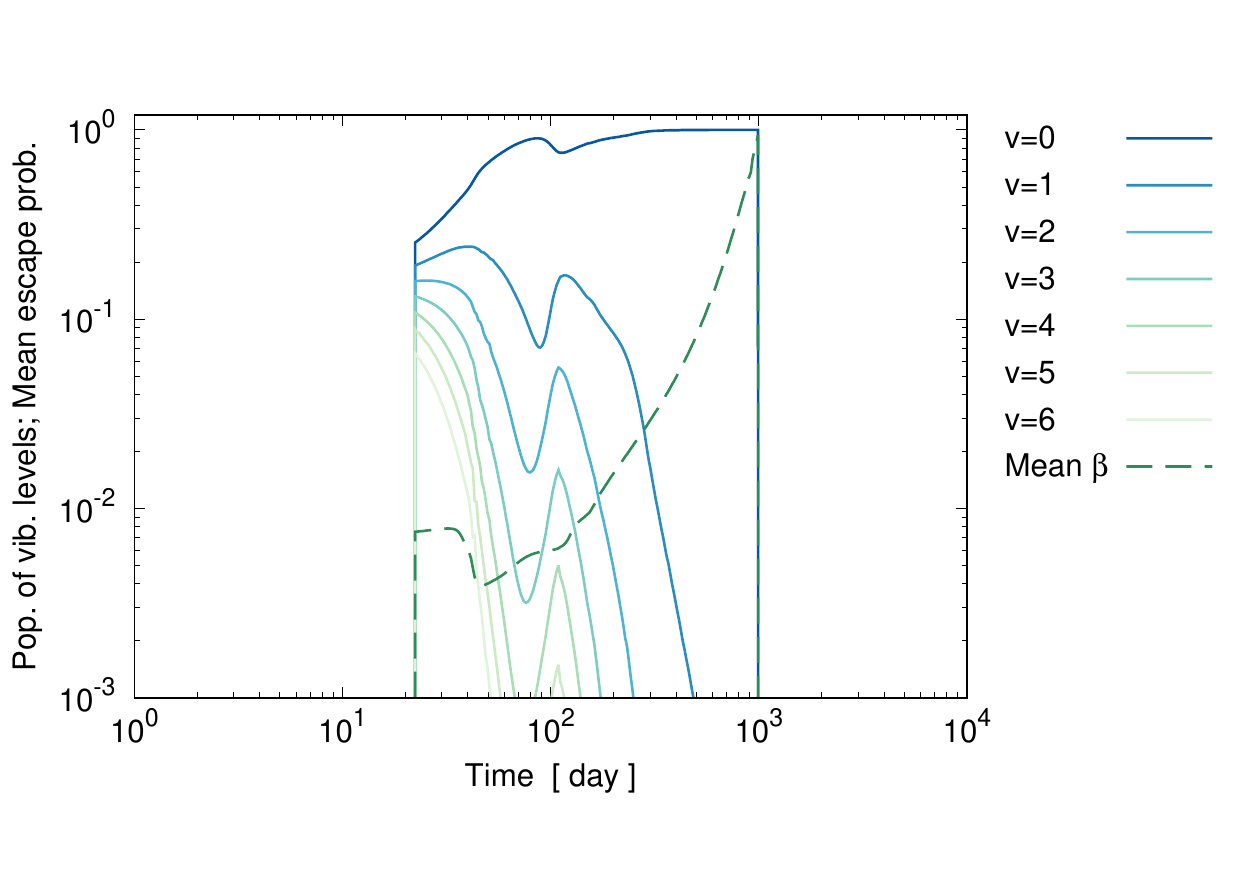}
\end{center}
\end{minipage}
\\
\begin{minipage}{0.5\hsize}
\vs{-0.5}
\begin{center}
\includegraphics[width=6.7cm,keepaspectratio,clip]{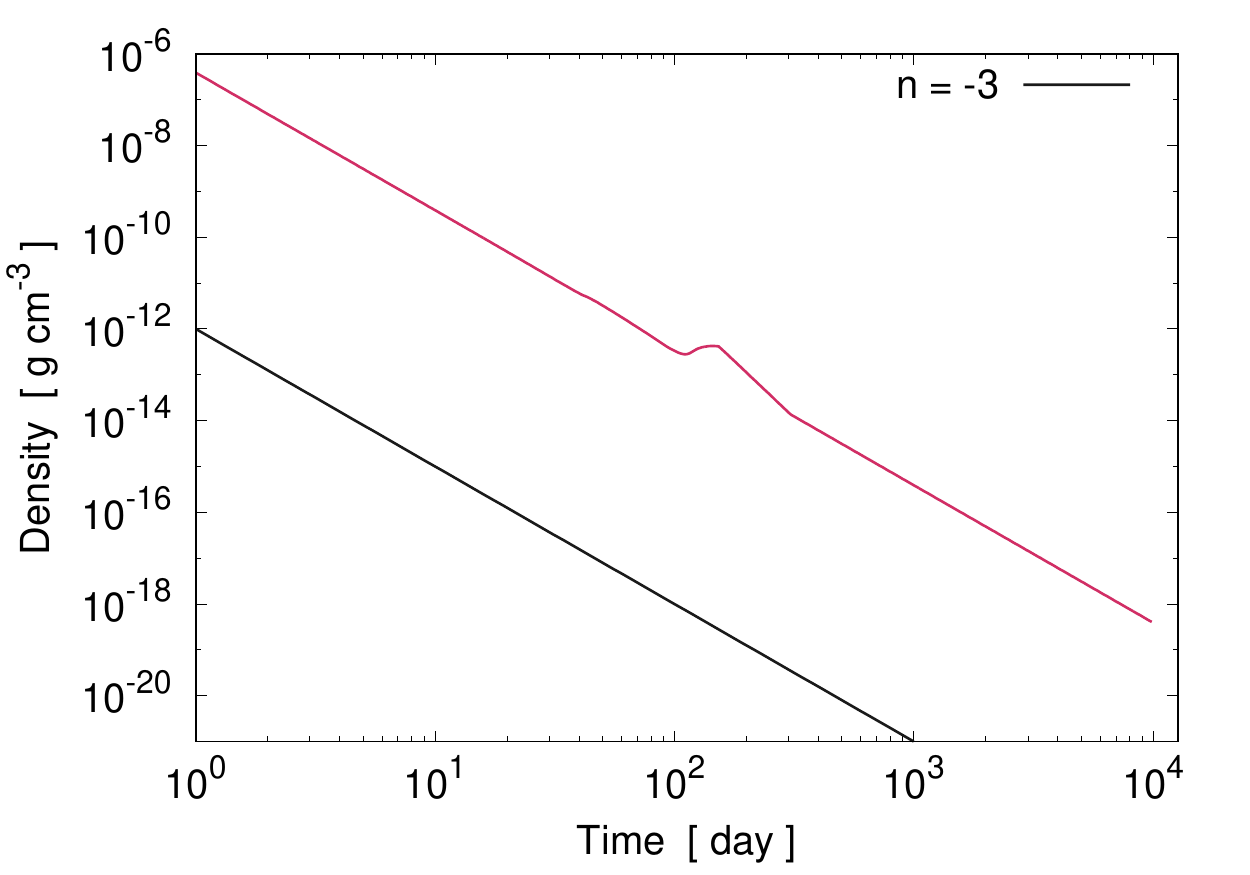}
\end{center}
\end{minipage}
\caption{One-zone calculation results for the case of $f_{\rm h} = 10^{-4}$, $f_{\rm d} = 5 \times 10^{-2}$, and $t_{\rm s} = 200$ days. %
Top panel: the time evolution of the population of CO vibrational levels $X_i$ and the mean escape probability $\overline{\beta}$. %
Bottom panel: the time evolution of the gas density; for reference, a power-law evolution with the power of $-$3 is also shown. For the details, see the text.} %
\label{fig:vib_pop_rho_mod}
\end{figure}

\hspace{\parindent} 
In the top panels in Figure~\ref{fig:one_zone_param1}, the results with the parameter values of $f_{\rm h} = 10^{-4}$, $f_{\rm d} = 5 \times 10^{-2}$, and $t_{\rm s} = 200$ days are shown. %
The gas temperature evolution is plotted in the left panel (violet line). %
The temperature has a break at around 14 days and soon after ($\sim$ 20 days) the temperature goes down to $\sim$ 10$^{4}$ K at which molecules start to form as seen in the amounts of CO and SiO in the same panel. %
In this case, the gas temperature quickly goes down to below $\sim$ 100 K at $\sim$ 200 days due to the CO line cooling after around 100 days. %

By using this case as an example, the population of vibrational levels of CO and the evolution of the gas density are described in Figure~\ref{fig:vib_pop_rho_mod}. %

In the top panel, the time evolution of the population of vibrational levels of CO is shown. %
Initially, the population seems to be determined almost as an equilibrium state because of the high density and temperature at this phase. %
Throughout the evolution, the lower the vibrational level is, the higher the populational level is; %
the level $v = 0$ is dominated at all phases, but higher levels are also populated in early phases until a few hundred days. %
The time evolution of the population of the levels is sensitively determined by the balance of the excitation (absorption) and de-excitation (emission) processes, which depends on the evolution of the gas density and temperature, the background radiation field, and the escape probabilities (line optical depths). %
The line optical depths depend on the gas density and the population of the levels; the evolution is complicated due to the dependencies. %
In the same panel, the mean escape probability ${\overline \beta}$ obtained by the summation of the ``effective" escape probabilities $f_{\rm red} \,\beta_{\kappa \lambda}$, see, Equations~(\ref{eq:beta}) and (\ref{eq:fred}), for all ro-vibrational lines with the weight of the population of the initial state is also shown. %
The value of the reduction factor $f_{\rm red}$ at 100 days is approximately 10$^{-2}$ for $t_{\rm s} = 200$ days. %
Therefore, the escape probabilities at early phases are effectively reduced by the reduction factor $f_{\rm red}$ to restrict CO line emissions and the resultant cooling. %
As mentioned above, the gas temperature quickly goes down to $\sim$ 100 K at $\sim$ 200 days even with the effective reduction through the factor $f_{\rm red}$. %

It is found that in one-zone calculations, overall, $f_{\rm red} =$ 200 days results in better fluxes of emission via CO ro-vibrational 
transitions. %
The impact of the reduction factor $f_{\rm red}$ through the parameter $t_{\rm s}$ is discussed in Appendix~\ref{app:fred} by taking several one-zone calculations as a reference. %

As seen in the time evolution of the gas density in the bottom panel of Figure~\ref{fig:vib_pop_rho_mod}, the value increases after 100 days due to the effect described in Equation~(\ref{eq:thermal_i}), i.e., the density evolution deviates from the power-law evolution with the power of $n = -3$ due to the deflation that stems from the cooling. %
The timing of the density increase corresponds to the rapid temperature decrease due to the cooling by CO ro-vibrational transitions around 100 days. %
It is noted that after 200 days, the density evolution again gradually approaches the original power-law evolution. This is because, in the calculation, the minimum temperature (specific internal energy) is set. %
Then, after the gas temperature reaches the minimum temperature (100 K), without effective heating due to the decay of $^{56}$Ni, $e$ becomes equal to $e_{\rm ad}$ in Equation~(\ref{eq:thermal_i}) at about 300 days, at which point, the evolution of the gas density becomes consistent with the original power-law evolution by the assumption. %

In the top right panel in Figure~\ref{fig:one_zone_param1}, the fluxes of CO vibrational bands derived from the calculation and observed ones (open and closed points: the former points are in the fundamental band and the latter ones are in the first overtone band) are plotted. %
Only before about 100 days, the calculated flux of the first overtone band (${\it \Delta}v=2$) is slightly higher than that of the fundamental band (${\it \Delta}v=1$). %
The calculation underestimates the fluxes of the fundamental and the first overtone bands a bit after 100 days compared with the observed fluxes. %
In particular, it is difficult to explain the peaks around 200--250 days and the comparable flux levels between the two bands around the peaks seen in the observed fluxes. %
Such difficulties are also common for other calculation models in this paper. %
Therefore, we would expect that supra-thermal electrons supplied by the decay of $^{56}$Ni may play a role in the excitation. %
Actually, the timings of the peaks of the observed fluxes are consistent with one at which the gas temperatures peak due to the heating by the decay of $^{56}$Ni$\,$\footnote{The half-lives (lifetimes) of $^{56}$Ni and $^{56}$Co are 6.10 and 77.12 days (8.80 and 111.3 days), respectively \citep{1994ApJS...92..527N}. Therefore, practically, $^{56}$Co contributes to the peak at around 200 days; $^{57}$Co \citep[the half-time is 271.74 days:][]{1998NDS....85..415B} may partly contribute as well as $^{56}$Co, although $^{57}$Co is not taken into account in this study.} (the feature of the peaks in the gas temperatures due to the decay of $^{56}$Ni can be more recognized in the 1D calculation results seen e.g., in the left panel in Figure~\ref{fig:rho_dens_1d}). %
It is noted that as mentioned in Section~\ref{sec:intro}, the observed CO vibrational bands are well reproduced by a one-zone model with time-dependent thermal-chemical calculations \citep{1995ApJ...454..472L}, in which the thermal evolution was calculated with a more realistic method including a time-dependent heating efficiency of gas and the line cooling not only from CO ro-vibrational transitions but also from ones of ions with non-thermal excitations, although an arbitrary time-dependent reduction factor for the energy deposition was introduced for the reproduction \citep{1995ApJ...454..472L}. %
However, since the reproduction of the observed CO light curves is beyond the scope of this paper, we leave these issues related to CO ro-vibrational transitions for future works (see, Section~\ref{subsec:limitation}). %

The time evolution of the amounts of CO and SiO and the seed atoms, carbon, oxygen, and silicon, is shown in the top left panel in Figure~\ref{fig:one_zone_param1} with the estimations (including theoretical calculations) for CO and SiO by the previous studies \citep[closed points for CO and open points for SiO:][]{1992ApJ...396..679L,1994ApJ...428..769L,1995ApJ...454..472L,2020A&A...642A.135L} and ones from the ALMA observations \citep[vertical bars with errors:][]{2017MNRAS.469.3347M}. %
It is noted that some of the estimated values in the previous study \citep{1992ApJ...396..679L} are based on the fitting of observed spectra but in the fitting, some assumptions (e.g., steady states, LTE, and NLTE) and models are involved. %
Therefore, the estimated values in the previous studies may not necessarily be the actual amounts as can be seen from the fact that the estimated values for CO are not fully consistent among the studies. %
Nevertheless, those values are useful for reference. %
The results of the molecules other than CO and SiO are presented later in some of 1D calculation results (see, Sections~\ref{subsubsec:1d_comp_sphel} and \ref{subsubsec:1d_angle}). %

As seen in the top left panel, CO and SiO molecules start to form at about 30 days quickly after the gas temperature reaches $\sim$ 10$^4$ K and the amounts of both CO and SiO reach over 0.1 $M_{\odot}$ at 40 days. %
Since there were no observations (at least for SN~1987A) before 100 days, unfortunately, it is difficult to set a constraint on the amounts of molecules before 100 days. %
After 100 days, some fractions of CO and SiO are destructed until 300 days (the major formation and destruction reactions of CO and SiO are discussed later in Section~\ref{subsubsec:chemi_reac}). %
The amount of CO is consistent in particular with the values of LD95 \citep{1995ApJ...454..472L} and the amount of SiO is also consistent with the estimations of LD94 \citep{1994ApJ...428..769L} relatively well. %
After 300 days, CO and SiO are recovered again. %
At the final phase of $\sim$ 10$^4$ days, the amount of CO is a bit less than the estimated range from ALMA; %
the amount of SiO is overestimated compared with the ALMA estimation. %
It is noted that in the ejecta of SN 1987A, the existence of a large mass of dust (0.3 $M_{\odot}$ amorphous carbon and 0.5 $M_{\odot}$ silicate) has been suggested by the observations by Herschel \citep[][]{2015ApJ...800...50M}. %
Thus, in reality, some fractions of molecules, in particular, SiO must have been converted to dust as partly supported by dust formation theories in CCSNe \citep{2013ApJ...776...24N,2003ApJ...598..785N} and the fact that the mass of CO estimated by the ALMA observations \citep{2017MNRAS.469.3347M} is higher than that of SiO. %
Therefore, some degree of overestimations of SiO compared with the ALMA estimation can be explained in this way. %
Then, the calculated amount of CO and SiO would be roughly consistent with the previous studies and the ALMA in this specific case. %

As mentioned above the gas temperature, however, quickly goes down to $\sim$ 100 K in spite of the underestimation of the fluxes of the CO vibrational bands. %
Actually, in the previous studies \citep[][]{1992ApJ...396..679L,1995ApJ...454..472L,2020A&A...642A.135L}\footnote{The former paper calculated the temperatures by spectral fitting under the assumptions of steady states and LTE. The middle one calculated time-dependent thermal and chemical evolutions with a more realistic thermal evolution model than that of this paper. The last one also calculated gas temperatures under the steady state assumption.}, the temperatures of carbon monoxide have been calculated as $> 500$ K before 1000 days. %
Since the adopted models and assumptions are different among the studies including this paper, such a rapid cooling would not be necessarily rejected. %
Even though, we prefer to find a milder gas cooling case (gas temperature evolution) as presented below. %

\paragraph{The case with $f_{\rm h} = 5 \times 10^{-4}$, $f_{\rm d} = 10^{-2}$, and 
$t_{\rm s} = 200$ days} \label{para:one_zone_param2}

%
\hspace{\parindent} 
In the middle panels in Figure~\ref{fig:one_zone_param1}, the results of the parameter values, $f_{\rm h} = 5 \times 10^{-4}$, $f_{\rm d} = 10^{-2}$, and $t_{\rm s} = 200$ days, are shown. %
In this case, the gas temperature (the left panel) remains greater than 100 K before 1000 days in contrast to the previous case. %
After approximately 300 days, the gas temperature increases by the heating due to the decay of radioactive $^{56}$Ni. %
As observed in the previous case, the flux of the first overtone band dominates that of the fundamental band before about 100 days. %
The monotonically increasing trend both for the fundamental and the first overtone bands before 100 days is different from the previous case with a dip around 70 days because of a bit different evolution of the gas temperature (density). %
Compared with the previous case, the fluxes (the right panel) of the CO fundamental and first overtone bands are slightly improved to be more consistent with the observations after 200 days, although at around 100 days the fluxes are a bit overestimated, and after 200 days the fluxes are still underestimated by more than one order of magnitude. %

The amount of CO (the left panel) before 100 days is not so different from the previous case but the amount after 100 days is apparently higher than the previous case and the estimations by the previous studies. %
This trend is partly due to the lower $f_{\rm d}$ value than that of the previous case. %
The amount of SiO also has a similar trend. %
However, as can be seen, the final amounts of CO and SiO are not so different from the ones in the previous case. %

\paragraph{The case with $f_{\rm h} = 10^{-3}$, $f_{\rm d} = 10^{-2}$, and 
$t_{\rm s} = 200$ days} \label{para:one_zone_param3}

%
\hspace{\parindent} 
In the bottom panels in Figure~\ref{fig:one_zone_param1}, the results of the parameter values of $f_{\rm h} = 10^{-3}$, $f_{\rm d} = 10^{-2}$, and $t_{\rm s} 
= 200$ days are shown. %
The behavior of the gas temperature (the left panel) is similar to the previous case but the value remains higher than those of the previous two cases after 100 days and before 1000 days due to the higher heating efficiency via the $f_{\rm h}$ value. %
The early trend (before 100 days) of the fluxes of CO vibrational bands is similar to the previous case; %
the high gas temperature, however, slightly increases the fluxes of CO vibrational bands (the right panel) compared with the previous two cases. %

The amount of CO (the left panel) before 100 days is not so different among the three cases including this case. %
However, the amount after 100 days is further shifted to higher values and the deviation from the estimations by the previous studies becomes significant. %
The amount of SiO is also a similar trend to the one of CO. %
Although the deviation of the amounts of CO and SiO from those estimated in the previous studies with data analysis and theoretical calculations is significant, the final amounts are more consistent with the ALMA estimations. %
The amount of CO at the final phase is at this time within the error bars of the ALMA estimation and the amount of SiO becomes close to the ALMA one. %

\paragraph{The criteria for selecting the best parameter set} \label{para:one_zone_selec}

\hspace{\parindent} 
In order to pick up reasonable parameter sets, the following three points are taken into account as the criteria to be satisfied. %
1.~The consistency of the amounts of CO and SiO with the previous estimations and the ALMA values, i.e., points and bars with errors shown in the left panels in Figure~\ref{fig:one_zone_param1}; %
2.~consistency of the calculated fluxes of CO vibrational bands with the observed fundamental and first overtone bands; %
3.~calculated gas temperatures should not be too low ($\sim$ 100 K) before 1000 days. 
Among the three cases presented above and the other models calculated, we arbitrarily select the most acceptable model to be the second case (the case in Section~\ref{para:one_zone_param2}) presented above according to the three points above. %
Hereafter, the impact of the parameters, $f_{\rm h}$ and $f_{\rm d}$, and trends of the other models are briefly described with a few representative/extreme cases. %

\begin{figure*}
\begin{minipage}{0.5\hsize}
\begin{center}
\includegraphics[width=8.5cm,keepaspectratio,clip]{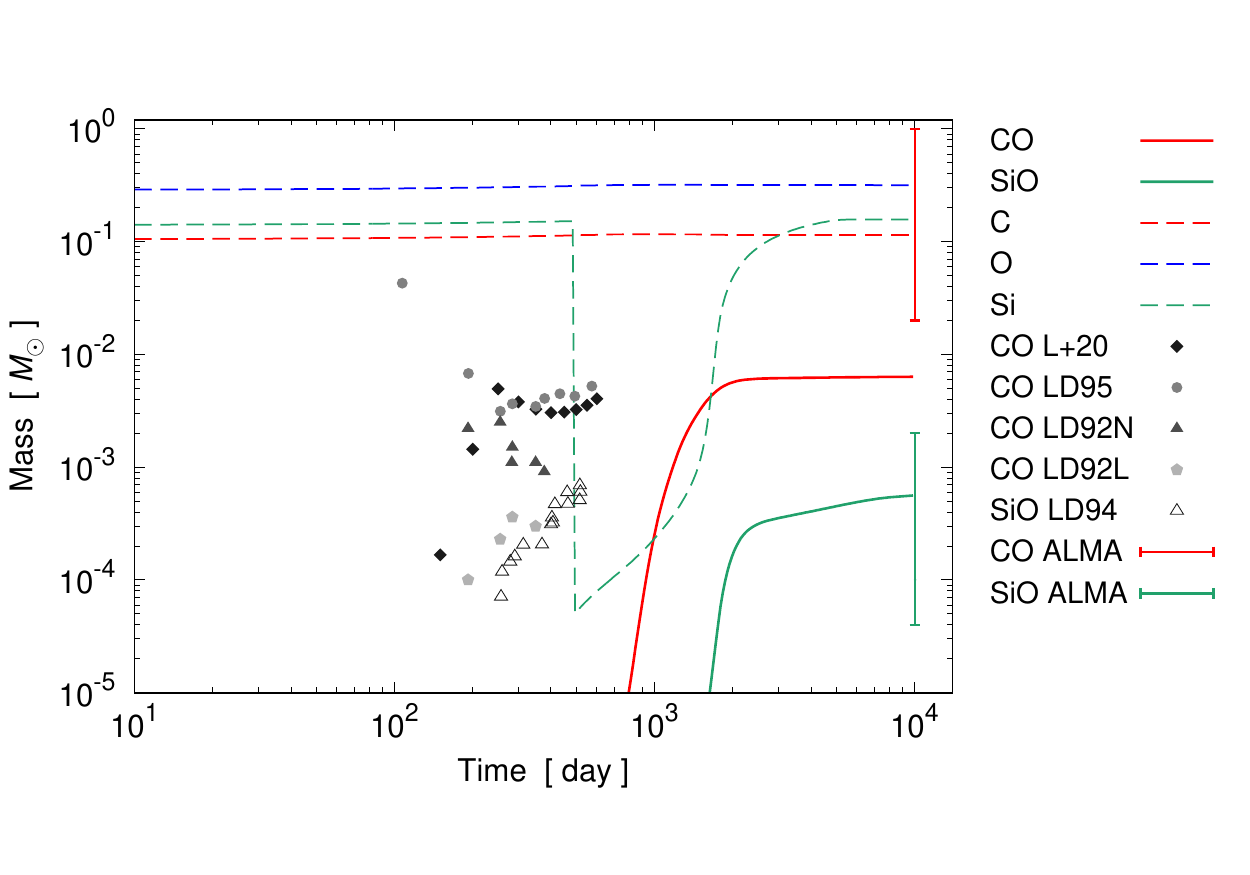}
\end{center}
\vs{-1.}
\end{minipage}
\begin{minipage}{0.5\hsize}
\begin{center}
\includegraphics[width=8.5cm,keepaspectratio,clip]{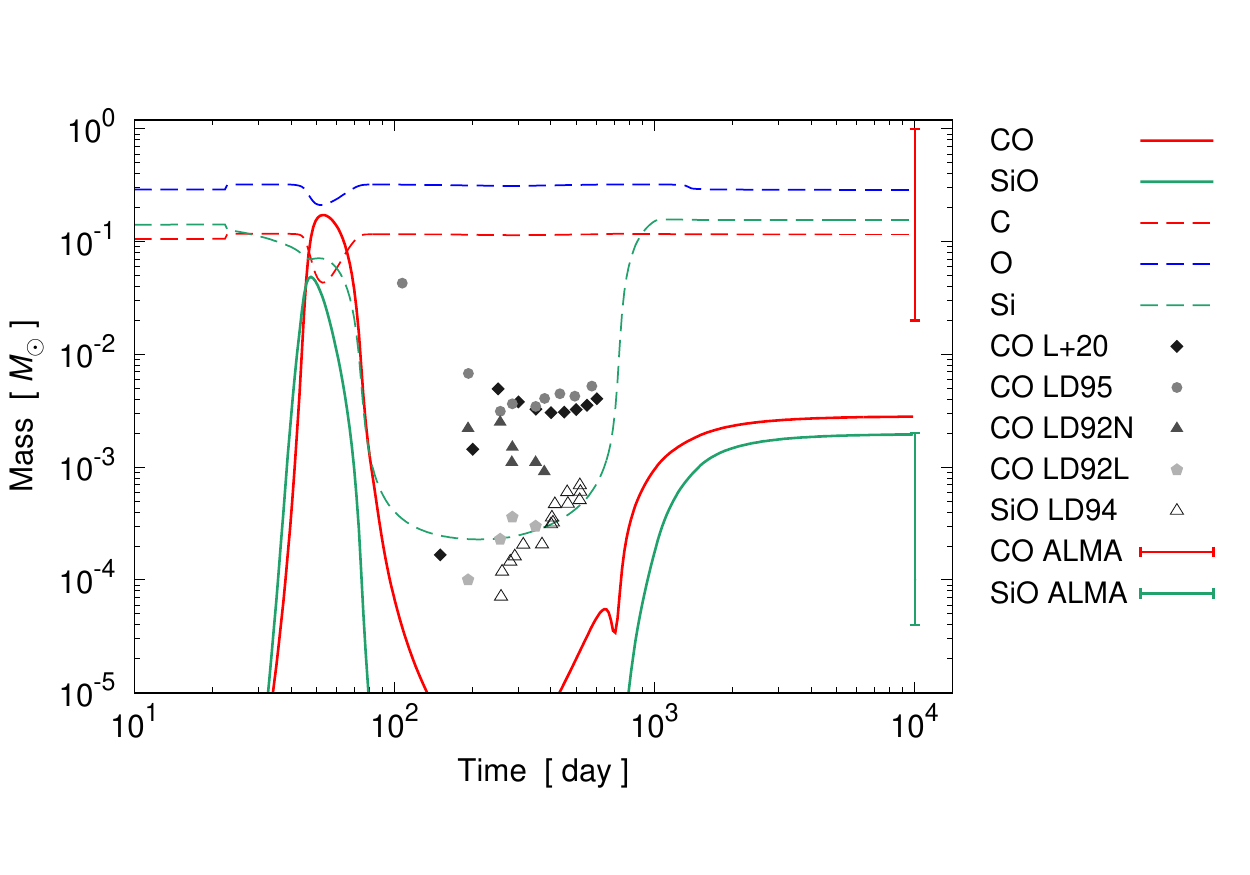}
\end{center}
\vs{-1.}
\end{minipage}
\caption{Time evolution of the amounts of CO and SiO and the seed atoms, carbon, oxygen, and silicon, compared with the estimations (points) for CO and SiO in previous studies (for the explanation of the points, see the caption of Figure~\ref{fig:one_zone_param1}). %
Left panel: $f_{\rm h} = 10^{-2}$, $f_{\rm d} = 10^{-2}$, and $t_{\rm s} = 200$ days. %
Right panel: $f_{\rm h} = 5 \times 10^{-4}$, $f_{\rm d} = 1.0$, and $t_{\rm s} = 200$ days.} %
\label{fig:one_zone_param2}
\end{figure*}

If the values of $f_{\rm h}$ are high ($f_{\rm h}$ = 5 $\times$ 10$^{-3}$ or 10$^{-2}$), i.e., the efficiency of the gas heating by the decay of $^{56}$Ni is high, the gas temperature goes down to $\sim$ 10$^{4}$ K at late phases ($\sim$ 1000 days). %
Then, molecules start to form at such late phases (the left panel in Figure~\ref{fig:one_zone_param2}), which is inconsistent with the fact that CO vibrational bands were observed as early as $\sim$ 100 days. %
Additionally, at the late phases, the gas densities are low ($\sim$ 10$^{-15}$ g~cm$^{-3}$) compared with ones at $\sim$ 100 days ($\sim$ 10$^{-12}$ g~cm$^{-3}$, see, the bottom panel in Figure~\ref{fig:vib_pop_rho_mod}); the amounts of molecules tend to be lower due to the low efficiency of two-body reactions at later phases compared with the cases of lower $f_{\rm h}$ values. %

In the case of high $f_{\rm d}$ values, i.e., the destruction efficiency is high, CO and SiO are destructed from $\sim$ 100 days to 300 days (the right panel in Figure~\ref{fig:one_zone_param2}) due to the ionizations by Compton electrons (\texttt{CM} reactions) and some reactions related to ionized atoms for this specific case (for the details of important chemical reactions for the formation and destruction of CO and SiO, see, Section~\ref{subsubsec:chemi_reac}), although after 300 days the amounts are recovered to some extent. %
In the destruction processes, ionized atoms could also play a role through ion-neutral reactions as described in the next section. %

It should be noted that in the one-zone approximation, the chemical composition is forced to be uniform throughout the ejecta core; %
matter mixing is then overestimated compared to the real environment. %
Since both the gas heating and the ionization and destruction by Compton electrons (\texttt{CM} reactions) depend on the local amount of $^{56}$Ni (see, Equations~(\ref{eq:epsilon_ni}), (\ref{eq:epsilon_co}), (\ref{eq:epsilon}), and (\ref{eq:kc})), these processes may also be overestimated with the effectively high local $^{56}$Ni abundance. %
Therefore, in the case of the one-zone approximation, lower values of $f_{\rm h}$ and $f_{\rm d}$ may be selected as reasonable parameter values compared with the cases with more realistic matter mixing as discussed later with 1D calculation results. %

\subsubsection{Important chemical reactions for CO and SiO} \label{subsubsec:chemi_reac} 

\begin{figure*}
\begin{minipage}{0.5\hsize}
\begin{center}
\hs{-1.5}
\includegraphics[width=9.5cm,keepaspectratio,clip]{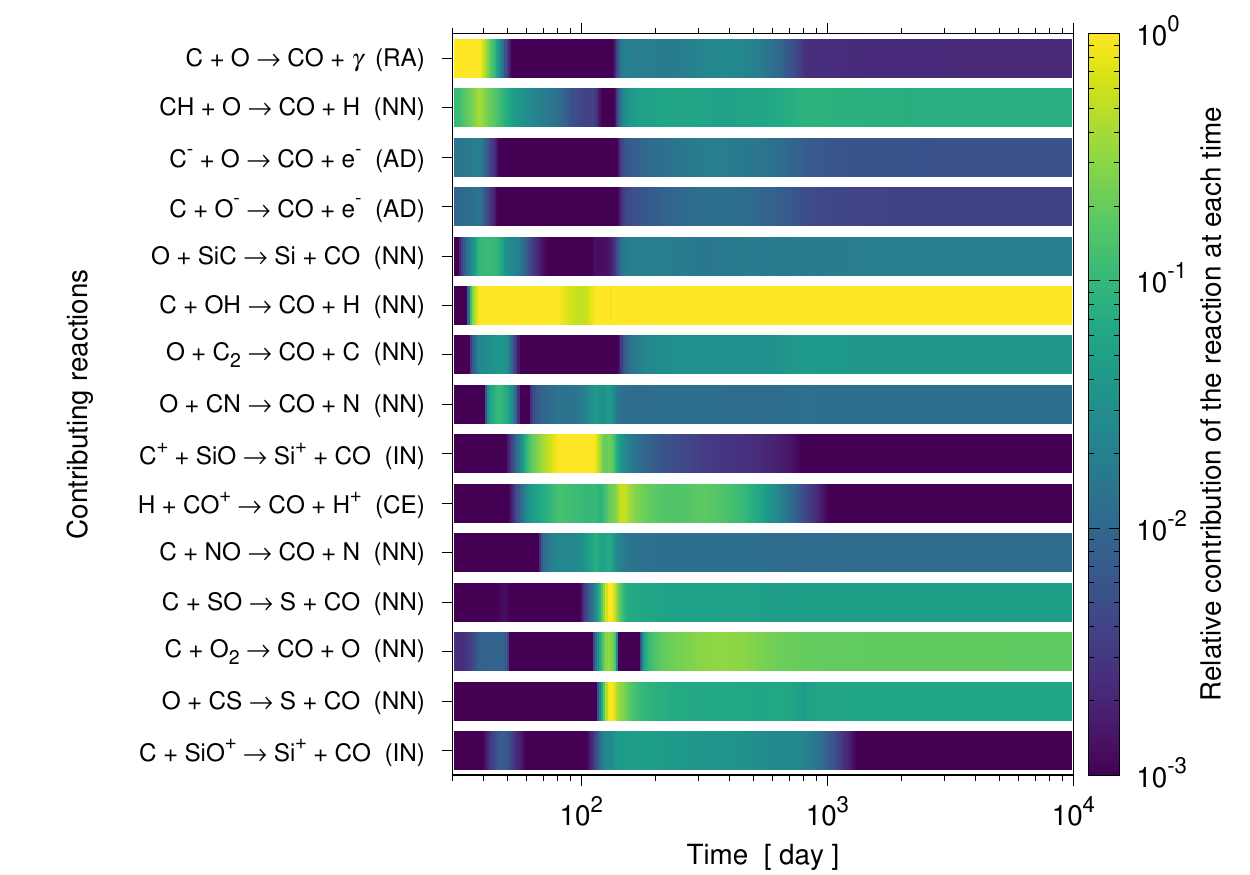}
\end{center}
\end{minipage}
\begin{minipage}{0.5\hsize}
\begin{center}
\hs{-1.5}
\includegraphics[width=9.5cm,keepaspectratio,clip]{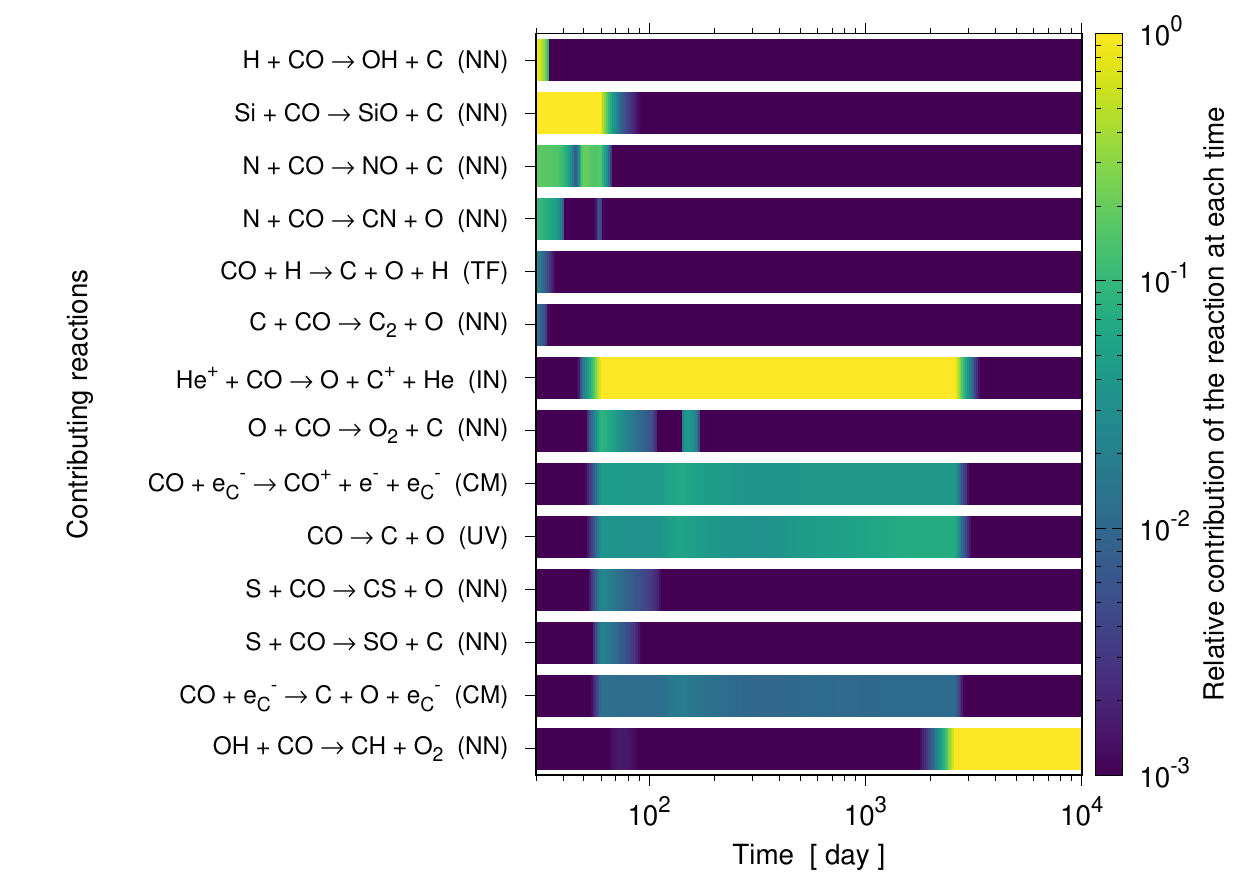}
\end{center}
\end{minipage}
\caption{Contributing chemical reactions for the formation (left) and destruction (right) of CO in which carbon monoxide (CO) is directly involved as a function of time (after 30 days) for the one-zone calculation of the case with $f_{\rm h} = 5 \times 10^{-4}$, $f_{\rm d} = 10^{-2}$, and $t_{\rm s} = 200$ days. %
The code inside the parentheses left side of each reaction denotes the corresponding reaction type listed in Table~\ref{table:types}. %
Colors denote the relative contribution of each reaction, which is proportional to the reaction flow $F_i$ ($D_i$) in Equation~(\ref{eq:rate_eq}), normalized as the maximum $F_i$ ($D_i$) among CO formation (destruction) reactions at each time to be unity. %
For neutral-neutral (\texttt{NN}) reactions, net contributions are counted by subtracting the reaction flow of the corresponding inverse reaction. Reactions are picked up if the relative contribution once becomes greater than 10$^{-3}$.} %
\label{fig:co_reac}
\end{figure*}

\hspace{\parindent} 
In this section, with the most acceptable case (the case in Section~\ref{para:one_zone_param2}; the middle panels in  Figure~\ref{fig:one_zone_param1}), i.e., $f_{\rm h} = 5 \times 10^{-4}$, $f_{\rm d} = 10^{-2}$, and $t_{\rm s} = 200$ days, as a representative, important chemical reactions for the formation and destruction of molecules are described focusing on CO and SiO. %
To pick up important reactions of which the contribution to the formation/destruction is high, the first and second terms in the right-hand side in Equation~(\ref{eq:rate_eq}), more specifically $F_i$ and $D_i$ in Equation~(\ref{eq:form_dest}), are monitored. %
Molecules start to form as early as approximately 30 days after the explosion (see, the middle left panel in Figure~\ref{fig:one_zone_param1}). %
The important reactions are picked up in chronological order for CO and SiO, separately (first CO, and later SiO). %
As seen in the evolution of the amounts of CO and SiO, the evolution can roughly be separated into four epochs, i.e., from the start of the formation of molecules to 40 days (the initial rapid rise), from 40 to 100 days (a plateau phase), from 100 days to 300 days (a destruction dominant phase), and after 300 days (a recovery phase). %
In Figure~\ref{fig:co_reac}, contributing chemical reactions for the formation and destruction of CO in which CO is directly involved are shown as a function of time (after 30 days) for reference. %

At the early phase from the start of the molecule formation ($\lesssim$ 30 days), basically, formation processes are dominant. %
The most dominant formation process of CO is initially the radiative association (\texttt{RA}) reaction below. %
\begin{equation}
{\rm C} + {\rm O} \lra {\rm CO} + \gamma.
\label{eq:co_rad}
\end{equation}
The secondary reactions are the neutral-neutral (\texttt{NN}) reactions, 
\begin{align}
{\rm O} + {\rm CH} \lra {\rm CO} + {\rm H}, \label{eq:o_ch} \\
{\rm C} + {\rm OH} \lra {\rm CO} + {\rm H}, \label{eq:c_oh}
\end{align}
where CH is dominantly formed by a radiative association reaction of the same form of Equation~(\ref{eq:co_rad}) and OH is dominantly produced by the \texttt{NN} reaction, O $+$ H$_2$ $\lra$ OH $+$ H, where H$_2$ is produced by the associative electron detachment (\texttt{AD}) reaction, H$^-$ + H $\lra$ H$_2$ $+$ e$^-$; %
H$^-$ can be produced by the radiative electron attachment (\texttt{REA}) reaction, H + e$^-$  $\lra$ H$^-$ + $\gamma$. %
It is noted that in most of the \texttt{NN} reactions adopted in this paper, there are corresponding inverse reactions, and the forward and inverse reactions compete with each other. %
Therefore, for \texttt{NN} reactions, net contributions are taken into account in judging whether the reaction is important or not. %
Then, the contributions of the \texttt{RA} reaction in Equation~(\ref{eq:co_rad}) and the \texttt{NN} reaction in Equation~(\ref{eq:o_ch}) are gradually reduced and the contribution of the \texttt{NN} reaction in Equation~(\ref{eq:c_oh}) instead increases (actually, the contribution of the \texttt{NN} reaction is initially negative, i.e., the inverse reaction is dominant) to be comparable with the \texttt{RA} reaction at $\sim$ 40 days. %
On the other hand, the main destruction processes before 40 days are the \texttt{NN} reactions below. %
\begin{align}
&{\rm H} + {\rm CO} \lra {\rm OH} + {\rm C}, \label{eq:h_co} \\
&{\rm Si} + {\rm CO} \lra {\rm SiO} + {\rm C}. \label{eq:si_co} 
\end{align}
Initially, the \texttt{NN} reaction in Equation~(\ref{eq:h_co}) dominates the reaction in Equation~(\ref{eq:si_co}); %
the latter, however, becomes the primary destruction reaction after 30 days as the contribution of the former decreases to be negative after 35 days. %
The latter reaction is also the formation reaction of SiO. %
As seen in the middle left panel in Figure~\ref{fig:one_zone_param1}, by approximately 40 days, a large fraction of the seed carbon and oxygen atoms are consumed. %

Then, after about 40 days, the dominant formation process of CO is replaced by the \texttt{NN} reaction in Equation~(\ref{eq:c_oh}) (until 80 days); %
the \texttt{RA} reaction in Equation~(\ref{eq:co_rad}) becomes no longer the primary in the formation processes. %
One of the secondary formation reactions is the \texttt{NN} reaction in Equation~(\ref{eq:o_ch}). %
As secondary reactions, the \texttt{NN} reactions below also partake in. %
\begin{align}
{\rm O} + {\rm CN} \lra {\rm CO} + {\rm N}, \label{eq:o_cn} \\
{\rm O} + {\rm SiC} \lra {\rm CO} + {\rm Si}, \label{eq:o_sic}
\end{align}
where SiC is mainly produced by the \texttt{RA} reaction, Si + C $\lra$ SiC + $\gamma$, at an early phase and later, the \texttt{NN} reaction, C + SiH $\lra$ SiC + H; %
SiH is produced by the \texttt{NN} reaction, Si + H$_2$ $\lra$ SiH + H. %
The formation processes of CN will be mentioned later. %
On the other hand, the primary destruction process becomes the \texttt{NN} reaction in Equation~(\ref{eq:si_co}) after 40 days. %
As one of the secondary destruction processes, the \texttt{NN} reaction below starts to contribute to the destruction. %
\begin{align}
{\rm N} + {\rm CO} \lra {\rm NO} + {\rm C} \label{eq:n_co}.
\end{align}
It is noted that CN seen in the reactants in Equation~(\ref{eq:o_cn}) is mainly produced by the \texttt{NN} reaction, C + NO $\lra$ CN + O; %
NO can also be seen in the products in Equation~(\ref{eq:n_co}). %
Therefore, through \texttt{NN} reactions, several molecules are complexly involved in the formation and destruction of CO, i.e., one of the products of the destruction process of CO is indirectly involved in the formation processes of CO. %
After 50 days, the contributions of the ion-neutral (\texttt{IN}) reactions below 
\begin{align}
&{\rm C^+} + {\rm SiO} \lra {\rm CO} + {\rm Si^+}, \label{eq:c+_sio} \\
&{\rm He^+} + {\rm CO} \lra {\rm O} + {\rm C^+} + {\rm He}, \label{eq:he+_co}
\end{align}
become gradually larger and larger as one of the formation and destruction processes, respectively. %
The former reaction is also a destruction process of SiO. The ions C$^+$ and He$^+$ can be produced by ionization of the corresponding atoms by Compton electrons in Equation~(\ref{eq:ion_x}) (\texttt{CM} reactions). %
After 60 days the former reaction becomes the secondary formation process and the latter process becomes the primary destruction process. %
The contributions of \texttt{CM} reactions, i.e., the ionization and destruction by Compton electrons due to the decay of $^{56}$Ni also gradually increase as destruction processes. %
In particular, the ionization process of CO by Compton electrons in Equation~(\ref{eq:ion_ab}) and the destruction by UV photons (\texttt{UV} reaction) become the secondary destruction processes. %
The \texttt{NN} reaction below also partakes in the destruction processes. %
\begin{align}
{\rm O} + {\rm CO} \lra {\rm O_2} + {\rm C}. \label{eq:o_co}
\end{align}
As CO$^+$ increases by the ionization in Equation~(\ref{eq:ion_ab}), the charge exchange (\texttt{CE}) reaction below becomes one of the secondary formation processes. %
\begin{align}
{\rm H} + {\rm CO^+} \lra {\rm CO} + {\rm H^+}. \label{eq:h_co+} 
\end{align}
After 80 days, the \texttt{IN} reaction in Equation~(\ref{eq:c+_sio}), i.e. conversion of SiO to CO, overtakes the \texttt{NN} reaction in 
Equation~(\ref{eq:c_oh}). %
The situation does not change until about 100 days. %

After 100 days, destruction processes are dominant as seen in the middle left panel in Figure~\ref{fig:one_zone_param1}. %
The primary destruction process is still the \texttt{IN} reaction via He$^+$ in Equation~(\ref{eq:he+_co}) and the ionization by Compton electrons in Equation~(\ref{eq:ion_ab}) and the dissociation by UV photons are secondary (the situation on the destruction processes continues until a few thousand days). %
The \texttt{NN} reaction in Equation~(\ref{eq:c_oh}) again becomes the primary formation process by overtaking the \texttt{IN} reaction in Equation~(\ref{eq:c+_sio}). %
After approximately 130 days, the \texttt{NN} reactions below also partake in the secondary formation processes as well as the \texttt{NN} reaction in Equation~(\ref{eq:o_ch}). %
\begin{align}
&{\rm C} + {\rm O_2} \lra {\rm CO} + {\rm O}, \label{eq:c_o2} \\
&{\rm O} + {\rm CS} \lra {\rm CO} + {\rm S}, \label{eq:o_cs} \\
&{\rm C} + {\rm SO} \lra {\rm CO} + {\rm S}. \label{eq:c_so} 
\end{align}
Here, O$_2$, CS, and SO are mainly produced by the \texttt{NN} reactions, O + OH $\lra$ O$_2$ + H, C + SO $\lra$ CS + O, and S + OH $\lra$ SO + H, respectively. %
It is interpreted that after 100 days, those molecules are converted to CO through the \texttt{NN} reactions in Equations~(\ref{eq:c_o2}), (\ref{eq:o_cs}), and (\ref{eq:c_so}). %
Among the secondary formation reactions, i.e., the \texttt{IN} reaction in Equation~(\ref{eq:c+_sio}), the \texttt{CE} reaction in Equation~(\ref{eq:h_co+}), and the \texttt{NN} reactions in Equations~(\ref{eq:o_ch}), (\ref{eq:c_o2}), (\ref{eq:o_cs}), and (\ref{eq:c_so}), the significance of each reaction depends on time. %
However, overall the situation continues until 300 days. %

After 300 days, the formation processes overwhelm the destruction processes. %
The primary formation process is still the \texttt{NN} reaction in Equation~(\ref{eq:c_oh}) and the secondary formation reactions are the \texttt{CE} reaction in Equation~(\ref{eq:h_co+}) (until about 600 days) and the \texttt{NN} reactions in Equations~(\ref{eq:o_ch}), (\ref{eq:c_o2}), (\ref{eq:o_cs}), and (\ref{eq:c_so}). %
The primary destruction process is the \texttt{IN} reaction in Equation~(\ref{eq:he+_co}) and the secondary destruction processes are the dissociation by UV photons and the ionization and destruction by Compton electrons in Equations~(\ref{eq:ion_ab}) and (\ref{eq:destruction}) until a few thousand days. After that, mostly the \texttt{NN} reaction below contributes to the destruction, although the amount of CO does not vary so much at this later phase. %
\begin{align}
{\rm OH} + {\rm CO} \lra {\rm CH} + {\rm O}_2. \label{eq:oh_co}
\end{align}

\begin{figure*}
\begin{minipage}{0.5\hsize}
\begin{center}
\hs{-1.5}
\includegraphics[width=9.5cm,keepaspectratio,clip]{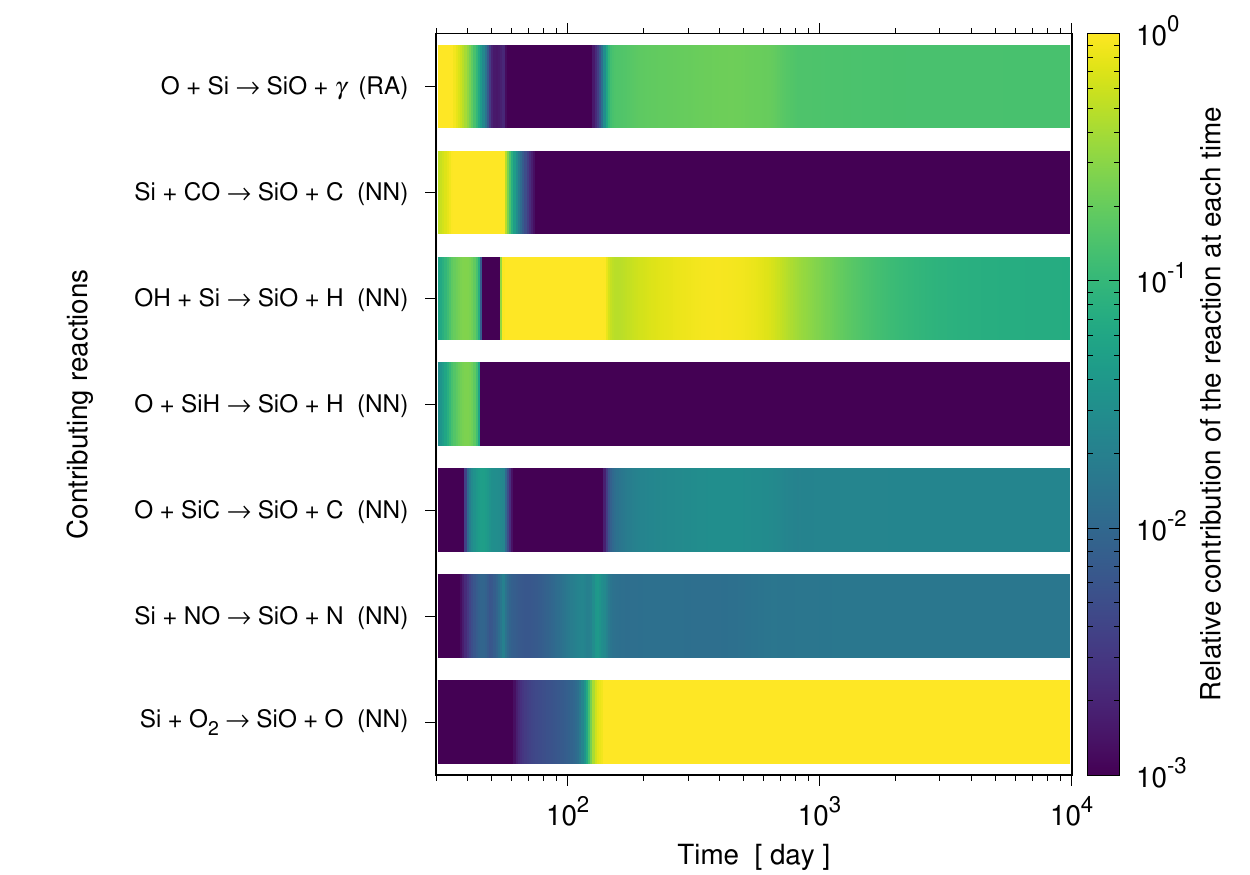}
\end{center}
\end{minipage}
\begin{minipage}{0.5\hsize}
\begin{center}
\hs{-1.5}
\includegraphics[width=9.5cm,keepaspectratio,clip]{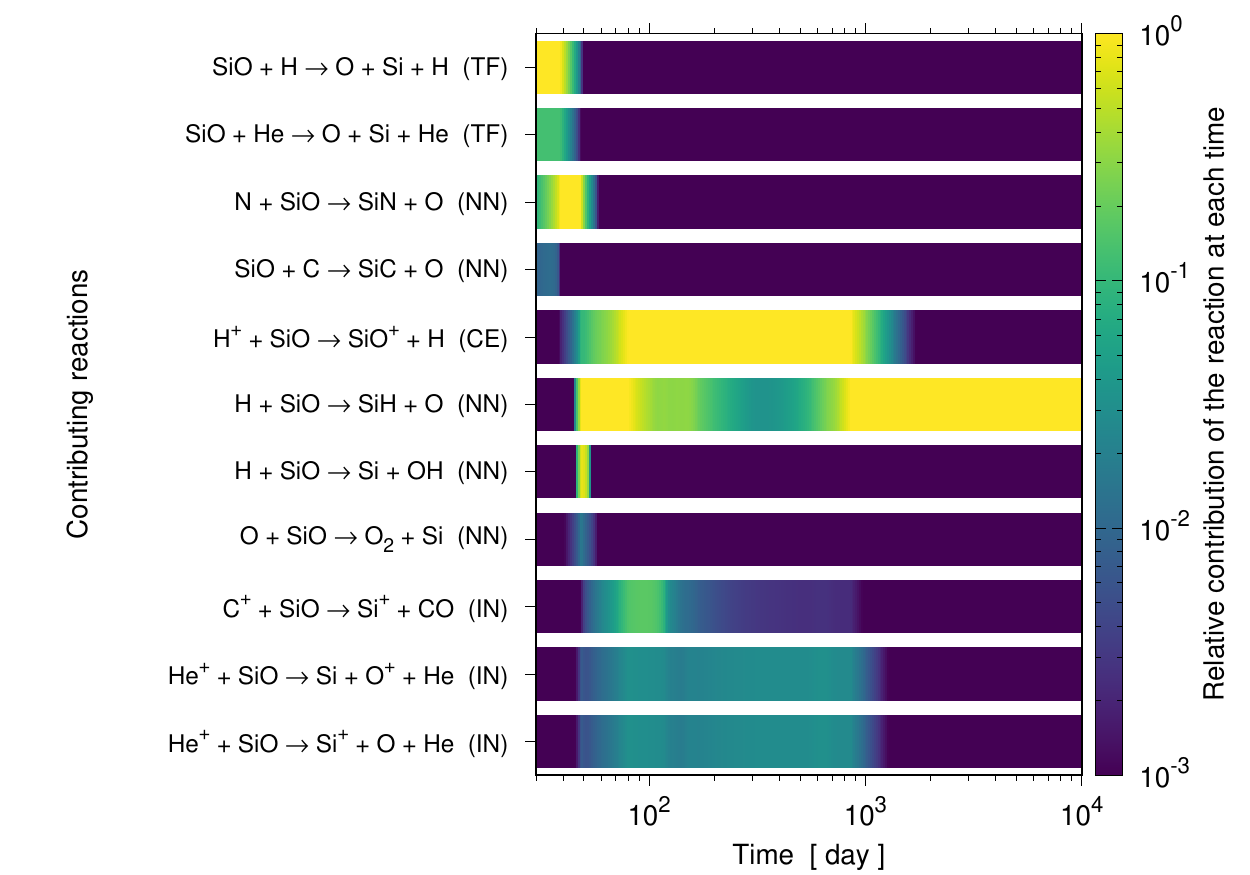}
\end{center}
\end{minipage}
\caption{Same as Figure~\ref{fig:co_reac} but for silicon monoxide (SiO).}
\label{fig:sio_reac}
\end{figure*}

Hereafter, similarly to CO, the formation and destruction processes of SiO are described. %
In Figure~\ref{fig:sio_reac}, contributing chemical reactions for the formation and destruction of SiO in which SiO is directly involved are shown as a function of time for reference. As seen in the left panel, the contributing SiO formation reactions are limited compared with those for CO. %
 
From the start of the molecule formation ($\sim$ 30 days) to 40 days, formation processes are dominant as seen in the middle left panel in Figure~\ref{fig:one_zone_param1}. Initially, the primary formation process is the \texttt{RA} reaction below. %
\begin{equation}
{\rm Si} + {\rm O} \lra {\rm SiO} + \gamma. \label{eq:sio_rad}
\end{equation}
The secondary formation reactions are the \texttt{NN} reaction in Equation~(\ref{eq:si_co}) mentioned as a destruction process of CO and the \texttt{NN} reaction below. %
\begin{align}
{\rm Si} + {\rm OH} \lra {\rm SiO} + {\rm H}, \label{eq:si_oh} 
\end{align}
where OH is mostly produced by the \texttt{NN} reaction, O $+$ H$_2$ $\lra$ OH $+$ H, as mentioned above. %
The destruction processes of the initial phase are thermal fragmentation reactions (\texttt{TF} reactions) with hydrogen and helium described in Section~{\ref{subsubsec:thermal_frag}}. %
The secondary destruction process is the \texttt{NN} reaction below. %
\begin{align}
&{\rm N} + {\rm SiO} \lra {\rm SiN} + {\rm O}. \label{eq:n_sio}
\end{align}
After 35 days, the \texttt{NN} reaction in Equation~(\ref{eq:si_co}) (CO destruction process) overtakes the \texttt{RA} reaction in Equation~(\ref{eq:sio_rad}) to become the primary formation process and the \texttt{NN} reaction below is added to the secondary formation reactions. %
\begin{align}
&{\rm O} + {\rm SiH} \lra {\rm SiO} + {\rm H}, \label{eq:o_sih}
\end{align}
where SiH is primarily produced by the \texttt{NN} reaction, Si + H$_2$ $\lra$ SiH + H, at this phase. %

After 40 days, the formation and destruction processes compete with each other. %
The primary formation process is the \texttt{NN} reaction in Equation~(\ref{eq:si_co}) for a while (until about 55 days) and the contribution of the \texttt{RA} reaction in Equation~(\ref{eq:sio_rad}) becomes gradually small. %
The destruction process (\texttt{NN} reaction) in Equation~(\ref{eq:n_sio}) overtakes \texttt{TF} reactions of SiO with hydrogen and helium. %
After 50 days, the \texttt{NN} reaction below partakes in the secondary formation process. %
\begin{align}
{\rm O} + {\rm SiC} \lra {\rm SiO} + {\rm C}, \label{eq:o_sic_sio}
\end{align}
where SiC is mainly formed by the \texttt{RA} reaction between silicon and carbon and the \texttt{NN} reaction, C + SiH $\lra$ SiC + H, as mentioned above. %
As the primary destruction processes, the \texttt{NN} reactions below are added by overtaking the \texttt{NN} reaction in Equation~(\ref{eq:n_sio}). %
\begin{align}
{\rm H} + {\rm SiO} \lra {\rm SiH} + {\rm O}, \label{eq:h_sio_sih} \\
{\rm H} + {\rm SiO} \lra {\rm OH} + {\rm Si}. \label{eq:h_sio_oh} 
\end{align}
Simultaneously, the contribution from the \texttt{IN} reaction in Equation~(\ref{eq:c+_sio}) (formation process of CO) and the \texttt{CE} and \texttt{IN} reactions below gradually increase as secondary destruction processes. %
\begin{align}
&{\rm H^+} + {\rm SiO} \lra {\rm SiO^+} + {\rm H}, \label{eq:h+_sio} \\
&{\rm He^+} + {\rm SiO} \lra {\rm Si} + {\rm O^+} + {\rm He}, \label{eq:he+_sio_o+} \\
&{\rm He^+} + {\rm SiO} \lra {\rm Si^+} + {\rm O} + {\rm He}. \label{eq:he+_sio_si+}
\end{align}
Soon after that, the contribution of the \texttt{NN} reaction in Equation~(\ref{eq:h_sio_oh}) as the destruction process becomes small (the inverse reaction in turn dominates the forward reaction). %
After 60 days, the \texttt{NN} reaction in Equation~(\ref{eq:si_oh}) overtakes the \texttt{NN} reaction in Equation~(\ref{eq:si_co}) to be the primary formation process. %
The contributions of the \texttt{NN} reactions below also gradually increase as the secondary formation processes. %
\begin{align}
&{\rm Si} + {\rm O_2} \lra {\rm SiO} + {\rm O}, \label{eq:si_o2} \\
&{\rm Si} + {\rm NO} \lra {\rm SiO} + {\rm N}, \label{eq:si_no}
\end{align} 
where O$_2$ is mainly produced by the \texttt{NN} reaction, O + OH $\lra$ O$_2$ + H, and NO is formed by the \texttt{NN} reaction, O + SiN $\lra$ NO + Si ; %
SiN is produced by the \texttt{NN} reaction in Equation~(\ref{eq:n_sio}). %
After 80 days, the \texttt{CE} reaction in Equation~(\ref{eq:h+_sio}) overtakes the \texttt{NN} reaction in Equation~(\ref{eq:h_sio_sih}) to become the primary destruction process. %
Overall the situation continues until 100 days. %

After 100 days, destruction processes are dominant as seen in the middle left panel in Figure~\ref{fig:one_zone_param1}. %
The primary formation process is the \texttt{NN} reaction in Equation~(\ref{eq:si_oh}) (until 140 days). %
The primary destruction process is the \texttt{CE} reaction in Equation~(\ref{eq:h+_sio}); %
the \texttt{NN} reaction in Equation~(\ref{eq:h_sio_sih}) and the \texttt{IN} reactions in Equations~(\ref{eq:c+_sio}), (\ref{eq:he+_sio_o+}), and (\ref{eq:he+_sio_si+}) are followed as the secondary destruction processes. %
After 140 days, the \texttt{NN} reaction in Equation~(\ref{eq:si_o2}) overtakes the \texttt{NN} reaction in Equation~(\ref{eq:si_oh}) to be the primary formation process. %
The contribution of the formation processes, the \texttt{RA} reaction in Equation~(\ref{eq:sio_rad}) and the \texttt{NN} reaction in Equation~(\ref{eq:o_sic_sio}), recover again to some extent. %
The contribution of the destruction reactions, the \texttt{NN} reaction in Equation~(\ref{eq:h_sio_sih}) and the \texttt{IN} reaction in Equation~(\ref{eq:c+_sio}), becomes smaller. %
After that, the situation overall continues until 300 days. %

After 300 days, the formation processes become dominant until the reactions are settled ($\sim$ 1000 days). %
The contributing formation and destruction processes, however, are overall the same as the situation at $\lesssim$ 300 days. %
After $\sim$ 1000 days, the contribution of the formation process, the \texttt{NN} reaction in Equation~(\ref{eq:si_oh}), becomes a bit small. %
The destruction process, the \texttt{NN} reaction in Equation~(\ref{eq:h_sio_sih}), becomes the primary process by overtaking the \texttt{CE} reaction in Equation~(\ref{eq:h+_sio}). %

In this section, by taking the representative case, important reactions for CO and SiO were described. %
For some of the reactions, in particular, the \texttt{NN} and \texttt{IN} reactions, listed above, hydrogen and helium atoms and the corresponding ions, H$^+$ and He$^+$, are involved. %
Such reactions play a non-negligible role in the formation and destruction processes of CO and SiO. %
In the one-zone approximation, as partly mentioned in the previous section, the mixing of hydrogen and helium into the ejecta core and the mixing of $^{56}$Ni, which is important for the ionization and destruction of molecules, into outer layers would be overestimated. %
Therefore, the description in this section may not be universal but as a reference, it may still be useful. %
Actually, most of the reactions related to the CO and SiO formation and destruction referred to in the later sections are covered above. %
The impact of matter mixing with more realistic 1D models on molecule formation is discussed in the next section. %

\subsection{1D calculation results} \label{subsec:1d_results} 

In this section, 1D calculation results are presented according to the strategy described in Section~\ref{sec:strategy}. %
In Section~\ref{subsubsec:1d_param}, as in the one-zone calculations, with several cases, the evolution of physical quantities is described and the most reasonable parameter set is selected. %
In Section~\ref{subsubsec:1d_single_star}, the results of the angle-averaged 1D profiles of the 3D models with the binary merger and the single-star progenitor models are presented including the description of the molecules other than CO and SiO. %
Section~\ref{subsubsec:1d_comp_sphel} is devoted to the comparison with the spherical cases. %
Finally, in Section~\ref{subsubsec:1d_angle}, differences in the specified directions in the 3D models are presented. %

\subsubsection{The fiducial case and the dependence on the parameters, 
$f_{\rm h}$, $f_{\rm d}$, and $t_{\rm s}$} \label{subsubsec:1d_param} 

As done in Section~\ref{subsec:one_zone_results} for one-zone calculations, to find a reasonable set (reasonable sets) of parameters, $f_{\rm h}$, $f_{\rm d}$, and $t_{\rm s}$, for 1D calculations, based on the angle-averaged 1D profiles of the model b18.3-high \citep{2020ApJ...888..111O}, in total 100 cases are calculated with the combinations of the parameters, $f_{\rm h} = 10^{-4}$, 5~$\times$~10$^{-4}$, 10$^{-3}$, 5~$\times$~10$^{-3}$, 10$^{-2}$, $f_{\rm d} = 10^{-2}$, 5~$\times$~10$^{-2}$, 10$^{-1}$, 5~$\times$~10$^{-1}$, 1.0, $t_{\rm s} =$ 200, 300, 500, and $\infty$ (practically $f_{\rm red} = 1.0$) days. %

In the following subsections, the results of two representative cases are presented in Sections~\ref{para:1d_param1} and \ref{para:1d_param2}, respectively, as candidates for the best parameter set to be fixed for the latter discussion. %
Then, in Section~\ref{para:1d_selec}, with a few additional cases, the parameter dependence is described, and among the two cases, the most reasonable parameter set is selected. %
For the selection of the best parameter set (the two candidates), the three criteria listed in Section~\ref{para:one_zone_selec} are taken into account. %
Section~\ref{para:1d_chemi_reac} is devoted to the description of the important chemical reactions for CO and SiO for the reasonable parameter set as a representative. %

\begin{figure*}
\begin{minipage}{0.5\hsize}
\begin{center}
\includegraphics[width=8.5cm,keepaspectratio,clip]{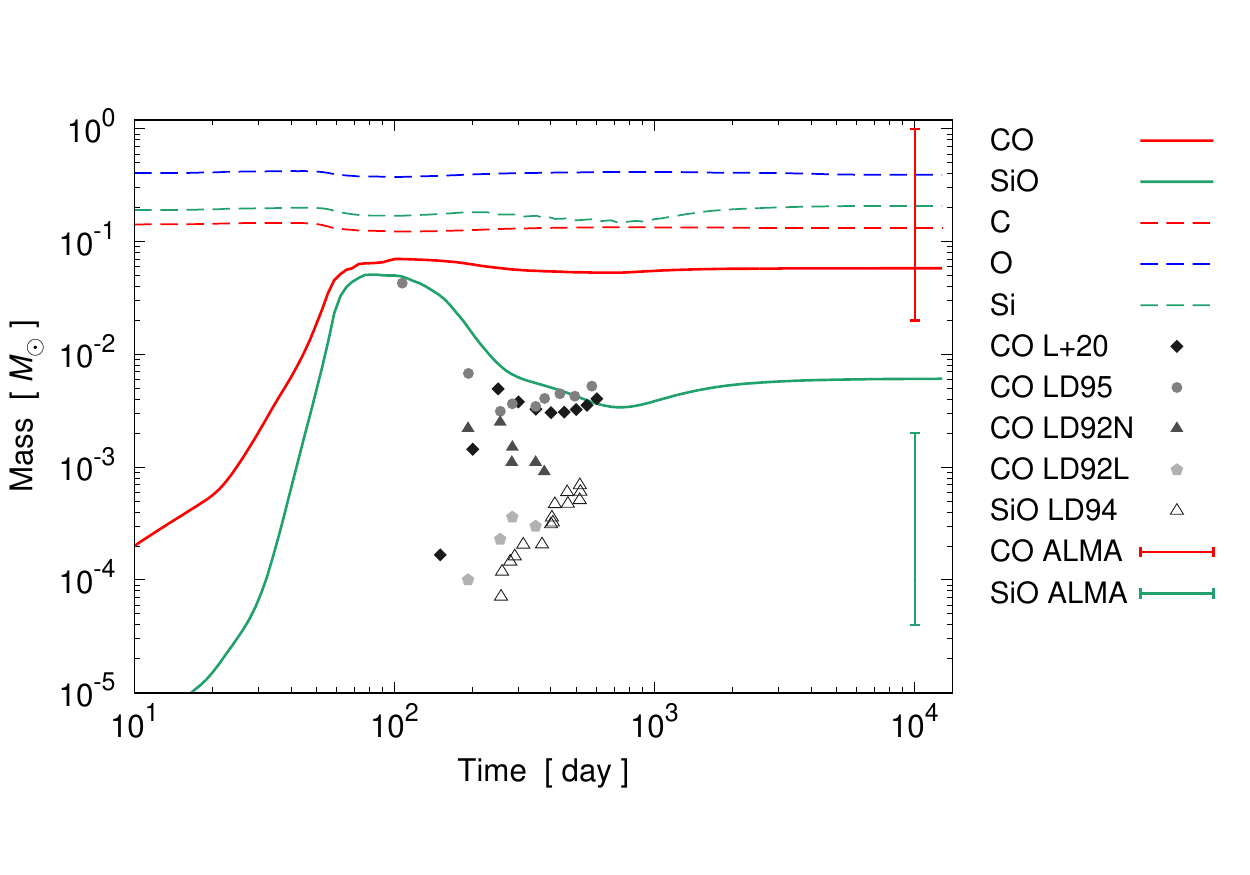}
\end{center}
\end{minipage}
\begin{minipage}{0.5\hsize}
\begin{center}
\includegraphics[width=8.5cm,keepaspectratio,clip]{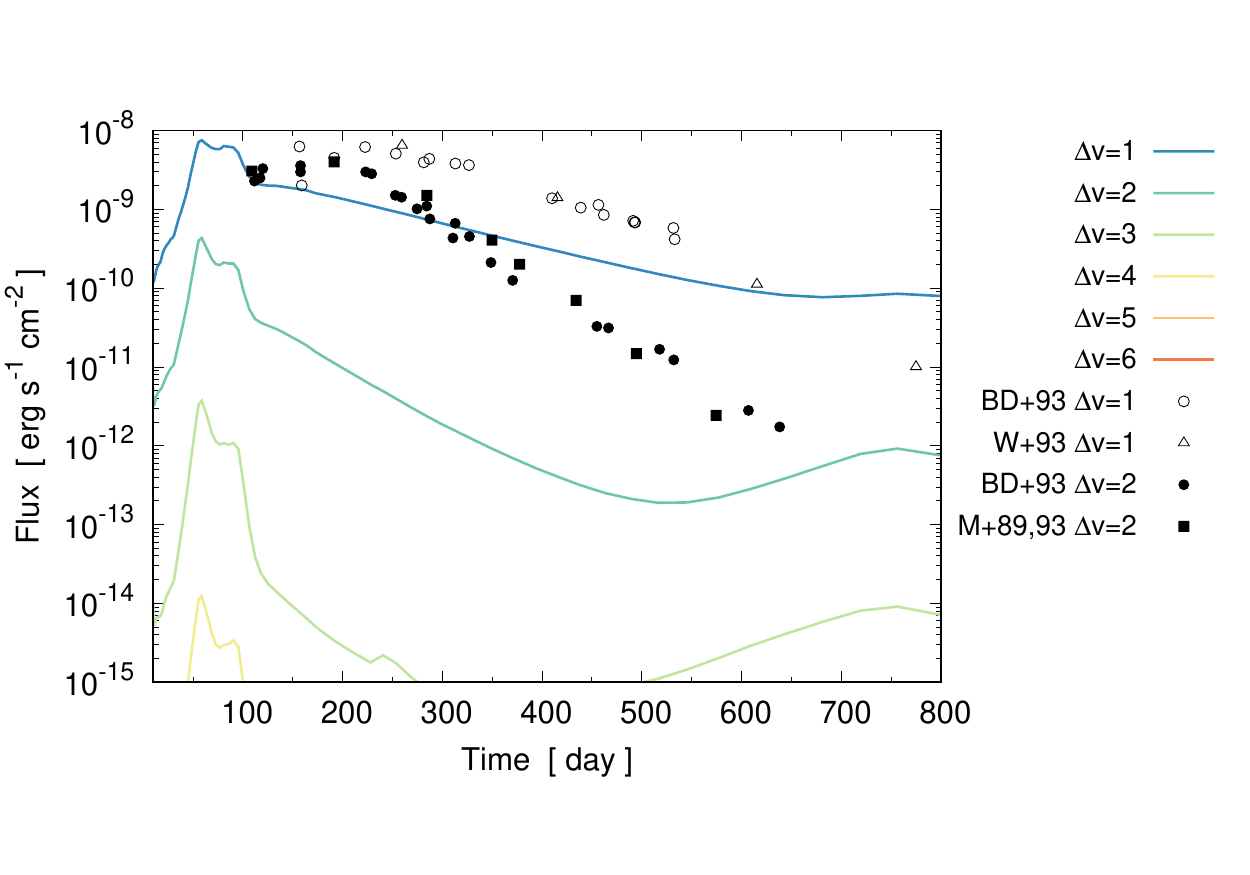}
\end{center}
\end{minipage}
\\
\begin{minipage}{0.5\hsize}
\vs{-1.}
\begin{center}
\includegraphics[width=8.5cm,keepaspectratio,clip]{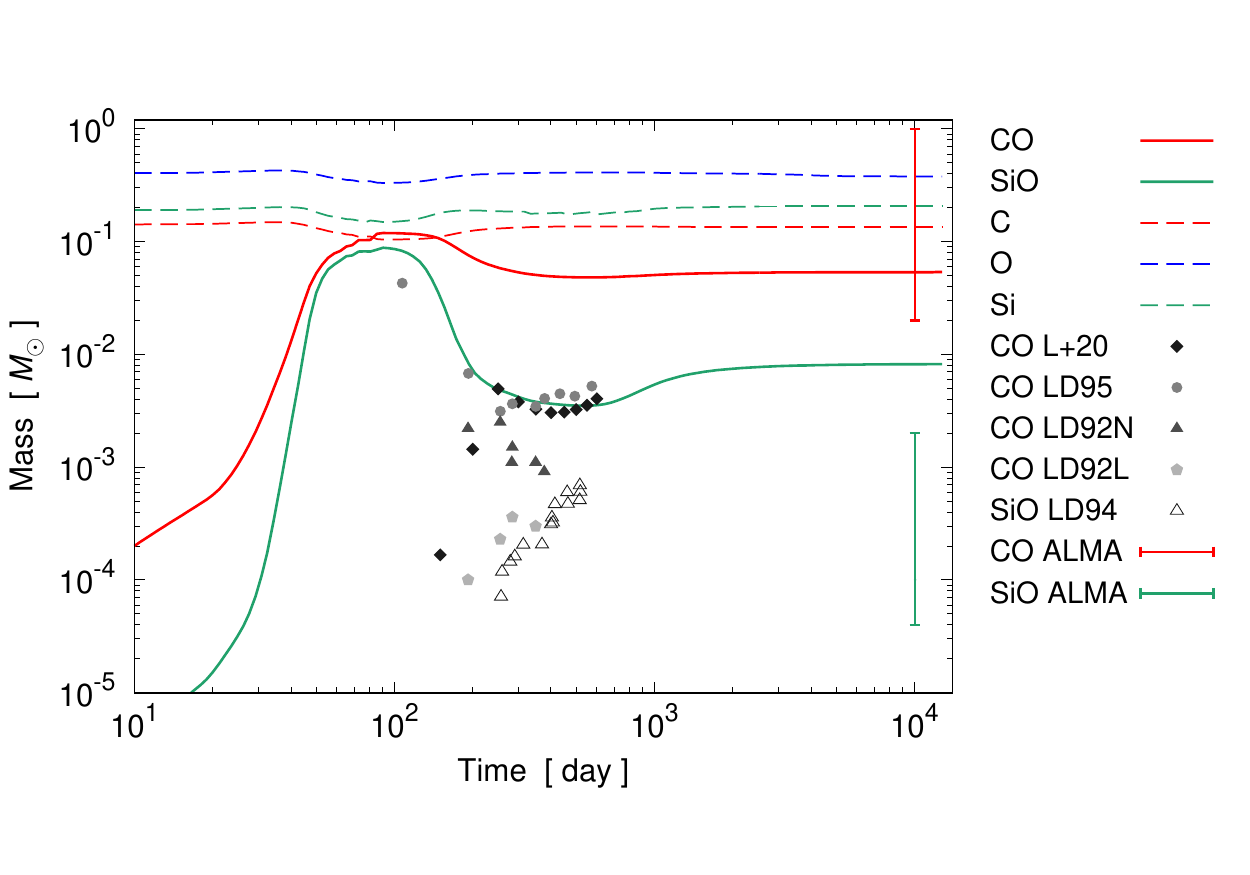}
\end{center}
\vs{-1.}
\end{minipage}
\begin{minipage}{0.5\hsize}
\vs{-1.}
\begin{center}
\includegraphics[width=8.5cm,keepaspectratio,clip]{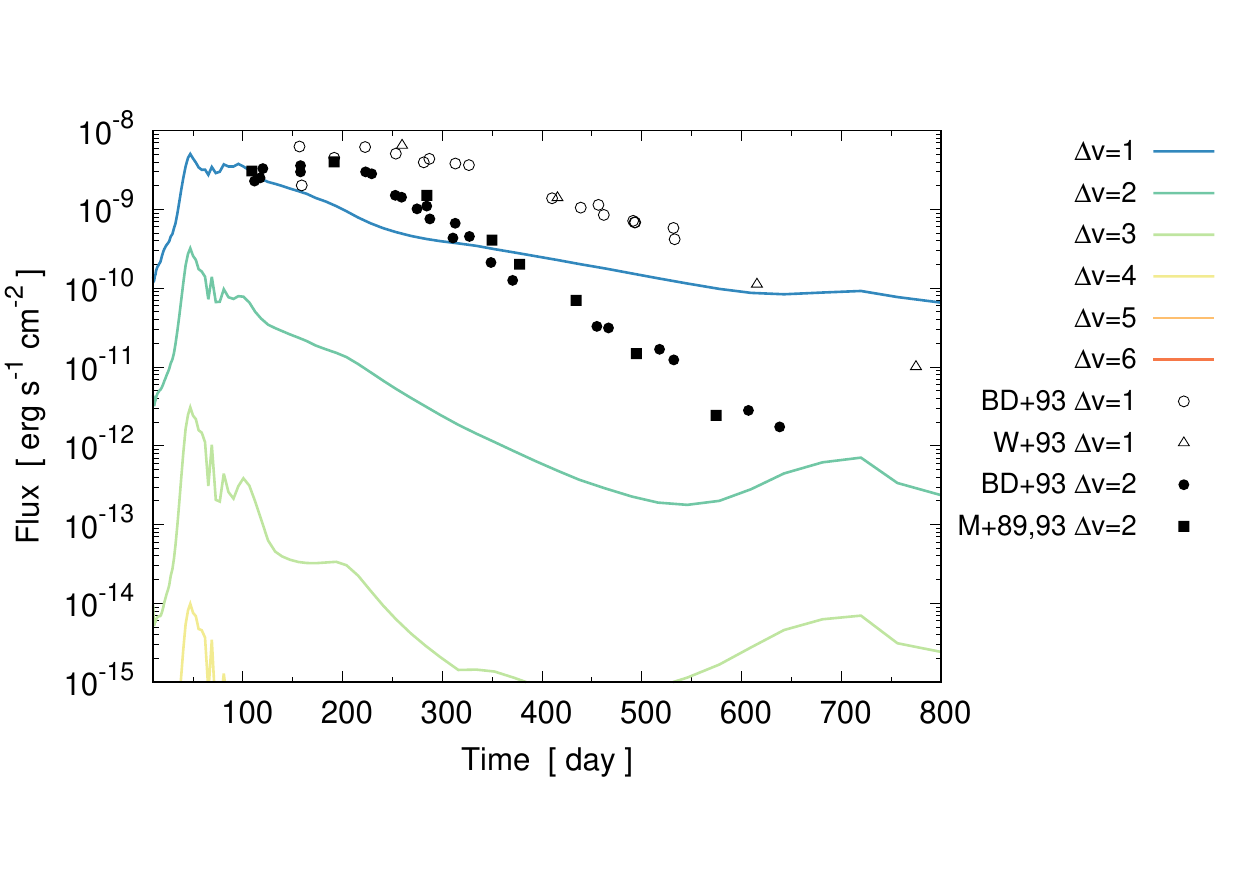}
\end{center}
\vs{-1.}
\end{minipage}
\caption{1D calculation results with the angle-averaged profiles for the model b18.3-high \citep{2020ApJ...888..111O} with the parameters, 
$f_{\rm h} = 10^{-2}$, $f_{\rm d} =$ 1.0, and $t_{\rm s} =$ 500 days (top panels) and $f_{\rm h} = 5 \times 10^{-3}$, $f_{\rm d} =$ 1.0, and $t_{\rm s} =$ 500 days (bottom panels: corresponding to the model b18.3-mean). %
Left panels: time evolution of the total amounts of CO and SiO and the seed atoms, carbon, oxygen, and silicon, compared with the estimations (including theoretical calculations) for CO and SiO in previous studies: %
LD92 \citep{1992ApJ...396..679L}, LD94 \citep{1994ApJ...428..769L}, LD95 \citep{1995ApJ...454..472L}, ALMA \citep{2017MNRAS.469.3347M}, and L+20 \citep{2020A&A...642A.135L}. %
Right panels: time evolution of the fluxes for CO vibrational bands, ${\it \Delta}v=1$ (fundamental), ${\it \Delta}v=2$ (first overtone), \ldots, ${\it \Delta}v=6$, compared with the observed light curves for ${\it \Delta}v=1$ \citep[BD93; W$+$93:][respectively]{1993A&A...273..451B,1993ApJS...88..477W} and ${\it \Delta}v=2$ \citep[M+89, 93; BD93:][respectively]{1989MNRAS.238..193M,1993MNRAS.261..535M,1993A&A...273..451B}.}
\label{fig:1d_param1}
\end{figure*}

\paragraph{The case with $f_{\rm h} = 10^{-2}$, $f_{\rm d} = 10^{-2}$, and $t_{\rm s} = 500$ days} \label{para:1d_param1}

\begin{figure*}
\begin{minipage}{0.5\hsize}
\begin{center}
\includegraphics[width=8.cm,keepaspectratio,clip]{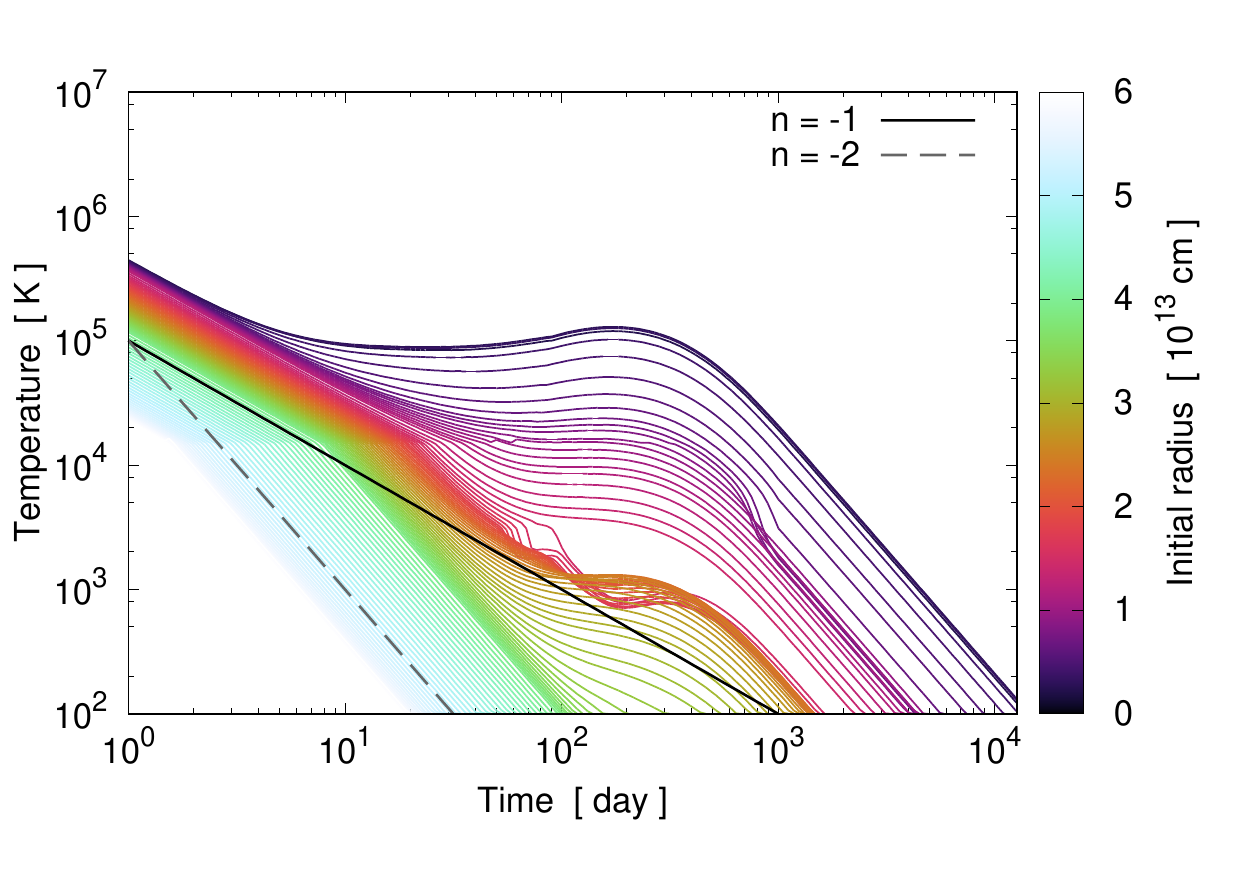}
\end{center}
\vs{-0.5}
\end{minipage}
\begin{minipage}{0.5\hsize}
\begin{center}
\includegraphics[width=8.cm,keepaspectratio,clip]{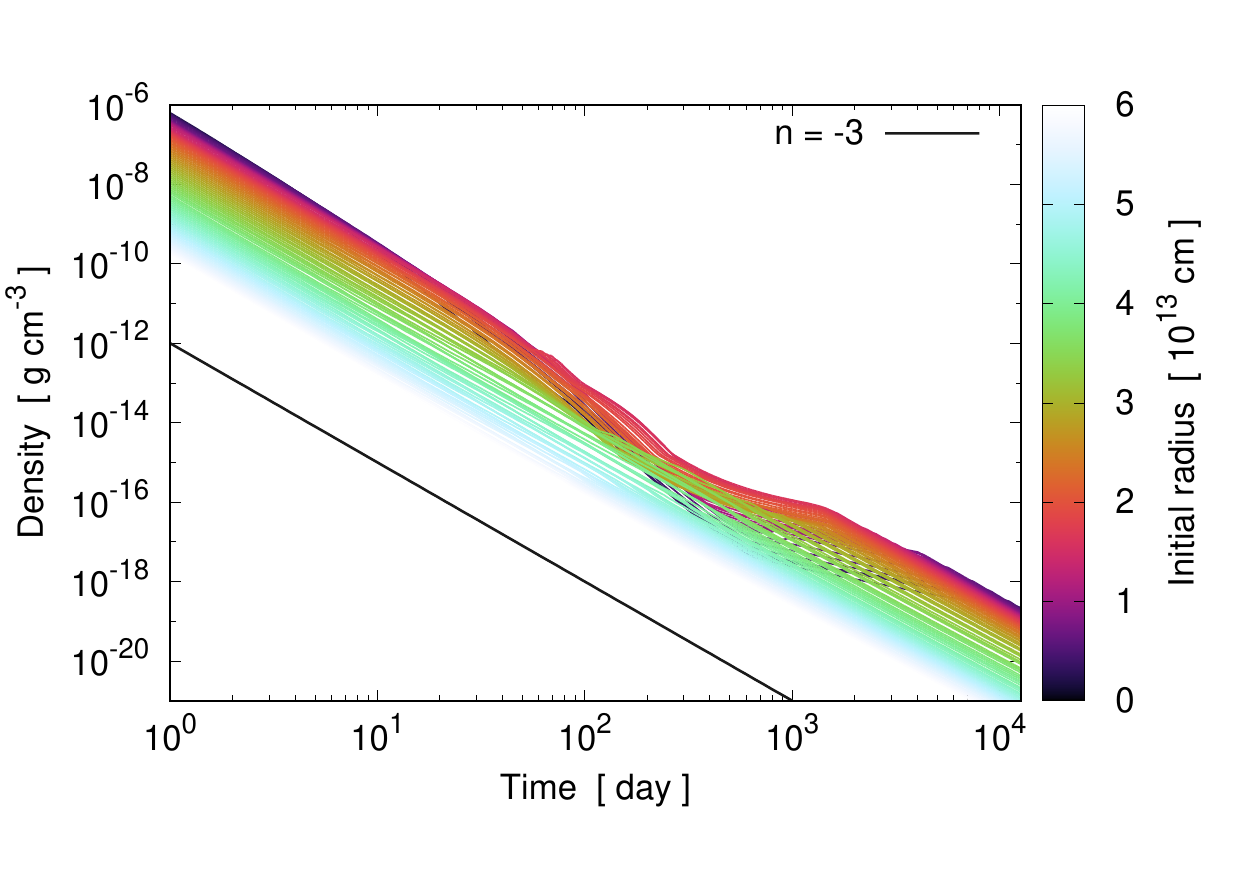}
\end{center}
\vs{-0.5}
\end{minipage}
\caption{1D calculation results with the angle-averaged profiles for the model b18.3-high \citep{2020ApJ...888..111O} with the parameters, $f_{\rm h} = 10^{-2}$, $f_{\rm d} =$ 1.0, and $t_{\rm s} =$ 500 days. %
Left (right) panel: time evolution of the gas temperatures (gas densities) of the tracer particles; the colors denote the initial positions of the particles. As a reference, power-law evolutions of the powers of $-1$, $-2$ (left), and $-3$ (right) are also shown.} %
\label{fig:rho_dens_1d}
\end{figure*}

\hspace{\parindent} 
In the upper panels in Figure~\ref{fig:1d_param1}, the results with the parameters, $f_{\rm h} = 10^{-2}$, $f_{\rm d} =$ 10$^{-2}$, and $t_{\rm s} =$ 500 days are shown as one of the candidates for the reasonable parameter set. %

As seen in the calculated fluxes of CO vibrational bands (the right panel), the peak flux level ($\sim$ 10$^{-8}$ erg s$^{-1}$ cm$^{-2}$) of the fundamental band (${\it \Delta}v=1$) is comparable with the observed peak fluxes but the calculation fails to reproduce the timing (200--250 days) and the dome-like feature of the observed peaks same as in the one-zone calculations. %
The qualitative feature of the fluxes of the other bands is similar to that of the fundamental band but the higher ${\it \Delta}v$ is, the lower the flux; the flux of the first overtone band (${\it \Delta}v=2$) is rather underestimated compared with the observed one even in early phases before 100 days. %
In the one-zone calculation results presented in Section~\ref{subsubsec:one_zone_trend} (see, the right panels in Figure~\ref{fig:one_zone_param1}), the flux of the first overtone band can be comparable or even higher compared with that of the fundamental band before 100 days. %
The discrepancy between the one-zone and 1D calculations may partly be attributed to the more effective mixing of $^{56}$Ni in the one-zone calculations than that in the 1D calculations and the resultant higher electron densities 
due to more effective ionization by Compton electrons, which may enhance the excitation of higher CO vibrational levels. %
The increase of fluxes after 500 days is the contribution from some inner particles in which the formation of CO and CO ro-vibrational transitions start at such later phases. %

The calculated flux levels depend on the parameter $t_{\rm s}$ as demonstrated in Appendix~\ref{app:fred} for one-zone calculations. %
The higher (lower) $t_{\rm s}$ value, the higher (lower) the peak flux levels. %
Among the calculated cases, the parameter value of $t_{\rm s} =$ 500 days results in better flux levels not only for this case but also for the other 1D calculations (the better value is different from one for the one-zone calculations, i.e., $t_{\rm s} =$ 200 days). %
Then, hereafter only the cases of the parameter of $t_{\rm s} =$ 500 days are focused (if not specifically mentioned, $t_{\rm s} =$ 500 days). %

In Figure~\ref{fig:rho_dens_1d}, the time evolution of the gas temperatures (left) and densities (right) are shown by taking this specific parameter set as one of the representatives. %
The gas temperatures of inner (tracer) particles are efficiently heated due to the decay of $^{56}$Ni making peaks around 200 days. %
The particles whose initial positions (radii) are less than 3.5 $\times$ 10$^{13}$ cm are more or less affected by the heating at some point. %
Some particles around 1.5 $\times$ 10$^{13}$ cm undergo the cooling by CO ro-vibrational transitions around 100 days; the gas temperature evolution of such particles is a bit complicated since the heating and cooling are competing with each other. %
Some of the inner particles also go through the cooling at later phases ($\lesssim$ 1000 days). %
The evolution of the gas densities fluctuates on top of the power-law evolution of the power of $-3$ due to the effect described in Equation~(\ref{eq:thermal_i}). %
The behavior of the particles affected by the cooling is a bit complex reflecting the temperature evolution. The gas densities recover the original power-law evolution after the gas temperatures reach the minimum (100 K) by the assumption (see, Section~\ref{subsec:one_zone_calc}). %

The time evolution of the total amounts (contribution from all the particles) of CO and SiO (the left panel) shows early ($\sim$ 10 days) formation of CO and SiO and smoother increase of the amounts than those in the one-zone calculation models (see, e.g., the left panels in Figure~\ref{fig:one_zone_param1}). %
This is because of the variations in physical quantities among the particles as seen in Figure~\ref{fig:rho_dens_1d}. %
Both the amounts of CO and SiO reach about 10$^{-2}$--10$^{-1}$ $M_{\odot}$ by 100 days in contrast to the representative one-zone calculations (Figure~\ref{fig:one_zone_param1}), in which both the amounts exceed 10$^{-1}$ $M_{\odot}$ at the peak. %
Actually, the seed atoms for CO and SiO are not so consumed after 40 days compared with the representative one-zone cases (Figure~\ref{fig:one_zone_param1}). %
The calculated amounts of CO and SiO are overestimated roughly by one order of magnitude compared with the previous studies (points) and the decreasing trend of CO from 100 days to a few hundred days seen in LD95 (filled circles) is not well reproduced. %
After 100 days, in particular, SiO decreases by one order of magnitude by 600 days; CO slightly decreases during this phase. After that, the amount of SiO recovers by a factor of a few. The increasing and decreasing trends are overall milder than those of the representative one-zone cases. %
This is probably attributed to the wide range of physical quantities among the particles as mentioned and less effective matter mixing in the 1D calculations. %

\paragraph{The case with $f_{\rm h} = 5 \times 10^{-3}$, $f_{\rm d} = 1.0$, and $t_{\rm s} = 500$ days} \label{para:1d_param2}

\hspace{\parindent} 
In the lower panels in Figure~\ref{fig:1d_param1}, the 1D calculation results with the parameters, $f_{\rm h} = 5 \times 10^{-3}$, $f_{\rm d} =$ 1.0, and $t_{\rm s} =$ 500 days, are shown as another candidate for the reasonable parameter set. %

The qualitative features of the fluxes of CO vibrational bands are similar to the previous case; the fluxes peak before 100 days in contrast to the observed fundamental and the first overtone bands; the higher ${\it \Delta}v$ is, the lower the flux is throughout the evolution; fluxes increase to some extent after 500 days. %
The peak flux of the fundamental band is slightly lower than that of the previous case; instead, the flux is a bit higher and smoother at around 100 days. %

The qualitative features of the amounts of CO and SiO (the left panel) are similar to the previous case but quantitative differences can be recognized. %
The initial increase of CO and SiO is rather similar to the previous case. %
However, the amount of CO becomes greater than 10$^{-1}$ $M_{\odot}$ at the peak and the amount of SiO nearly reaches 10$^{-1}$ $M_{\odot}$. %
After 100 days both CO and SiO have a clear decreasing trend compared with the previous case; by 300 days, the amount of CO decreases by a factor of five in contrast to the previous case and the amount of SiO is smaller than that of the previous case at this phase. %
After 300 days, the amount of CO only slightly recovers, and the amount of SiO recovers by a factor of more than a few. The final amount of CO is comparable with the previous case, and the amount of SiO is slightly greater than that of the previous case. %

\paragraph{The selection of the best parameter set and the parameter dependence} \label{para:1d_selec}

\begin{figure}
\begin{minipage}{0.5\hsize}
\vs{-0.5}
\begin{center}
\includegraphics[width=8.cm,keepaspectratio,clip]{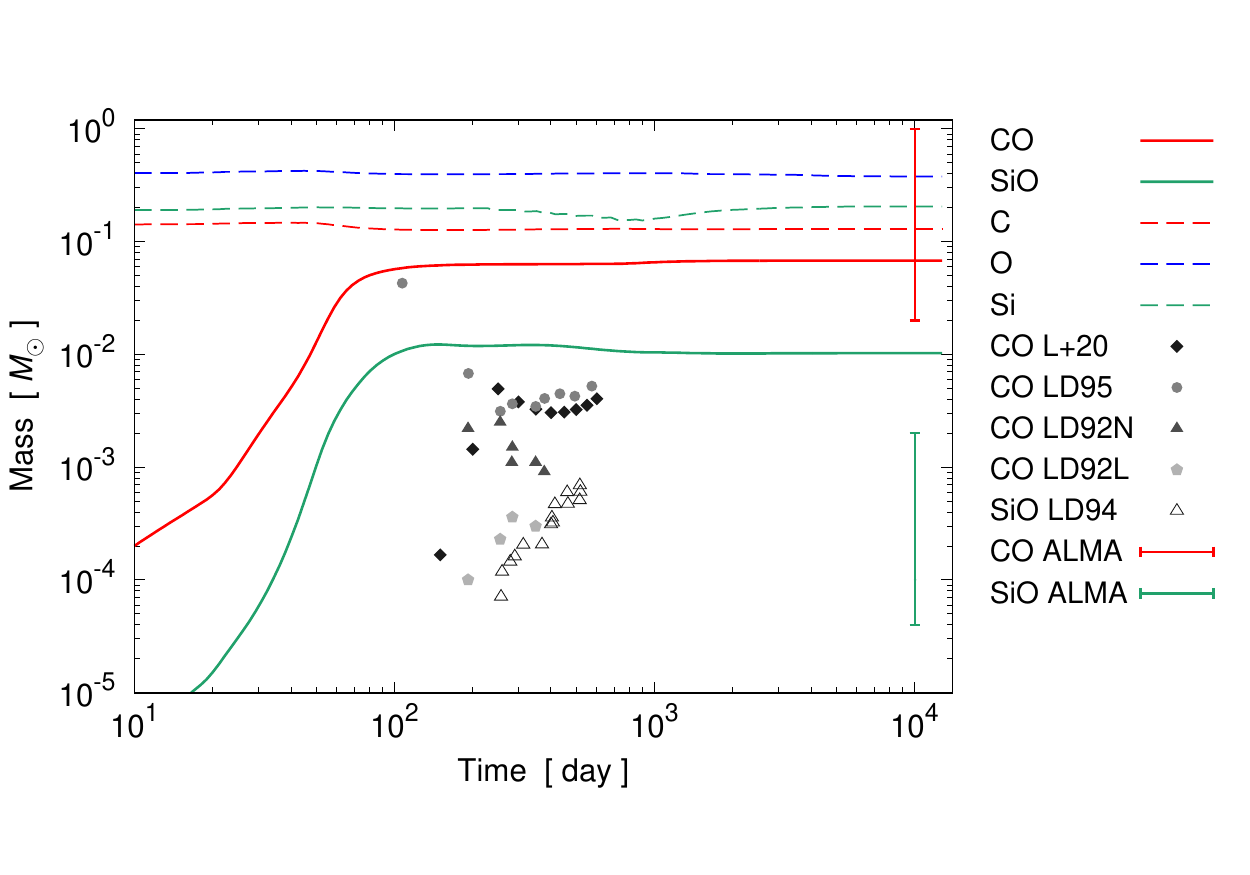}
\end{center}
\vs{-0.5}
\end{minipage}
\\
\begin{minipage}{0.5\hsize}
\vs{-0.7}
\begin{center}
\includegraphics[width=8.cm,keepaspectratio,clip]{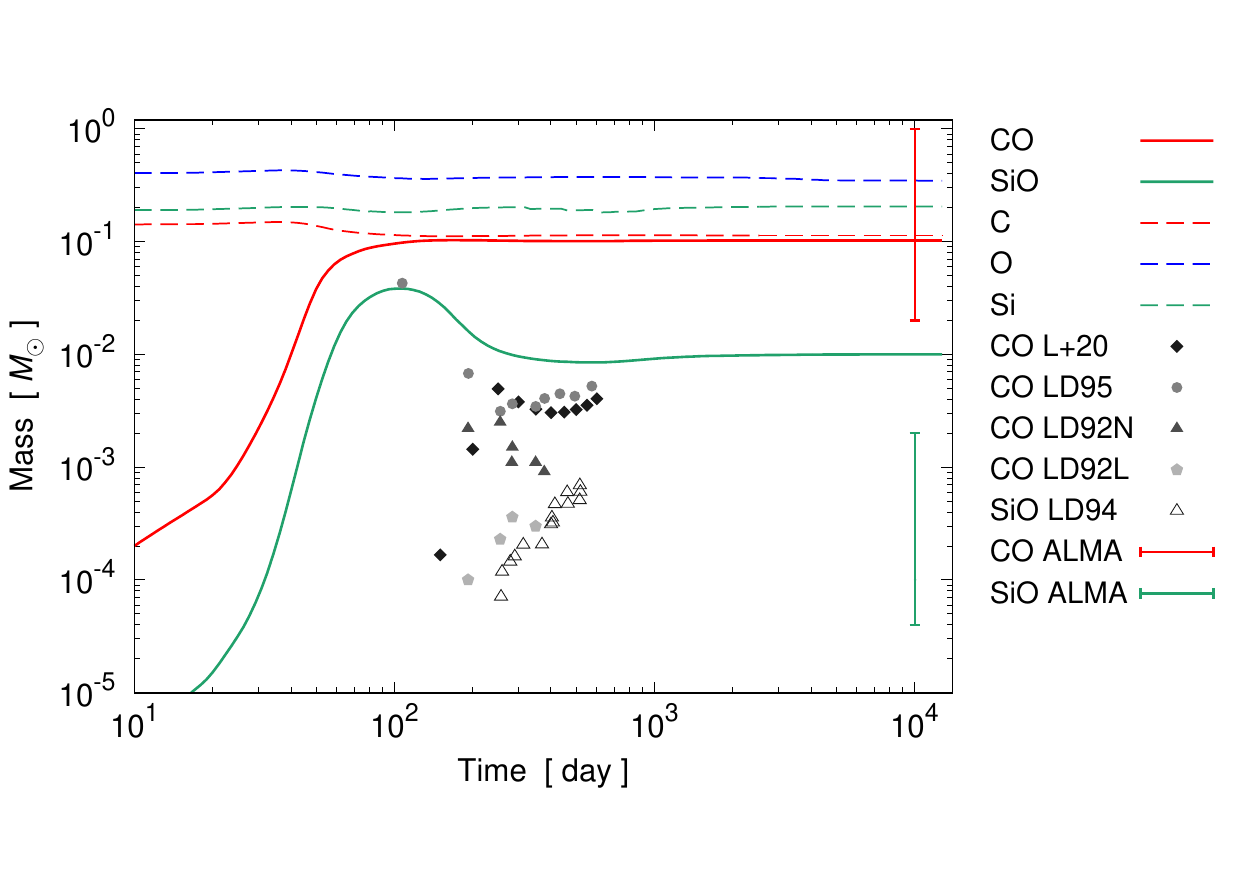}
\end{center}
\vs{-0.5}
\end{minipage}
\caption{1D calculation results with the angle-averaged profiles for the model b18.3-high \citep{2020ApJ...888..111O} with the parameters, $f_{\rm h} = 10^{-2}$, $f_{\rm d} =$ 10$^{-2}$, and $t_{\rm s} =$ 500 days (upper panel) and $f_{\rm h} = 5 \times 10^{-3}$, $f_{\rm d} =$ 10$^{-2}$, and $t_{\rm s} =$ 500 days (lower panel). %
The time evolution of the total amounts of CO and SiO, and the seed atoms, carbon, oxygen, and silicon, are shown. %
The points are the estimations for CO and SiO in the previous studies; for the details, see, the caption of Figure~\ref{fig:1d_param1}.} %
\label{fig:1d_param2}
\end{figure}

\begin{figure*}
\begin{minipage}{0.5\hsize}
\begin{center}
\includegraphics[width=8.cm,keepaspectratio,clip]{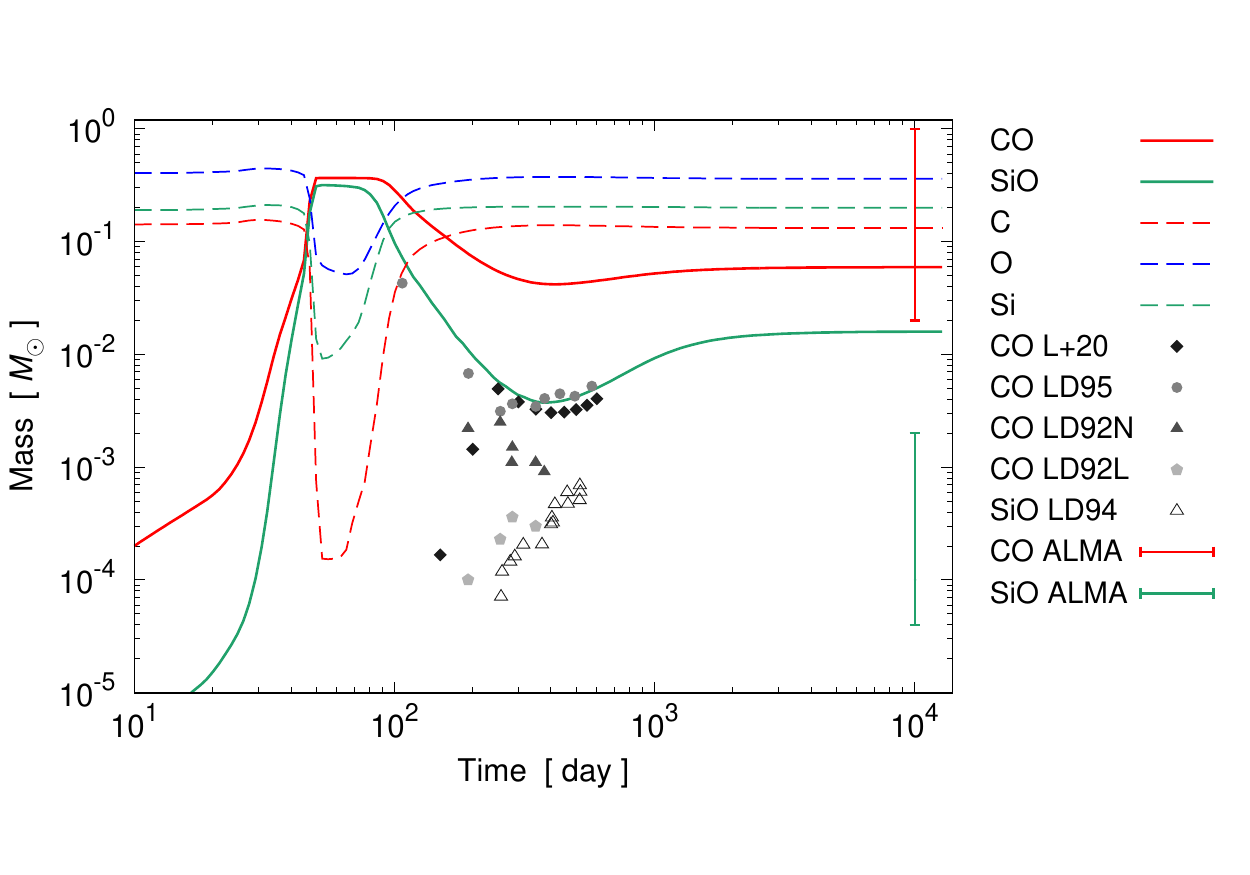}
\end{center}
\vs{-0.5}
\end{minipage}
\begin{minipage}{0.5\hsize}
\begin{center}
\includegraphics[width=8.cm,keepaspectratio,clip]{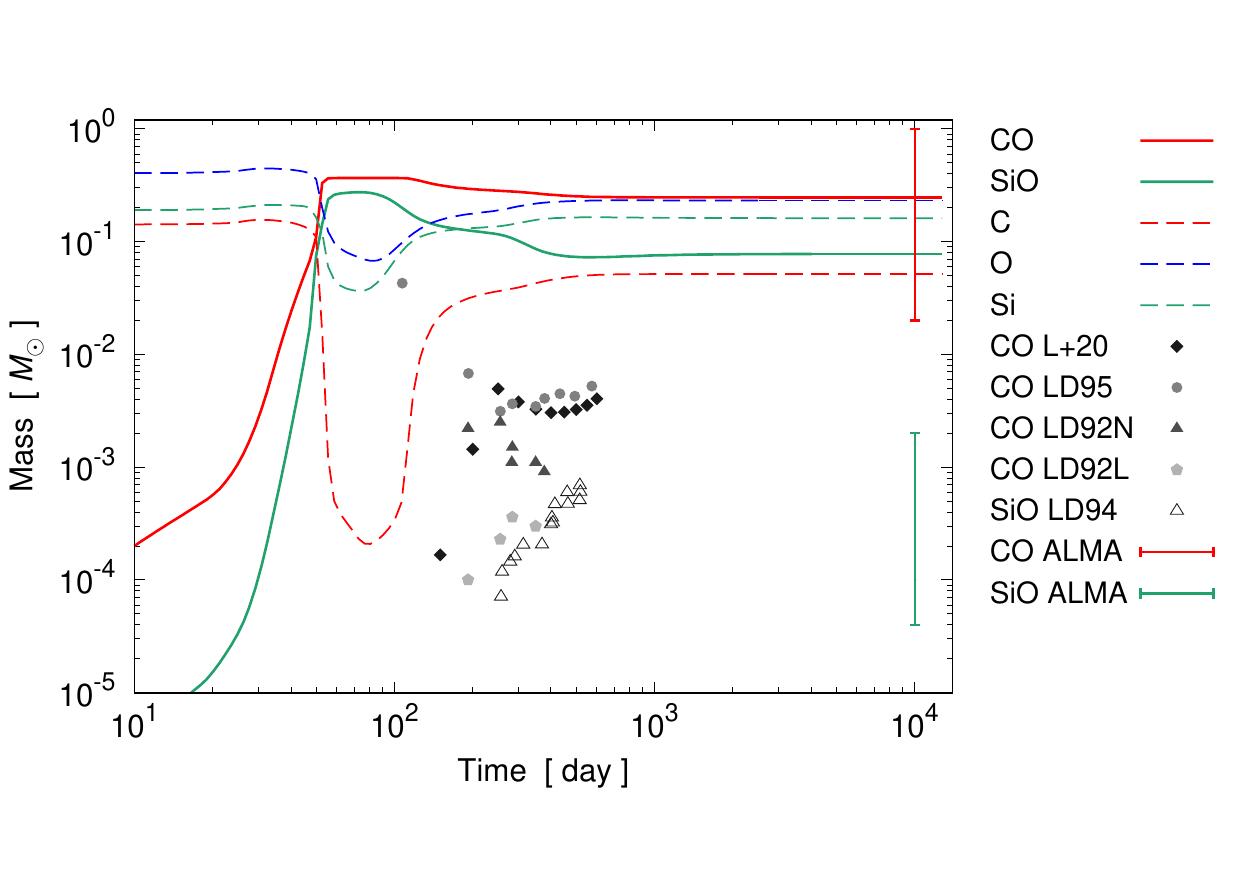}
\end{center}
\vs{-0.5}
\end{minipage}
\caption{Same as Figure~\ref{fig:1d_param2} but for the parameters of $f_{\rm h} = 10^{-4}$, $f_{\rm d} =$ 1.0, and $t_{\rm s} =$ 500 days (left panel) and $f_{\rm h} = 10^{-4}$, $f_{\rm d} =$ 10$^{-2}$, and $t_{\rm s} =$ 500 days (right panel).}
\label{fig:1d_param3}
\end{figure*}

\hspace{\parindent} 
To find the reasonable parameter set, the three criteria to be satisfied mentioned in Section~\ref{para:one_zone_selec} are taken into account for 1D calculations too. %
In the 1D calculations, the third point, which is related to the gas temperature, is not so problematic as long as the calculations are performed with the reduction factor $f_{\rm red}$ with $t_{\rm s} =$ 500 days. %
Actually, outer particles tend to reach such a low temperature at a few hundred days even without the cooling by CO ro-vibrational transitions; %
unless the majority of the particles have such an immediate low temperature, we do not make an issue out of the model. %
Then, the first and second points matter. %
In this regard, among the calculated 1D cases, the two cases presented above in this section would be candidates for the reasonable parameter set to be fixed for later discussion. %

Before fixing the parameter set, the overall features for other calculated cases and the dependence on the values of the parameters, $f_{\rm h}$ and $f_{\rm d}$, are described. %
To see the impact of the parameter $f_{\rm d}$, as the counterparts of the two cases above, the results with the $f_{\rm d}$ value of the opposite end of the two cases, i.e., $f_{\rm d} = 10^{-2}$, are shown in Figure~\ref{fig:1d_param2}, respectively. %
As seen in the results with the parameters, $f_{\rm h} = 10^{-2}$, $f_{\rm d} =$ 10$^{-2}$, and $t_{\rm s} =$ 500 days (upper panel), there is no distinct decrease (destruction) of CO and SiO around 100--300 days compared with the counterpart case with $f_{\rm d} =$ 1.0 (upper left panel in Figure~\ref{fig:1d_param1}). %
This trend can simply be understood as the consequence of the inefficient ionization and destruction by Compton electrons (\texttt{CM} reactions) due to the low value of $f_{\rm d} =$ 10$^{-2}$ and several ion-involving reactions, e.g., the \texttt{CE} reaction in Equation~(\ref{eq:h+_sio}). %

In the case of $f_{\rm h} = 5 \times 10^{-3}$, $f_{\rm d} =$ 10$^{-2}$, and $t_{\rm s} =$ 500 days (lower panel), the destruction of SiO after 100 days can be recognized in contrast to the previous case; the destruction is, however, less evident compared with the counterpart case (lower left panel in Figure~\ref{fig:1d_param1}). %
The destruction of SiO after 100 days is not due to \texttt{CM} reactions but due to a few \texttt{NN} reactions. The decreasing trend of CO seen in the counterpart case after 100 days disappears in this case. %

Figure~\ref{fig:1d_param3} shows the results for additional two cases of $f_{\rm d} =$ 1.0 and $f_{\rm d} = 10^{-2}$ with $f_{\rm h} = 10^{-4}$ to see the impact of the parameter $f_{\rm h}$; the value of the parameter $f_{\rm h}$ is the low end in this study. %
Then, the gas heating is inefficient compared with the cases shown above. %
As expected, the gas temperatures reach about 10$^{4}$ K at earlier phases and molecule formation starts in a high-density environment compared with the cases with higher $f_{\rm h}$ values. %
In the case of $f_{\rm h} = 10^{-4}$ and $f_{\rm d} = 1.0$ (left panel in Figure~\ref{fig:1d_param3}), apparently the amounts of CO and SiO at 40--100 days are the highest compared with the previous cases and the amounts are almost comparable with the initial amounts of the seeds. %
However, after 100 days, due to the efficient \texttt{CM} reactions (destruction and/or ionization by Compton electrons), the amounts are reduced to some extent and the final amounts are not dramatically different from those of the previous cases, e.g., one with $f_{\rm h} = 5 \times 10^{-3}$ and $f_{\rm d} =$ 1.0 (lower left panel in Figure~\ref{fig:1d_param1}), although the amount of SiO is the highest than that of the previous cases. %
On the other hand, in the case of $f_{\rm h} = 10^{-4}$ and $f_{\rm d} = 10^{-2}$ (right panel in Figure~\ref{fig:1d_param3}), basically the inefficient \texttt{CM} reactions with the low $f_{\rm d}$ value results in the highest amounts of CO and SiO throughout the evolution among at least all the presented 1D cases above. %

As a summary of the impacts of the parameters $f_{\rm h}$ and $f_{\rm d}$ on the amounts of CO and SiO, the parameter $f_{\rm h}$ mainly affects the amounts in early phases before 100 days, and the parameter $f_{\rm d}$ affects the destruction of CO and SiO from 100 days to a few hundred days. %
The lower $f_{\rm h}$ value is, the higher the amounts of CO and SiO before 100 days. %
The higher $f_{\rm d}$ value is, the more efficient destruction of CO and SiO at 100--300 days is. %
These qualitative features are the same as observed in one-zone calculations (see, Section~\ref{subsec:one_zone_results}).

In view of the first criterion (described in Section~\ref{para:one_zone_selec}), the cases of the $f_{\rm h}$ value less than 5 $\times$ 10$^{-3}$ would result in too high amounts of CO and SiO at 40--600 days compared with the previous estimations. %
Then, among all the calculated 1D cases (for the model b18.3-high), again, the first and second cases (the cases in Sections~\ref{para:1d_param1} and \ref{para:1d_param2}, respectively) would be candidates for the fiducial case to be fixed. %
Among the estimations of the amounts of CO in the previous studies, only LD95 \citep{1995ApJ...454..472L} solved the time-dependent chemical evolution with a similar method for CO ro-vibrational transitions with an optically thick regime as described in Section~\ref{subsec:cooling}. %
Therefore, we would arbitrarily prefer to reproduce the estimations by LD95, in particular, the trend with a valley (i.e., the initial rise, the middle destruction, and the final recovery). %
In this regard, the second case has a clearer decreasing trend of the amounts of both CO and SiO, although the amount of CO at 100 days is a bit higher than the value by LD95 compared with the first case. %
The estimations in the previous studies are not consistent with each other, and in the estimations, more or less some theoretical models are involved. %
Therefore, a strict quantitative consistency of the amounts of CO and SiO with the previous estimations may not be indispensable as mentioned. %
In view of the second criterion (described in Section~\ref{para:one_zone_selec}), the fluxes of CO vibrational bands for the two cases are qualitatively similar to each other and it is difficult to find an apparently better case than those in the two cases among all the 1D calculations. %
Thus, we would select the second case presented in Section~\ref{para:1d_param2}, i.e., $f_{\rm h} = 5 \times 10^{-3}$, $f_{\rm d} =$ 1.0, and $t_{\rm s} =$ 500 days as the fiducial case to be fixed for later discussion on the impact of effective matter mixing. %
Hereafter, if the parameter values of $f_{\rm h}$, $f_{\rm d}$, and $t_{\rm s}$ are not specifically mentioned, the values are $5 \times 10^{-3}$, 1.0, and 500 days, respectively. %

\paragraph{Important chemical reactions for CO and SiO for the fiducial 1D calculation model} \label{para:1d_chemi_reac}

\begin{figure*}
\begin{minipage}{0.5\hsize}
\begin{center}
\hs{-1.5}
\includegraphics[width=9.5cm,keepaspectratio,clip]{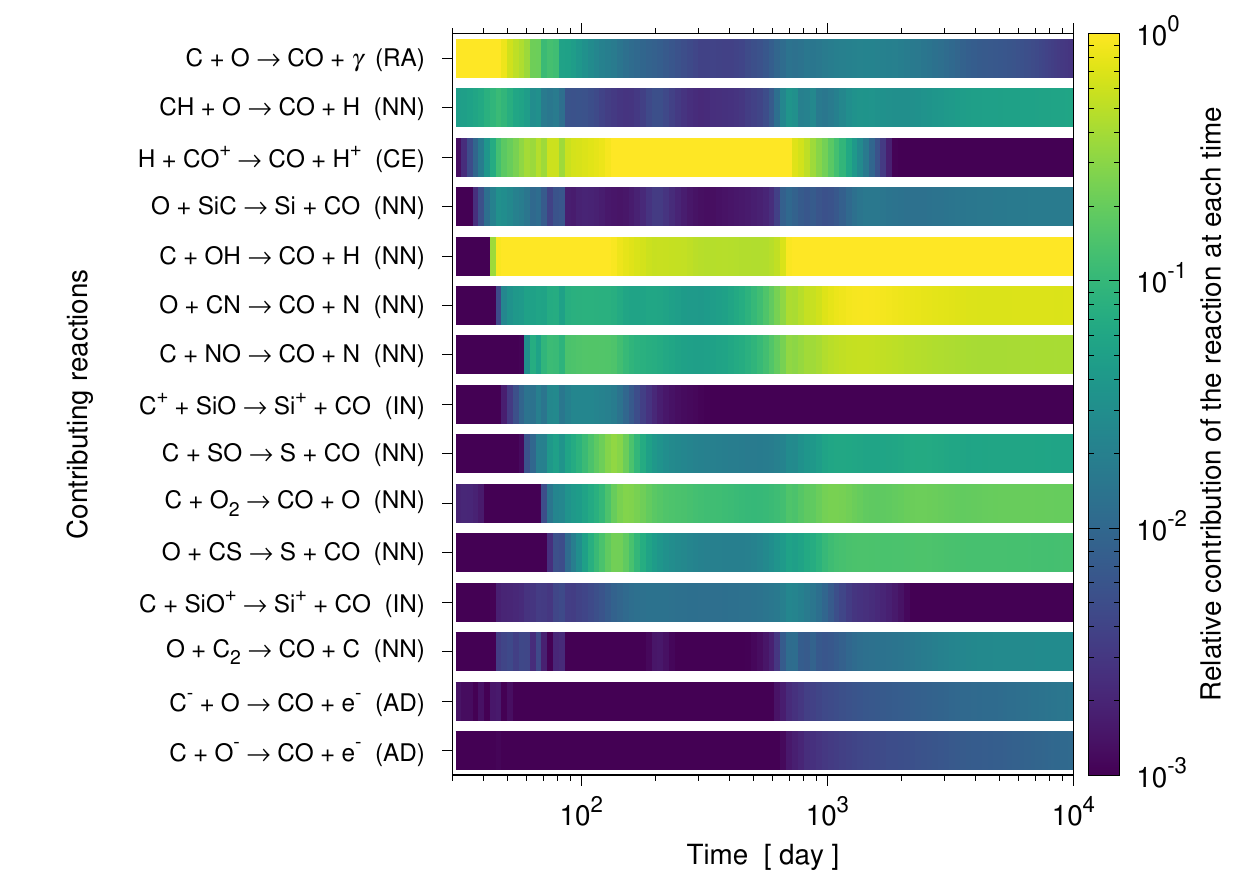}
\end{center}
\end{minipage}
\begin{minipage}{0.5\hsize}
\begin{center}
\hs{-1.5}
\includegraphics[width=9.5cm,keepaspectratio,clip]{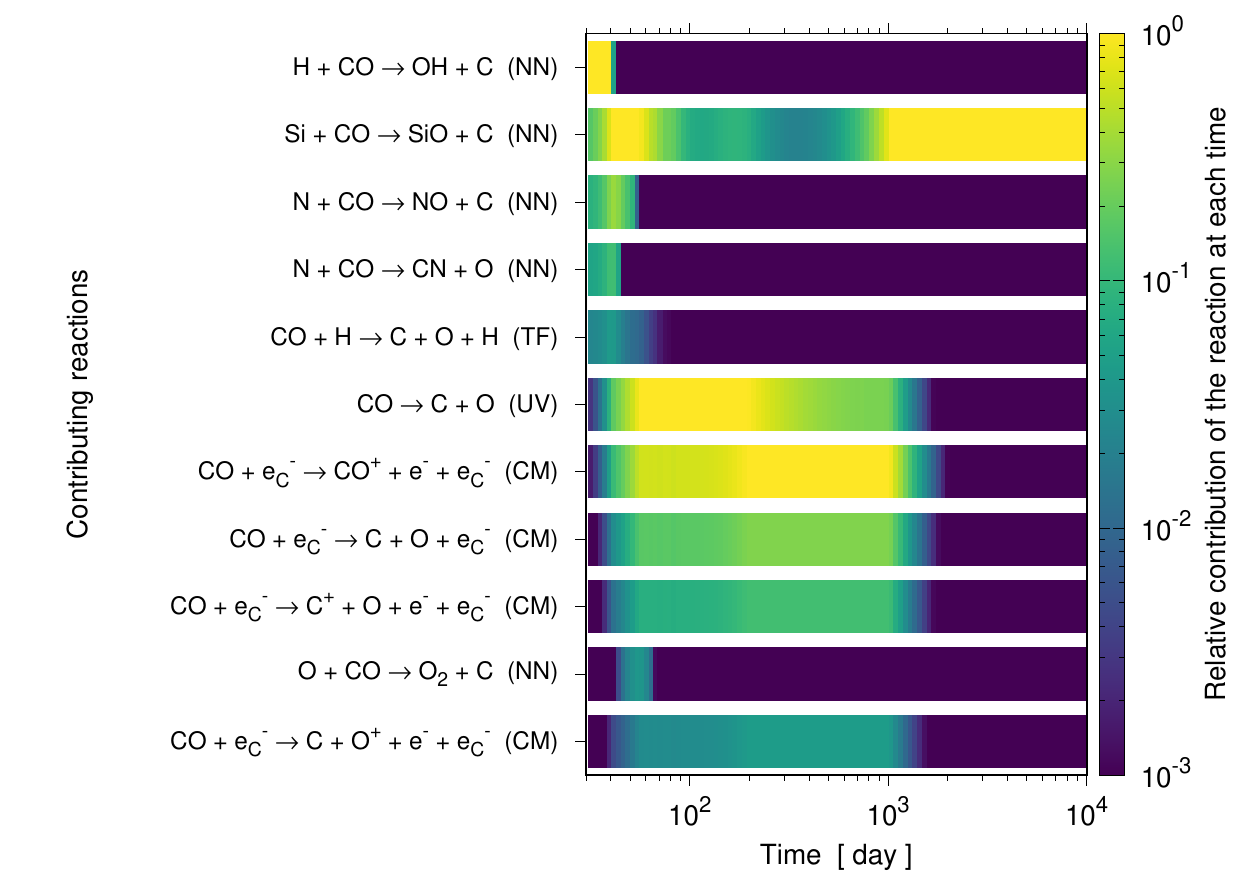}
\end{center}
\end{minipage}
\caption{Contributing chemical reactions for the formation (left) and destruction (right) of CO in which carbon monoxide (CO) is directly involved as a function of time (after 30 days) for the model b18.3-mean. %
The code inside the parentheses left side of each reaction denotes the corresponding reaction type listed in Table~\ref{table:types}. %
Colors denote the relative contribution of each reaction (contribution from all the tracer particles are counted by taking the masses as a weight), which is proportional to the reaction flow $F_i$ ($D_i$) in Equation~(\ref{eq:rate_eq}), normalized as the maximum $F_i$ ($D_i$) among CO formation (destruction) reactions at each time to be unity. %
For neutral-neutral (\texttt{NN}) reactions, net contributions are counted by subtracting the reaction flow of the corresponding inverse reaction. %
Reactions are picked up if the relative contribution once becomes greater than 10$^{-3}$.} %
\label{fig:co_reac_b18.3-mean}
\end{figure*}

\begin{figure*}
\begin{minipage}{0.5\hsize}
\begin{center}
\hs{-1.5}
\includegraphics[width=9.5cm,keepaspectratio,clip]{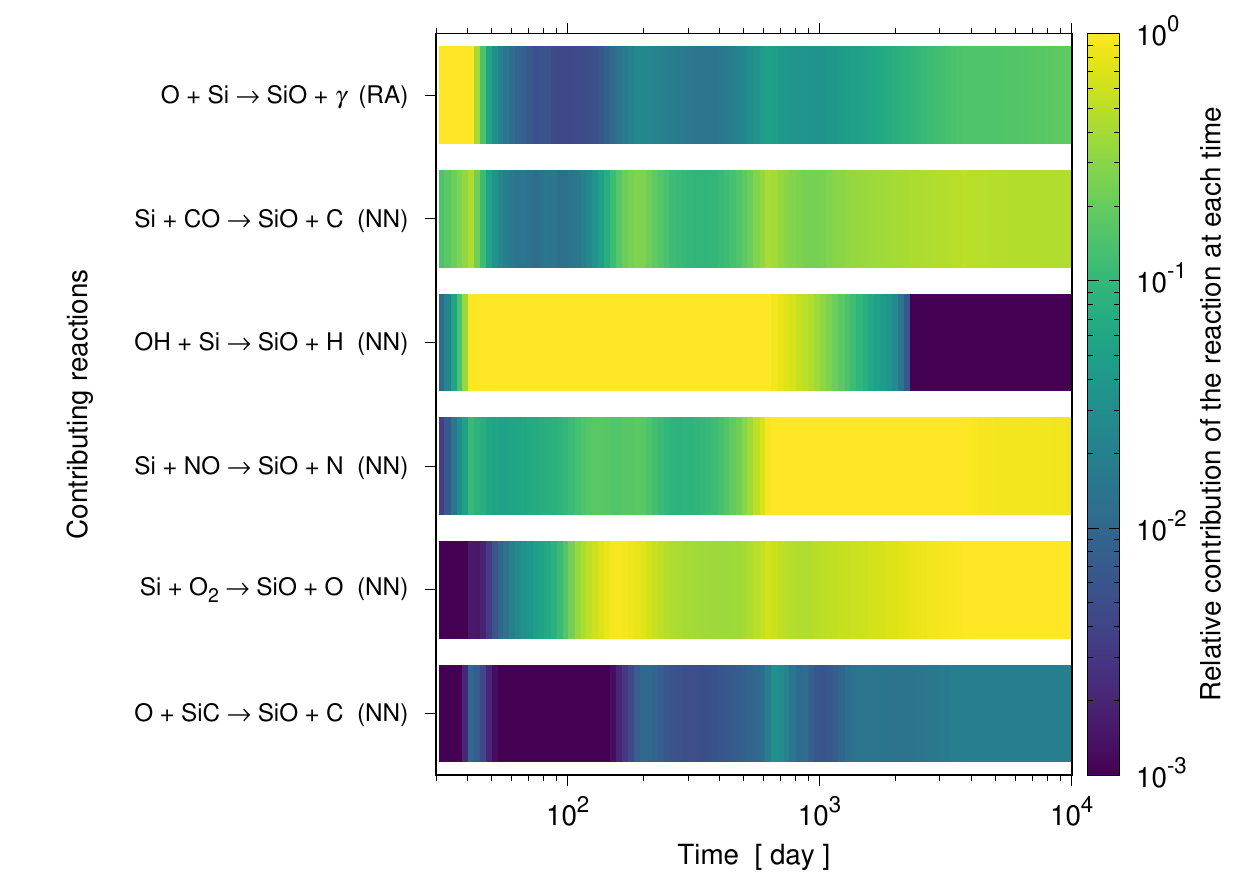}
\end{center}
\end{minipage}
\begin{minipage}{0.5\hsize}
\begin{center}
\hs{-1.5}
\includegraphics[width=9.5cm,keepaspectratio,clip]{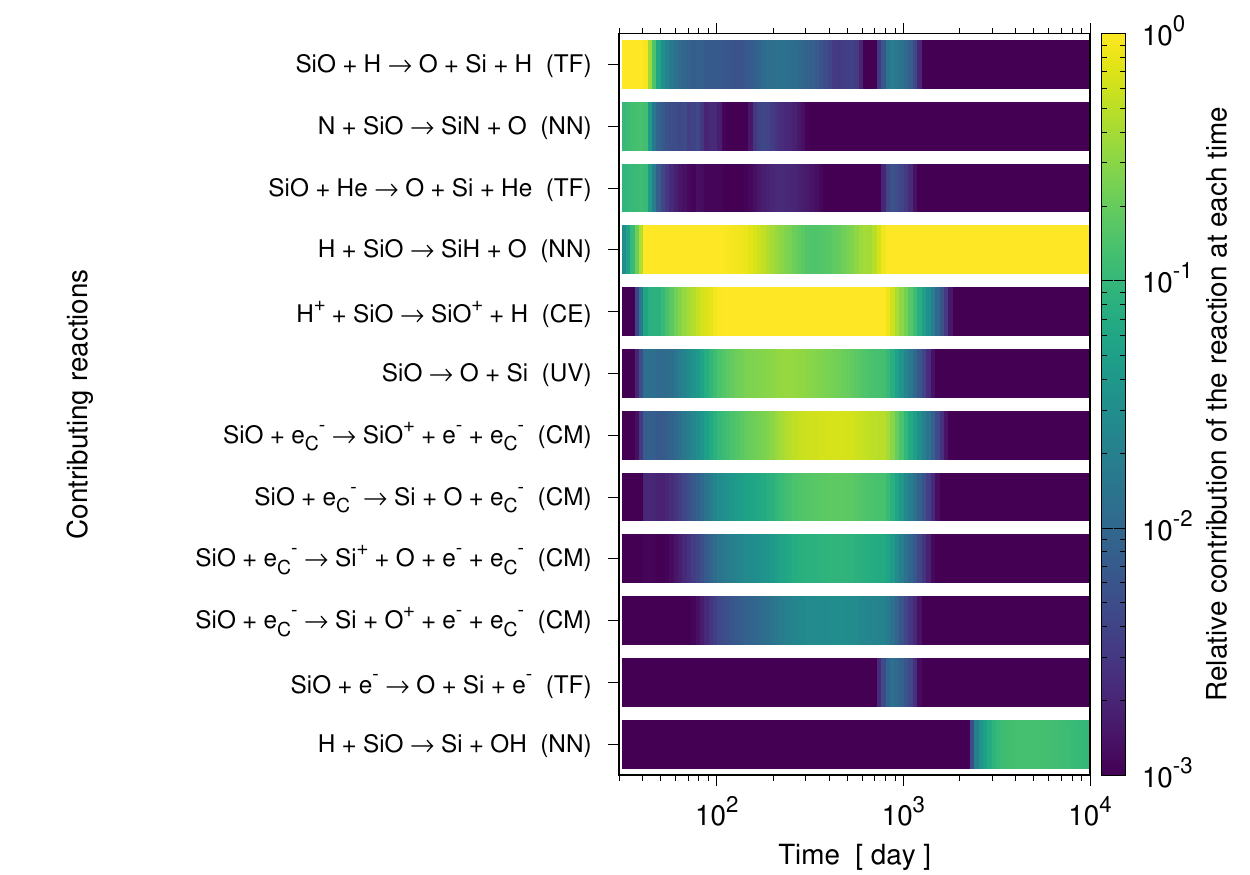}
\end{center}
\end{minipage}
\caption{Same as Figure~\ref{fig:co_reac_b18.3-mean} but for silicon monoxide (SiO).} %
\label{fig:sio_reac_b18.3-mean}
\end{figure*}

\hspace{\parindent}
As done in Section~\ref{subsubsec:chemi_reac} for the fiducial one-zone calculation model (middle panels in Figure~\ref{fig:one_zone_param1} in Section~\ref{para:one_zone_param2}), important chemical reactions for the formation and destruction of CO and SiO are described briefly for the fiducial 1D calculation model (b18.3-mean), i.e., the calculation with the angle-averaged 1D profile based on the 3D model b18.3-high \citep{2020ApJ...888..111O} with the parameters, $f_{\rm h} = 5 \times 10^{-3}$, $f_{\rm d} =$ 1.0, and $t_{\rm s} =$ 500 days, comparing with the fiducial one-zone case. %

Since the contributing chemical reactions depend on the time and each particle, it is difficult to describe the formation and destruction processes of molecules in detail; %
nonetheless, by integrating the contributions from all the particles with the particle's mass as a weight, a similar discussion is presented here. %
In Figure~\ref{fig:co_reac_b18.3-mean} (Figure~\ref{fig:sio_reac_b18.3-mean}), contributing chemical reactions for the formation and destruction of CO (SiO) are shown for reference. %
As for CO formation (left panel in Figure~\ref{fig:co_reac_b18.3-mean}), the listed contributing reactions in the figure are the same as the corresponding ones for the fiducial one-zone calculation (left panel in Figure~\ref{fig:co_reac}); differences, however, can be recognized in the relative contributions between the fiducial one-zone and 1D calculations. %
As for the CO destruction and SiO formation and destruction, the listed reactions are a bit different from those in the fiducial one-zone model, and the relative contributions are also different to some extent. %
Hereafter, important chemical reactions are described for each representative epoch in chronological order, focusing on the epochs when the changes in the quantities of CO and SiO are large; %
formation (destruction) reactions are focused for formation (destruction) dominant phases. %

From the start of the molecule formation to about 100 days (initial increasing phase), the contributing formation processes of CO are the \texttt{RA} reaction in Equation~(\ref{eq:co_rad}), the \texttt{NN} reactions in Equations~(\ref{eq:o_ch}) and (\ref{eq:c_oh}), the \texttt{CE} reaction in Equation~(\ref{eq:h_co+}), and the \texttt{NN} reaction in Equation~(\ref{eq:o_cn}); %
compared with the one-zone case (see, the left panel in Figure~\ref{fig:co_reac}), the contribution of \texttt{CE} reaction in Equation~(\ref{eq:h_co+}) becomes large, and the contributions of the \texttt{NN} reaction in Equation~(\ref{eq:o_sic}) and the \texttt{IN} reaction in Equation~(\ref{eq:c+_sio}) become small. %
The reactions contributing to the formation of SiO are the \texttt{RA} reaction in Equation~(\ref{eq:sio_rad}) and the \texttt{NN} reactions in Equations~(\ref{eq:si_co}), (\ref{eq:si_oh}), (\ref{eq:si_no}), and (\ref{eq:si_o2}); compared with the one-zone case, the contribution of the \texttt{NN} reactions in Equations~(\ref{eq:si_no}) and (\ref{eq:si_o2}) become large, and the \texttt{NN} reactions in Equations (\ref{eq:si_co}) and (\ref{eq:o_sih}) become small. %

The amounts of CO and SiO decrease from 100 days to approximately 300 days. %
In this phase, the contributing destruction processes of CO are the ionization and destruction by Compton electrons (\texttt{CM} reactions) described in Equations~(\ref{eq:ion_ab}) and (\ref{eq:destruction}), the dissociation by UV photons (\texttt{UV} reaction), and the \texttt{NN} reaction in Equation~(\ref{eq:si_co}). %
It is noted that in the one-zone case, the \texttt{IN} reaction with He$^+$ in Equation~(\ref{eq:he+_co}) is the primary destruction process in contrast to the 1D case here. %
This feature in the one-zone case may be attributed to the efficient mixing of helium and $^{56}$Ni compared with the 1D case. %
The \texttt{IN} reaction above does not contribute in the 1D case. %
The \texttt{NN} reaction in Equation~(\ref{eq:si_co}) becomes effective in contrast to the one-zone case. %
The contributing destruction processes of SiO are the \texttt{NN} reaction in Equation~(\ref{eq:h_sio_sih}) and the \texttt{CE} reaction in Equation~(\ref{eq:h+_sio}), the ionization by Compton electrons (\texttt{CM} reaction), and the destruction by UV photons (\texttt{UV} reaction). %
The contribution of the \texttt{NN} reaction in Equation~(\ref{eq:h_sio_sih}) becomes large compared with the one-zone case. %
The contributions of the \texttt{IN} reactions with He$^+$ in Equations~(\ref{eq:he+_sio_o+}) and (\ref{eq:he+_sio_si+}) are minor in the 1D case in contrast to the one-zone case. %

After 300 days, the amount of CO slightly increases. %
On the other hand, the amount of SiO is distinctively recovered to some extent. %
The contributing formation processes of CO at this phase are the \texttt{CE} reaction in Equation~(\ref{eq:h_co+}), the \texttt{NN} reactions in Equations~(\ref{eq:c_oh}), (\ref{eq:o_cn}), (\ref{eq:c_o2}), and the \texttt{NN} reaction below. %
\begin{align}
{\rm C} + {\rm NO} \lra {\rm CO} + {\rm N}. \label{eq:c_no}
\end{align}
The \texttt{NN} reactions in Equations~(\ref{eq:o_cs}) and (\ref{eq:c_so}) also contribute as the secondaries. %
The contributions of the \texttt{CE} reaction in Equation~(\ref{eq:h_co+}) and the \texttt{NN} reactions in Equations~(\ref{eq:o_cn}) and (\ref{eq:c_no}) become large compared with the one-zone case. %
The major contributing formation processes of SiO at this phase are the \texttt{NN} reaction in Equations~(\ref{eq:si_oh}), (\ref{eq:si_no}), (\ref{eq:si_o2}), and (\ref{eq:si_co}). %
The contributions of the \texttt{NN} reactions in Equations~(\ref{eq:si_co}) and (\ref{eq:si_no}) become large compared with the one-zone case. %

As a summary, compared with the fiducial one-zone model, the less efficient mixing of $^{56}$Ni, hydrogen, and helium in the 1D model could decrease the contributions of the reactions involved with ionized hydrogen and helium atoms, i.e., the \texttt{CE} reactions in Equations~(\ref{eq:c+_sio}) and (\ref{eq:h+_sio}) and the \texttt{IN} reactions in Equations~(\ref{eq:he+_co}), (\ref{eq:he+_sio_o+}), and (\ref{eq:he+_sio_si+}). %

\subsubsection{The results including molecules other than CO and SiO with the angle-averaged 1D profiles based on the 3D models; the comparison between the binary merger and single-star progenitor models} \label{subsubsec:1d_single_star}

\begin{figure*}

\begin{minipage}{0.5\hsize}
\vs{-0.5}
\begin{center}
\includegraphics[width=9.5cm,keepaspectratio,clip]{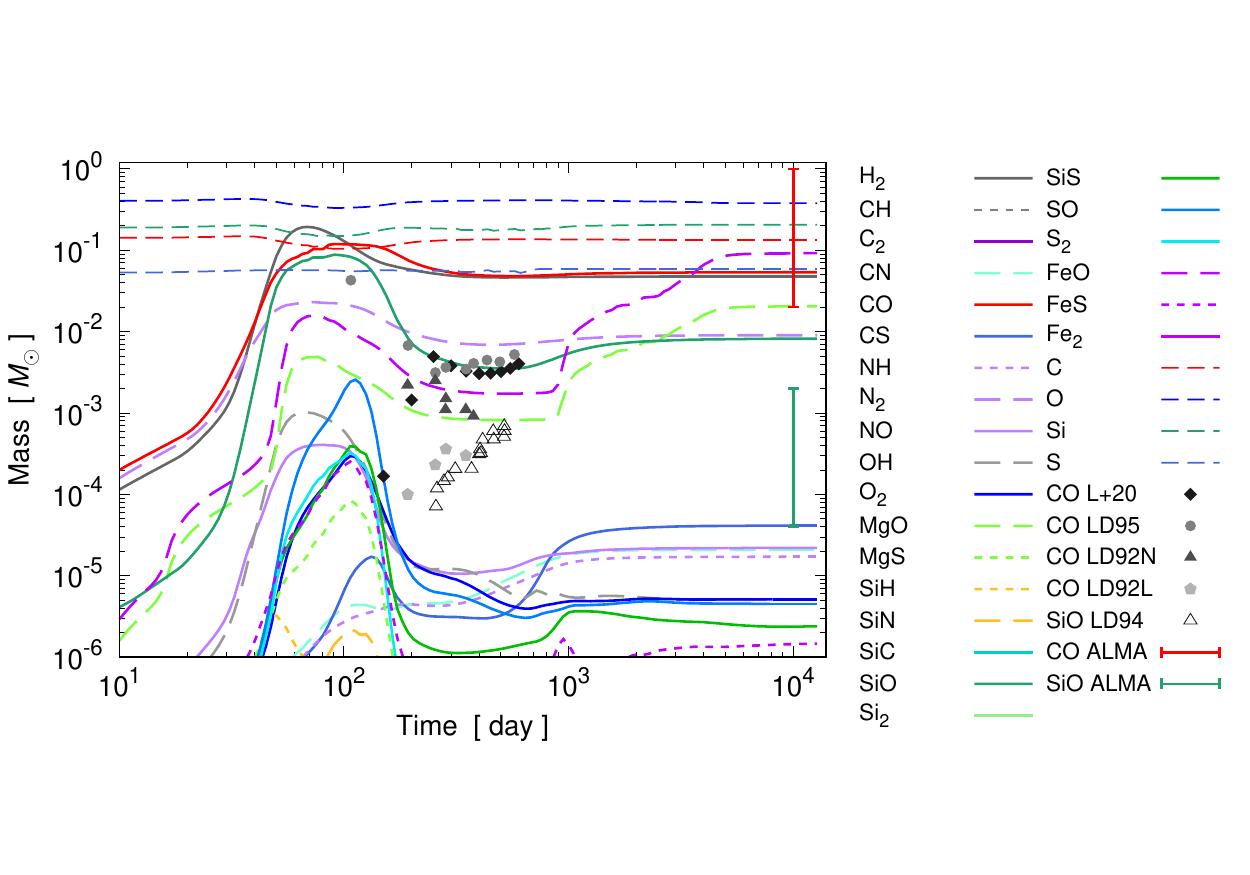}
\end{center}
\end{minipage}
\begin{minipage}{0.5\hsize}
\vs{-0.5}
\begin{center}
\includegraphics[width=7.4cm,keepaspectratio,clip]{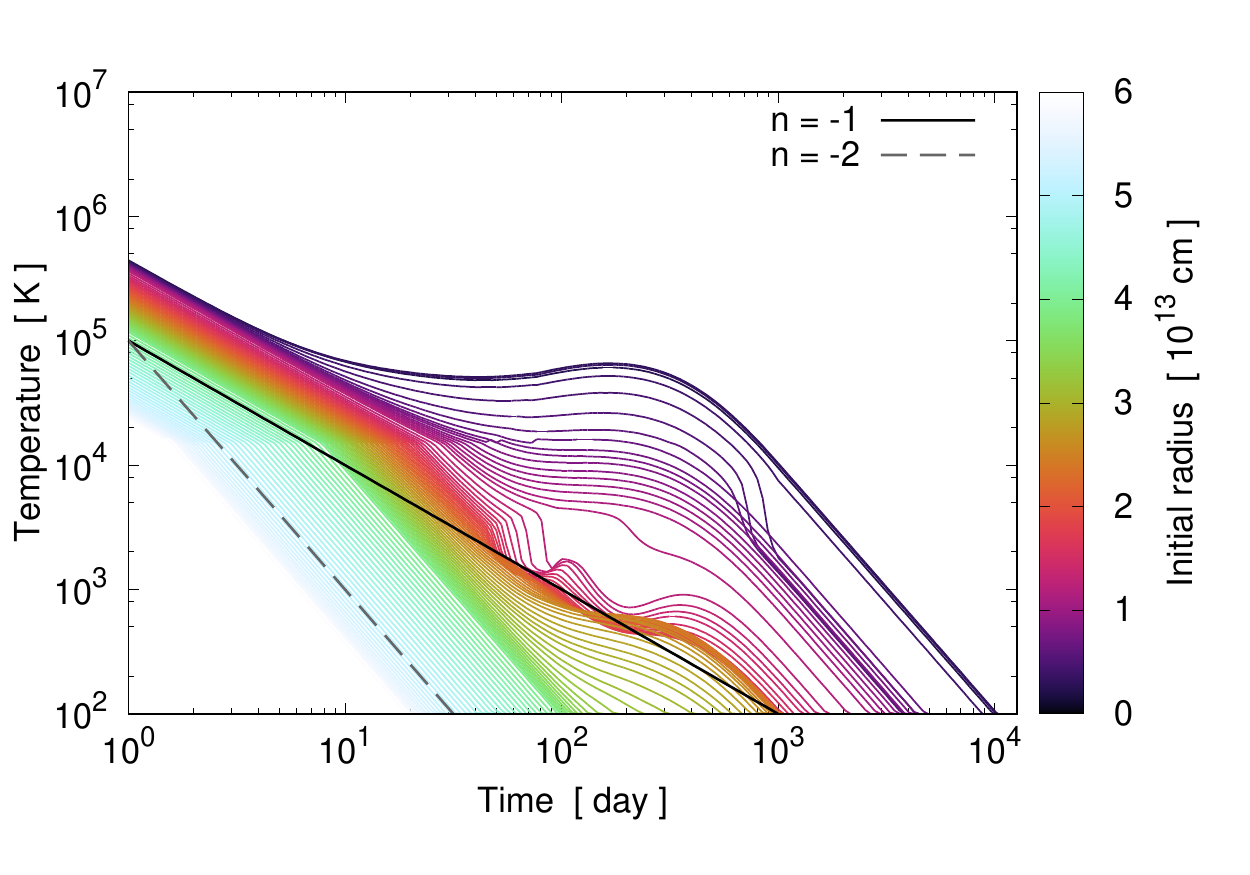}
\end{center}
\end{minipage}
\\
\begin{minipage}{0.5\hsize}
\vs{-1.7}
\begin{center}
\includegraphics[width=9.5cm,keepaspectratio,clip]{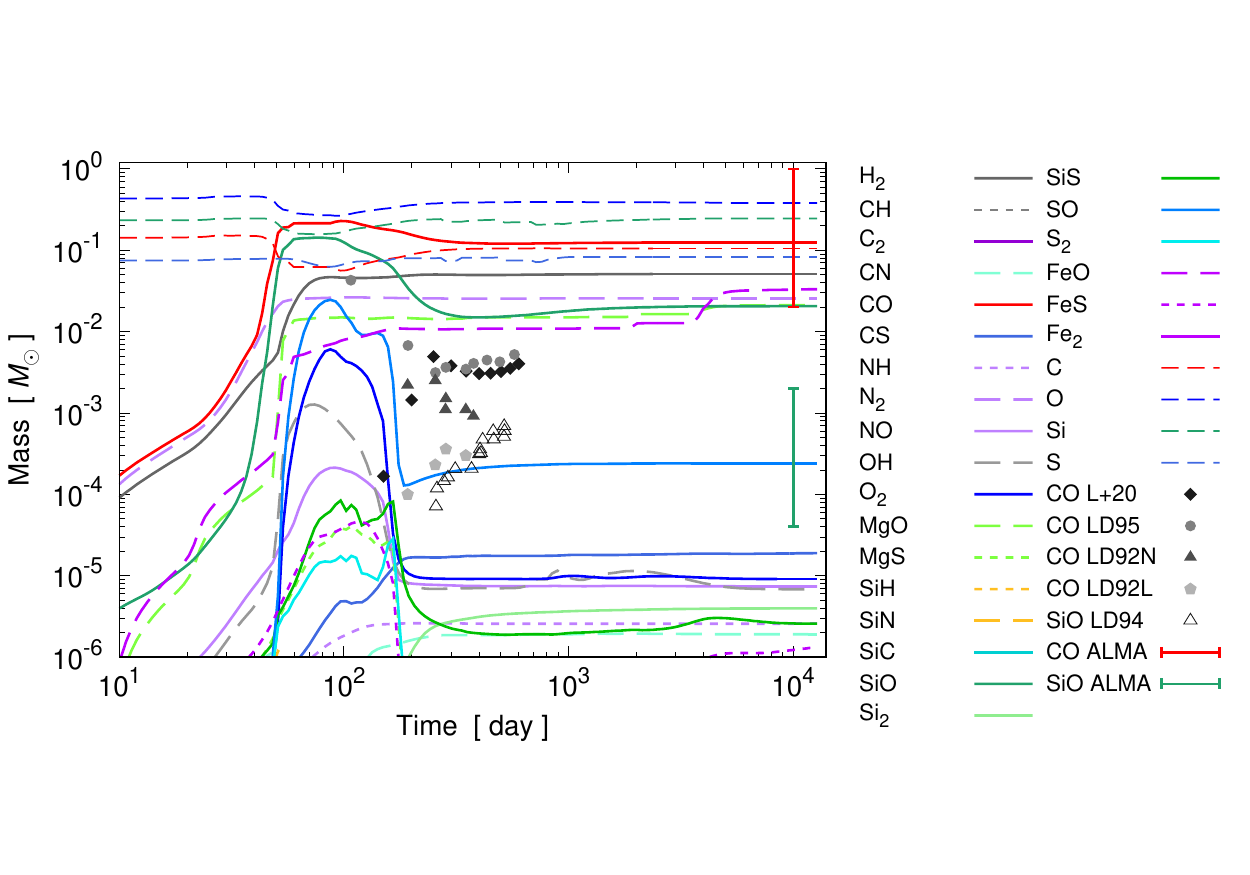}
\end{center}
\end{minipage}
\begin{minipage}{0.5\hsize}
\vs{-1.7}
\begin{center}
\includegraphics[width=7.4cm,keepaspectratio,clip]{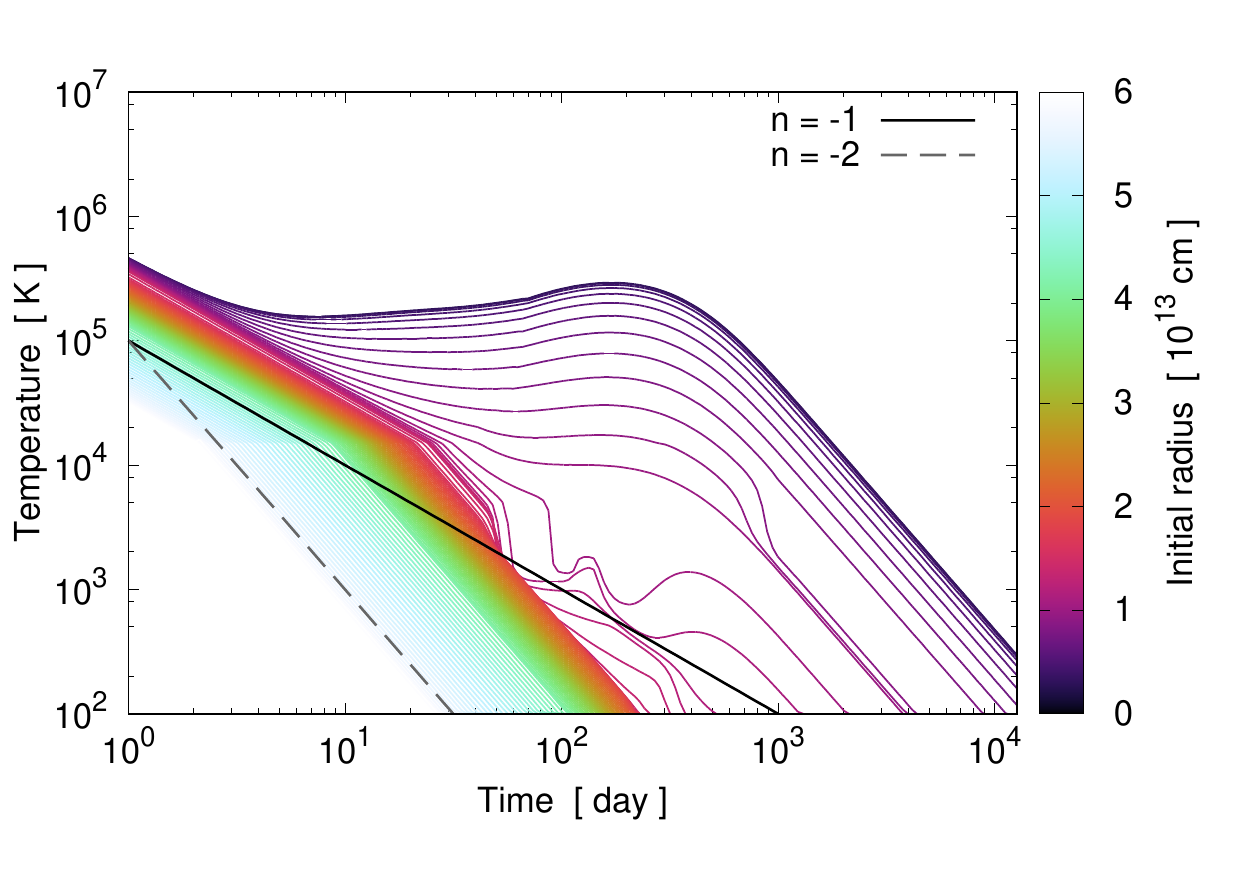}
\end{center}
\end{minipage}
\\
\begin{minipage}{0.5\hsize}
\vs{-1.7}
\begin{center}
\includegraphics[width=9.5cm,keepaspectratio,clip]{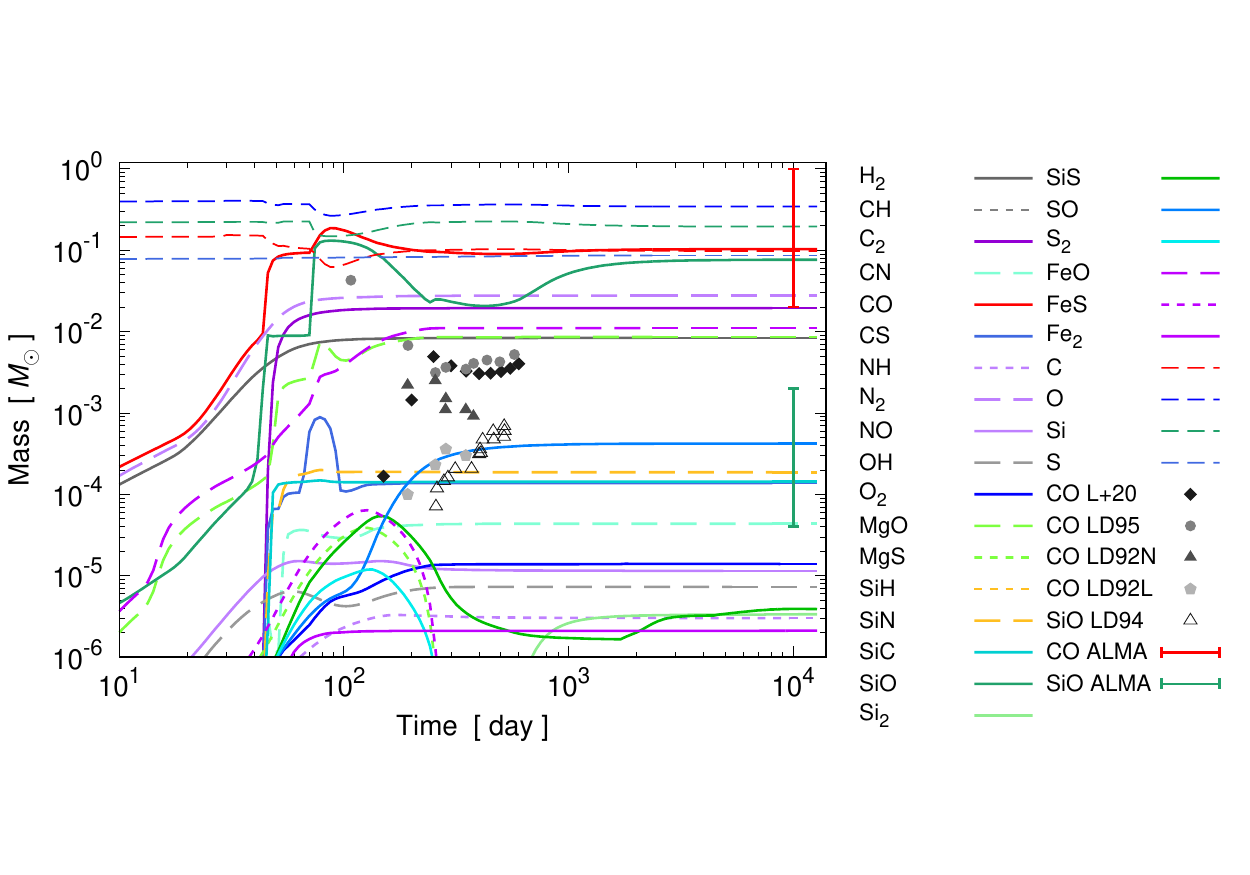}
\end{center}
\vs{-1.2}
\end{minipage}
\begin{minipage}{0.5\hsize}
\vs{-1.7}
\begin{center}
\includegraphics[width=7.4cm,keepaspectratio,clip]{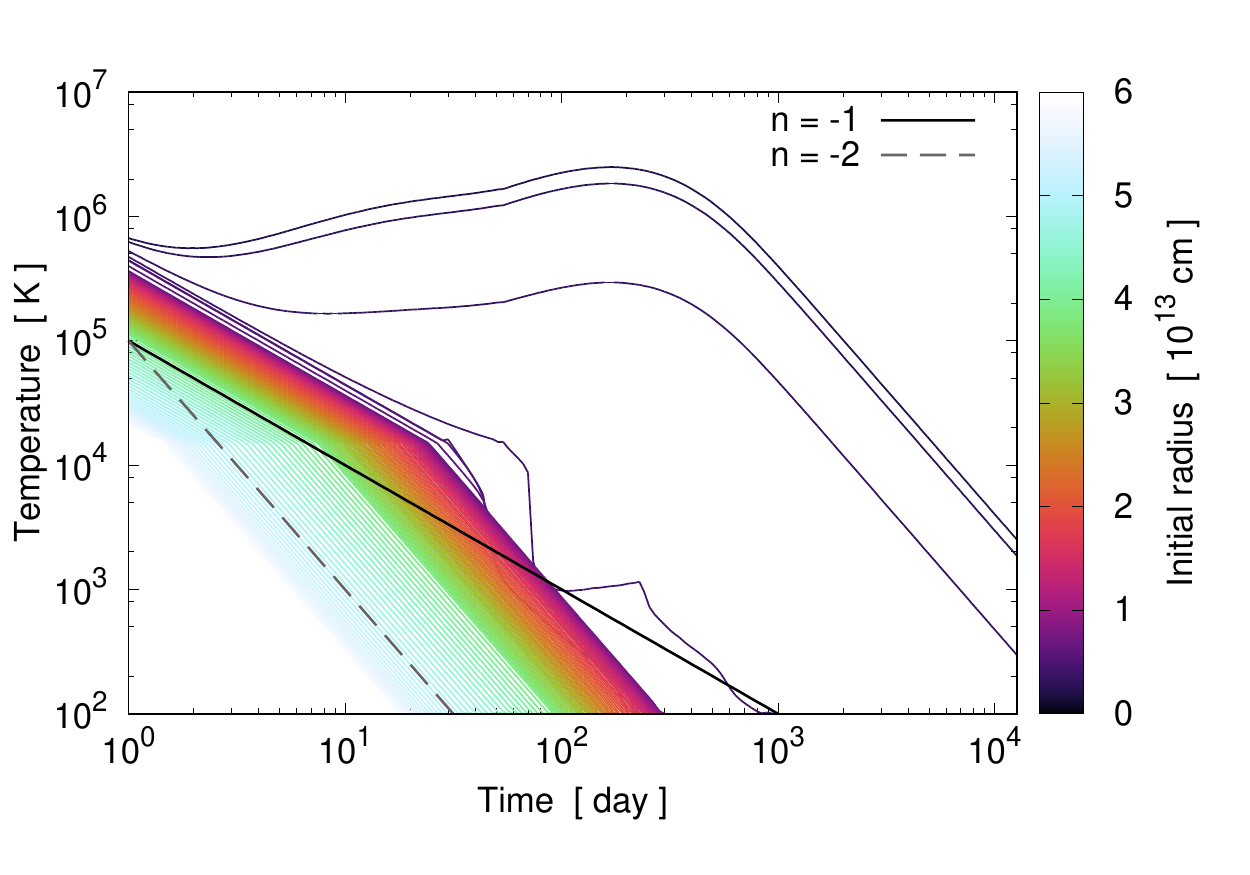}
\end{center}
\vs{-1.2}
\end{minipage}
\caption{1D calculation results of the models, b18.3-mean (top panels), b18.3-sphel (middle panels), and b18.3-sphel-pure (bottom panels). %
Left panels: the time evolution of the amounts of diatomic molecules and several seed atoms compared with the estimation for CO and SiO (points) in the previous studies (for the details, see, the caption of Figure~\ref{fig:1d_param1}). %
Right panels: time evolution of the gas temperatures of the tracer particles; the colors denote the initial positions of the particles. As a reference, power-law evolutions of the powers of $-1$ and $-2$ are shown.} %
\label{fig:1d_mass_temp_binary}
\end{figure*}

\begin{figure}
\begin{minipage}{0.5 \hsize}
\vs{-0.5}
\begin{center}
\includegraphics[width=8.5cm,keepaspectratio,clip]{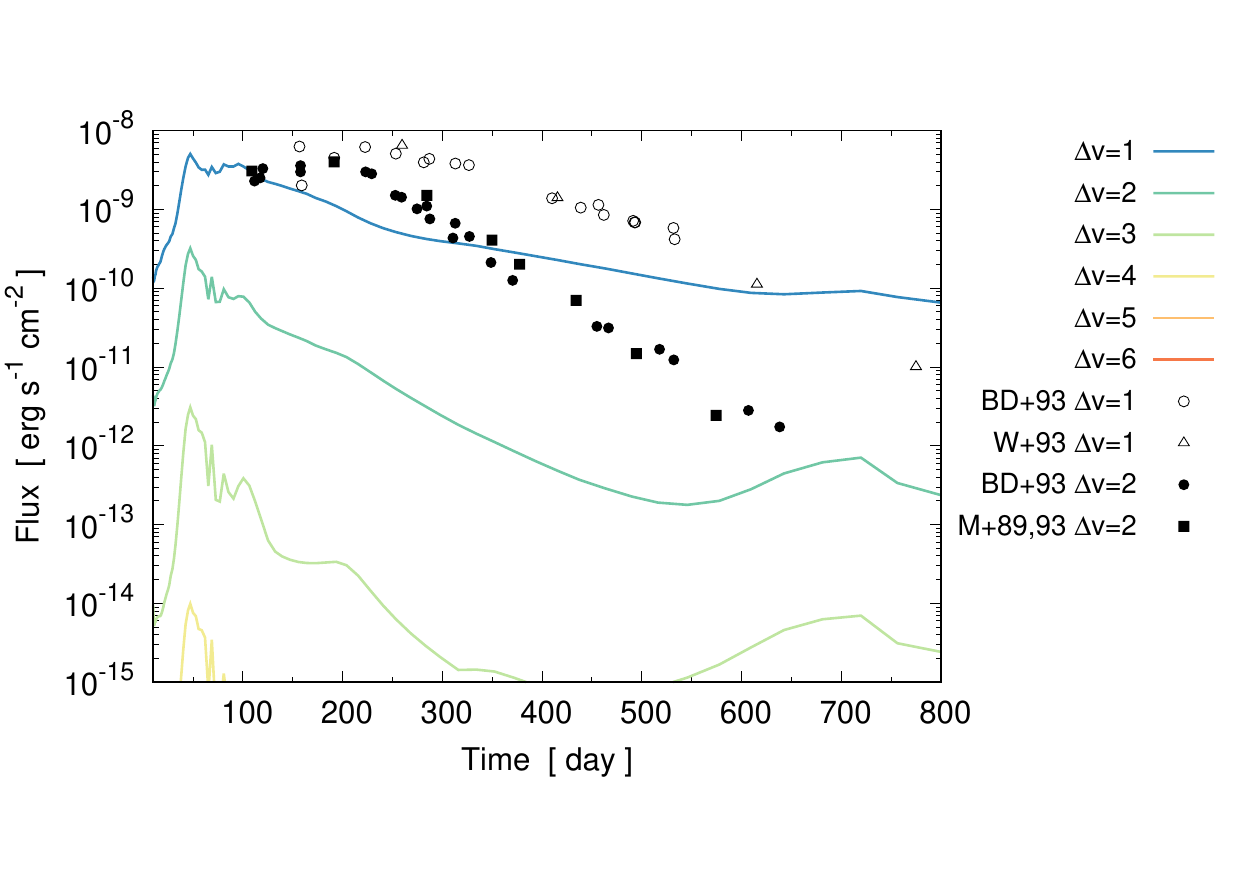}
\end{center}
\end{minipage}
\\
\begin{minipage}{0.5 \hsize}
\vs{-1.}
\begin{center}
\includegraphics[width=8.5cm,keepaspectratio,clip]{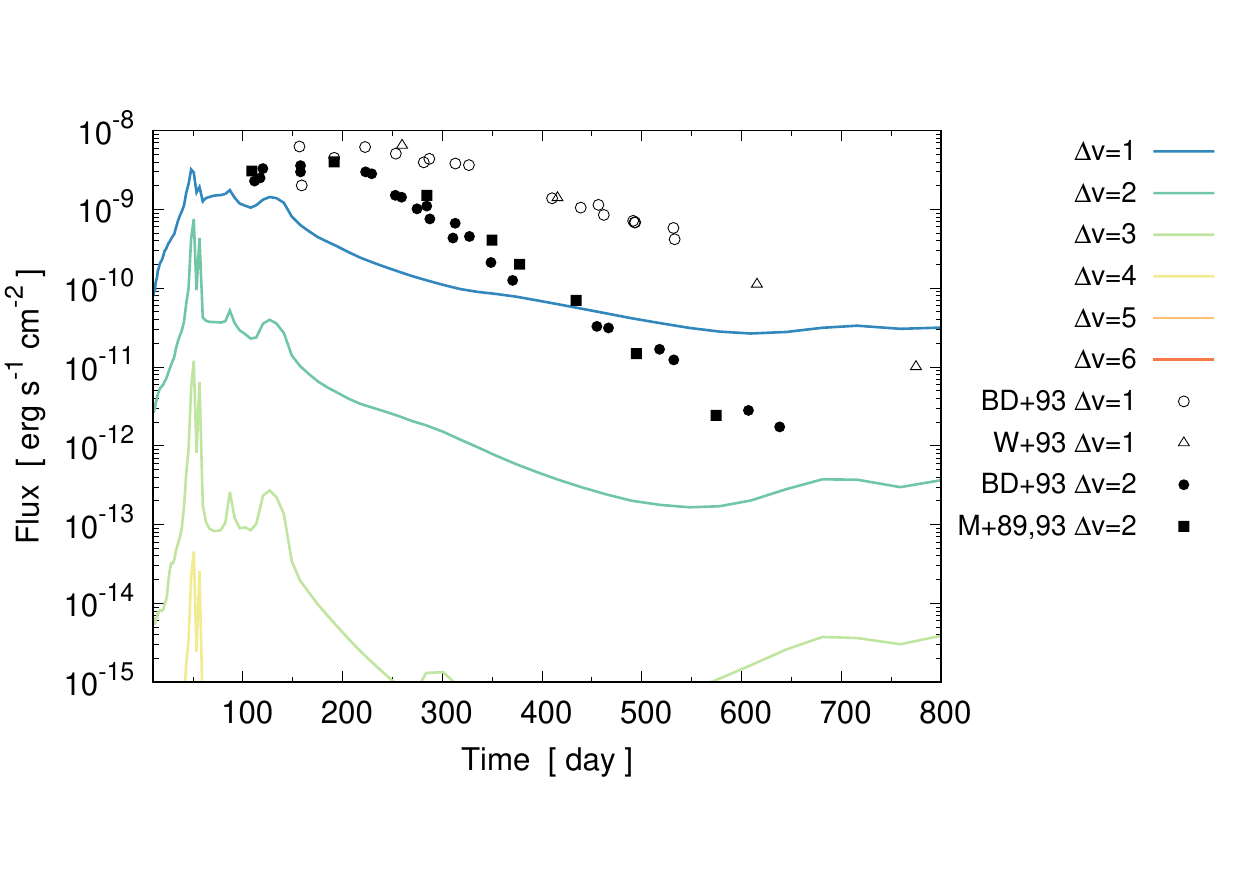}
\end{center}
\end{minipage}
\\
\begin{minipage}{0.5 \hsize}
\vs{-1.}
\begin{center}
\includegraphics[width=8.5cm,keepaspectratio,clip]{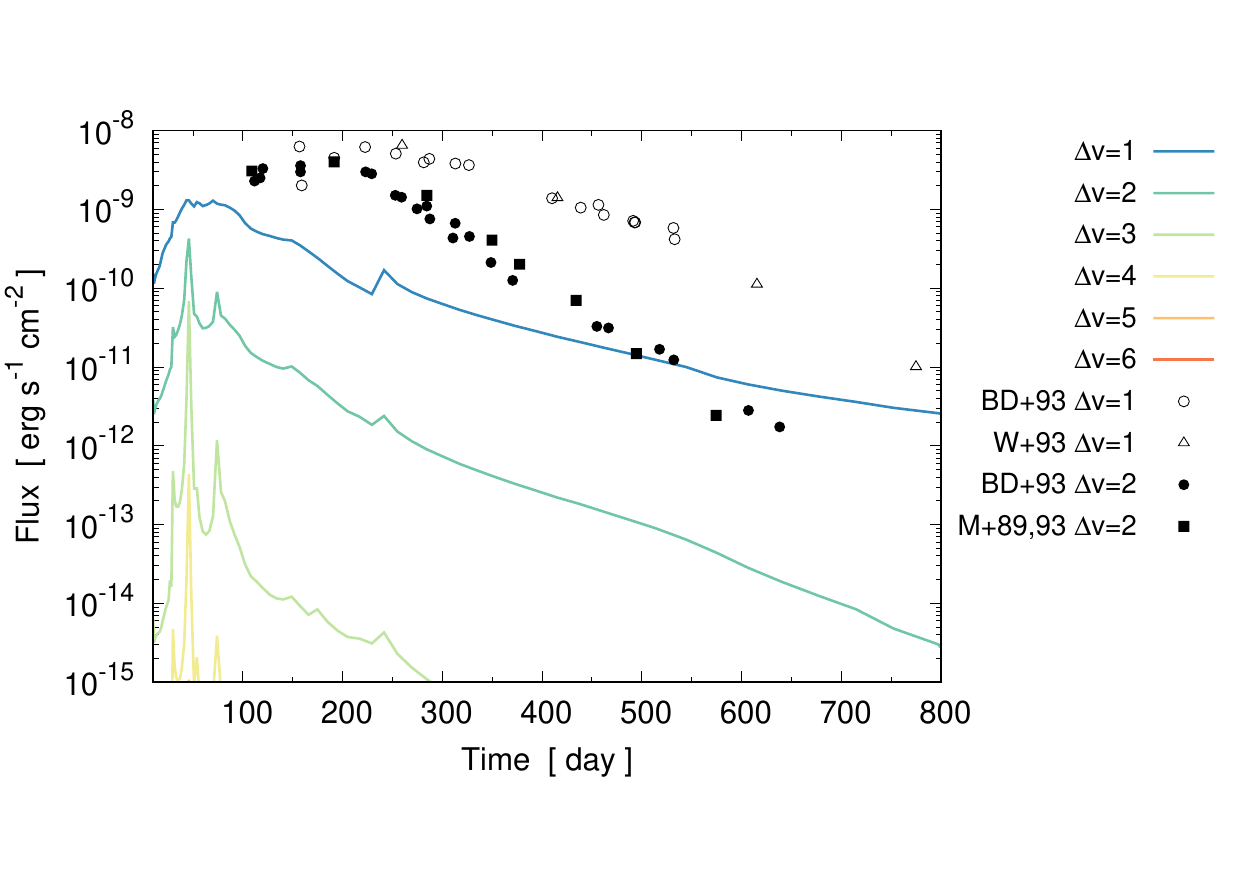}
\end{center}
\vs{-0.7}
\end{minipage}
\caption{Time evolution of the fluxes for CO vibrational bands, ${\it \Delta}v=1$ (fundamental), ${\it \Delta}v=2$ (first overtone), \ldots, ${\it \Delta}v=6$, compared with the observed light curves for ${\it \Delta}v=1$ \citep[BD93; W$+$93:][respectively]{1993A&A...273..451B,1993ApJS...88..477W} and ${\it \Delta}v=2$ \citep[M+89, 93; BD93:][respectively]{1989MNRAS.238..193M,1993MNRAS.261..535M,1993A&A...273..451B}, for the models b18.3-mean (top), b18.3-sphel (middle), and b18.3-sphel-pure (bottom). %
The top panel is the same as the lower right panel in Figure~\ref{fig:1d_param1}.} %
\label{fig:1d_engy_loss_binary}
\end{figure}

Having determined the fiducial values for the parameters ($f_{\rm h} = 5 \times 10^{-3}$, $f_{\rm d} =$ 1.0, and $t_{\rm s} =$ 500 days), here the results including molecules other than CO and SiO with the angle-averaged 1D profiles based on the 3D explosion models \citep[the models b18.3-high and n16.3-high:][]{2020ApJ...888..111O} with the binary merger (b18.3) and the single-star (n16.3) progenitor models, i.e., the results of the models b18.3-mean and n16.3-mean  are described in Sections~\ref{para:b18.3-mean} and \ref{para:n16.3-mean}, respectively. %

\paragraph{The results of the model b18.3-mean} \label{para:b18.3-mean}

\hspace{\parindent}
In the top panels in Figure~\ref{fig:1d_mass_temp_binary}, the results of the model b18.3-mean are shown. %
The qualitative features of the evolution of the gas temperatures (the right panel) are similar to the case of the parameters, $f_{\rm h} = 10^{-2}$, $f_{\rm d} =$ 1.0, and $t_{\rm s} =$ 500 days (see, the left panel in Figure~\ref{fig:rho_dens_1d}), although some differences can be recognized due to the smaller $f_{\rm h}$ value; %
inner particles are preferentially heated by the decay of $^{56}$Ni making peaks around 200 days in the gas temperatures; %
some of the particles whose initial positions are around 1.5 $\times$ 10$^{13}$ cm and a few inner particles experience both the heating and the cooling by CO ro-vibrational transitions. %
The highest peak temperature around 200 days ($< 10^{5}$ K) is lower than that of the case of $f_{\rm h} = 10^{-2}$ ($> 10^{5}$ K) as expected. %
The time evolution of the fluxes of CO vibrational bands was already described in Section~\ref{para:1d_param2} (see, the lower right panel in Figure~\ref{fig:1d_param1}; %
for the comparison in Section~\ref{subsubsec:1d_comp_sphel}, the same panel is shown in the top panel in Figure~\ref{fig:1d_engy_loss_binary}). %

In the left panel, the time evolution of the total amounts of the molecules including those other than CO and SiO is shown.\footnote{In the similar plots hereafter, for completeness, all the diatomic molecules taken into account in this study (24 molecules) are plotted. The amounts of some of the molecules that do not appear in the plots are less than 10$^{-6}$ $M_{\odot}$ throughout the evolution.} %
The time evolution of the amounts of CO and SiO and the important chemical reactions for their formation and destruction were already described in Section~\ref{para:1d_param2} (see, the lower left panel in Figure~\ref{fig:1d_param1}) and \ref{para:1d_chemi_reac}, respectively. %
As with CO and SiO, the molecules other than CO and SiO have qualitatively similar trends in the time evolution, i.e., the amounts increase until about 100 days; after that, those decrease to some extent; after a few hundred days, some of them increase again. %
Hereafter, the time evolution of the amounts and primary formation and destruction reactions are described for some of the molecules other than CO and SiO focusing on epochs when the changes in the amounts are remarkable; %
formation (destruction) reactions are focused during the formation (destruction) dominant phases. %

At the end of the calculation, the molecules whose amounts are greater than 10$^{-3}$ $M_{\odot}$ are CO, SiO, H$_2$, N$_2$, MgO, and FeO; there are some gaps in the amounts between the molecules above and the others. %
The sequence of the reactions for H$_2$ formation is the \texttt{REA} reaction, H + e$^-$  $\lra$ H$^-$ + $\gamma$; the \texttt{AD} reaction, H$^-$ + H $\lra$ H$_2$ $+$ e$^-$ (as mentioned in Section~\ref{subsubsec:chemi_reac}). %
The destruction processes of H$_2$ after 60 days are basically the \texttt{UV} reaction, the ionization by Compton electrons (one of \texttt{CM} reactions), and the \texttt{NN} reaction of oxygen with H$_2$ (products: OH and helium). %
The primary formation processes of N$_2$ are the \texttt{NN} reaction of nitrogen with NO before about 200 days and the \texttt{NN} reaction of nitrogen with CN after 200 days. %
The destruction processes of N$_2$ after about 100 days are the \texttt{CM} reactions (ionization, dissociation, dissociative ionization) and the \texttt{UV} reaction. %
The formation and destruction processes of MgO and FeO are similar to each other. The formation process of MgO (FeO) before about 200 days is the \texttt{NN} reaction between magnesium (iron) and O$_2$ and the ones after that are three-body (\texttt{3B}) reactions with hydrogen and helium, e.g., Mg (Fe) + O + H $\lra$ MgO (FeO) + H. %
The major destruction processes are the \texttt{CM} and \texttt{UV} reactions. %

Among the molecules other than those mentioned above, the amounts of OH, NO, SO, O$_2$, and SiS become greater than 10$^{-4}$ $M_{\odot}$ by 200 days and those remain greater than 10$^{-6}$ $M_{\odot}$ after the destruction dominant phase around a few hundred days. 
The primary formation reactions of OH are the \texttt{RA} reaction between oxygen and hydrogen atoms before 50 days and the \texttt{NN} reaction between oxygen and H$_2$ after 50 days. %
The destruction process of OH after 60 days is the \texttt{NN} reaction of OH with silicon (products: SiO + H). %
NO is primarily formed by the \texttt{NN} reactions of nitrogen with CO and OH before 200 days and the \texttt{NN} reaction of oxygen with NH after 200 days. %
The destruction processes of NO after 100 days are the \texttt{NN} reactions of NO with silicon (products: SiO and nitrogen) and with carbon (products: CO and nitrogen). %
The major formation processes of SO before 100 days are the \texttt{RA} reaction of sulfur and oxygen atoms and the \texttt{NN} reaction between oxygen and S$_2$. The formation process after 100 days is the \texttt{NN} reaction of sulfur with OH. %
The primary destruction reactions of SO after 100 days are the \texttt{NN} reactions of SO with nitrogen (products: NO and sulfur) and with carbon (products: CO and sulfur; CS and oxygen). %
The primary formation reactions of O$_2$ before 100 days are the \texttt{RA} reaction between two oxygen atoms and the \texttt{NN} reaction of oxygen and OH. %
The destruction processes of O$_2$ after 100 days are the \texttt{NN} reactions of O$_2$ with silicon (products: SiO and oxygen) and with carbon (products: CO and oxygen). %
SiS is primarily formed by the \texttt{RA} reaction between silicon and sulfur atoms (before about 100 days and after 200 days) and the \texttt{NN} reaction between silicon and S$_2$ (from 100 to 200 days). %
The destruction processes of SiS after 100 days are the \texttt{NN} reactions of SiS with nitrogen (products: SiN and sulfur) and with sulfur (products: S$_2$ and silicon). %
The \texttt{TF} reaction of SiS with hydrogen also contributes to the destruction. %

It is noted that the products of some of the destructive \texttt{NN} reactions of the molecules above are CO and/or SiO. Therefore, such molecules are somehow converted to CO and/or SiO through the \texttt{NN} reactions. %

The amount of S$_2$ becomes greater than 10$^{-4}$ $M_{\odot}$ around 100 days but the one becomes less than 10$^{-6}$ $M_{\odot}$ due to destruction processes by 200 days. %
The primary formation processes of S$_2$ before 100 days are the \texttt{NN} reactions of sulfur with SiS and with FeS. %
The destruction processes of S$_2$ after 100 days are the \texttt{NN} reactions of S$_2$ with oxygen (products: SO and sulfur) and with carbon (products: CS and sulfur). %

The amounts of the molecules, CS, CN, and NH, increase after about 100 days and those become greater than 10$^{-5}$ $M_{\odot}$ by the end of the calculation. %
The primary formation processes of CS after 100 days are the \texttt{NN} reaction of carbon with SO and the \texttt{NN} reaction of sulfur with CN (only after 600 days). %
The destruction process of CS after 130 days is the \texttt{NN} reaction of CS with oxygen (products: CO and sulfur). %
CN is primarily formed by the \texttt{NN} reactions of carbon with NO and with NH. %
The primary formation process of NH is the \texttt{NN} reaction of nitrogen with H$_2$. %

Other molecules that are not mentioned above, CH, C$_2$, SiH, SiN, SiC, and Fe$_2$ except for MgS and FeS, are minor in the amounts throughout the evolution. %

The amounts of MgS and FeS are qualitatively similar; the amount of MgS (FeS) once becomes greater than 10$^{-5}$ (10$^{-4}$) $M_{\odot}$ around 100 days, although later those become less than 10$^{-6}$ (10$^{-5}$) $M_{\odot}$. %
The primary formation process of MgS (FeS) is the \texttt{NN} reaction of magnesium (iron) with SO; the primary destruction process is the \texttt{NN} reaction of MgS (FeS) with sulfur. %

\paragraph{The results of the model n16.3-mean} \label{para:n16.3-mean}

\begin{figure*}
\begin{minipage}{0.5\hsize}
\vs{-0.5}
\begin{center}
\includegraphics[width=9.5cm,keepaspectratio,clip]{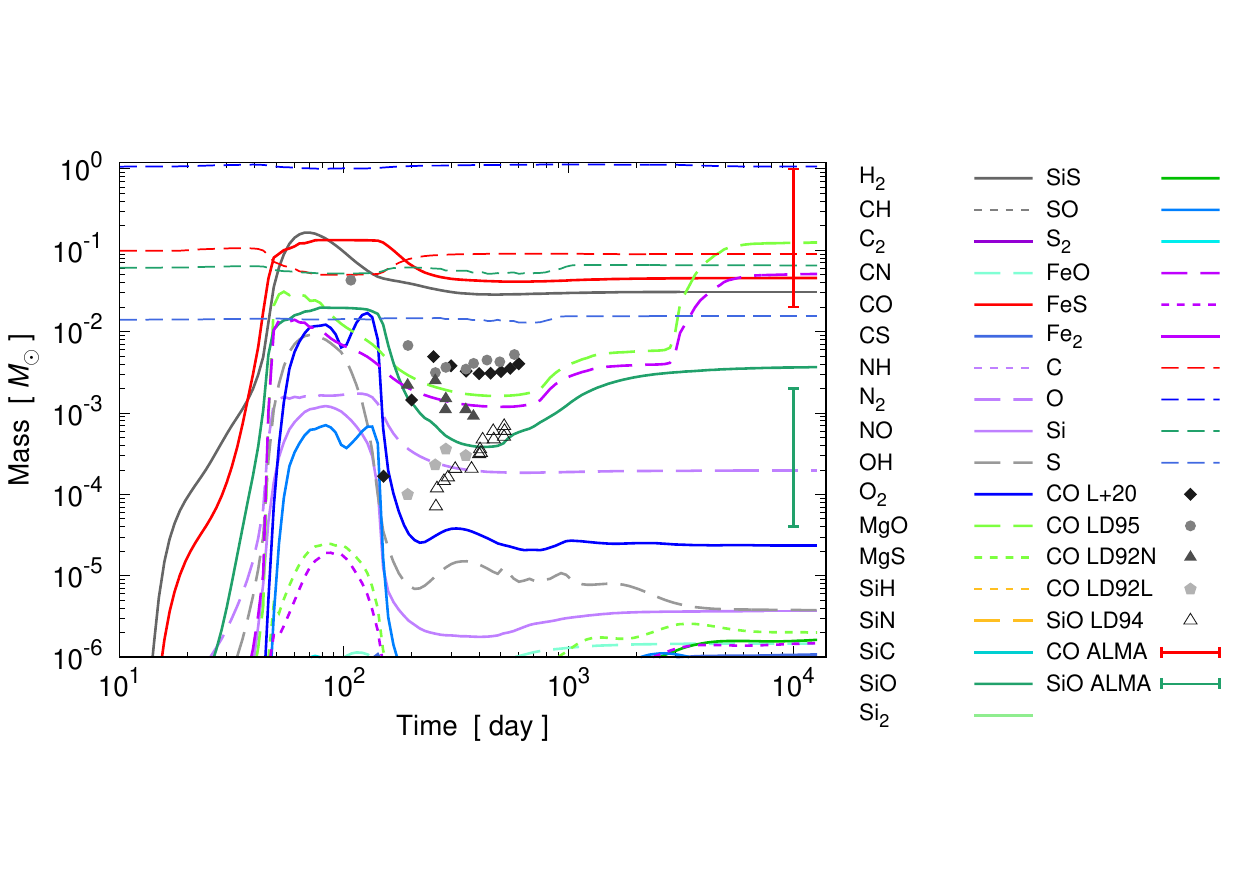}
\end{center}
\end{minipage}
\begin{minipage}{0.5\hsize}
\vs{-0.5}
\begin{center}
\includegraphics[width=7.4cm,keepaspectratio,clip]{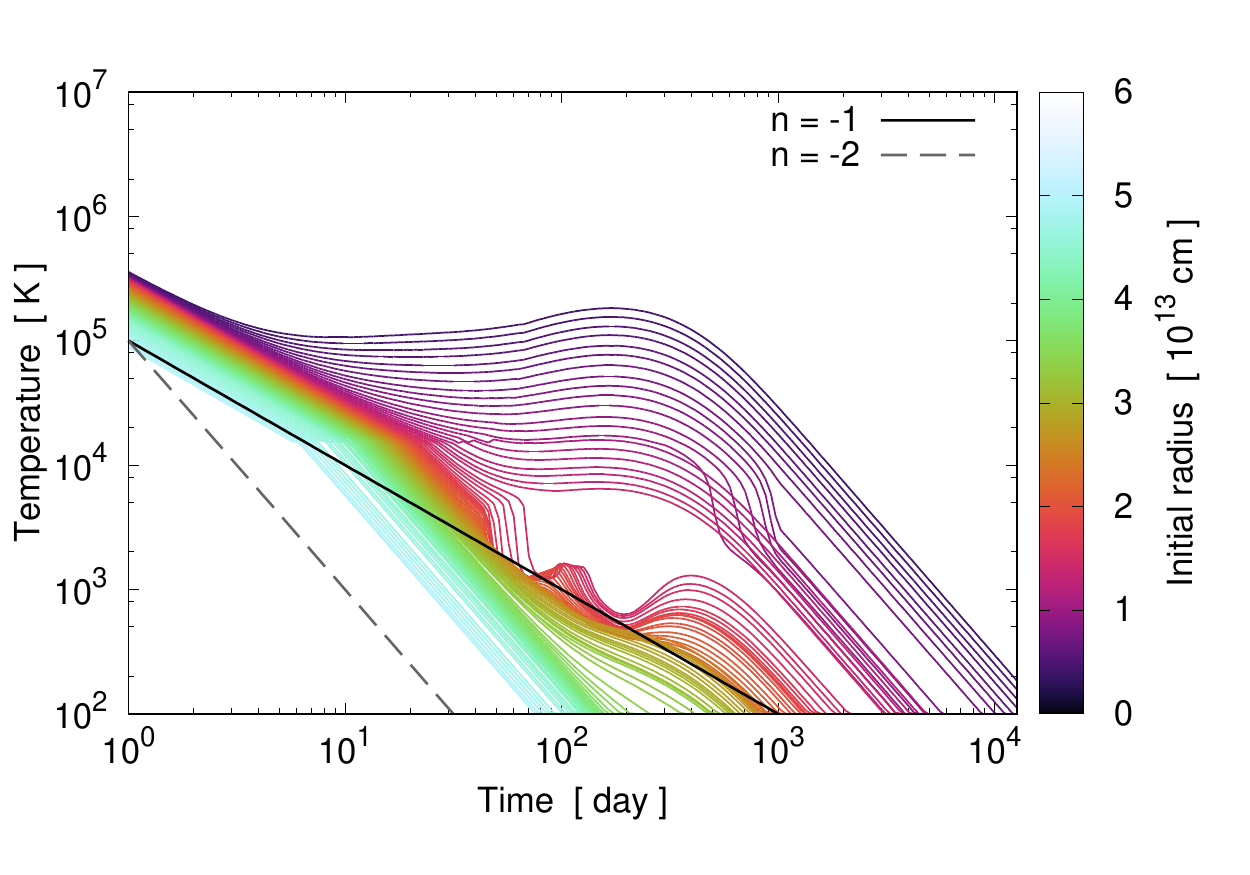}
\end{center}
\end{minipage}
\\
\begin{minipage}{0.5\hsize}
\vs{-1.7}
\begin{center}
\includegraphics[width=9.5cm,keepaspectratio,clip]{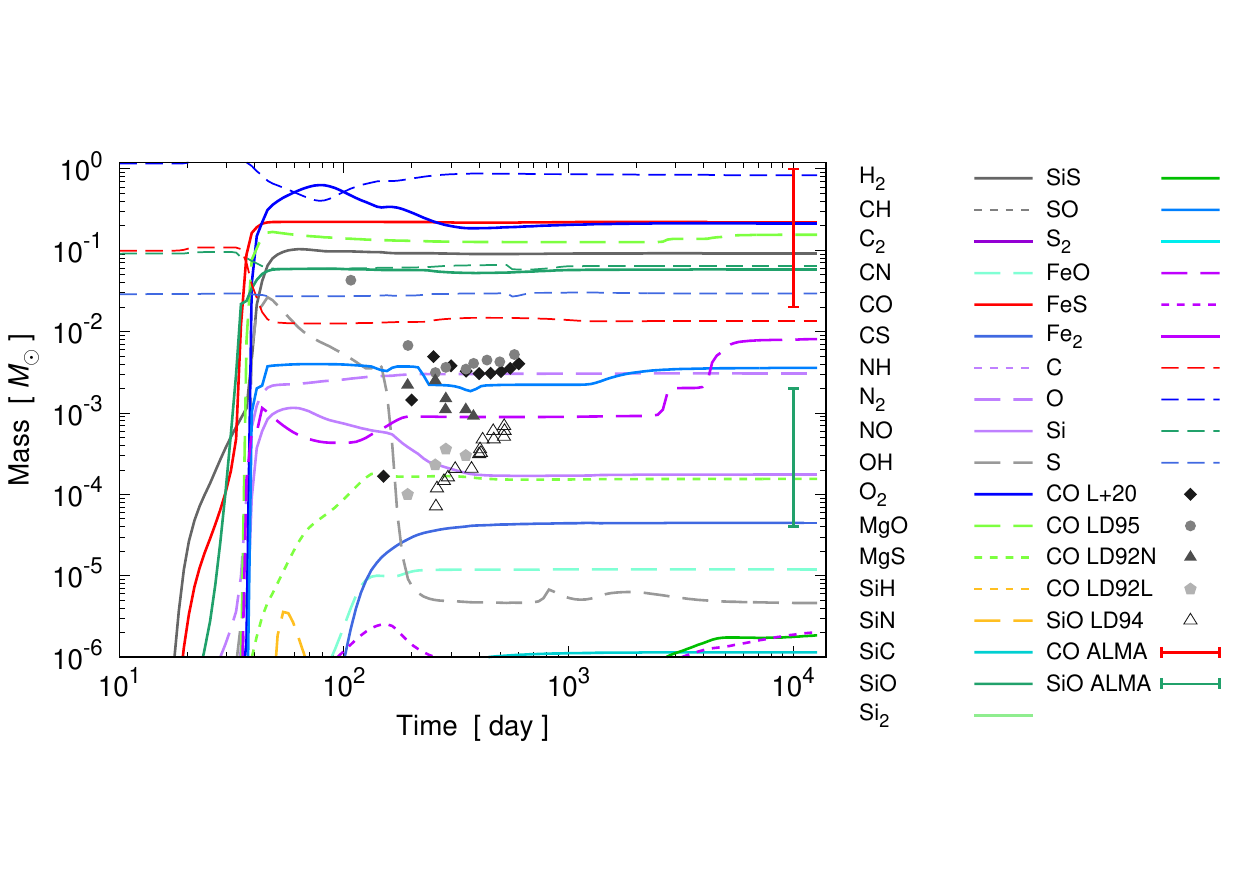}
\end{center}
\end{minipage}
\begin{minipage}{0.5\hsize}
\vs{-1.7}
\begin{center}
\includegraphics[width=7.4cm,keepaspectratio,clip]{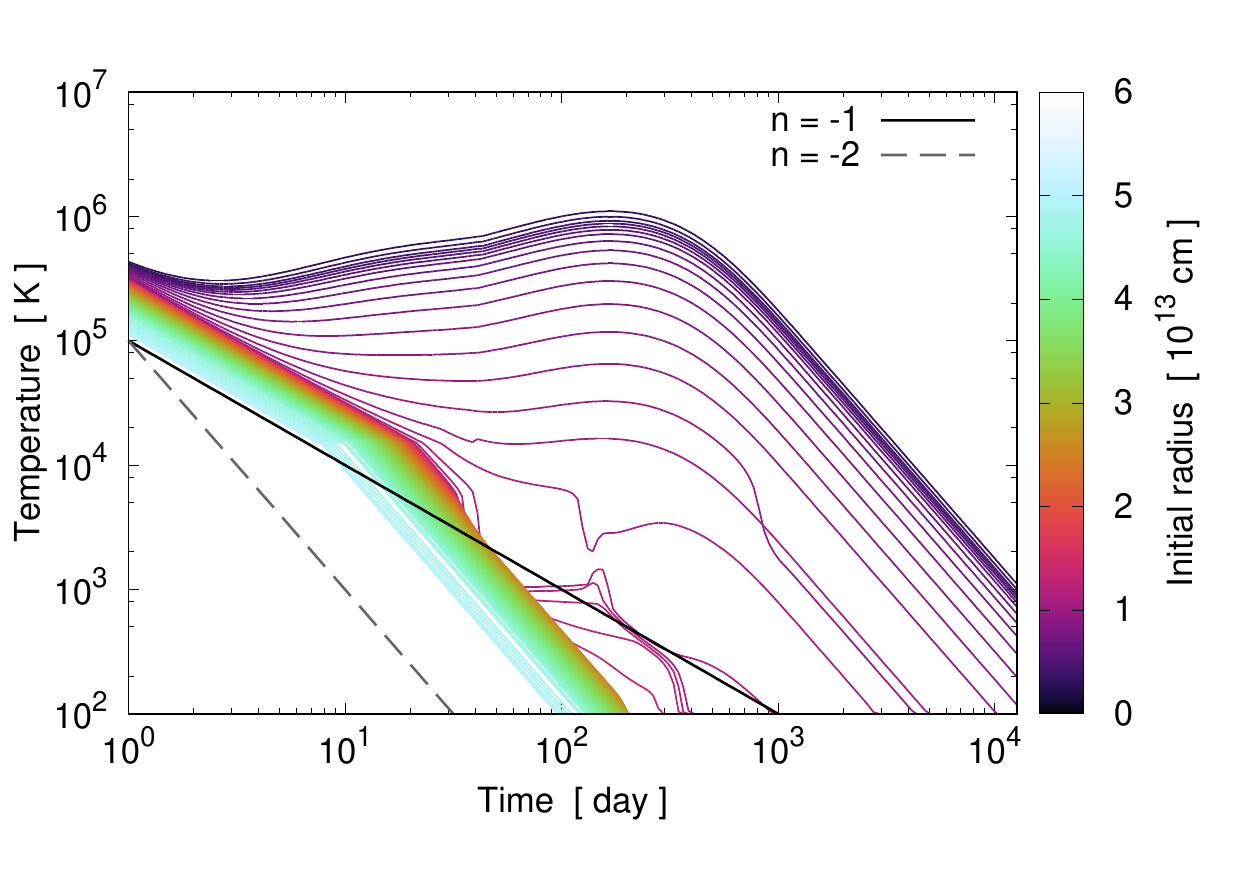}
\end{center}
\end{minipage}
\\
\begin{minipage}{0.5\hsize}
\vs{-1.7}
\begin{center}
\includegraphics[width=9.5cm,keepaspectratio,clip]{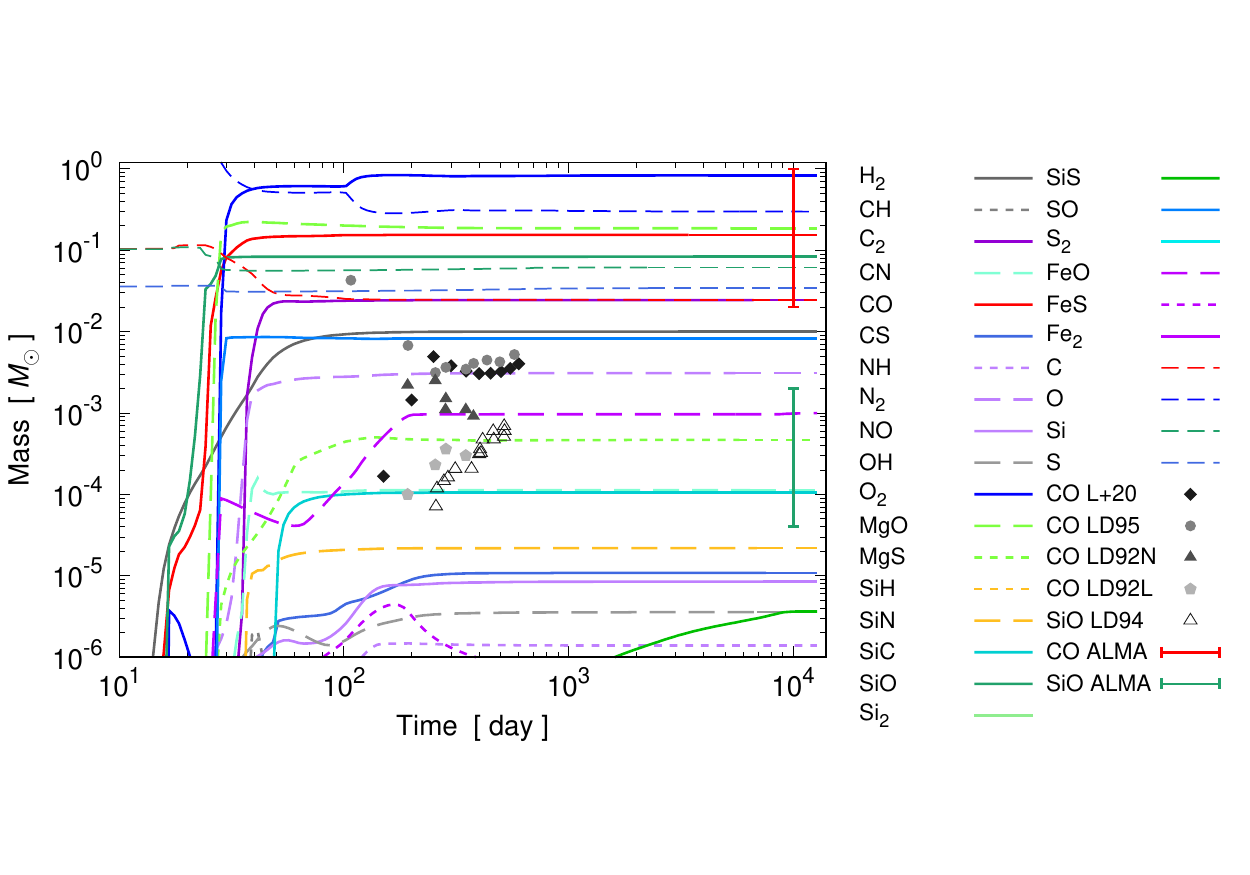}
\end{center}
\vs{-1.2}
\end{minipage}
\begin{minipage}{0.5\hsize}
\vs{-1.7}
\begin{center}
\includegraphics[width=7.4cm,keepaspectratio,clip]{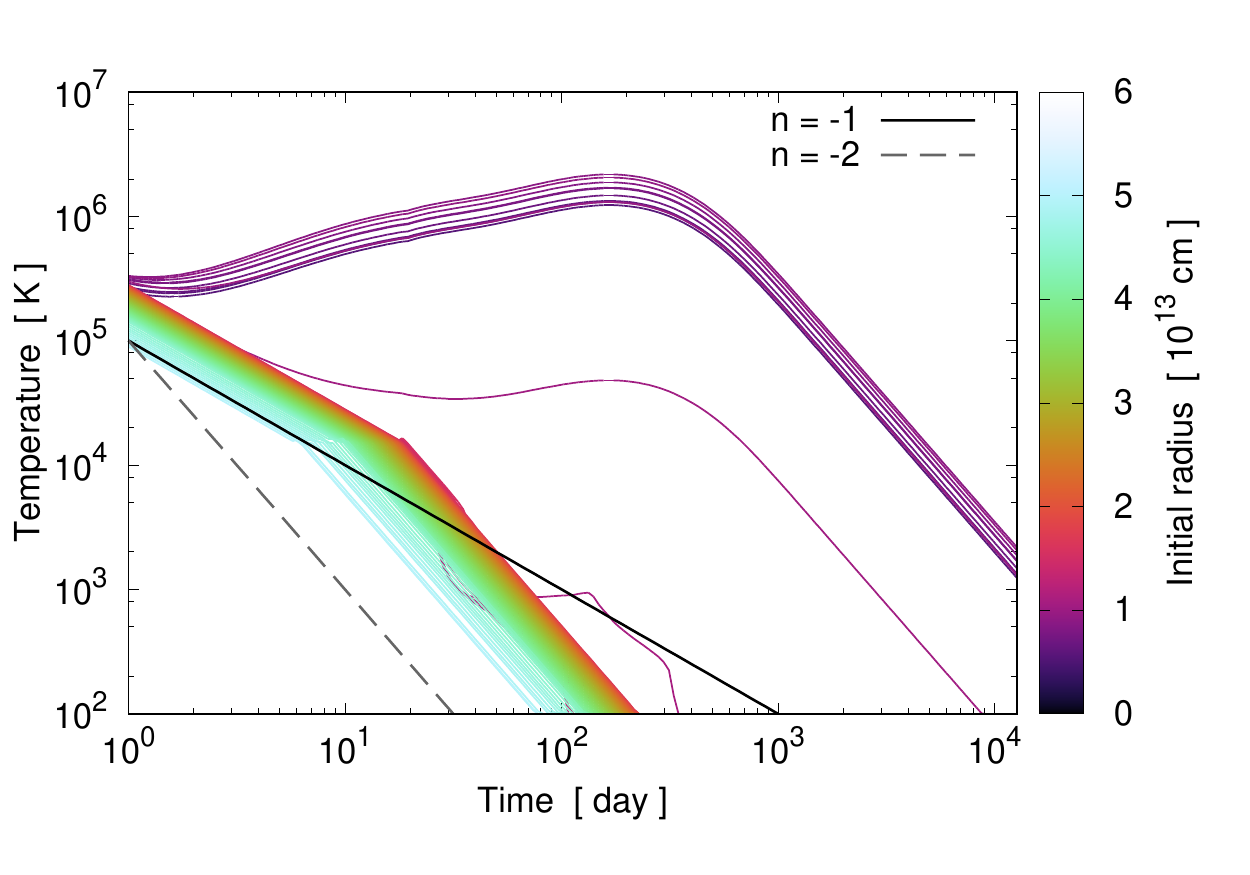}
\end{center}
\vs{-1.2}
\end{minipage}
\caption{Same as Figure~\ref{fig:1d_mass_temp_binary} but for the models n16.3-mean (top panels), n16.3-sphel (middle panels), and n16.3-sphel-pure (bottom panels).} %
\label{fig:1d_mass_temp_single}
\end{figure*}

\begin{figure}
\begin{minipage}{0.5 \hsize}
\vs{-0.5}
\begin{center}
\includegraphics[width=8.5cm,keepaspectratio,clip]{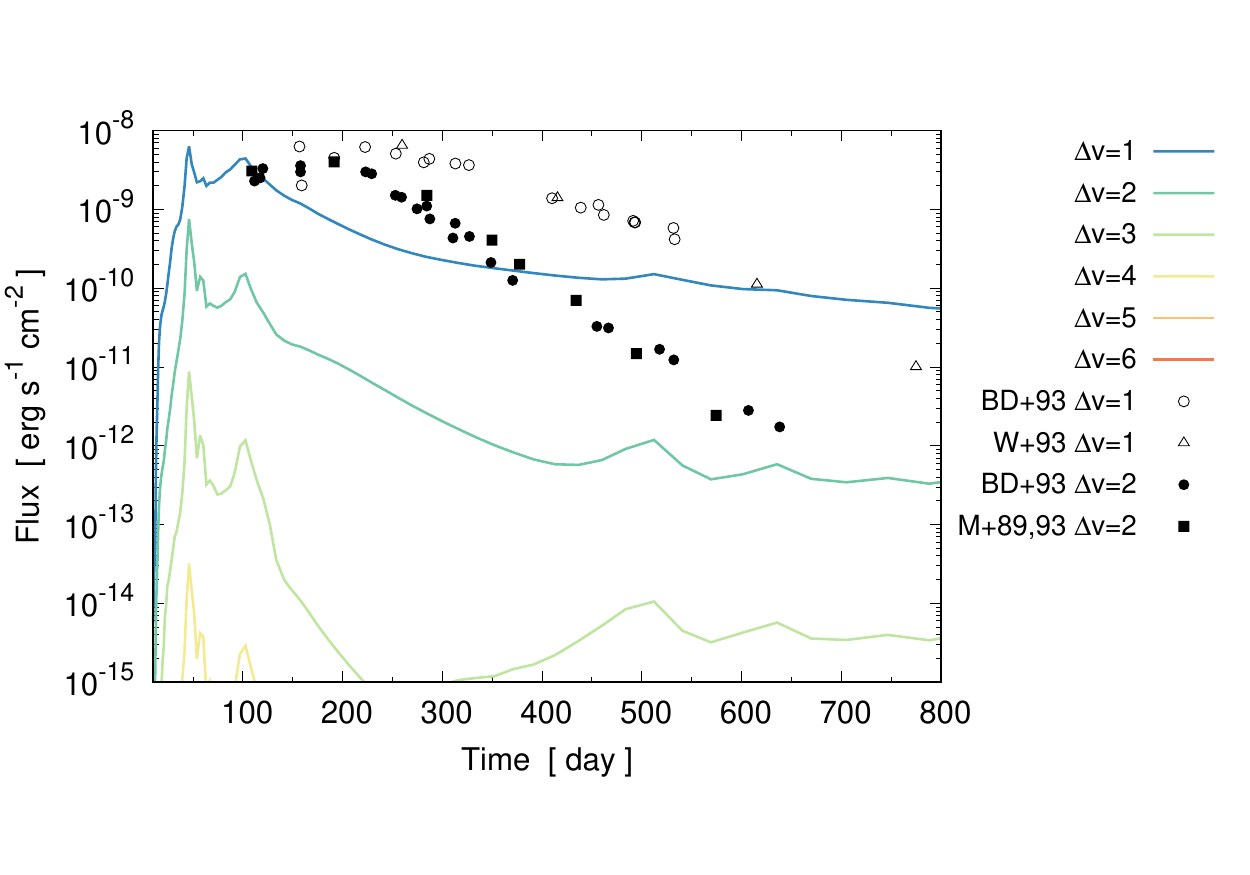}
\end{center}
\end{minipage}
\\
\begin{minipage}{0.5 \hsize}
\vs{-1.}
\begin{center}
\includegraphics[width=8.5cm,keepaspectratio,clip]{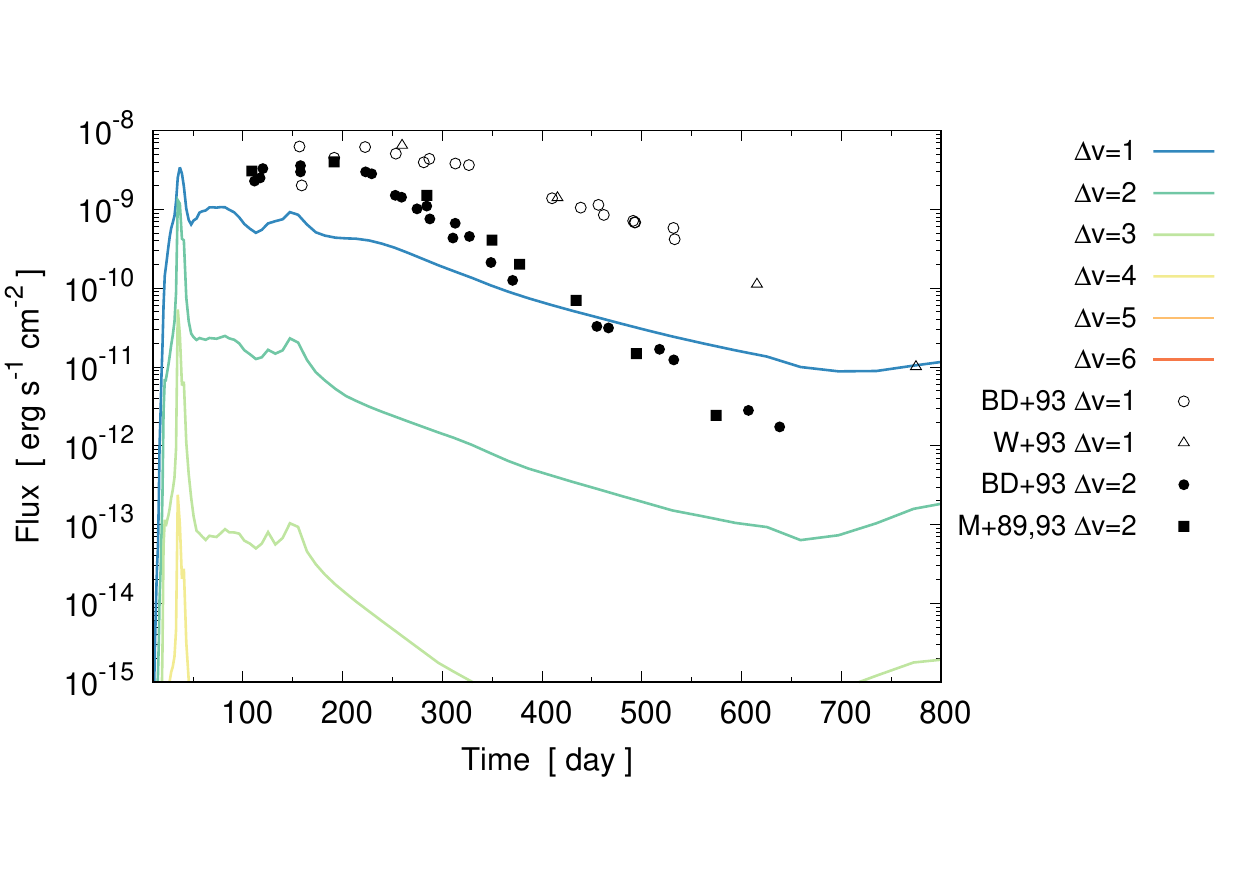}
\end{center}
\end{minipage}
\\
\begin{minipage}{0.5 \hsize}
\vs{-1.}
\begin{center}
\includegraphics[width=8.5cm,keepaspectratio,clip]{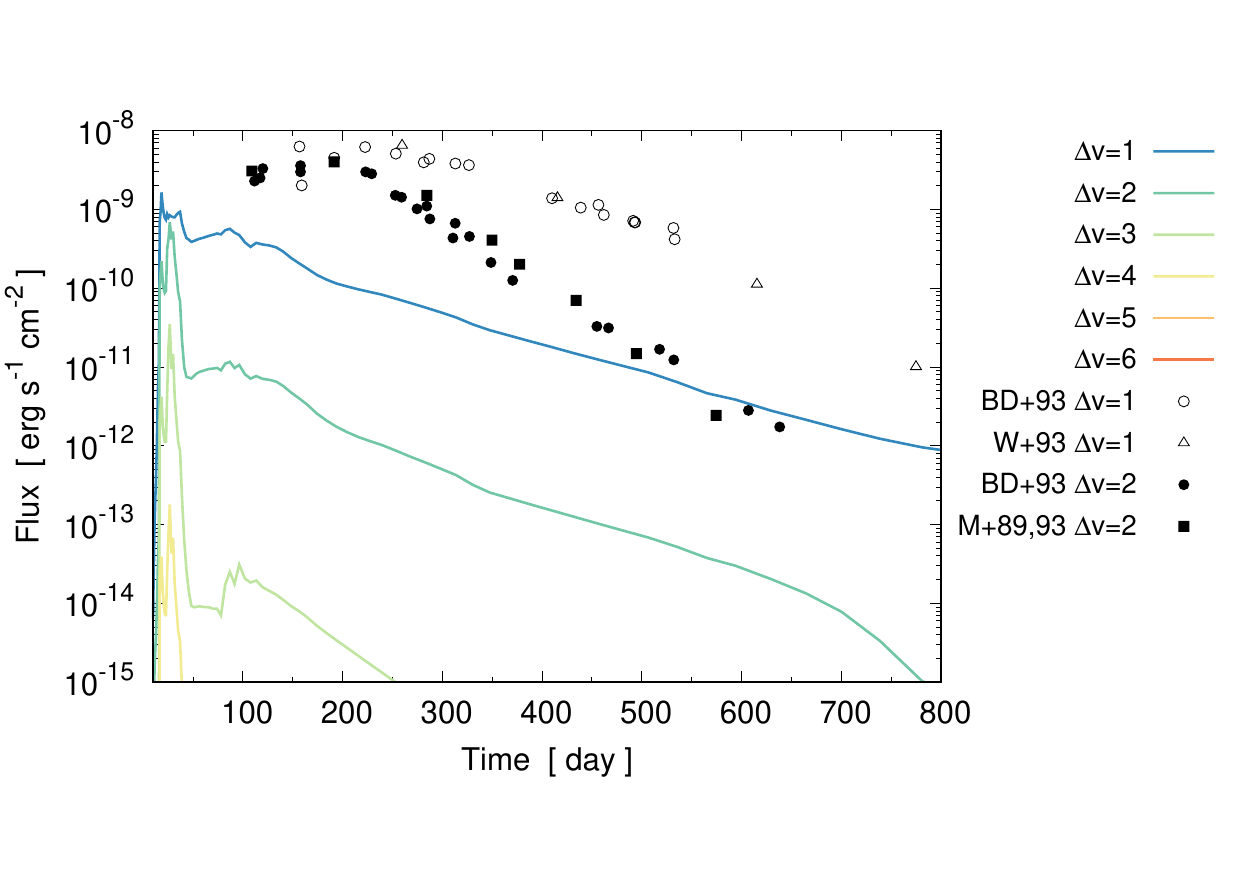}
\end{center}
\vs{-0.7}
\end{minipage}
\caption{Same as Figure~\ref{fig:1d_mass_temp_binary} but for the models n16.3-mean (top), n16.3-sphel (middle), and n16.3-sphel-pure (bottom).} %
\label{fig:1d_engy_loss_single}
\end{figure}

\hspace{\parindent}
In this section, the results of the model n16.3-mean are presented by comparing with those of the model b18.3-mean mainly focusing on the differences. %
In the top panels in Figure~\ref{fig:1d_mass_temp_single}, the results of the model n16.3-mean are shown. %

As can be seen in the right panel, the gas temperatures of outer particles ($\gtrsim 5 \times 10^{13}$ cm in the bottom panels in Figure~\ref{fig:prof_mean}) are overall higher than those of the model b18.3-mean (top right panel in Figure~\ref{fig:1d_mass_temp_binary}). %
Then, the timing when the gas temperatures of the outer particles go down to $\sim$ 10$^4$ K, i.e., when the molecule formation starts, is delayed compared with one for the model b18.3-mean; %
actually, the molecule formation starts after 10 days (before 10 days for the case of b18.3-mean) as seen in the left panel. %
Overall, other qualitative features of the gas temperature evolution are similar to the case of the b18.3-mean; %
due to the gas heating by the decay of $^{56}$Ni, in particular, inner particles are heated up and the gas temperatures peak at about 200 days; %
some particles initially positioned at (1.5--2.5) $\times$ 10$^{13}$ cm go through the cooling via CO ro-vibrational transitions after about 60 days and some inner particles also cooled at later phases ($\lesssim$ 1000 days). %

The fluxes of CO vibrational bands (the top panel in Figure~\ref{fig:1d_engy_loss_single}) peak before 100 days and the calculation fails to reproduce the peaks of the observed fundamental and first overtone bands same as other results shown above. %
The peak flux levels are similar to those of the b18.3-mean case but slightly lower around the observed peaks. %
The increases in the fluxes at later phases (after 300 days for the fundamental and first overtone bands) are attributed to the cooling of inner particles. %

Related to the formation of molecules, the amounts of initial seed atoms are different from those of the model b18.3-mean as seen in the top left panel in Figure~\ref{fig:1d_mass_temp_single}. %
Apparently, the amount of oxygen atoms ($\sim$ 1 $M_{\odot}$) is higher than that of the model b18.3-mean (see, the top left panel in Figure~\ref{fig:1d_mass_temp_binary}). %
On the other hand, the amounts of carbon and silicon atoms are lower than those of the model b18.3-mean. %
Then, the abundance ratios of oxygen atoms to carbon and silicon are higher than the case of the model b18.3-mean. %
Additionally, in the model n16.3-mean, the amount of carbon atoms is higher than that of silicon in contrast to the model b18.3-mean. %
The initial amount of sulfur is also different from the case of b18.3-mean. %
Such differences in the initial abundance ratios of seed atoms could potentially affect the formation of molecules. %

The qualitative trends of the evolution of the amount of CO is similar to the case of b18.3-mean except for the initial rise. %
On the other hand, the amount of SiO is rather different between the two models. %
In the model n16.3-mean, the amount of SiO at around 40--200 days is lower compared with the model b18.3-mean roughly by a factor of five. %
The amount of SiO at about 200--600 days is rather lower (at most one order of magnitude) than that of the model b18.3-mean. %
The final amount of SiO is also lower than that of the model b18.3-mean approximately by a factor of a few. %

The contributing formation and destruction processes of SiO at about 40--200 days are not so different between the models b18.3-mean and n16.3-mean, although the order of the significance of those processes is a bit different between the two models depending on time. %
After 200 days, the primary destruction process of SiO is the \texttt{CE} reaction in Equation~(\ref{eq:h+_sio}) for both the models b18.3-mean and n16.3-mean. %
The secondary destruction processes are the \texttt{NN} reaction in Equation~(\ref{eq:h_sio_sih}), the ionization by Compton electrons (one of \texttt{CM} reactions), and the \texttt{UV} reaction; %
in the model n16.3-mean, among the secondary processes, the latter processes dominate the \texttt{NN} reaction (in the model b18.3-mean, the \texttt{NN} reaction dominates the processes related to Compton electrons). %
The difference in the amount of SiO between the two models may be attributed to the different initial abundance ratios of the seed atoms, in particular, the ratio of carbon to silicon, and/or the differences in gas densities and temperatures. %

As for other molecules, for example, probably due to the differences in the initial abundance ratios of sulfur to oxygen, the model n16.3-mean leads to a higher amount of O$_2$ and a lower amount of SO than those of the model b18.3-mean. %
Similarly, in the model n16.3-mean, the amount of SiS is apparently lower than that of the model b18.3-mean due to the lower silicon and sulfur abundances. %

Here, the primary formation processes of O$_2$ is the \texttt{RA} reaction between two oxygen atoms and the \texttt{NN} reaction of oxygen with OH as the same as the model b18.3-mean. %
SO is mainly formed by the \texttt{RA} reaction between sulfur and oxygen atoms and the \texttt{NN} reaction of sulfur with O$_2$. %
It is noted that in the model b18.3-mean, the \texttt{NN} reaction of oxygen with S$_2$ is dominant instead of the \texttt{NN} reaction above. %
The discrepancy is probably due to the high initial abundance ratio of oxygen to sulfur in the model n16.3-mean. %
SiS is mainly produced by the \texttt{RA} reaction between silicon and sulfur atoms. 
The \texttt{NN} reaction of silicon with S$_2$, which is effective in the model b18.3-mean, however, does not contribute to the formation of SiS in the model n16.3-mean. %
This can also be attributed to the low initial abundance ratio of sulfur to the others. %

After about 130 days, for both two models, O$_2$, SO, and SiS are significantly destructed (in the model n16.3-mean, SiS does not appear in the top left panel in Figure~\ref{fig:1d_mass_temp_single} before 2000 days). %
The main destruction processes of O$_2$ at this phase are the \texttt{NN} reactions of O$_2$ with silicon and with carbon. %
The \texttt{CE} reaction between H$^+$ and O$_2$ also contributes to some extent in contrast to the model b18.3-mean. %
SO is primarily destructed by the \texttt{NN} reactions of carbon with SO (two cases of products: CS and oxygen; %
CO and sulfur), the \texttt{NN} reaction of nitrogen with SO, and the \texttt{CE} reaction between H$^+$ and SO. %
The contribution of the \texttt{CE} reaction above is limited in the model b18.3-mean. %
The destruction processes of SiS are a bit complicated. %
SiS is destructed mainly by the \texttt{NN} reactions of SiS with nitrogen (products: SiN and sulfur) and with hydrogen (products: SiH and sulfur). %
The thermal fragmentation reaction of SiS with helium also primarily contributes to the destruction. %
The \texttt{CE} reaction between H$^+$ and SiS contributes to some extent in contrast to the model b18.3-mean. %
Among those destruction processes, the significance varies depending on time and the models. %
It is noted that since the \texttt{CE} reactions mentioned above involve ions, the mixing of $^{56}$Ni, which influences the ionization by Compton electrons, may affect the significance. %
On the other hand, the primary destruction processes of O$_2$, SO, and SiS are not always ion-related reactions like the \texttt{CE} reactions above but are rather \texttt{NN} reactions; %
the amounts would be determined by the complicated balance between the \texttt{NN} reactions depending on the gas temperatures. %

Among the other molecules, the trends of H$_2$, OH, MgO, and FeO are qualitatively similar to those of the model b18.3-mean. %
The amount of MgO is, however, larger than that of FeO in the model n16.3-mean in contrast to the model b18.3-mean. %
The contributing formation and destruction processes of MgO and FeO are consistent with the model b18.3-mean (see, Section~\ref{para:b18.3-mean}). %
This feature is probably attributed to the initial high abundance of magnesium in the model n16.3-mean (see, e.g., the left panels in Figures~\ref{fig:prof_mean} and \ref{fig:prof_sphel}). %

A distinct difference between the models b18.3-mean and n16.3-mean is the amount of N$_2$; %
the amount of N$_2$ in the model n16.3-mean is significantly lower than that of the model b18.3-mean. %
The contributing formation and destruction reactions are the same as the model b18.3-mean. %
The feature above is likely attributed to the fact that the nitrogen abundance, in particular, at the outer layers in the progenitor model b18.3 is higher than that in n16.3. %
As mentioned, since the progenitor model b18.3 is based on a binary merger, the CNO cycle additionally triggered during the merger process increases the CNO cycle-processed materials including nitrogen ($^{14}$N) into the envelope. %

As a summary, it is difficult to clearly figure out what is the primary reason for the differences in the amounts of those molecules between the two models. %
Even though, the initial abundance ratios of seed atoms, $^{56}$Ni, hydrogen, and helium, which reflect the matter mixing, likely affect the results as presented above. %

\subsubsection{Comparison with the spherical cases} \label{subsubsec:1d_comp_sphel} 

In this section, in order to see the impact of the effective matter mixing, by comparing the models b18.3-mean and n16.3-mean, the results based on the initial profiles for spherical cases (spherical explosion and purely spherical cases), i.e., the results for the models b18.3-sphel, b18.3-sphel-pure, n16.3-sphel, and n16.3-sphel-pure are presented; %
the former two and the latter two are described in Sections~\ref{para:1d_sphel_binary} and \ref{para:1d_sphel_single}, respectively. %

\paragraph{The cases with the binary merger progenitor model} \label{para:1d_sphel_binary}

\hspace{\parindent} 
In the middle panels in Figure~\ref{fig:1d_mass_temp_binary}, the results of the model b18.3-sphel (the spherical explosion case) are shown. %
Inner particles (initially less than about 1 $\times$ 10$^{13}$ cm) are heated by the decay of $^{56}$Ni (the right panel) since $^{56}$Ni is initially distributed only inside $\sim$ 1 $\times$ 10$^{13}$ cm in contrast to the model b18.3-mean, in which particles initially inside $\sim 3$ $\times$ 10$^{13}$ cm are a matter of heating. %
As a result, the peak temperatures of the inner particles at around 200 days become higher than those of the model b18.3-mean due to the high local $^{56}$Ni abundance. %
Some particles initially at (1--2) $\times$ 10$^{13}$ cm and a few inner particles go through the cooling by CO ro-vibrational transitions. %

In the middle panel in Figure~\ref{fig:1d_engy_loss_binary}, the time evolution of the fluxes of CO vibrational bands is shown. The calculated fluxes of CO vibrational bands are qualitatively similar to those of the b18.3-mean (the top panel) but quantitatively the flux levels are slightly lower than those of the b18.3-mean. %

The qualitative features of the evolution of the amounts of CO and SiO (the middle left panel in Figure~\ref{fig:1d_mass_temp_binary}) are roughly consistent with that of the model b18.3-mean; %
some differences, however, can also be recognized. %
The amounts of CO and SiO are higher than those of the model b18.3-mean throughout the evolution. %

The initial rise of the amounts of molecules is steeper than that of the model b18.3-mean. %
At approximately 50 days the amounts of both CO and SiO exceed 0.1 $M_{\odot}$. %
Then, later the amounts are reduced by destruction processes; %
the destruction of some of the molecules, in particular SiO, is, however, milder than that of the model b18.3-mean. %
This is because $^{56}$Ni is concentrated only in inner regions compared with the model b18.3-mean. %
In the inner particles heated by the decay of $^{56}$Ni, the molecule formation is delayed. %
On the other hand, in the outer particles, with the relatively low local $^{56}$Ni abundance, molecules start to form earlier in higher-density environment and the destruction processes caused by the decay of $^{56}$Ni, i.e., the \texttt{CE} reaction in Equation~(\ref{eq:h+_sio}) and the \texttt{CM} reactions (ionization, dissociation, and dissociative ionization by Compton electrons), are less effective compared with the model b18.3-mean. %
In total, less efficient mixing of $^{56}$Ni in the model b18.3-sphel results in earlier effective formation and less effective destruction of molecules compared with the model b18.3-mean. %
Actually, it is confirmed that in the model b18.3-sphel, the primary destruction processes of SiO after about 100 days are the \texttt{NN} reactions in Equation~(\ref{eq:h_sio_sih}) and the \texttt{CE} reaction in Equation~(\ref{eq:h+_sio}) is the secondary in contrast to the model b18.3-mean (in the model b18.3-mean, the \texttt{CE} reaction is the primary), which supports the statement above. %

Additionally, the evolution of the amounts of other molecules such as SO and O$_2$ is distinctively different. %
The amounts of SO and O$_2$ at around 50--130 days are rather higher (at most more than one order of magnitude) than those of the model b18.3-mean and the final amounts are also affected. %
At about 50--100 days (SO and O$_2$ increasing phase), the primary formation processes of SO are the \texttt{NN} reactions of sulfur with O$_2$ and with OH. %
The primary formation process of O$_2$ is the \texttt{NN} reaction of oxygen with OH. %
For both SO and O$_2$, the major formation processes at this phase are consistent with the model b18.3-mean. %

After 100 days, in the model b18.3-sphel, the amounts of SO and O$_2$ start to decrease in contrast to the model b18.3-mean. %
The major destruction reactions of SO are the \texttt{NN} reactions of SO with carbon (products: CO and sulfur; CS and oxygen). %
The primary destruction process of O$_2$ is the \texttt{NN} reaction of O$_2$ with silicon. %
Therefore, the destruction processes of SO and O$_2$ are also consistent with the model b18.3-mean. %
The abundances of SO and O$_2$ are probably determined by the balance of the \texttt{NN} reactions depending on the gas temperatures. %

The behavior of H$_2$ is different from that of b18.3-mean. %
In the model b18.3-sphel, the early formation of H$_2$ is limited, and the ratios of the amount of H$_2$ to CO and SiO are apparently smaller than those in the model b18.3-mean. %
Since the primary formation process is the sequence, the \texttt{REA} reaction, H + e$^-$  $\lra$ H$^-$ + $\gamma$; the \texttt{AD} reaction, H$^-$ + H $\lra$ H$_2$ $+$ e$^-$, the smaller amount of H$_2$ in the model b18.3-sphel may be attributed to the less effective ionization due to less effective mixing of $^{56}$Ni. %

Among the other molecules, N$_2$, MgO, and FeO have a different feature from that of the model b18.3-mean; %
after the formation of those molecules before 100 days, there is no distinct destruction of those species. %
Since the destruction processes of those are basically the \texttt{CM} reactions, the less effective destruction in the model b18.3-sphel may be attributed to the less effective mixing of $^{56}$Ni, again.

In the bottom panels in Figure~\ref{fig:1d_mass_temp_binary}, the results of the model b18.3-sphel-pure (the purely spherical case) are shown. %
First, as seen in the right panel, only several inner particles are heated by the decay of $^{56}$Ni, since the $^{56}$Ni is initially distributed at the innermost region ($\lesssim$ 0.5 $\times$ 10$^{13}$ cm; see, the top left panel in Figure~\ref{fig:prof_sphel_pure}). %
The peak temperatures of the heated particles are significantly higher than those of the models b18.3-mean and b18.3-sphel due to the high local $^{56}$Ni abundance. %
Only several inner particles are affected by the cooling due to the CO ro-vibrational transitions. %
The gas temperature evolution of the other particles not affected by the heating and cooling follows the broken power-law. %

In the bottom panel in Figure~\ref{fig:1d_engy_loss_binary}, the fluxes of CO vibrational bands are shown. %
The fluxes at early phases (before 300 days) are distinctively lower than those of the models b18.3-mean and b18.3-sphel. %
It may be understood that due to the inefficient ionization by Compton electrons (because of the less effective mixing of $^{56}$Ni), the number density of (thermal) electrons for exciting the vibrational levels of CO is lower than that of efficient ionization cases (efficient mixing cases). %
Actually, this tendency, i.e., less efficient ionization results in lower CO vibrational fluxes, is also seen in between the models b18.3-mean and b18.3-sphel; %
the fluxes of CO vibrational transitions in the model b18.3-mean (the top panel in Figure~\ref{fig:1d_engy_loss_binary}), in which the mixing of $^{56}$Ni and ionization by Compton electrons are efficient compared with the model b18.3-sphel, are overall higher than that of the model b18.3-sphel (the middle panel in the same figure). %
The sudden rise in the fluxes at around 300 days is apparently attributed to the particle that has a rapid cooling (see, the bottom right panel in Figure~\ref{fig:1d_mass_temp_binary}); %
this feature would be artificial due to the insufficient number of particles around the regions where the particle with the rapid cooling is (the gas temperatures have huge gaps between adjacent particles). %
Although higher resolution (number) of tracer particles may result in smoother light curves of CO vibrational bands and the radial distributions of molecules, it should not affect the total amounts of molecules much. %

The evolution of the amounts of molecules has noticeable differences from the models b18.3-mean and b18.3-sphel. %
CO and SiO molecules are sharply formed after about 40 days when the gas temperatures of most of the particles go down to $\sim$ 10$^4$ K. %
After the rapid formation of CO and SiO, plateaus of the amounts of them can be recognized. %
This phase corresponds to the timing when molecules start to form in the tracer particles inside the helium-rich shell ($\lesssim$ 0.8 $\times$ 10$^{13}$ cm in the top left panel in Figure~\ref{fig:prof_sphel_pure}), i.e., their gas temperatures cool down to $\sim$ 10$^4$ K. %
The amount of SiO has a plateau with sharp edges compared with that of CO. %
After SiO enters the plateau phase, the amount of CO continues to increase rapidly for a while. %
Then, the increase becomes slow. %
The plateau of SiO with the sharp edges is probably due to the abundance ratios among the seed atoms, carbon, oxygen, and silicon, inside the helium-rich shell; %
the high abundance ratio of carbon to oxygen (actually, $N({\rm C}) > N({\rm O})$ in the helium-rich shell, where $N({\rm C})$ and $N({\rm O})$ denote the number density of the seed carbon and oxygen atoms, respectively) and the low abundance ratios of silicon to carbon and oxygen may inhibit the formation of SiO. %
After 70 days, the amounts of CO and SiO again start to increase. %
This phase corresponds to the timing when molecules start to form in the tracer particles inside oxygen and silicon-rich shells ($\lesssim$ 0.5 $\times$ 10$^{13}$ cm in the top left panel in Figure~\ref{fig:prof_sphel_pure}). %
The high abundance ratios of oxygen and silicon allow SiO to form and contribute to increasing both the amounts of CO and SiO. %
After 100 days, the amounts of CO and SiO, in particular, SiO, decrease. %
The primary destruction processes of CO are the \texttt{NN} reaction in Equation~(\ref{eq:si_co}) (SiO formation reaction) and the \texttt{CM} and \texttt{UV} reactions. %
SiO is primarily destructed by the \texttt{CM} and \texttt{UV} reactions. %
For the destruction by \texttt{CM} and \texttt{UV} reactions, several particles affected by the decay of $^{56}$Ni (deviated from the broken power-low evolution) except for ones whose peak temperatures after 100 days are greater than 10$^5$ K contribute to the destruction. %
After 500 days, SiO distinctly increases again, and the amount becomes comparable with that of CO. %
The contributing formation processes of SiO are the \texttt{NN} reactions in Equations~(\ref{eq:si_co}) and (\ref{eq:si_o2}). %

It is noted that in the innermost particles, due to the strong heating with the high local $^{56}$Ni abundance, the gas temperatures go down to 10$^{4}$ K only at rather late phases after a few thousand days when the gas densities are very low. %
Then, the contributions of the innermost particles to the amounts of molecules may be relatively small. %

The properties of the formation of other molecules are also different from the models b18.3-mean and b18.3-sphel. %
In the models b18.3-mean and b18.3-sphel, the formation of SO and O$_2$ is recognized at about 40--200 days; in the model b18.3-sphel-pure, the two molecules are, however, not distinctly formed at least before 100 days. %
Instead, CS and SiC are formed just after 40 days, and the amount of CS changes remarkably at around 70 days. %
CS is mainly formed by the \texttt{RA} reaction between carbon and sulfur atoms, the \texttt{NN} reaction of sulfur with C$_2$, and the \texttt{NN} reaction of carbon with SO before 70 days. %
SiC is primarily formed by the \texttt{NN} reaction of carbon with SiO, the \texttt{RA} reaction between silicon and carbon atoms, and the \texttt{NN} reaction of silicon with C$_2$ before 70 days. %
The major destruction processes of CS after 70 days are the \texttt{NN} reaction of hydrogen with CS (products: CH and sulfur) and the \texttt{NN} reaction of oxygen with CS (products: CO and sulfur). %
SiC is mainly destructed by the \texttt{NN} reactions of oxygen with SiC (products: SiO and carbon; CO and silicon) after 70 days, although the decrease of the amount is not distinctively recognized. %
It is noted that in both of the formation processes of CS and SiC, C$_2$ plays a major role. %
In the model b18.3-sphel-pure, C$_2$ is significantly formed after 40 days in contrast to the models b18.3-mean and b18.3-sphel; %
the amount exceeds 10$^{-2}$ $M_{\odot}$. %
C$_2$ is mainly formed by the \texttt{RA} reaction between two carbon atoms and the \texttt{NN} reactions of carbon with CN and CH. %
Similarly, for example, SiN is also markedly formed in contrast to the models b18.3-mean and b18.3-sphel, where SiN is primarily formed by the \texttt{NN} reaction of nitrogen with SiO. %

On the other hand, as observed in the model b18.3-sphel, the formation of H$_2$ is further limited compared with that of the model b18.3-mean due to the less effective mixing of $^{56}$Ni. %
Additionally, the formation of OH is also significantly reduced compared with the models b18.3-mean and b18.3-sphel. %
The primary formation processes of OH before 100 days are the \texttt{RA} reaction between oxygen and hydrogen atoms and the \texttt{NN} reaction of oxygen with H$_2$. %
Therefore, the formation of OH is also affected by the mixing of $^{56}$Ni through the latter reaction above. %

As a summary, the distinct features found in the model b18.3-sphel-pure are probably attributed to the initial stratified profiles of the seed atoms and the abundance ratios of each layer (see the top left panel in Figure~\ref{fig:prof_mean}) owing to the purely spherical hydrodynamical evolution and the less effective mixing of $^{56}$Ni into outer layers. %

\paragraph{The cases with the single-star progenitor model} \label{para:1d_sphel_single}

\hspace{\parindent} 
In this subsection, the results of spherical cases based on the single-star progenitor model (n16.3), i.e., ones for the models n16.3-sphel and n16.3-sphel-pure, are presented mainly focusing on the differences from those based on the binary merger progenitor model (b18.3). %

In the middle and bottom panels in Figure~\ref{fig:1d_mass_temp_single}, the results of the models n16.3-sphel and n16.3-sphel-pure are shown, respectively. %
The qualitative features of the gas temperatures are similar to the models b18.3-sphel and b18.3-sphel-pure. %
Some recognizable differences are as follows. %
The temperatures of the innermost particles in the model n16.3-sphel are higher than those in the model b18.3-sphel. %
In the models n16.3-sphel and n16.3-sphel-pure, the number of significantly heated particles whose peak 
temperatures at around 200 days greater than 10$^{5}$~K is higher than those of the corresponding counterpart models b18.3-sphel and b18.3-sphel-pure. %
This feature may partly be attributed to the differences in the initial distribution of $^{56}$Ni at the innermost region between the spherical models with the progenitor models b18.3 and n16.3 (see, the left panels in Figures~\ref{fig:prof_sphel} and \ref{fig:prof_sphel_pure}). %

In the middle and bottom panels in Figure~\ref{fig:1d_engy_loss_single}, the fluxes of CO vibrational bands for the models n16.3-sphel and n16.3-sphel-pure are shown, respectively. %
As seen in the cases of the binary merger progenitor model (b18.3), the more spherical the evolution is, the lower the fluxes of CO vibrational bands are; %
the fluxes of CO vibrational bands in the models n16.3-mean, n16.3-sphel, and n16.3-sphel-pure decrease in this order. %
Again, this may be attributed to the inefficient ionization by Compton electrons produced by the decay of $^{56}$Ni in the spherical cases. %
In the model n16.3-sphel-pure, there is no distinct increase of fluxes at later phases (after 300 days) in contrast to the model b18.3-sphel-pure. This is because no particle goes through rapid cooling at such late phases. %

The evolution of the amounts of molecules in the models n16.3-sphel and n16.3-sphel-pure is different from the corresponding counterpart models based on the binary merger progenitor model (b18.3) in several points. %

As seen in the model n16.3-mean, in both models n16.3-sphel and n16.3-sphel-pure, molecules start to form a bit later time (after 10 days) compared with the models based on the progenitor model b18.3 by reflecting the fact that the gas temperatures of outer particles are relatively higher than those of the corresponding counterpart models with the progenitor model b18.3. %
In the models n16.3-sphel and n16.3-sphel-pure, the amounts of CO and SiO are more stable after 40 days compared with the models b18.3-sphel and b18.3-sphel-pure. %
In the model b18.3-sphel-pure, in particular, the amount of SiO varies in time partly due to the stratified distributions of the seed atoms. %
The more stable feature in the amounts of CO and SiO in the spherical models based on the single-star progenitor model (n16.3) might partly be attributed to the less contribution from the significantly heated particles, in which molecule formation starts only at later phases in low-density environment due to the heating, (the number of those particles is larger than the counterpart models with the binary merger model as mentioned) and the relatively larger contribution from the particles that are not significantly affected by the decay of $^{56}$Ni, in which the destruction processes induced by the decay of $^{56}$Ni are not so significant. %

As mentioned in Section~\ref{para:n16.3-mean}, the initial abundance ratios of the seed atoms in the model n16.3-mean are different from those of the counterpart model b18.3-mean; %
the abundance ratios of oxygen to silicon and carbon are higher than that of the model b18.3-mean and the amount of oxygen in n16.3-mean is initially as high as $\sim$ 1 $M_{\odot}$. %
This feature does not change in the spherical cases, i.e., the models n16.3-sphel and n16.3-sphel-pure. %
By reflecting this feature, in the models n16.3-sphel and n16.3-sphel-pure, the amount of O$_2$ is significantly higher than those of the models based on the binary merger progenitor (b18.3) and the model n16.3-mean. %
In the model n16.3-sphel, the amount of O$_2$ is as high as 0.6 $M_{\odot}$ and is higher than that of CO at approximately 40--200 days. %
Later, it becomes almost comparable to that of CO after the destruction from 100 days to a few hundred days. %
O$_2$ is primarily formed by the \texttt{NN} reaction of oxygen with OH and the \texttt{NN} reaction between two OH molecules. %
The contributing destruction processes after 100 days are the \texttt{NN} reactions of O$_2$ with silicon (products: SiO and oxygen), with sulfur (products: SO and oxygen), and with carbon (products: CO and oxygen), the \texttt{UV} reaction, and the ionization by Compton electrons (one of the \texttt{CM} reactions). %

In the model n16.3-sphel-pure, the amount of O$_2$ is the highest among the molecules after 30 days and it is as high as $\sim$ 1 $M_{\odot}$ at the last phase. %
The primary formation process of O$_2$ in the model n16.3-sphel-pure is the \texttt{AD} reaction, O$^-$ + O $\lra$ O$_2$ + e$^-$, where O$^-$ can be formed by the \texttt{REA} reaction, O + e$^-$ $\lra$ O$^-$ + $\gamma$. %
The \texttt{NN} reaction of oxygen with MgO also contributes to the increase of O$_2$ after 100 days. %
Therefore, the formation processes of O$_2$ in the n16.3-sphel-pure are peculiar compared with the other models. %
In the model n16.3-sphel-pure, the seed oxygen is initially concentrated in the oxygen-rich shell (see, the lower left panel in Figure~\ref{fig:prof_sphel_pure}), and the abundance ratios of oxygen to others, in particular, the ratio of oxygen to carbon is higher than those of the other models. %
Additionally, around the inner edge of the oxygen-rich shell, $^{56}$Ni is overlapped, where the number density of electrons is enhanced by the ionization by Compton electrons. %
Therefore, in such regions, the former formation sequence above would preferentially work compared with the other models. %

As seen between the models n16.3-mean and b18.3-mean, the amount of MgO is higher than that of FeO in the models n16.3-sphel and n16.3-sphel-pure. %
On the other hand, the amount of FeO dominates that of MgO in the models based on the binary merger progenitor (b18.3) model at least in the last phase. %
As mentioned (in Section~\ref{para:n16.3-mean}), this feature is probably attributed to the higher initial magnesium abundance in the models based on the single-star progenitor model (n16.3). %
Another distinct feature in the models n16.3-sphel and n16.3-sphel-pure is that the amount of MgO is significantly higher than that of FeO compared with the model n16.3-mean. %
The primary formation processes of MgO (FeO) are consistent with the models based on the binary merger model and the model n16.3-mean, i.e., the \texttt{NN} reaction of magnesium (iron) with O$_2$ and the \texttt{3B} reaction among magnesium (iron), oxygen, and hydrogen. %
The feature of the high MgO abundance in the models n16.3-sphel and n16.3-sphel-pure is probably attributed to the higher significance of the former reaction compared with the other models due to the high O$_2$ abundance. %

N$_2$ is also formed in the models n16.3-sphel and n16.3-sphel-pure, and the amount is smaller than that in the counterpart models b18.3-sphel and b18.3-sphel-pure. %
This is because of the initial smaller abundance ratios of the seed nitrogen atoms to the others as mentioned. %

\subsubsection{Differences among the specified directions} \label{subsubsec:1d_angle} 

In Sections~\ref{subsubsec:1d_param}, \ref{subsubsec:1d_single_star}, and \ref{subsubsec:1d_comp_sphel}, the impact of matter mixing on the molecule formation was described by using the angle-averaged 1D profiles. %
In reality, radial profiles could, however, vary depending on the direction. %
Such angle-averaged profiles may overestimate the effects of matter mixing compared with the case of the direct application to the 3D models. %
Additionally, in the 3D models b18.3-high and n16.3-high \citep{2020ApJ...888..111O}, the explosion is globally asymmetric (bipolar-like) as mentioned. %
It is worth seeing the dependence of the molecule formation on the specified direction \citep[see,][for discussion on the light curve dependence on different directions in the highly asymmetric explosion]{2021MNRAS.503..797K}. %
In this section, the molecule formation results for the specified directions, $+Z$, $-Z$, and $+Y$, are presented. %
It is reminded that the $+Z$ and $-Z$ axes are pointed to the stronger and weaker bipolar explosion directions, respectively; %
the $+Y$ axis is perpendicular to the bipolar explosion axis. It should be noted that the initial energy injections (depositions) in the $-Z$ and $+Y$ directions are lower than that in the $+Z$ direction by factors of 2.2 and 3.7 $\times$ 10$^2$, respectively \citep{2020ApJ...888..111O}. %
The different energy depositions significantly affect the amounts and the distribution (radial extension) of $^{56}$Ni as can be seen in Figures~\ref{fig:b18.3_angle} and \ref{fig:n16.3_angle}; %
in the $+Z$ and $-Z$ directions, the amounts and the radial extension of $^{56}$Ni are much higher than those in the $+Y$ direction, which could affect the formation of molecules. %

In Sections~\ref{para:1d_direc_binary} and \ref{para:1d_direc_single}, the time evolution of physical quantities for the models b18.3-zp, b18.3-zn, b18.3-yp, n16.3-zp, n16.3-zn, and n16.3-yp are presented (the former three and the latter three models are in Sections~\ref{para:1d_direc_binary} and \ref{para:1d_direc_single}, respectively). %
Radial distributions of the molecules for the former three and the latter three models are described in Sections~\ref{para:1d_radial_binary} and \ref{para:1d_radial_single}, respectively. %

\paragraph{The time evolution of quantities in the models with the binary merger progenitor model} \label{para:1d_direc_binary}

\begin{figure*}
\begin{minipage}{0.5\hsize}
\begin{center}
\includegraphics[width=9.cm,keepaspectratio,clip]{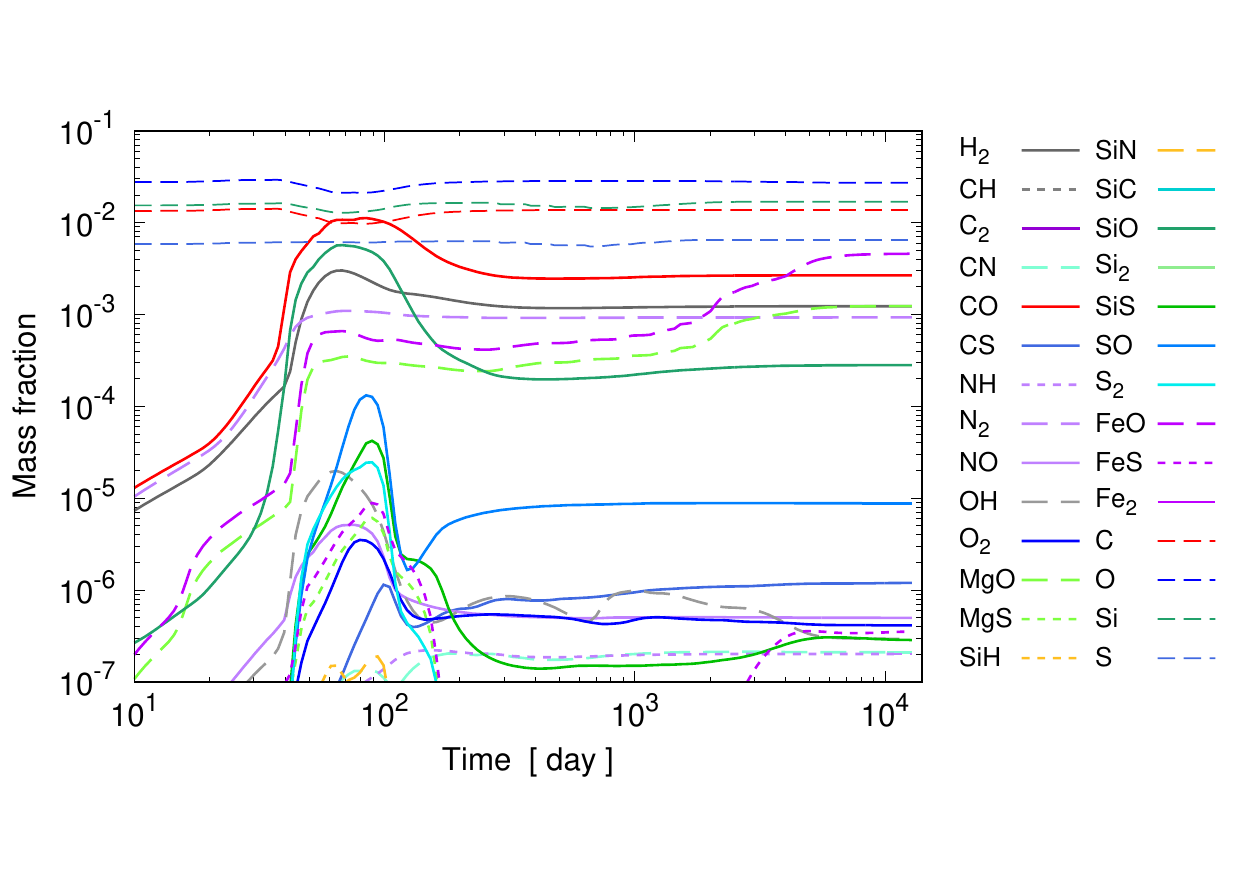}
\end{center}
\end{minipage}
\begin{minipage}{0.5\hsize}
\begin{center}
\includegraphics[width=7.5cm,keepaspectratio,clip]{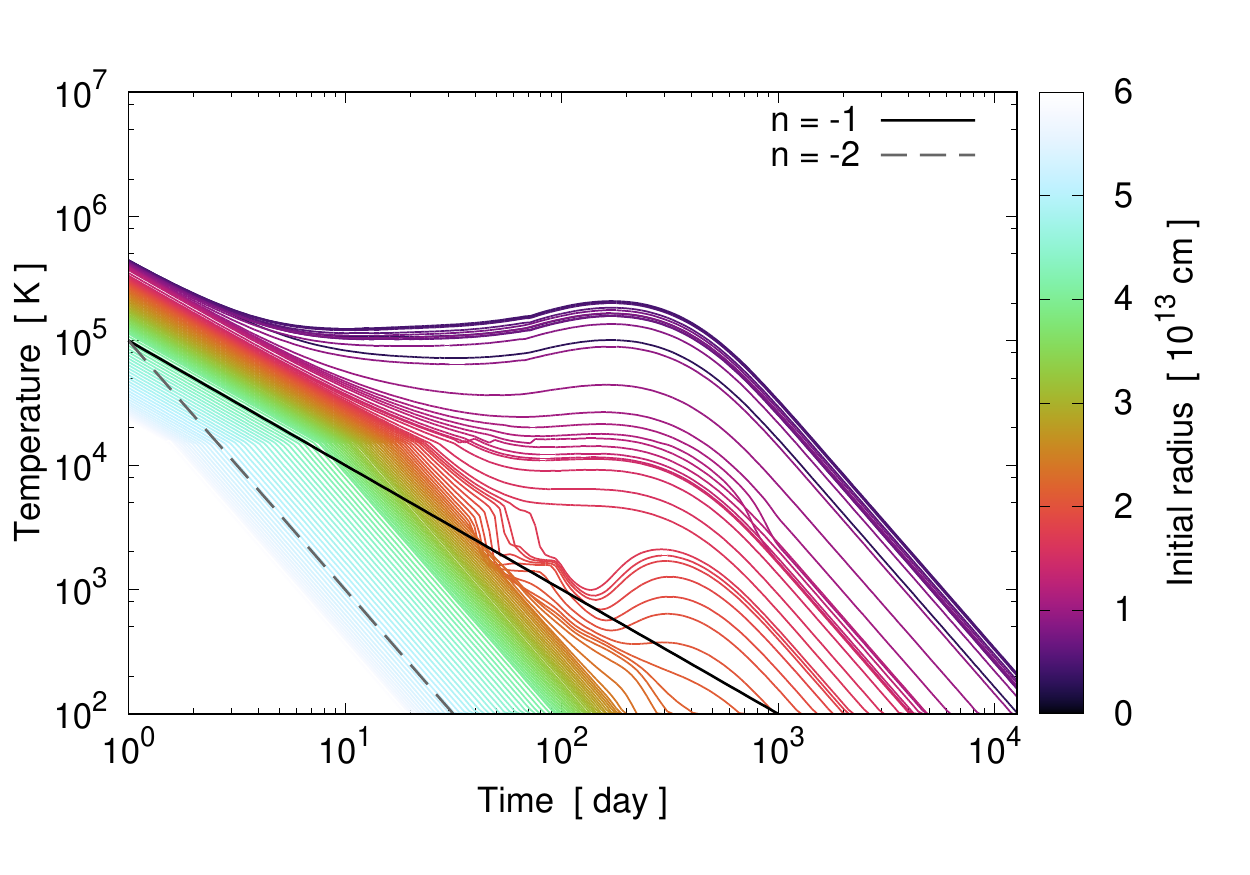}
\end{center}
\end{minipage}
\\
\begin{minipage}{0.5\hsize}
\vs{-1.3}
\begin{center}
\includegraphics[width=9.cm,keepaspectratio,clip]{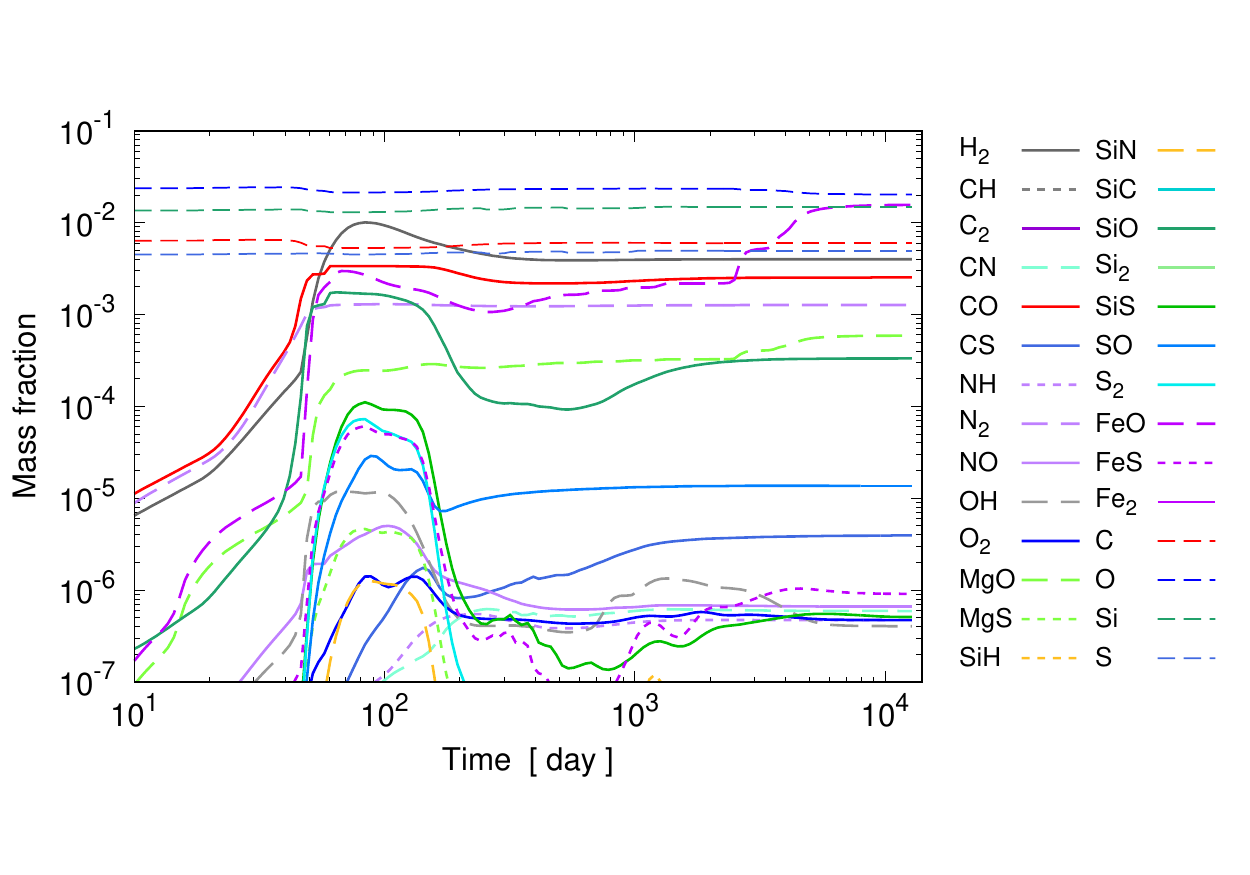}
\end{center}
\end{minipage}
\begin{minipage}{0.5\hsize}
\vs{-1.3}
\begin{center}
\includegraphics[width=7.5cm,keepaspectratio,clip]{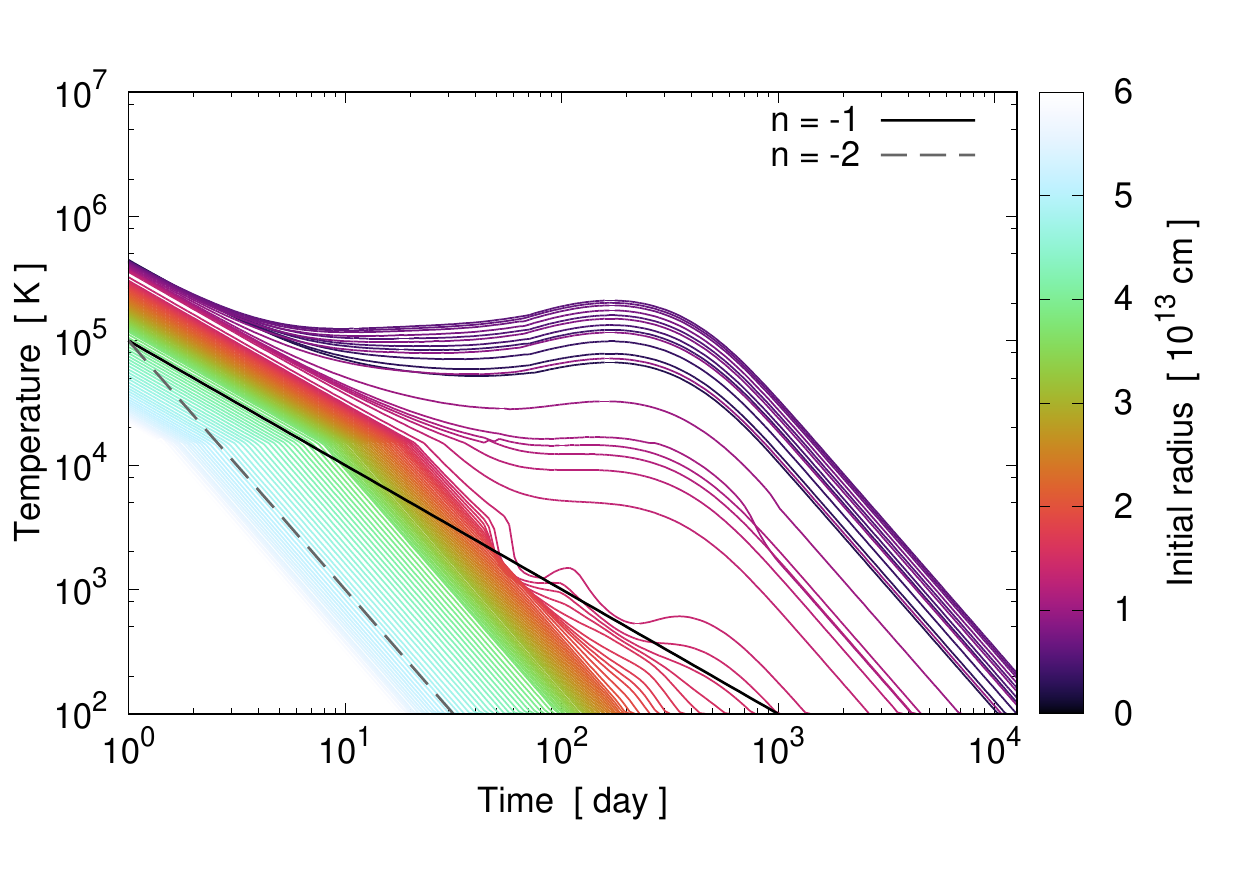}
\end{center}
\end{minipage}
\\
\begin{minipage}{0.5\hsize}
\vs{-1.3}
\begin{center}
\includegraphics[width=9.cm,keepaspectratio,clip]{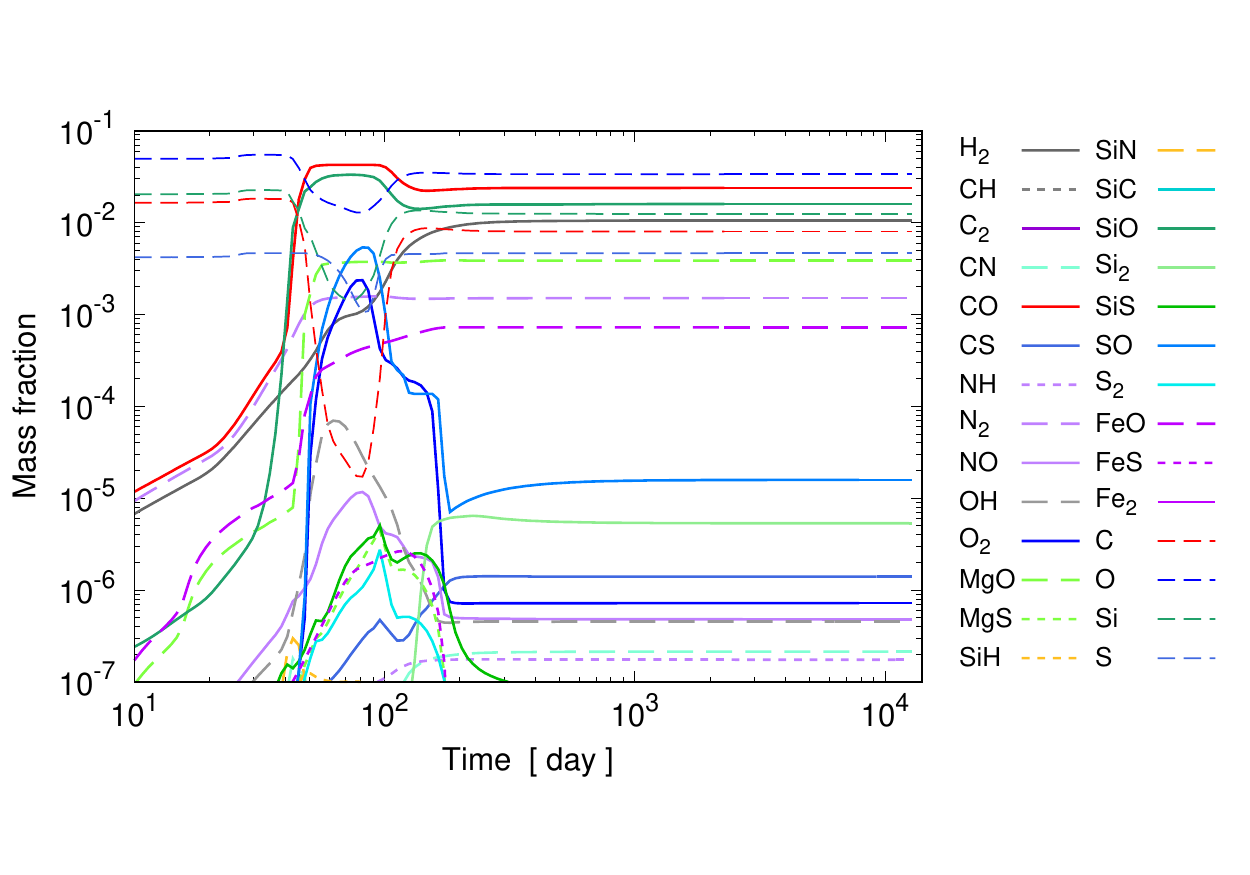}
\end{center}
\vs{-1.}
\end{minipage}
\begin{minipage}{0.5\hsize}
\vs{-1.3}
\begin{center}
\includegraphics[width=7.5cm,keepaspectratio,clip]{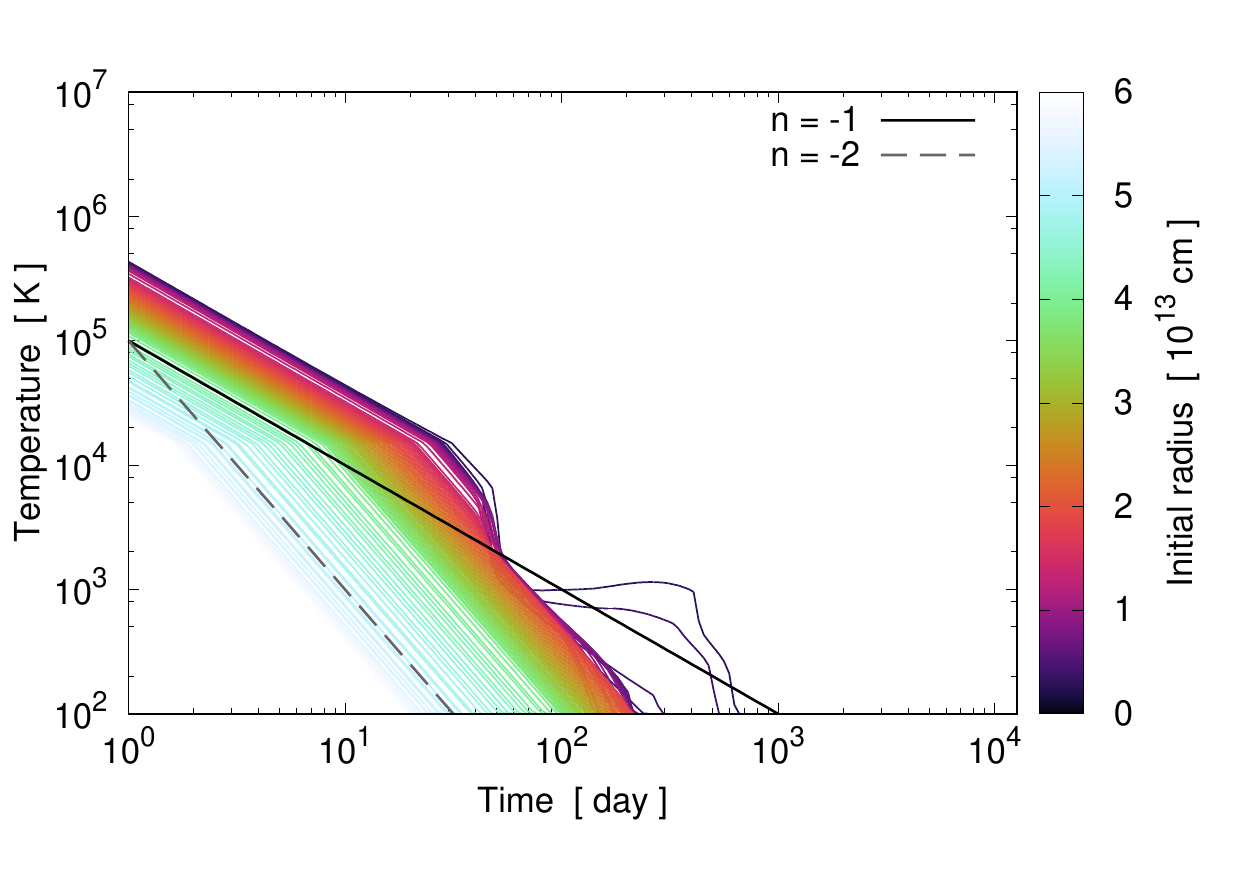}
\end{center}
\vs{-1.}
\end{minipage}
\caption{Time evolutions of the mass fractions (not the total amounts) of molecules and several seed atoms (left panels) and the gas temperatures (right panels). %
From top to bottom, the results for the models b18.3-zp, b18.3-zn, and b18.3-yp are shown, respectively. %
In the right panels, colors denote the initial positions of the particles.} %
\label{fig:1d_b18.3_angle}
\end{figure*}
\hspace{\parindent} 
Figure~\ref{fig:1d_b18.3_angle} shows the results of the models b18.3-zp, b18.3-zn, and b18.3-yp. %
First, from the evolution of gas temperatures (right panels), the evolution in the $+Y$ direction (bottom one) is apparently different from the ones in the $+Z$ and $-Z$ directions. %
This feature is clearly attributed to the differences in the deposited energies during the explosion, i.e., in the $+Y$ direction, the deposited energy is much lower (by a factor of a few hundred) than those in the other two directions (bipolar-like explosion axis); %
the amount of $^{56}$Ni synthesized in the $+Y$ direction is significantly lower than those in the other two directions. %
Hence, in the $+Y$ direction, only several inner particles are affected by the decay of $^{56}$Ni by reflecting the fact that $^{56}$Ni is initially located only in the innermost region (less than 2 $\times 10^{12}$ cm; %
see the distribution of Fe corresponding to $^{56}$Ni in the bottom left panel in Figure~\ref{fig:b18.3_angle}). %
On the other hand, $^{56}$Ni is more extensively distributed (up to $\sim 2 \times 10^{13}$ cm) in the $+Z$ and $-Z$ directions. %
Then, as seen in the top and middle right panels in Figure~\ref{fig:1d_b18.3_angle}, the numbers of heated particles are apparently higher than that in the $+Y$ direction; %
the particles initially at less than 2 $\times 10^{13}$ cm are distinctively heated and the temperatures peak at around 200 days. %
In the $+Z$ and $-Z$ directions, some of the particles initially located at approximately (1.5--2.5) $\times$ 10$^{13}$ cm go through the cooling by CO ro-vibrational transitions. %
In the $+Y$ direction, the inner particles initially located inside 1.5 $\times$ 10$^{13}$ cm are affected by the cooling. %
It is noted that in the inner regions where $^{56}$Ni is initially distributed, helium is also present (see, the left panels in 
Figure~\ref{fig:b18.3_angle}); %
part of helium in such a region may stem from the unburned alpha particles during the explosive nucleosynthesis. %

The time evolution of the mass fractions of molecules (not the amounts) is shown in the left panels. %
For all three directions, CO and SiO rapidly increase around 40 days but the mass fractions along the $+Z$ and $-Z$ directions are rather different from ones along the $+Y$ direction. %
In the $+Z$ and $-Z$ directions, the mass fractions of both CO and SiO at 50--100 days range over 10$^{-3}$--10$^{-2}$. %
On the other hand, in the $+Y$ direction, the mass fractions CO and SiO at this phase are as much as 10$^{-2}$--10$^{-1}$; %
actually, large fractions of the seed atoms, i.e., carbon, oxygen, and silicon, are consumed in the $+Y$ direction (see, the bottom left panel). %
In the $+Y$ direction, the mass fraction of SiO is comparable with that of CO at the final phase. %
This trend (the mass fractions of CO and SiO in the $+Y$ directions are higher than those in the other two directions) should be attributed to the different energy depositions during the explosion again, which result in higher initial abundances of the seed atoms in the $+Y$ direction and higher $^{56}$Ni abundances in the $+Z$ and $-Z$ directions (see, Figure~\ref{fig:b18.3_angle}). %
In the $+Y$ direction, molecules start to form at earlier phases with a high density because the inefficient gas heating due to the lower local $^{56}$Ni abundance makes the gas temperatures about 10$^{4}$ K at earlier phases compared with the other two directions. %

After 100 days, the mass fractions decrease due to destructive reactions. %
In the $+Z$ and $-Z$ directions, in particular, SiO is distinctively destructed compared with the $+Y$ direction. %
At this phase, CO is mainly destructed by the ionization, dissociation, and dissociative ionization due to Compton electrons (\texttt{CM} reactions) and the destruction by UV photons (\texttt{UV} reaction) in the $+Z$ and $-Z$ directions. %
On the other hand, the primary destruction process of CO is the \texttt{NN} reaction in Equation~(\ref{eq:si_co}) in the $+Y$ direction. %
SiO is mainly destructed by the \texttt{CE} reaction in Equation~(\ref{eq:h+_sio}) in the $+Z$ and $-Z$ directions; %
the \texttt{NN} reaction in Equation~(\ref{eq:h_sio_sih}), the ionization and dissociation through \texttt{CM} reactions,  and the \texttt{UV} reaction are followed as secondary destruction processes. %
In the $+Z$ ($-Z$) direction, the contribution from the \texttt{CM} and \texttt{UV} reactions are higher (lower) than that in the $-Z$ ($+Z$) direction. %
In the $+Y$ direction, the primary destruction processes of SiO are the \texttt{NN} reactions in Equations~(\ref{eq:h_sio_sih}) and (\ref{eq:h_sio_oh}) in contrast to the $+Z$ and $-Z$ directions. %
The differences in the destruction processes between the bipolar-like directions ($+Z$ and $-Z$) and the $+Y$ direction are because of the different local $^{56}$Ni abundances due to the different initial energy depositions, again. %
The differences in the significance of the SiO destruction processes between $+Z$ and $-Z$ directions may partly be attributed to the distribution of $^{56}$Ni; %
in the $-Z$ direction, the innermost oxygen- and silicon-rich region (6 $\times$ 10$^{12}$ cm) in the initial seed atom distributions (middle left panel in Figure~\ref{fig:b18.3_angle}), the mass fraction of $^{56}$Ni is smaller than those of oxygen and silicon in contrast to the $+Z$ direction. %

As for other molecules, from here, several features different among the directions are briefly described. %
For example, H$_2$ is also remarkably formed after 40 days in the $+Z$ and $-Z$ directions. %
On the other hand, in the $+Y$ direction, H$_2$ is formed at later phases after 100 days. %
In the $+Z$ and $-Z$ directions, H$_2$ is primarily formed by the \texttt{AD} reaction, H$^-$ + H $\lra$ H$_2$ + e$^-$ (H$^-$ can be formed by the \texttt{REA} reaction, H + e$^-$ $\lra$ H$^-$ + $\gamma$) after 40 days. %
Therefore, the ionization by Compton electrons plays a role in the formation of H$_2$. %
This means that the higher local $^{56}$Ni abundances that stem from the higher initial energy depositions in the $+Z$ and $-Z$ directions could also be important for the formation of H$_2$. %
Actually, in the $+Y$ direction, the main formation process of H$_2$ after 100 days is not the \texttt{AD} reaction above but the \texttt{NN} reaction of hydrogen with SiH. %
In the $-Z$ direction, the mass fraction of H$_2$ is higher than that of CO after approximately 40 days in contrast to the $+Z$ direction, which may be attributed to the fact that the mass fraction of hydrogen in the $-Z$ direction is initially higher than that in the $+Z$ direction in the outer layers (greater than 1 $\times 10^{13}$ cm) as seen in Figure~\ref{fig:b18.3_angle}. %

SO and O$_2$ are also markedly produced at around 40--200 days in the $+Y$ direction compared with the other two directions. %
However, after 200 days, the mass fractions are not so dramatically different among the directions. %
In the $+Z$ and $-Z$ directions, the primary formation processes of SO before 200 days are the \texttt{RA} reaction between sulfur and oxygen atoms (before 40--50 days), the \texttt{NN} reactions of oxygen with S$_2$, and the \texttt{NN} reaction of sulfur with OH. In the $+Y$ direction, the \texttt{NN} reaction of sulfur with O$_2$ also contributes as one of the primary formation reactions. %
The primary formation processes of O$_2$ are the \texttt{RA} reaction between two oxygen atoms (before 40--50 days) and the \texttt{NN} reaction of oxygen with OH in all three directions. %
In the $+Y$ direction, the \texttt{NN} reaction of oxygen with SiO also contributes temporarily. %
The contributing destruction processes of SO at the destruction dominant phase are the \texttt{NN} reaction of carbon with SO (products: oxygen and CS; CO and sulfur), the \texttt{NN} reaction of iron with SO (products: oxygen and FeS), and the \texttt{NN} reaction of magnesium with SO (products: oxygen and MgS). %
In the $-Z$ ($+Y$) direction, the contribution of the \texttt{NN} reaction of magnesium (iron) is limited compared with the other directions. %
The primary destruction process of O$_2$ is the \texttt{NN} reaction of silicon with O$_2$ (products: oxygen and SiO) in all three directions. %
Although the destruction processes of SO and O$_2$ are depending on the direction, basically, those are all \texttt{NN} reactions, i.e., the destructions are not at least directly related to the decay of $^{56}$Ni. %

As seen in the final phase, FeO and MgO are also formed to some extent. %
In the $+Z$ and $-Z$ directions, overall, FeO dominates MgO throughout the evolution. %
On the other hand, in the $+Y$ direction, MgO dominates FeO after approximately 40 days. %
This feature may be attributed to the low iron (decayed from $^{56}$Ni; $^{56}$Ni $\lra$ $^{56}$Co $\lra$ $^{56}$Fe) abundance in the $+Y$ direction. %
FeO is primarily formed by the \texttt{NN} reaction of iron with O$_2$ or the \texttt{3B} reaction, Fe + O + H $\lra$ FeO + H, depending on time before about 100 days. %
After 100 days FeO is primarily formed by the \texttt{3B} reaction above. %
Similarly, MgO is mainly produced by the \texttt{NN} reaction of magnesium with O$_2$ or the \texttt{3B} reaction, Mg + O + H $\lra$ MgO + H before 100 days. %
After 100 days MgO is primarily formed by the \texttt{3B} reaction. %
It is noted that in the $+Z$ and $-Z$ directions, in particular, FeO increases even after a few thousand days. %
It is interpreted that inner particles heated by the decay of $^{56}$Ni with high local iron (from the decayed $^{56}$Ni, again) abundance partake in the molecule formation at about a few thousand days; %
at this phase, only \texttt{3B} reactions play a role in the formation of FeO and MgO. %

\paragraph{The time evolution of quantities in the models with the single-star progenitor model} \label{para:1d_direc_single}

\begin{figure*}
\begin{minipage}{0.5\hsize}
\begin{center}
\includegraphics[width=9.cm,keepaspectratio,clip]{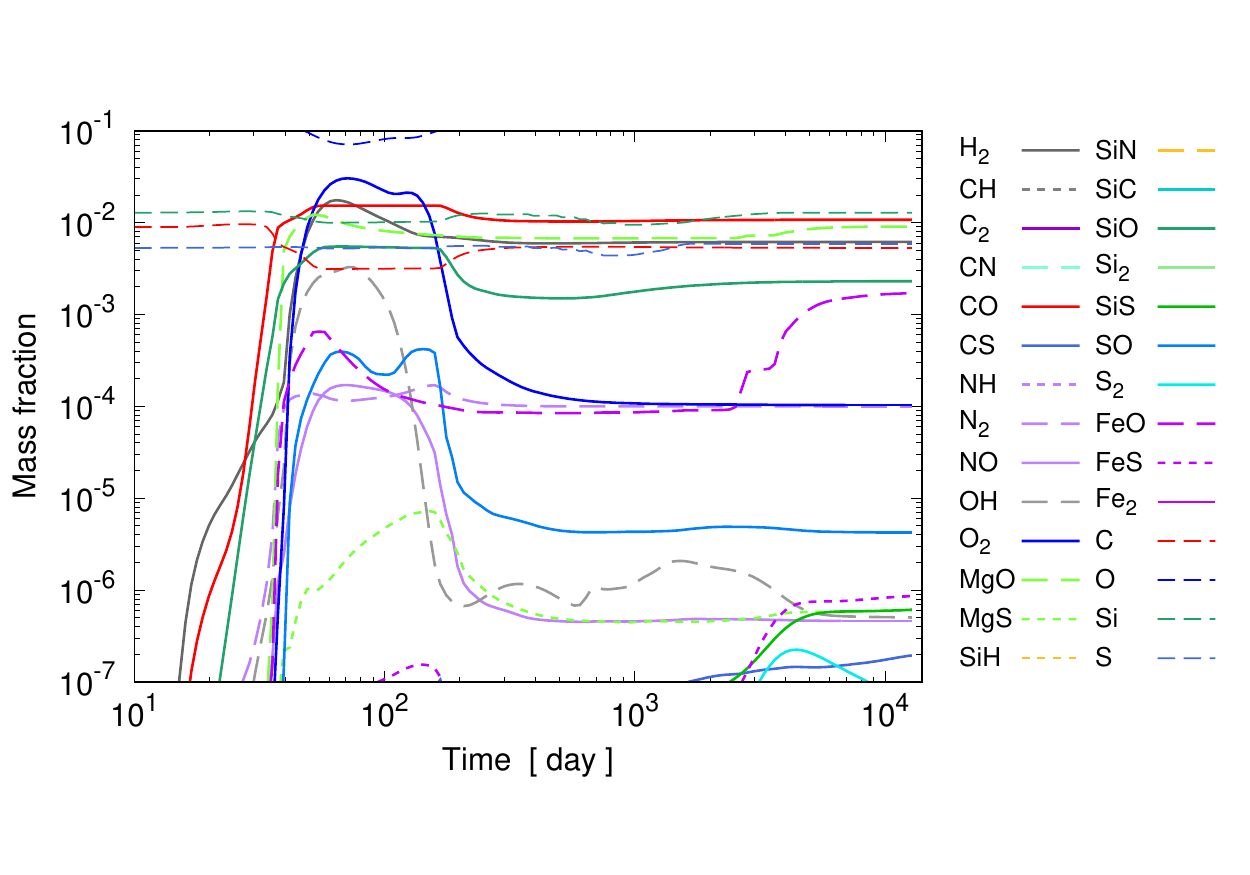}
\end{center}
\end{minipage}
\begin{minipage}{0.5\hsize}
\begin{center}
\includegraphics[width=7.5cm,keepaspectratio,clip]{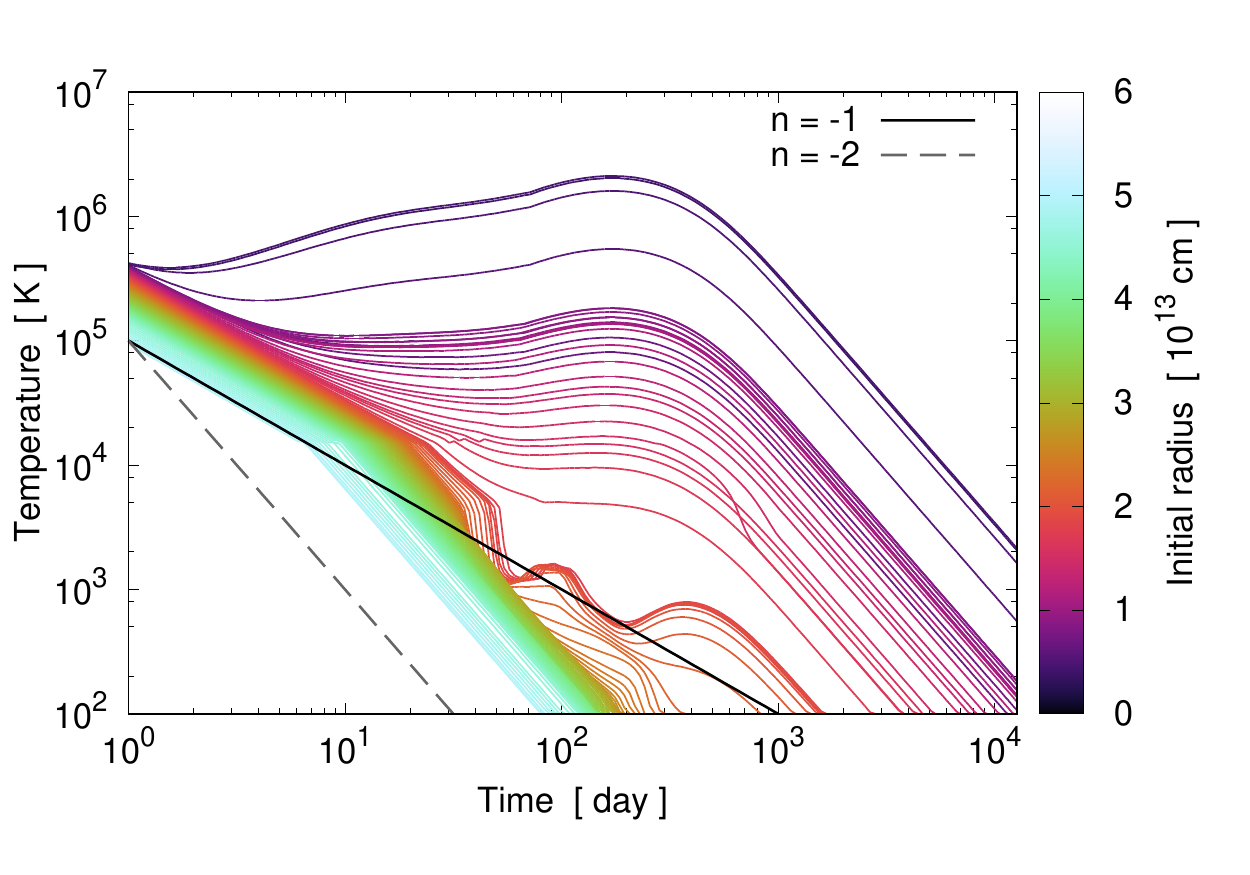}
\end{center}
\end{minipage}
\\
\begin{minipage}{0.5\hsize}
\vs{-1.3}
\begin{center}
\includegraphics[width=9.cm,keepaspectratio,clip]{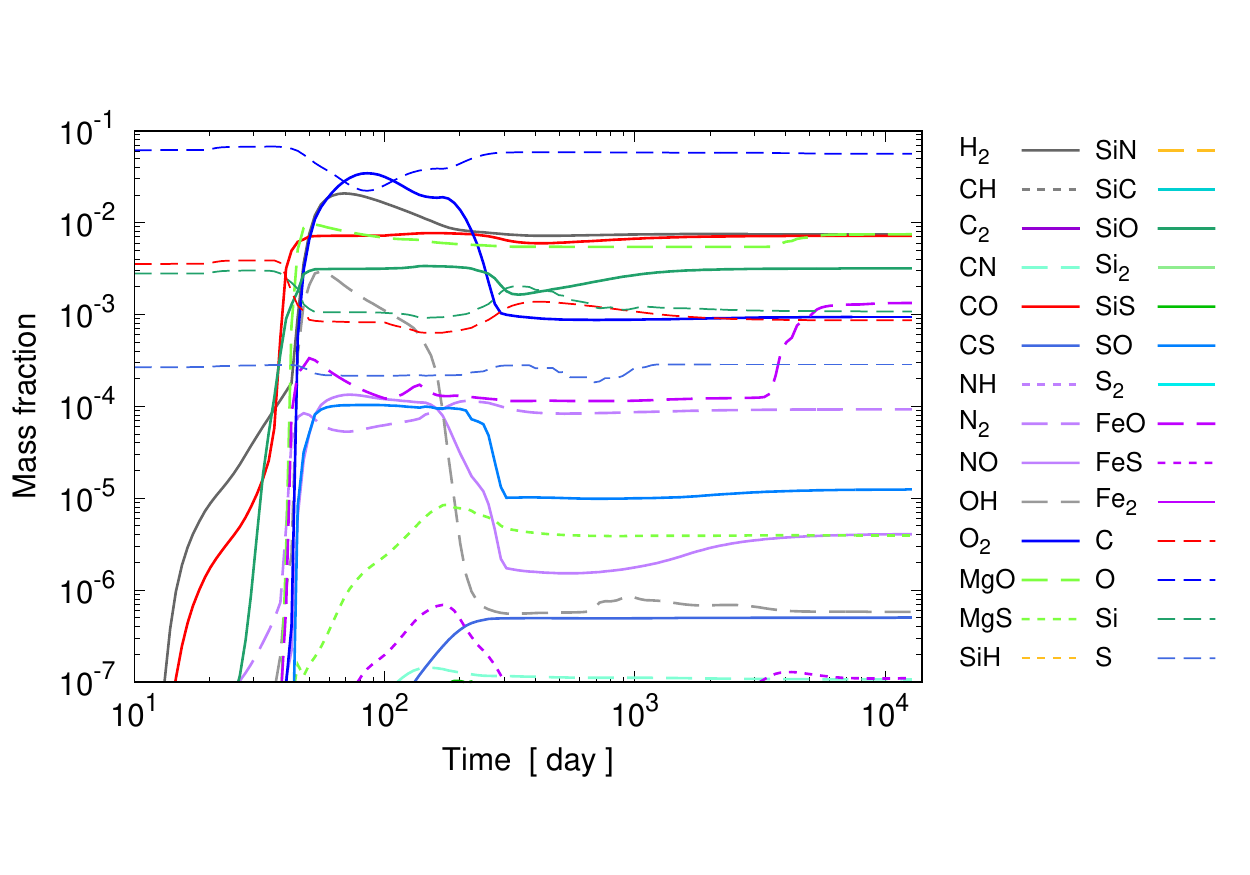}
\end{center}
\end{minipage}
\begin{minipage}{0.5\hsize}
\vs{-1.3}
\begin{center}
\includegraphics[width=7.5cm,keepaspectratio,clip]{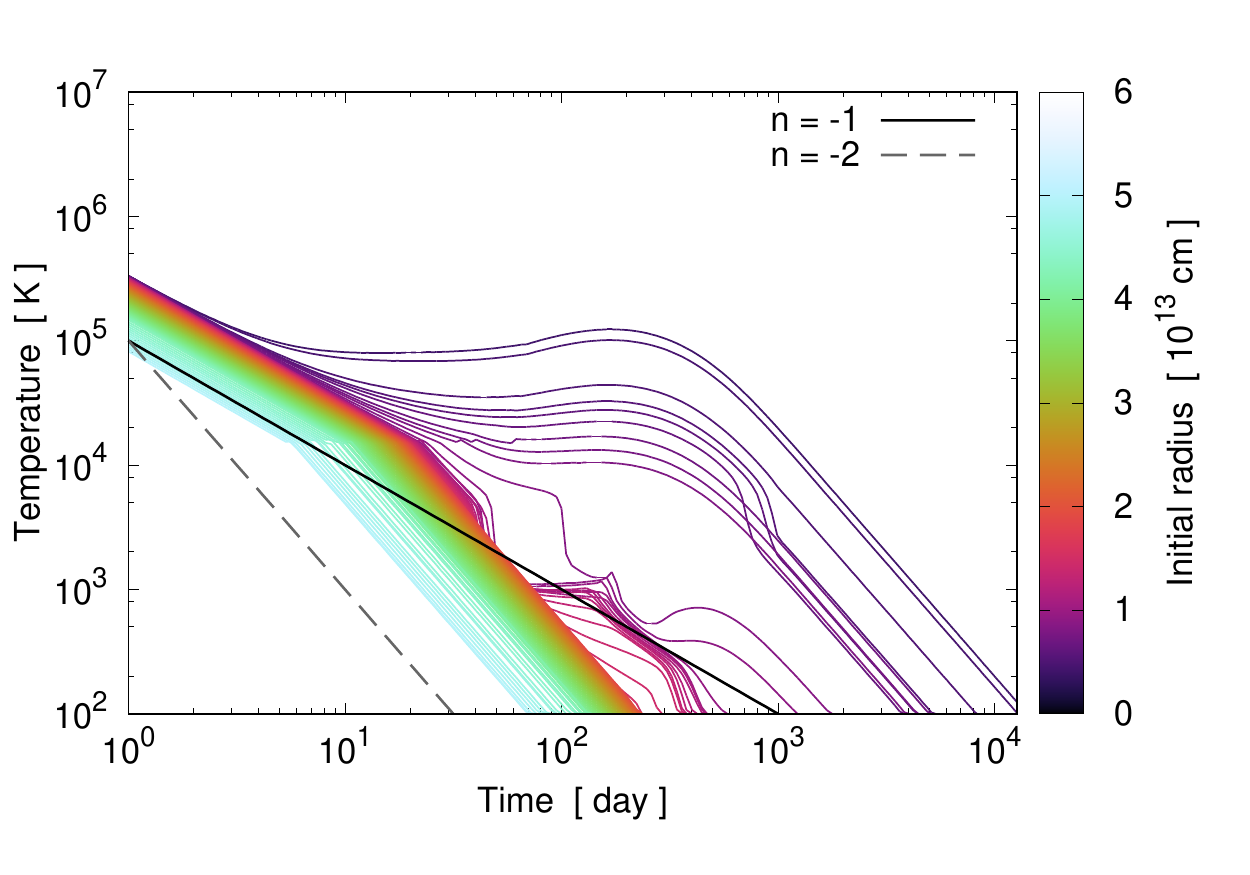}
\end{center}
\end{minipage}
\\
\begin{minipage}{0.5\hsize}
\vs{-1.3}
\begin{center}
\includegraphics[width=9.cm,keepaspectratio,clip]{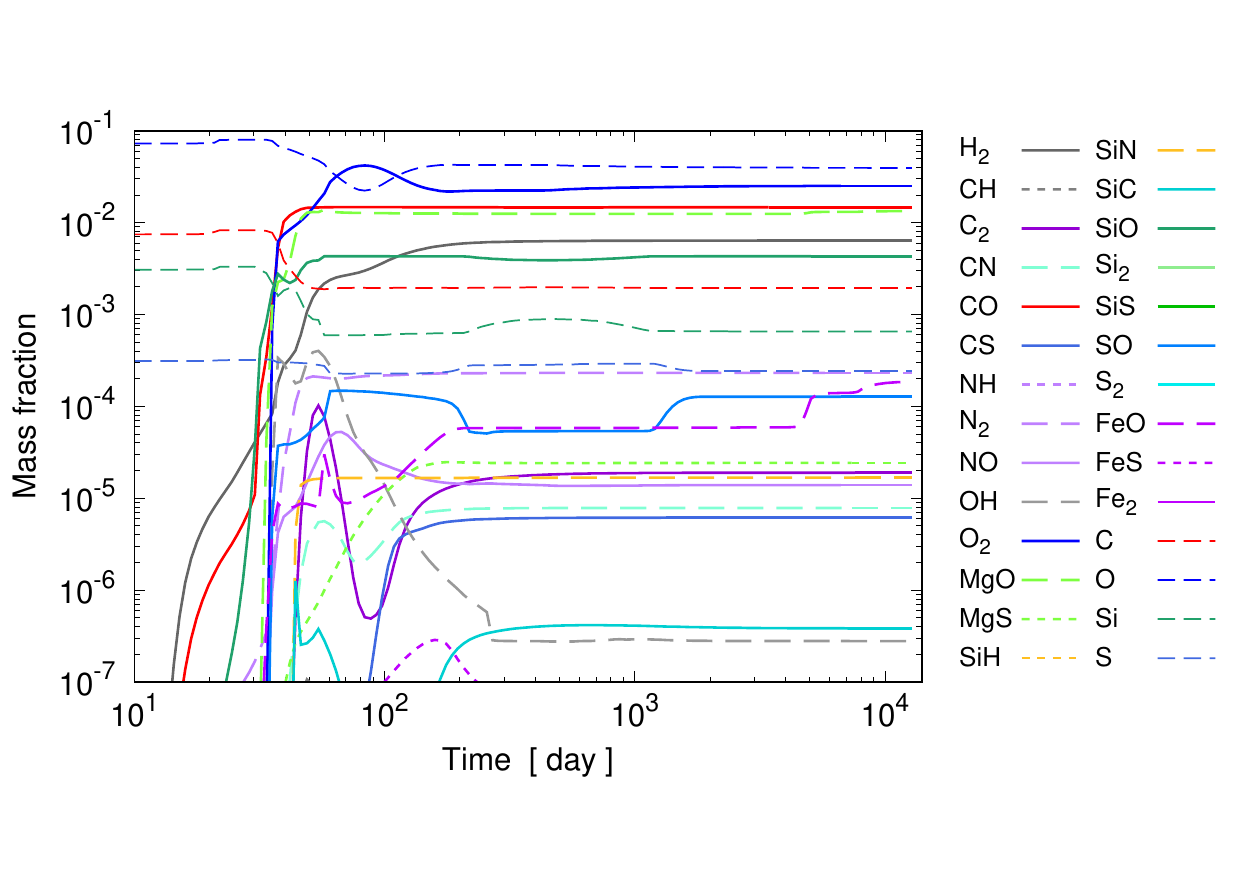}
\end{center}
\vs{-1.}
\end{minipage}
\begin{minipage}{0.5\hsize}
\vs{-1.3}
\begin{center}
\includegraphics[width=7.5cm,keepaspectratio,clip]{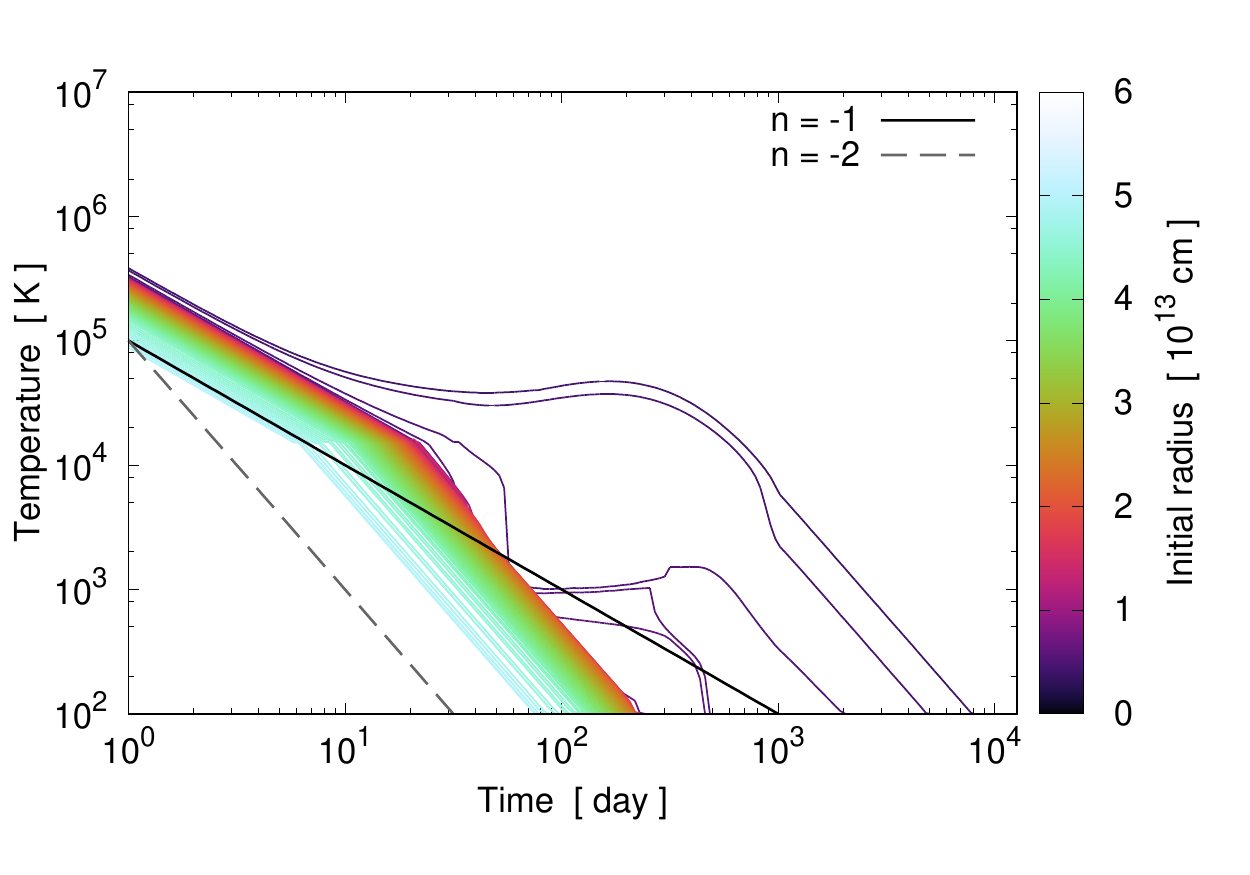}
\end{center}
\vs{-1.}
\end{minipage}
\caption{Same as Figure~\ref{fig:1d_b18.3_angle} but for the models, n16.3-zp, n16.3-zn, and n16.3-yp.} %
\label{fig:1d_n16.3_angle}
\end{figure*}
\hspace{\parindent} 
In Figure~\ref{fig:1d_n16.3_angle}, the results of the models based on the single-star progenitor model n16.3, i.e., n16.3-zp, n16.3-zn, and n16.3-yp, are shown. %
From the evolution of the gas temperatures (left panels), it is recognized that the number of heated particles is the largest in the $+Z$ direction and it is the smallest in the $+Y$ direction. %
In the $-Z$ direction, the number is in between the former two directions. %
This feature should be attributed to the different initial energy depositions, again. %
Actually, as can be seen in Figure~\ref{fig:n16.3_angle}, in the $+Z$, $-Z$, and $+Y$ directions, $^{56}$Ni is initially distributed extensively in this order. %
The number of heated particles reflects how extensively $^{56}$Ni is initially distributed. %
The temperatures of the heated particles peak at around 200 days. %
In the $+Z$ direction, the peak temperatures of some of the heated particles are higher than 10$^6$ K, which is higher than the highest peak temperature e.g., in the model b18.3-zp. %
As seen in the top left panel in Figure~\ref{fig:n16.3_angle}, there is a region (at around 5 $\times$ 10$^{12}$ cm) where the mass fraction of Fe (practically corresponding to $^{56}$Ni) is as high as $\sim$ 0.7; %
such a high mass fraction of $^{56}$Ni can not be seen in the model b18.3-zp. %
The particles that have peak temperatures higher than 10$^{6}$ K come from such a high local $^{56}$Ni abundance region. %
In the $+Z$ direction, some of the particles initially at about (1.5--2.5) $\times$ 10$^{13}$ cm go through the cooling by CO ro-vibrational transitions. %
In the $-Z$ direction, some of the particles initially at (1--2) $\times$ 10$^{13}$ cm are affected by the cooling. %
In the $+Y$ direction, only several inner particles are affected by the cooling. %

As seen in the evolution of the mass fractions of molecules (left panels), the decreases in the values of CO and SiO are limited compared with the models based on the progenitor model b18.3. %
In contrast to the models based on b18.3, molecules start to form after 10 days. %
In the $+Z$ and $-Z$ directions, CO is only slightly destructed after 200 days and SiO decreases by a factor of a few. %
In the $+Y$ direction, there is no recognizable change in the values of CO and SiO after about 50 days. %

In the $+Z$ and $-Z$ directions, CO is primarily formed by the \texttt{RA} reaction in Equation~(\ref{eq:co_rad}) before approximately 30--40 days; %
after that, the \texttt{NN} reaction in Equation~(\ref{eq:c_oh}) primarily contributes until the value enters a plateau (50--60 days). %
The \texttt{CE} reaction in Equation~(\ref{eq:h_co+}) also contributes to some extent as the secondary reaction. %
In the $+Y$ direction, the primary formation processes are the same as the $+Z$ and $-Z$ directions except the \texttt{CE} reaction does not contribute. %
In the $+Z$ and $-Z$ directions, SiO is primarily produced by the \texttt{RA} reaction in Equation~(\ref{eq:sio_rad}) before approximately 30 days; %
later SiO is mainly produces by \texttt{NN} reactions in Equations~(\ref{eq:si_oh}), (\ref{eq:si_o2}), and (\ref{eq:si_co}) until the values have a plateau. %
In the $+Y$ direction, the situation is similar but the contribution from the \texttt{RA} reaction is more limited; %
the \texttt{RA} reaction forms SiO until 25 days as the primary formation process. %
The \texttt{NN} reaction in Equation~(\ref{eq:si_oh}) also does not contribute distinctively. %
In the $+Z$ and $-Z$ directions, after approximately 200 days, CO and SiO are destructed to some extent as mentioned above. %
The primary destruction processes of CO at this phase are the ionization and dissociation by Compton electrons (\texttt{CM} reactions), and the \texttt{UV} reaction is followed as the secondary process. %
In both the $+Z$ and $-Z$ directions, the primary destruction process of SiO after 200 days is the \texttt{CE} reaction in Equation~(\ref{eq:h+_sio}); the \texttt{NN} reactions in Equations~(\ref{eq:h_sio_sih}) and (\ref{eq:h_sio_oh}) also contribute as the secondary processes. %

A distinct feature that can not be seen in the models based on the binary merger progenitor model b18.3 is that O$_2$ is remarkably formed compared with the models with b18.3. %
This feature is basically attributed to the high abundance ratios of the seed oxygen atom to the others as mentioned. %
In all three directions, the mass fraction of O$_2$ peaks at around 70--80 days. %
It is noted that the mass fraction of O$_2$ is higher than those of CO and SiO at the peaks. %
After the peak, in the $+Z$ ($-Z$) direction, the value decreases by roughly two orders (one order) of magnitude. %
In the $+Y$ direction, the value decreases only by a factor of a few. %
In all the three directions, initially, the \texttt{RA} reaction between two oxygen atoms is dominant as the primary formation process until 30--40 days. %
After that, the \texttt{NN} reactions between oxygen and OH and between two OH molecules overtake the \texttt{RA} reaction; the significance of the two \texttt{NN} reactions varies depending on the direction and time. %
It is noted that in the models with the progenitor model b18.3, the dominant \texttt{NN} reaction for the formation of O$_2$ is only the \texttt{NN} reaction of oxygen with OH at the formation dominant phase. %
As for the destruction processes after the peak ($\sim$ 100 days). %
In the $+Z$ and $-Z$ directions, the contributing destruction processes of O$_2$ are the \texttt{NN} reactions of O$_2$ with silicon and sulfur and the \texttt{CE} reaction between H$^+$ and O$_2$. %
In the $+Y$ direction, only the \texttt{NN} reaction of O$_2$ with silicon contributes in contrast to the other two directions. %
The contribution of the \texttt{CE} reaction in the $+Z$ and $-Z$ may be due to the efficient ionization by \texttt{CM} reactions owing to the high $^{56}$Ni abundance. %

As for other molecules, for example, SO qualitatively has a similar trend with O$_2$. %
The primary formation process of SO is initially the \texttt{RA} reaction between sulfur and oxygen until 20--40 days in all the three directions. %
After that, the \texttt{NN} reaction of sulfur with O$_2$ dominates other formation reactions. %
The primary destruction process after 100 days is the \texttt{NN} reaction of hydrogen with SO (products: sulfur and OH). %
Later, in the $+Z$ and $-Z$ directions, the \texttt{CE} reaction between H$^+$ and SO and/or the \texttt{NN} reaction of carbon with SO (products: oxygen and  CS; sulfur and CO) are dominant as the primary destruction process at around the rapidly decreasing phase (after 150 days and 200 days, respectively); %
in the $+Y$ direction, the \texttt{NN} reaction of hydrogen with SO above remains the primary. %
In the $+Y$ direction, after about 1000 days, the mass fraction of SO distinctively increases again. %
This is the contribution from the innermost particles heated by the decay of $^{56}$Ni, which partake in the molecule formation at such a later phase when the gas temperatures drop to 10$^4$ K. %
The primary formation process is the \texttt{NN} reaction of oxygen with S$_2$ at this phase. %

The behavior of H$_2$ is qualitatively similar compared with the counterpart model with the progenitor model b18.3. %
After 40 days, in all three directions, H$_2$ starts to form rapidly. %
In the $+Z$ and $-Z$ directions, the mass fraction of H$_2$ peaks at around 60 days. %
On the other hand, in the $+Y$ direction, the increase is slow and continues until 200 days. %
The primary formation processes of H$_2$ depend on the directions. %
In the $+Z$ direction, after 40 days, H$_2$ is primarily formed by the \texttt{AD} reaction, H$^-$ + H $\lra$ H$_2$ + e$^-$. %
In the $-Z$ direction, H$_2$ is mainly formed by the \texttt{NN} reaction of hydrogen with CH and the \texttt{NN} reaction between two OH molecules; %
the \texttt{AD} reaction above contributes as the secondary and overtakes the \texttt{NN} reactions after 60 days. %
In the $+Y$ direction, the primary formation process of H$_2$ is the \texttt{NN} reaction of hydrogen with CH and/or the \texttt{NN} reaction between two OH molecules before 60 days; %
later, the \texttt{NN} reaction of hydrogen with SiH becomes the primary process. %
In the $+Y$ direction, the \texttt{AD} reaction contributes to the formation after 40 days as one of the secondaries. %
Therefore, the more extensively $^{56}$Ni is initially distributed (see, Figure~\ref{fig:n16.3_angle}), the more significant the contribution of the \texttt{AD} reaction is. %
This is because the number density of one of the reactants of the \texttt{AD} reaction, i.e., H$^-$, increases due to the ionization by the \texttt{CM} reactions through the \texttt{REA} reaction, H + e$^-$ $\lra$ H$^-$ + $\gamma$ as mentioned. %

Similarly to the models with the progenitor model b18.3, FeO and MgO are also formed to some extent; %
the mass fraction of MgO, however, dominates that of FeO in all three directions in contrast to the models with b18.3. %
This is because the initial mass fractions of magnesium are overall higher than those in the models with b18.3 (see, Figures~\ref{fig:b18.3_angle} and~\ref{fig:n16.3_angle}). %
After a few thousand days, in all three directions, FeO and MgO (in particular, FeO in the $+Z$ and $-Z$ directions) increase, although the increase of MgO in the $+Y$ direction is not remarkable. %
As described in the models with the progenitor model b18.3, such a later increase of FeO and MgO is attributed to the fact that the particles heated by the decay of $^{56}$Ni with the initial high local $^{56}$Ni abundance partake in the formation of the molecules at such a later phase; the heated particles contain plenty of the seed iron to form FeO. 
The primary formation process of MgO approximately before 100 days is the \texttt{NN} reaction of magnesium with O$_2$. %
Later, the \texttt{3B} reaction, Mg + O + H $\lra$ MgO + H becomes the primary formation process until the end of the simulation in the $+Z$ and $-Z$ directions. %
In the $+Y$ direction, after the \texttt{NN} reaction-dominated phase, the \texttt{3B} reaction among magnesium, oxygen, and helium also contributes as one of the primary processes. %
In all the three directions, the primary formation process of FeO is the \texttt{NN} reaction of iron with O$_2$ before 100 days. %
After a few thousand days, FeO increases again in particular in the $+Z$ and $-Z$ directions. %
At the phase, in the $+Z$ and $-Z$ directions, the \texttt{3B} reaction, Fe + O + H $\lra$ FeO + H, is the primary formation process; %
in the $+Y$ direction, the \texttt{3B} reaction among iron, oxygen, and hydrogen (or second oxygen), is the primary depending on time. %

\paragraph{Radial distributions of molecules in the models with the binary merger progenitor model} \label{para:1d_radial_binary}

\begin{figure*}
\begin{minipage}{0.5\hsize}
\begin{center}
\includegraphics[width=9.cm,keepaspectratio,clip]{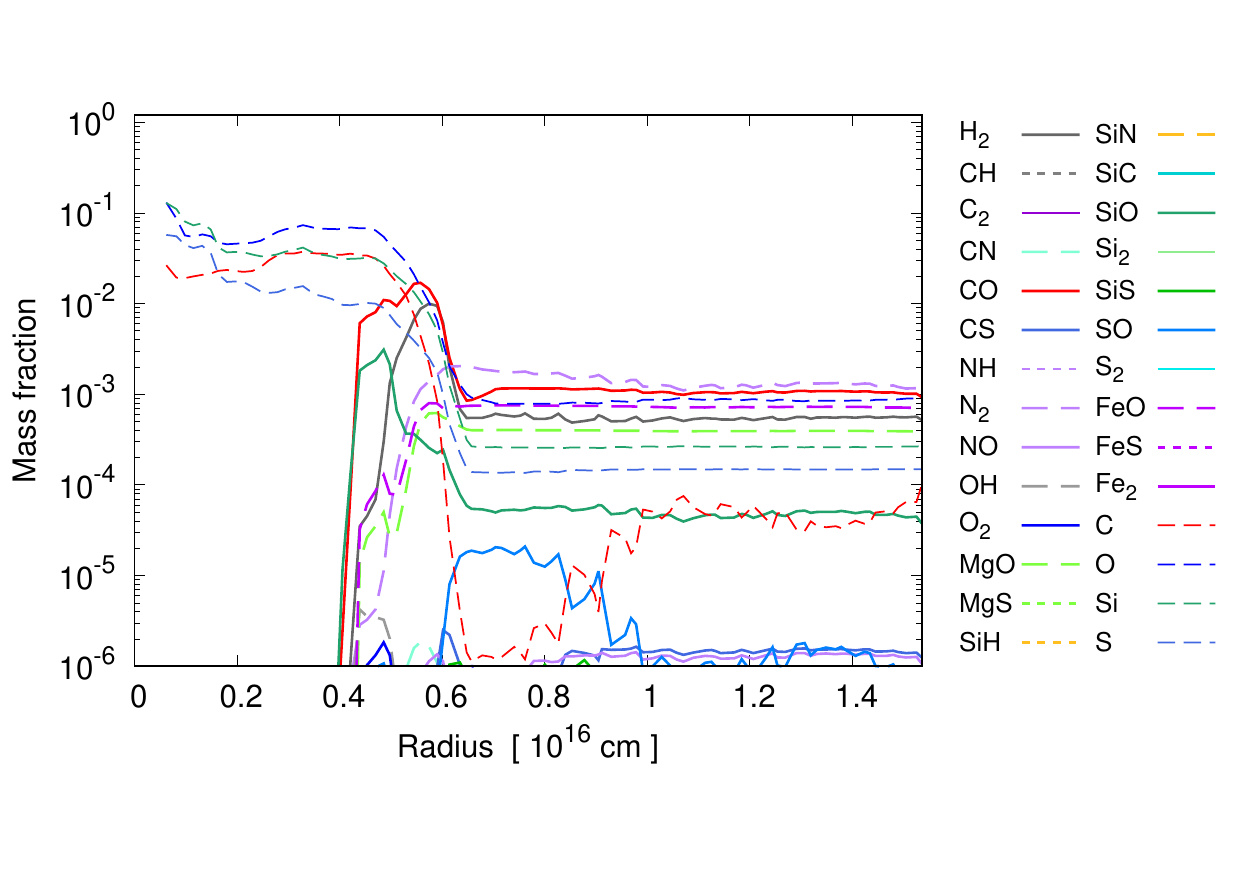}
\end{center}
\end{minipage}
\begin{minipage}{0.5\hsize}
\begin{center}
\includegraphics[width=9.cm,keepaspectratio,clip]{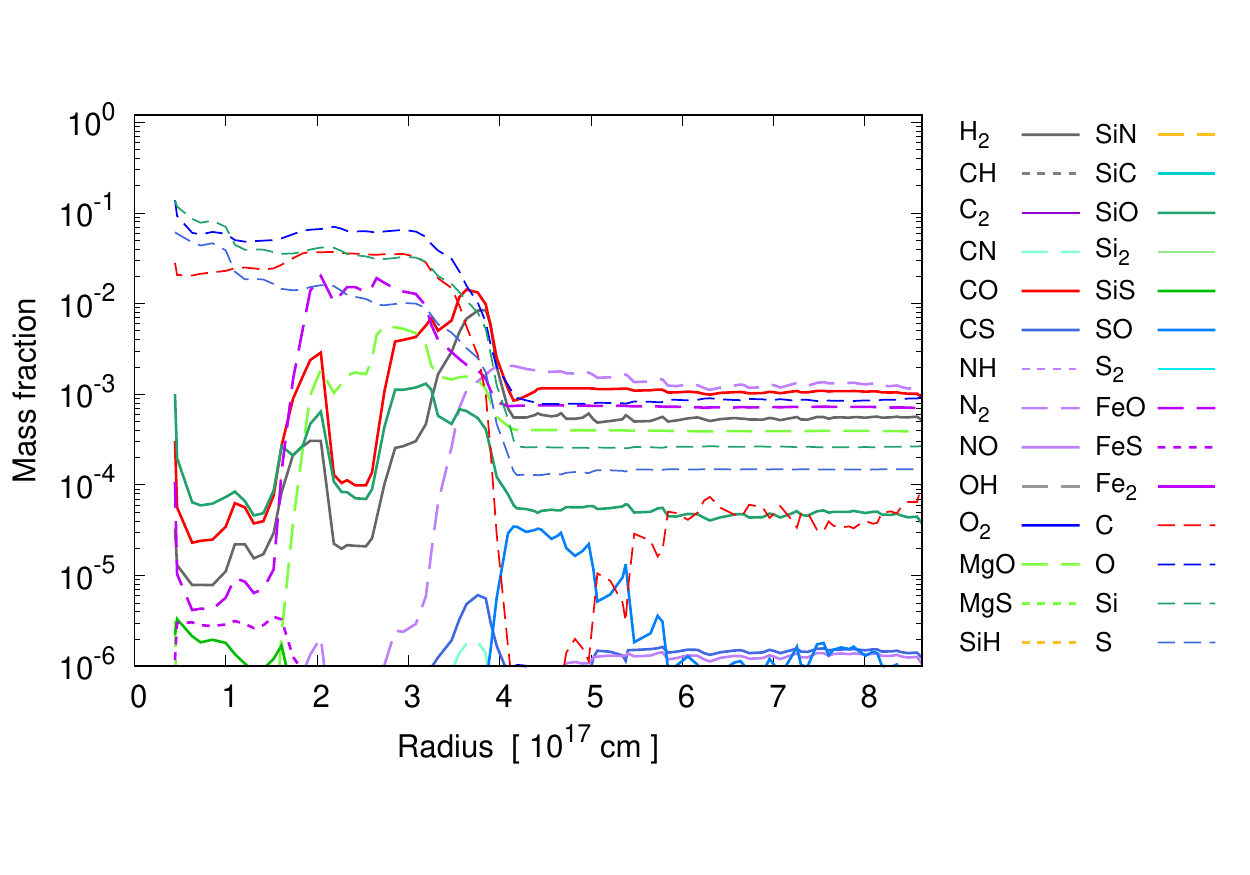}
\end{center}
\end{minipage}
\\
\begin{minipage}{0.5\hsize}
\vs{-1.3}
\begin{center}
\includegraphics[width=9.cm,keepaspectratio,clip]{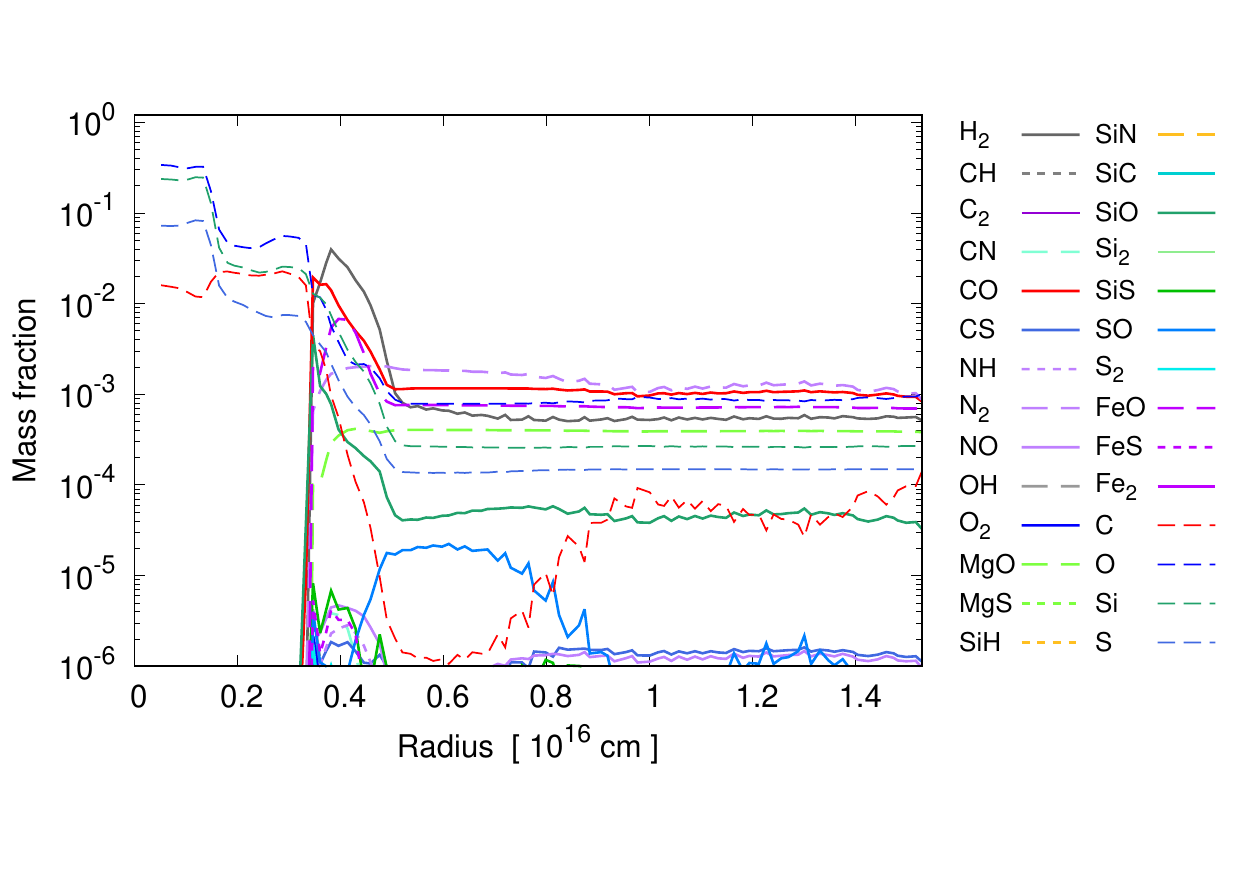}
\end{center}
\end{minipage}
\begin{minipage}{0.5\hsize}
\vs{-1.3}
\begin{center}
\includegraphics[width=9.cm,keepaspectratio,clip]{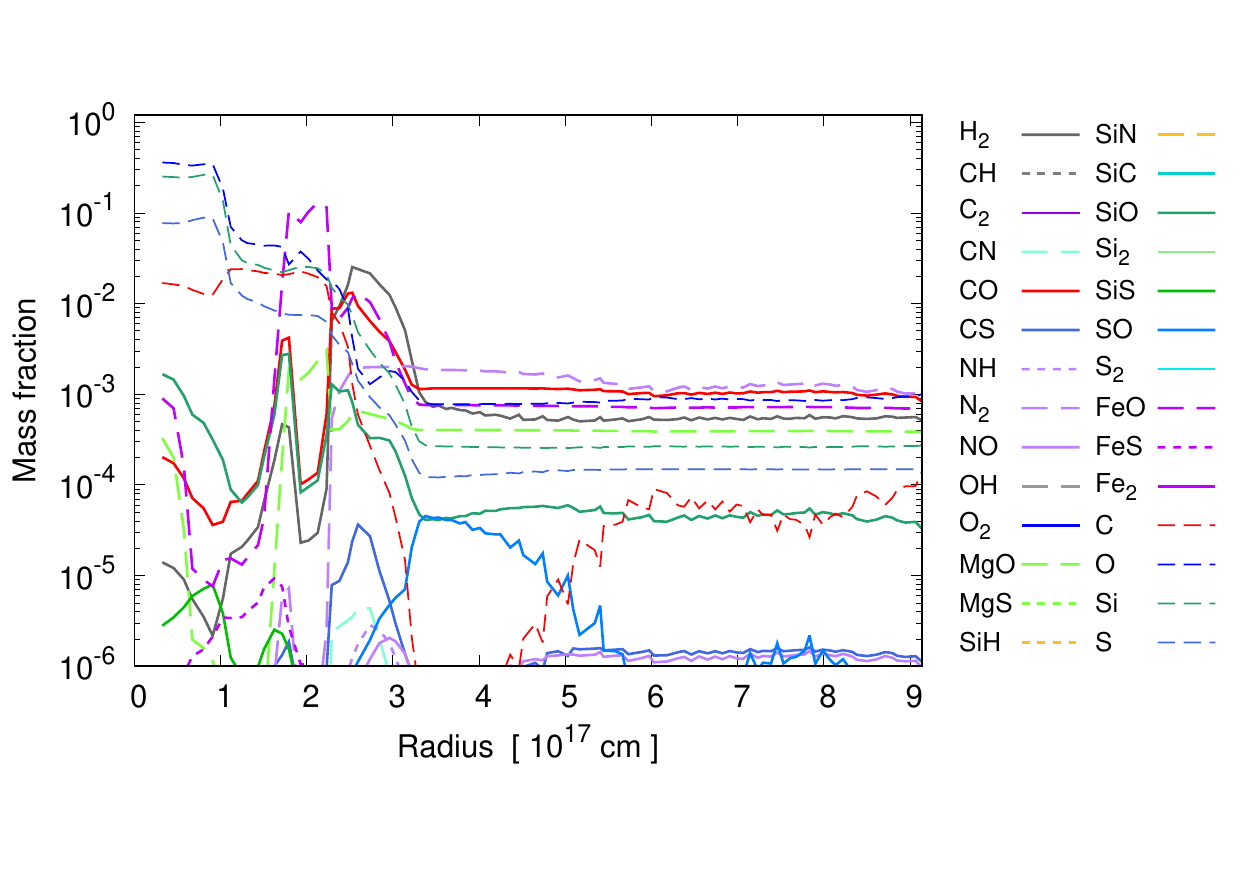}
\end{center}
\end{minipage}
\\
\begin{minipage}{0.5\hsize}
\vs{-1.3}
\begin{center}
\includegraphics[width=9.cm,keepaspectratio,clip]{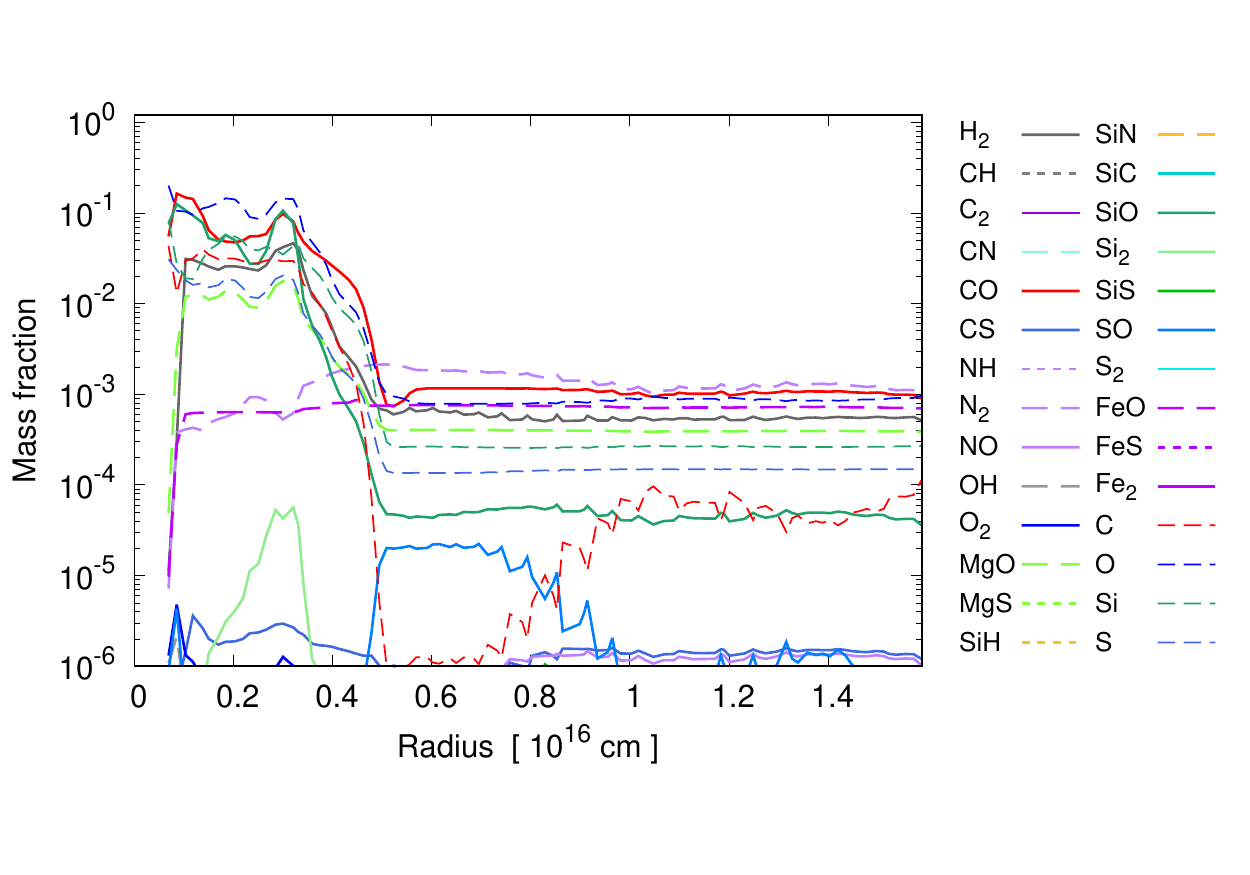}
\end{center}
\vs{-1.}
\end{minipage}
\begin{minipage}{0.5\hsize}
\vs{-1.3}
\begin{center}
\includegraphics[width=9.cm,keepaspectratio,clip]{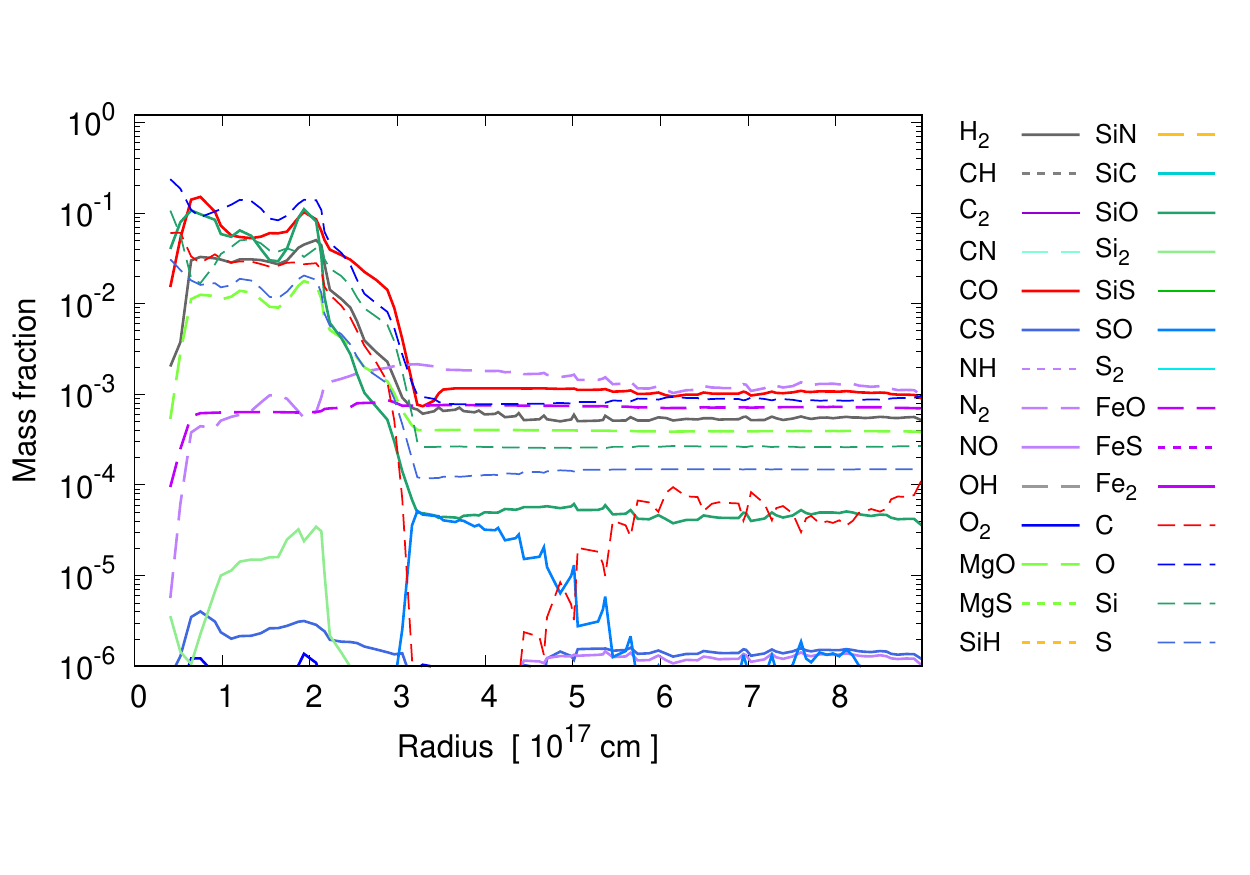}
\end{center}
\vs{-1.}
\end{minipage}
\caption{Radial profiles of the mass fractions of the molecules and several seed atoms for the models b18.3-zp (top), b18.3-zn (middle), and b18.3-yp (bottom). %
Left panels: the profiles at approximately 200 days. %
Right panels: the profiles at the end of the calculation.} %
\label{fig:1d_b18.3_prof}
\end{figure*}
\hspace{\parindent} 
So far, the directional dependence of the time evolution of the total mass fractions of molecules was discussed. %
Hereafter, the dependence of the radial distributions of molecules is described in detail. %

In Figure~\ref{fig:1d_b18.3_prof}, the radial profiles of the mass fractions of molecules and several seed atoms for the models b18.3-zp (top panels), b18.3-zn (middle panels), and b18.3-yp (bottom panels) are shown for two different epochs (approximately 200 days and the end of the calculation in the left and right panels, respectively). %
As mentioned above, during approximately 100-300 days, in particular, in the $+Z$ and $-Z$ directions, the destruction processes of some molecules are dominant (see, left panels in Figure~\ref{fig:1d_b18.3_angle}). %
As a representative snapshot of the destruction phase, ones at approximately 200 days are selected. %

At about 200 days (left panels), in the $+Z$ ($-Z$) direction, it is apparent that molecules are not formed at the regions inside $\sim$ 4 $\times$ 10$^{15}$ cm (3.5 $\times$ 10$^{15}$ cm). %
On the other hand, in the $+Y$ direction, molecules such as CO, SiO, and H$_2$ are remarkably formed even in inner regions. %
This is because in the $+Z$ and $-Z$ directions, at this epoch, the gas temperatures of the inner particles are too high for molecule formation due to the gas heating by the decay of $^{56}$Ni (see, the right panels in Figure~\ref{fig:1d_b18.3_angle}); %
on the other hand, in the $+Y$ direction, the local $^{56}$Ni abundance is overall low (gas heating is less), and the molecule formation is already developed because of the early starting due to the early decrease of the gas temperatures to $\sim$ 10$^4$ K. %
Again, the different initial $^{56}$Ni abundances are due to the different energy depositions during the bipolar-like explosion. %

In all three directions, outer layers where all the mass fractions shown are approximately less than 10$^{-3}$ and flat with some exceptions can be recognized. %
The transition points (radii) between the outer layers and the inner regions are similar in the three directions, although in the $+Z$ direction, the radius of the point is slightly larger than those in the other two directions. %
In the outer layers, CO and SiO are also formed, although the mass fractions are not so high. %

As for other molecules, in the outer layers, N$_2$, MgO, and FeO are distinctly formed; %
the formation of SO is also recognized, although the region is spatially limited (0.4--1.0 $\times$ 10$^{16}$ cm). %
In the $+Y$ direction, N$_2$, MgO, and FeO are formed even inner regions; %
the formation of Si$_2$ can also be recognized at a limited region (0.1--0.4 $\times$ 10$^{16}$ cm). %

After 200 days, the regions where molecules are formed gradually extend inside as inner particles heated by the decay of $^{56}$Ni partake in the molecule formation when the gas temperatures go down to $\sim$ 10$^{4}$ K. %

At the end of the calculation (right panels), in the $+Z$ and $-Z$ directions, the formation of molecules, e.g., CO, SiO, H$_2$, MgO, and FeO, are recognized at inner regions, although the mass fractions are not so high compared with those in the $+Y$ directions. %

As seen in the $+Z$ direction, the mass fractions of e.g., CO, SiO, and H$_2$ are not smoothly distributed in the inner regions (less than 4 $\times$ 10$^{17}$ cm). %
In the region of (2.0--2.8) $\times$ 10$^{17}$ cm (a dip region), the mass fractions of the three molecules above are apparently lower than those of the surroundings. %
It is noted that the transition point ($\sim$ 4 $\times$ 10$^{17}$ cm) at which the mass fractions of CO, SiO, and H$_2$ increase compared with those in the outer layers corresponds to the initial position of 2.4 $\times$ 10$^{13}$ cm (see, the top left panel in Figure~\ref{fig:b18.3_angle}) at which the mass fractions of the seed atoms increase. %

What corresponds to the upper bound of the dip region? As seen in the top right panel in Figure~\ref{fig:1d_b18.3_angle}, at 200 days, there is a gap in the gas temperatures between particles with and without a distinct gas cooling. %
Actually, the upper bound of the dip region (i.e., 2.8 $\times$ 10$^{17}$ cm) corresponds to the temperature gap. %
Inside the upper bound (corresponding to the upper region from the temperature gap in the top right panel in Figure~\ref{fig:1d_b18.3_angle}), particles go through the gas heating by the decay of $^{56}$Ni without a distinct gas cooling. %
Then, the molecule formation is delayed compared with the particles outside the upper bound; %
some fraction of formed molecules may also be destructed by the processes involved with Compton electrons (\texttt{CM} rections) and/or UV photons (\texttt{UV} reactions) and/or ions. %
On the other hand, outside the upper bound, the heating of the gas (ionization and/or destruction of molecules) is less effective and some of the particles go through the gas cooling by CO ro-vibrational transitions; %
the molecule formation may partly be enhanced by the earlier start of the formation and the higher density caused by the cooling through the effect described in Equation~(\ref{eq:thermal_i}). %

Inside the lower bound of the dip region (i.e., 2.0 $\times$ 10$^{17}$ cm), the mass fractions of CO, SiO, and H$_2$ are apparently higher than those in the dip region. What makes the difference? 

First, it is found that the lower bound roughly corresponds to the initial position of 1.1 $\times$ 10$^{13}$ cm at which the abundance of $^{56}$Ni is a bit sharply changed (see, the top left panel in Figure~\ref{fig:b18.3_angle}). %
Hence, inside the lower bound, the local abundance of $^{56}$Ni is higher than that in the dip region. %
The decay of $^{56}$Ni generally plays a role in the gas heating, ionization of atoms, and ionization and destruction of molecules. %
Therefore, considering the latter effect, it sounds a bit strange that inside the lower bound, the formation of CO, SiO, and H$_2$ is efficient compared with the dip region. %

The reason is probably that inside the lower bound, the decay of $^{56}$Ni actually ``enhances" the formation of the three molecules above through the sequences as follows. %
The decay of $^{56}$Ni increases the number density of electrons; %
H$^-$ is produced by the \texttt{REA} reaction, H + e$^-$ $\lra$ H$^-$ + $\gamma$; %
then, H$_2$ is produced by the \texttt{AD} reaction, H$^-$ + H $\lra$ H$_2$ + e$^-$; %
OH is formed by the \texttt{NN} reaction, O + H$_2$ $\lra$ OH + H; %
finally, CO and SiO are efficiently produced through the \texttt{NN} reactions in Equations~(\ref{eq:c_oh}) and (\ref{eq:si_oh}) with OH. %
Actually, it is confirmed that inside the lower bound, the contributions of the \texttt{NN} reactions in Equations~(\ref{eq:c_oh}) and (\ref{eq:si_oh}) on the formation of CO and SiO are higher than those in the dip region. %

What is the condition that the sequences that enhance the formation of CO and SiO dominate the destructive processes through the \texttt{CM} and \texttt{UV} reactions and/or ones involved with ions? %
Inside the lower bound of the dip region, the tracer particles are heated by the decay of $^{56}$Ni, and the start of the formation of molecules is delayed. %
In this case, before the start of the molecule formation, the number density of electrons increases due to the ionization of the seed atoms by Compton electrons through the reaction in Equation~(\ref{eq:ion_x}). %
Hence, the abundance ratios of electrons to the other species become higher than that of the cases where the formation of molecules starts at an early phase of the active decay of $^{56}$Ni, which may effectively increase the contribution from the formation processes in Equations~(\ref{eq:c_oh}) and (\ref{eq:si_oh}) through the sequences above with the high electron abundance. %

In the $-Z$ direction, similarly to the $+Z$ direction, at approximately (1.7--2.3) $\times$ 10$^{17}$ cm, a sharp dip of CO, SiO, and H$_2$ is recognized. %
In the $+Y$ direction, compared with the profiles at 200 days, there are not so large differences, which means that the molecule formation is overall settled before 200 days. %

As for other molecules, MgO and FeO are also formed in the inner regions in $+Z$ and $-Z$ directions. %
In all three directions, SO is formed at the bottom of the outer layers and the distributions are not so changed from those at 200 days. %
In the $+Z$ and $-Z$ directions, at the end of the calculation, the formation of CS is recognized just inside the region where SO is formed with some overlaps. %

\paragraph{Radial distributions of molecules in the models with the single-star progenitor model} \label{para:1d_radial_single}

\begin{figure*}
\begin{minipage}{0.5\hsize}
\begin{center}
\includegraphics[width=9.cm,keepaspectratio,clip]{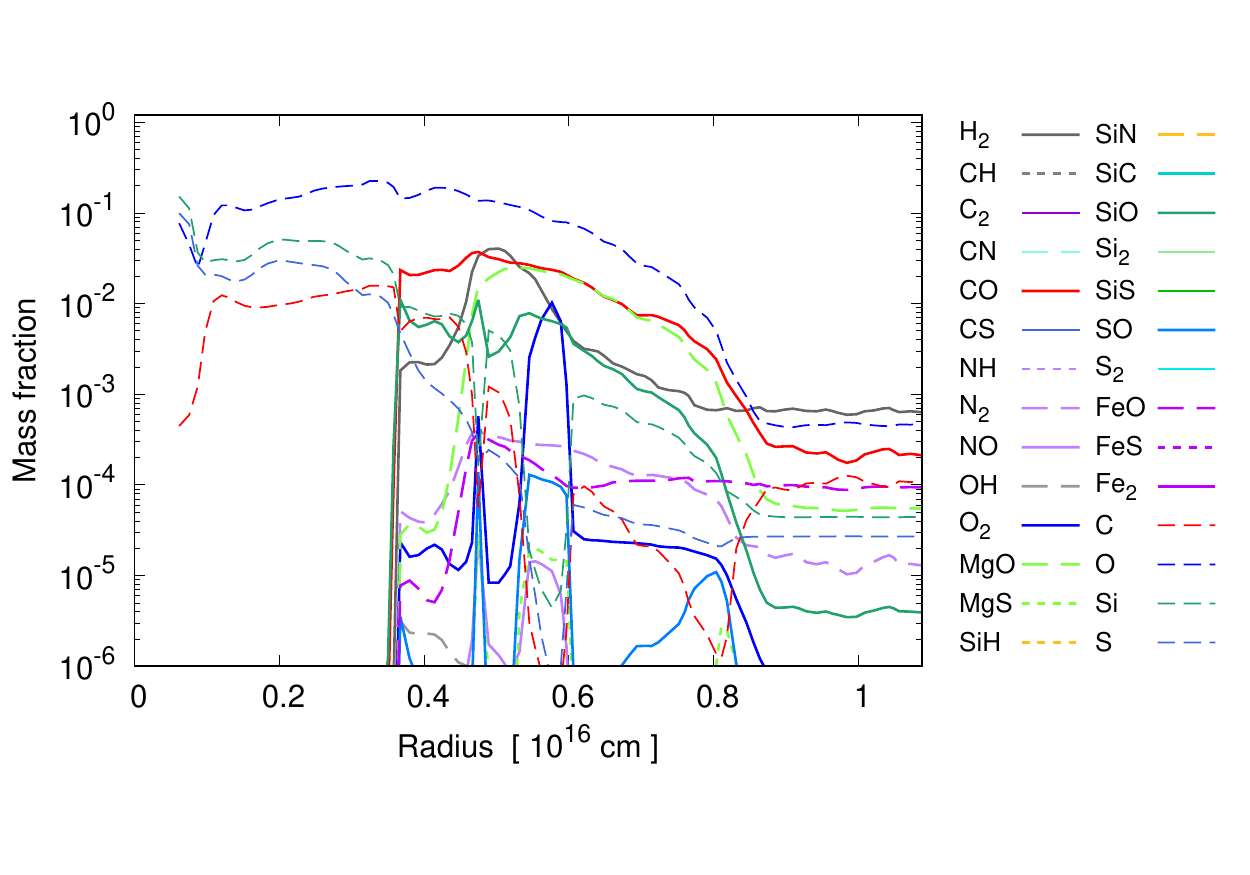}
\end{center}
\end{minipage}
\begin{minipage}{0.5\hsize}
\begin{center}
\includegraphics[width=9.cm,keepaspectratio,clip]{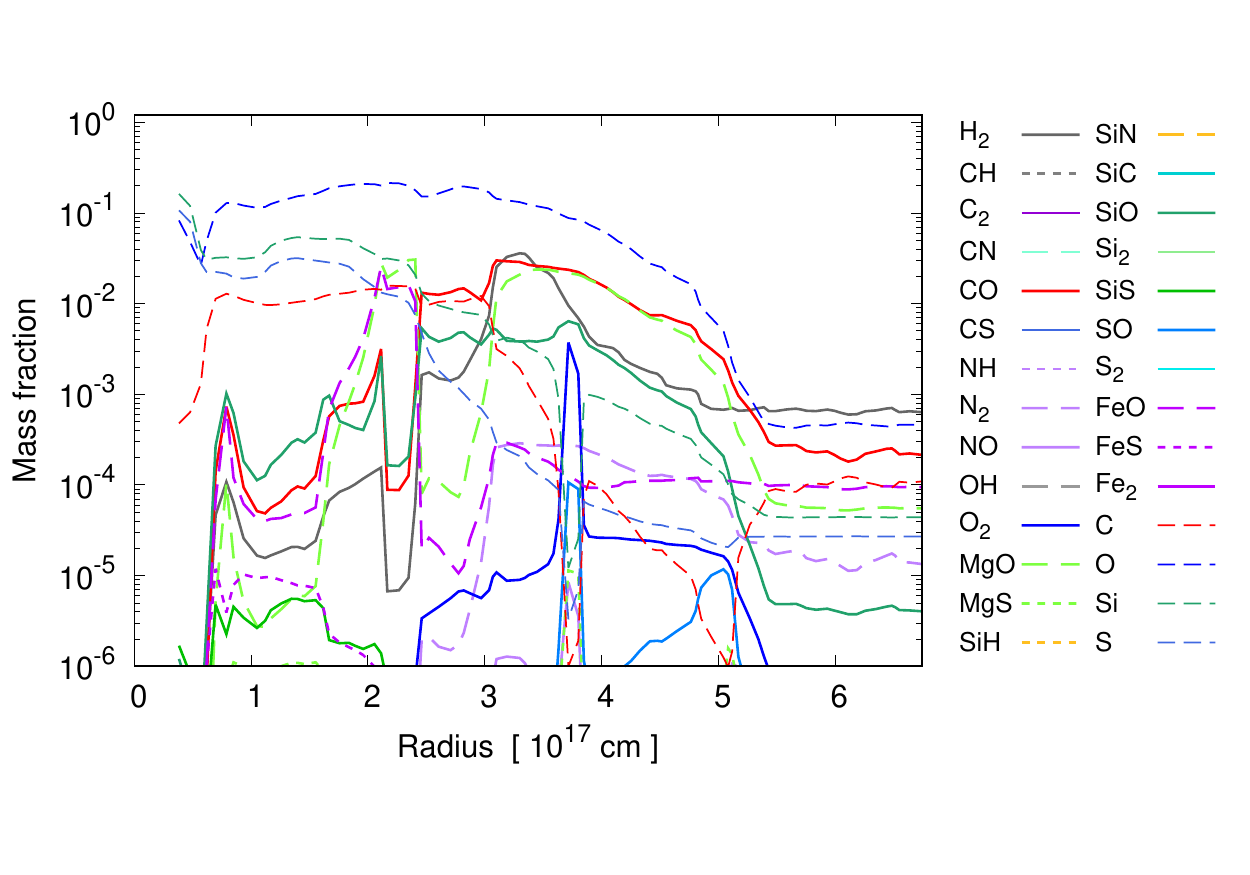}
\end{center}
\end{minipage}
\\
\begin{minipage}{0.5\hsize}
\vs{-1.3}
\begin{center}
\includegraphics[width=9.cm,keepaspectratio,clip]{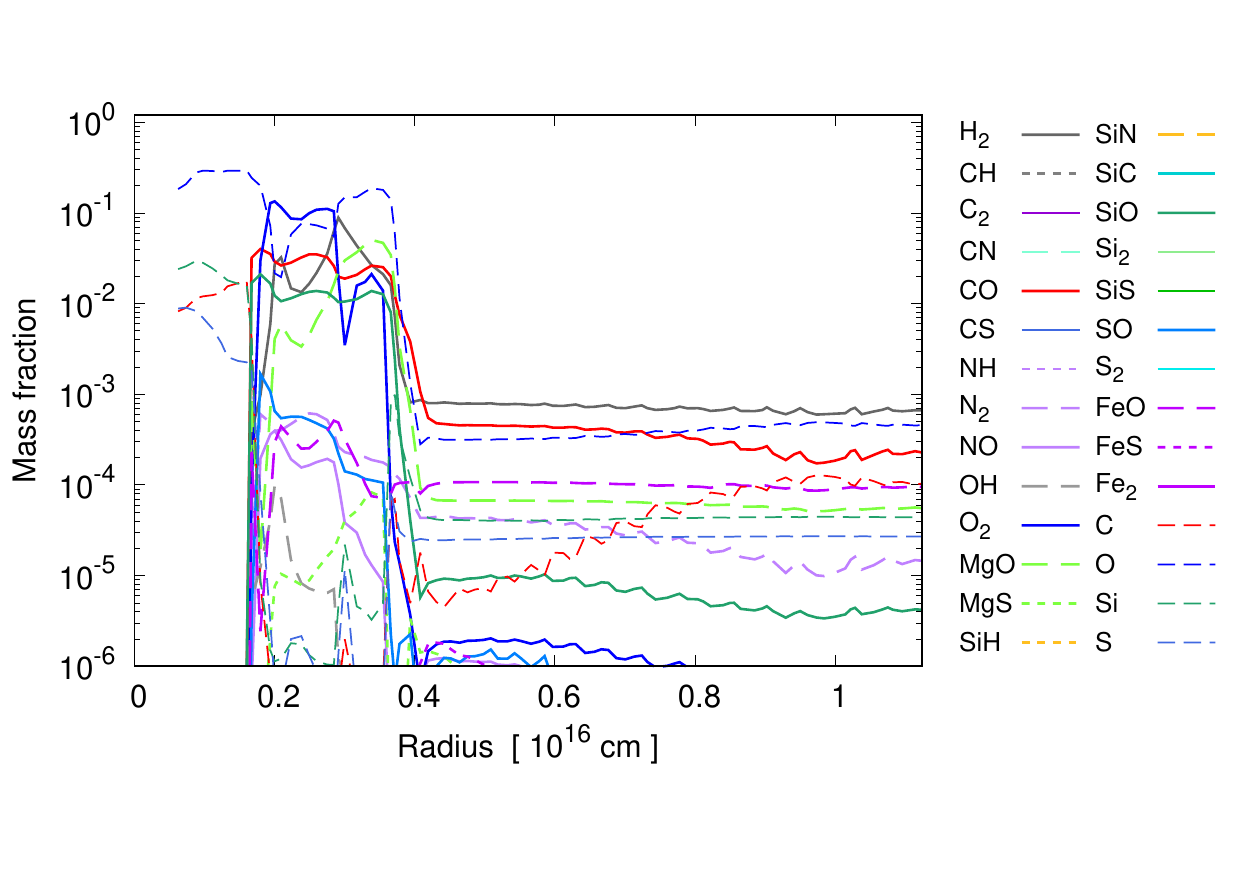}
\end{center}
\end{minipage}
\begin{minipage}{0.5\hsize}
\vs{-1.3}
\begin{center}
\includegraphics[width=9.cm,keepaspectratio,clip]{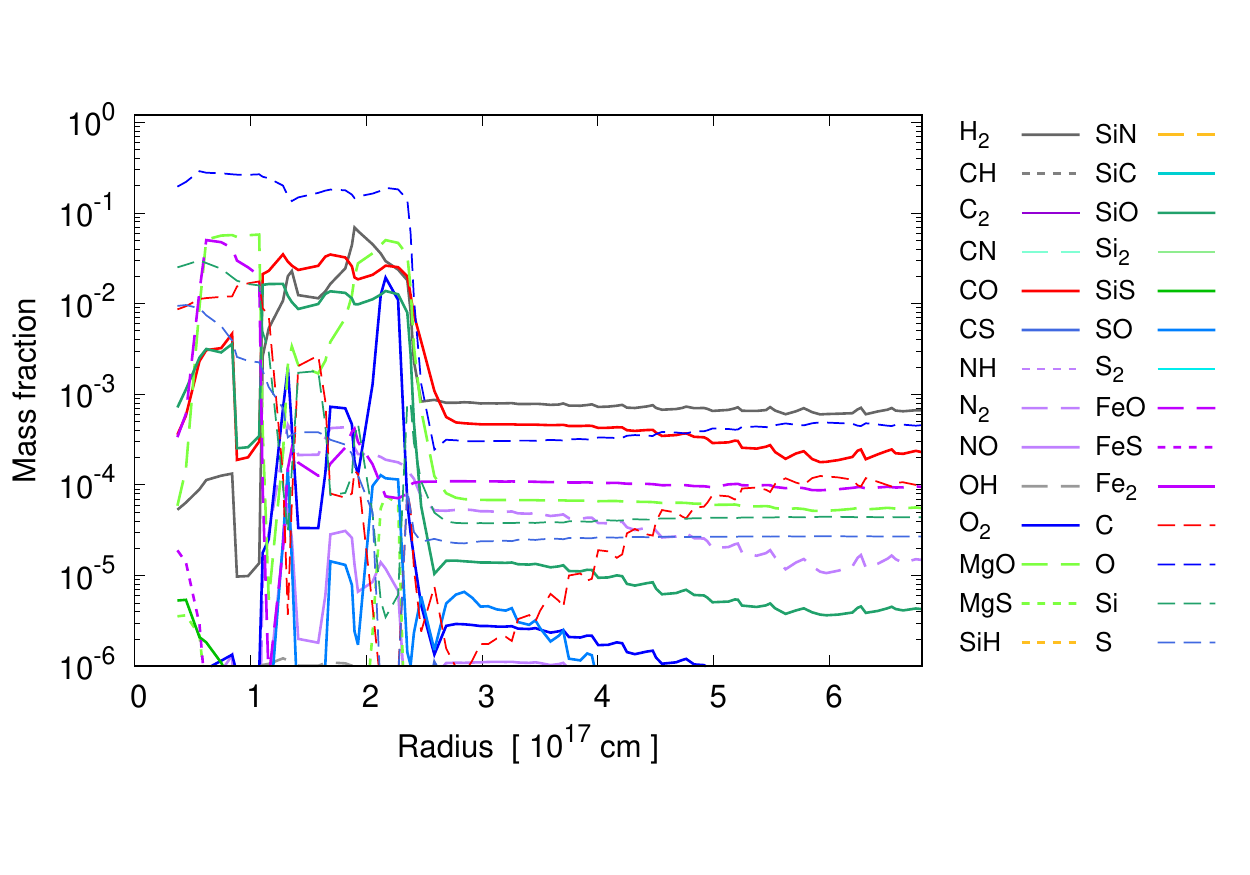}
\end{center}
\end{minipage}
\\
\begin{minipage}{0.5\hsize}
\vs{-1.3}
\begin{center}
\includegraphics[width=9.cm,keepaspectratio,clip]{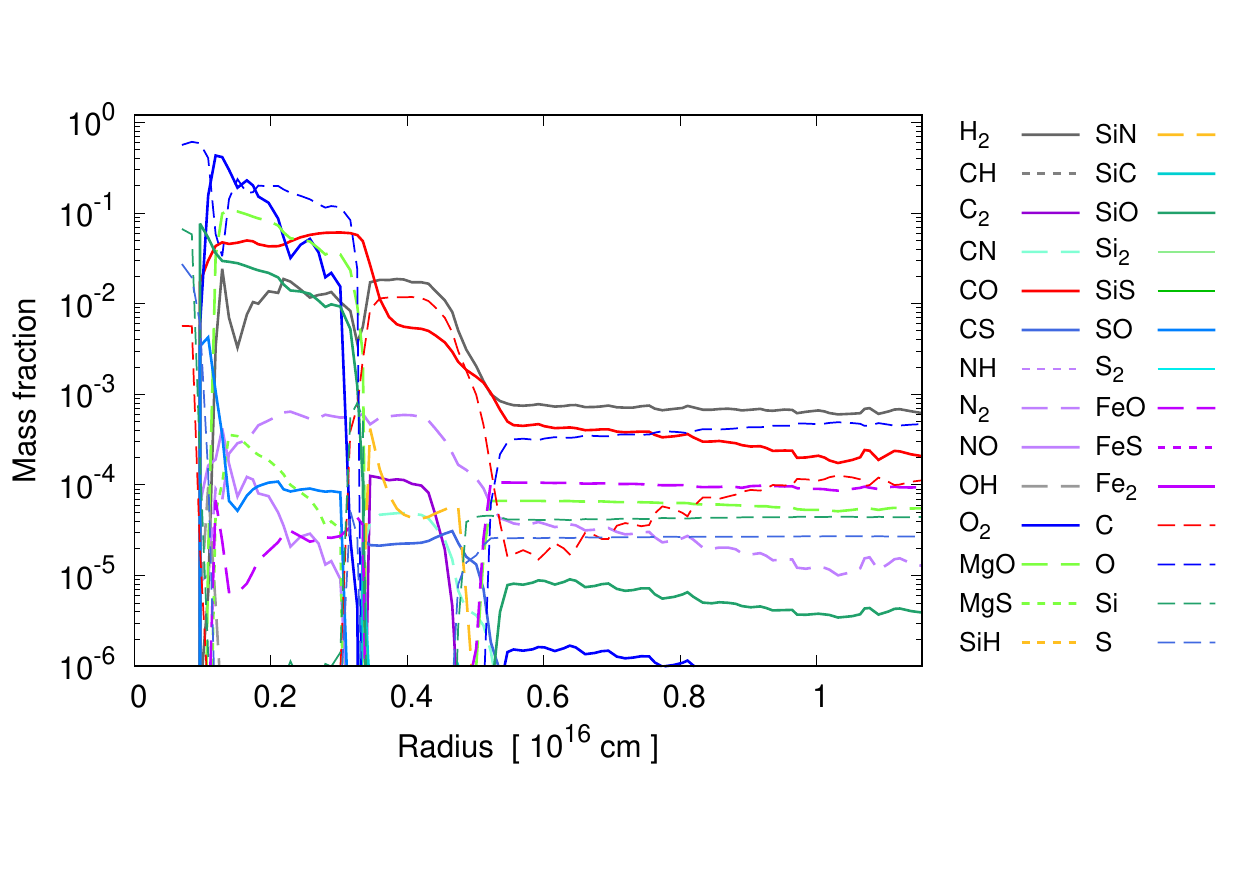}
\end{center}
\vs{-1.}
\end{minipage}
\begin{minipage}{0.5\hsize}
\vs{-1.3}
\begin{center}
\includegraphics[width=9.cm,keepaspectratio,clip]{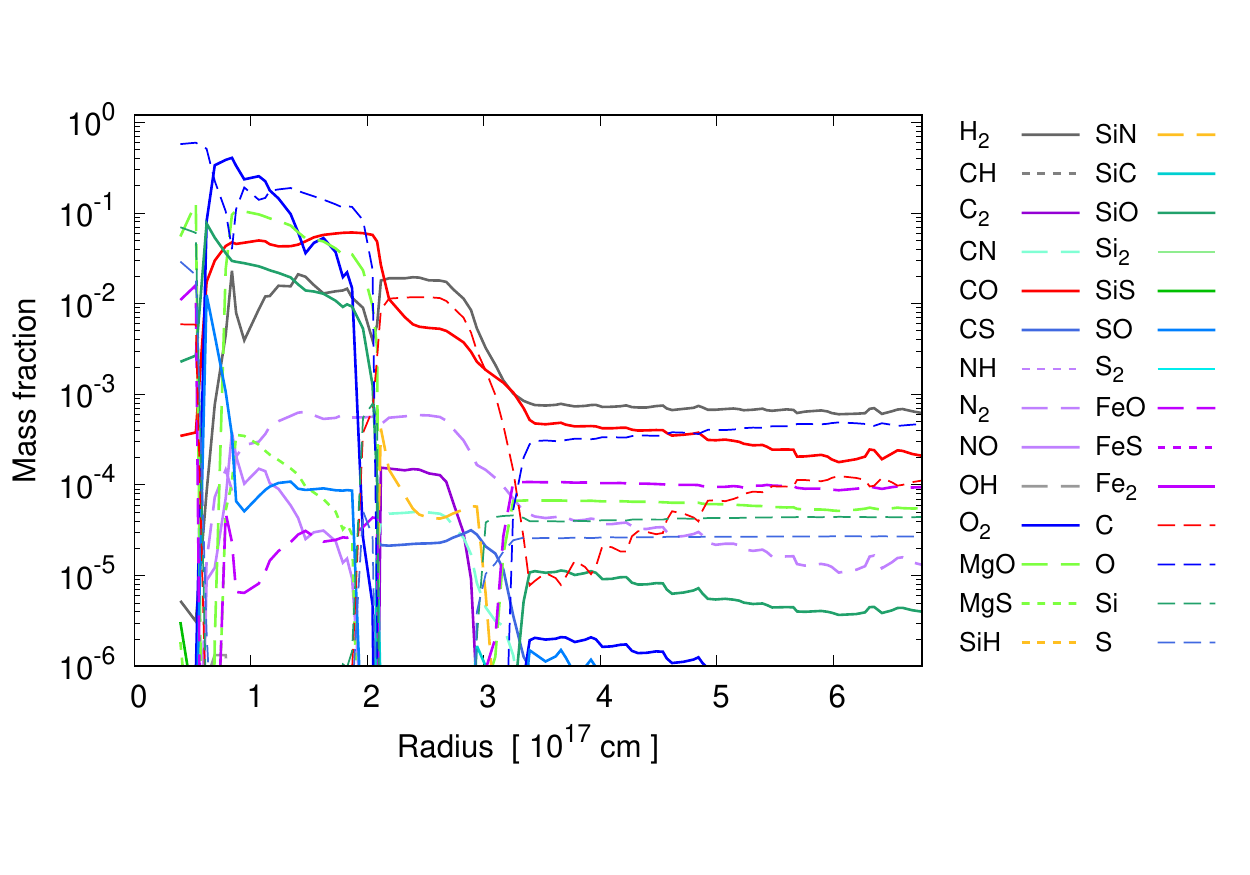}
\end{center}
\vs{-1.}
\end{minipage}
\caption{Same as Figure~\ref{fig:1d_b18.3_prof} but for the models n16.3-zp (top), n16.3-zn (middle), and n16.3-yp (bottom).} %
\label{fig:1d_n16.3_prof}
\end{figure*}
\hspace{\parindent} 
In Figure~\ref{fig:1d_n16.3_prof}, the profiles for the models, n16.3-zp (top panels), n16.3-zn (middle panels), and n16.3-yp (bottom panels) are shown. %
At about 200 days (left panels), it is apparent that in the $+Z$, $-Z$, and $+Y$ directions, the inner regions where molecules are remarkably formed with the mass fractions over $\sim$ 10$^{-3}$ are extended to $\sim$ 9 $\times$ 10$^{15}$ cm, 4 $\times$ 10$^{15}$ cm, and 5 $\times$ 10$^{15}$ cm, respectively. %
The radii roughly reflect the initial extensions of the regions where the seed atoms are abundant (see, Figure~\ref{fig:n16.3_angle}). %
In the $+Z$ and $-Z$ directions, there are inner regions where molecules are not distinctly formed; %
the regions are extended to roughly 4 $\times$ 10$^{15}$ cm and 2 $\times$ 10$^{15}$ cm, respectively. %
On the other hand, in the $+Y$ direction, molecules are remarkably formed except for the innermost region. %

Another distinct feature compared with the models b18.3-zp, b18.3-zn, and b18.3-yp is the formation of O$_2$. %
In the models with the progenitor model b18.3, the formation of O$_2$ is not evidently recognized. %
As seen in the remained seed atoms at 200 days in the $+Z$ and $-Z$ directions, the mass fractions of the remained oxygen atoms are apparently larger than those in the models with b18.3. %
The high mass fractions of O$_2$ compared with the models with b18.3 is basically attributed to the initial high oxygen abundance. %
Hereafter, for each direction, the properties of O$_2$ and other major molecules at about 200 days are described. %

In the $+Z$ direction, the mass fraction of O$_2$ peaks at approximately 5.8 (4.7) $\times$ 10$^{15}$ cm with the mass fraction of $\sim$ 10$^{-2}$ (10$^{-3}$) and in the other regions within (3.5--8.2) $\times$ 10$^{15}$ cm, the mass fraction is roughly the order of 10$^{-5}$. %
In contrast to O$_2$, CO, SiO, and H$_2$ are extensively formed within the regions of roughly (3.5--9.0) $\times$ 10$^{15}$ cm. %
In the regions where the mass fraction of O$_2$ is $\sim$ 10$^{-5}$, O$_2$ is actually more abundant at an earlier time. %
Then, O$_2$ is destructed (converted to other molecules) by the \texttt{NN} reactions, Si + O$_2$ $\lra$ SiO + O and C + O$_2$ $\lra$ CO + O, approximately from 60 days. %
Whether O$_2$ survives or not may sensitively be determined by a bit complicated balance of \texttt{NN} reactions depending on the thermal history. %
It is noted that in the evolution of the gas temperatures (the top right panel in Figure~\ref{fig:1d_n16.3_angle}), e.g., at 200 days, a gap of the gas temperatures (approximately from 1.5 $\times$ 10$^{3}$ K to 4.5 $\times$ 10$^3$ K) is recognized. %
Some of the particles whose temperatures at 200 days are below the lower bound have complicated temperature histories affected by both the heating and cooling. %
The regions of approximately (4.2--5.5) $\times$ 10$^{15}$ cm are affected by both the heating and cooling. %

In the $-Z$ direction, the formation of O$_2$ is more significant; %
the distribution of O$_2$ is roughly consistent with those of CO, SiO, and H$_2$ and there are regions where the mass fraction of O$_2$ is higher than that of CO with the peak value of $\sim$ 10$^{-1}$. %
It is noted that the regions of (1.7--4.0) $\times$ 10$^{15}$ cm correspond to the initial positions of (0.9--2.0) $\times$ 10$^{13}$ cm (see, the middle left panel in Figure~\ref{fig:n16.3_angle}); %
the radial fluctuations of the mass fractions of seed atoms, in particular, oxygen and silicon, seen in the initial profiles seem to be somehow reflected in the profiles of those of CO, SiO, and O$_2$ at 200 days. %

In the $+Y$ direction, O$_2$ is significantly formed inside $\sim$ 3 $\times$ 10$^{15}$ cm; %
the distribution is roughly consistent with that of SiO. %
The peak value of the mass fraction of O$_2$ is as high as 0.4 at approximately 1 $\times$ 10$^{15}$ cm. %
In contrast to O$_2$ and SiO, the regions where the mass fractions of CO and H$_2$ are high (greater than 10$^{-3}$) are more extended to 5 $\times$ 10$^{15}$ cm. %
The regions of (1.0--3.0) $\times$ 10$^{15}$ cm and (3.0--5.0) $\times$ 10$^{15}$ cm roughly correspond to the initial positions of (0.6--1.5) $\times$ 10$^{13}$ cm and (1.5--2.4) $\times$ 10$^{13}$ cm, respectively (see the bottom left panel in the Figure~\ref{fig:n16.3_angle}). %
In the latter region in the initial profiles, the mass fractions of oxygen and silicon are low compared with those in inner regions (the mass fraction of silicon is the order of 10$^{-5}$); %
instead, the mass fractions of hydrogen and helium are high. %
On the other hand, in the former regions, there is plenty of oxygen and silicon. %
The distinct different features across 3.0 $\times$ 10$^{15}$ cm seen in the profiles at 200 days are attributed to the initial distributions of seed atoms. %

As for other molecules, in the $+Z$ direction, MgO, FeO, and N$_2$ are also formed in the inner regions with similar distributions as CO and SiO at (4.5--9.0) $\times$ 10$^{15}$ cm. %
In particular, the mass fraction of MgO is comparable to that of CO in the regions. %
The formation of SO is also recognized at around the positions of the peaks of O$_2$. %
In the regions inside about 6.0 $\times$ 10$^{15}$ cm, FeO slightly increases inward. %
The increase is due to the seed iron originating from $^{56}$Ni. %
In the $-Z$ direction, the formation of MgO, FeO, N$_2$, and SO are recognized. %
Same as the $+Z$ direction, MgO dominates FeO. %
MgO is abundantly formed in the regions of (2.0--4.0) $\times$ 10$^{15}$ cm and peaks at around 3.5 $\times$ 10$^{15}$ cm with the value of 5 $\times$ 10$^{-2}$. %
N$_2$ and SO are a bit more abundantly formed in the regions of (2.0--3.0) $\times$ 10$^{15}$ cm. %
In the $+Y$ direction, first, in the regions of (3.0--5.0) $\times$ 10$^{15}$ cm, the formation of C$_2$, SiN, CN, and CS is recognized. %
In the regions of (1.0-3.0) $\times$ 10$^{15}$ cm, the formation of MgO and SO is evident. %
N$_2$ is formed in both regions above. %
In the former regions, the formation of the molecules mentioned is attributed to the initial abundances (abundance ratios) of the seed atoms, i.e., the relatively low oxygen and silicon abundances compared with the inner regions. %

After 200 days, the particles heated by the decay of $^{56}$Ni gradually partake in the molecule formation from the outer ones. %

At the end of the calculation, in the $+Z$ direction, the distributions of the molecules in the regions of approximately (2.4--5.3) $\times$ 10$^{17}$ cm are overall consistent with those in the regions of (3.5--9.0) $\times$ 10$^{15}$ cm at 200 days, although the mass fractions of SiO and O$_2$ in the regions of (2.4--3.5) $\times$ 10$^{17}$ cm are a bit different from those at 200 days. %
In the regions inside 2.4 $\times$ 10$^{17}$ cm, the formation of CO, SiO, and H$_2$ is recognized; %
the distributions, however, are not smooth. In these regions, particles are affected by the decay of $^{56}$Ni. %
As found in the model b18.3-zp, a dip region, i.e., (2.2--2.4) $\times$ 10$^{17}$ cm, where the mass fractions of CO, SiO, and H$_2$ are small compared with the surroundings is recognized. %
In the regions inside the upper bound, tracer particles are affected by the heating due to the decay of $^{56}$Ni without a distinct cooling through CO ro-vibrational transitions. %
As mentioned, the decay basically plays a role in the ionization and destruction of molecules; %
it, however, could enhance the formation of CO and SiO probably in certain cases if the molecule formation starts well after the decay becomes effective. %
The transition point whether the destruction or enhancement is dominant is at around 2.2 $\times$ 10$^{17}$ cm in the model n16.3-zp as also seen in the model b18.3-zp (2.0 $\times$ 10$^{17}$ cm for b18.3-zp; see, the top right panel in Figure~\ref{fig:1d_b18.3_angle}). %
In the regions inside the outer bound of the dip region ($\lesssim$ 2.4 $\times$ 10$^{17}$ cm), in particular, in the dip region, the formation of MgO and FeO is prominent. %
It is interpreted that the destruction of CO and SiO by the decay of $^{56}$Ni increases the seed oxygen abundance; then, the oxygen is converted to MgO and FeO. %
The high abundance ratio of FeO to MgO in this region than that in the outer layers may be due to the abundant seed iron atoms originating from $^{56}$Ni. %
Although the mass fraction is rather small (the order of 10$^{-6}$), the formation of SiS can also be recognized at the regions inside 2 $\times$ 10$^{17}$ cm. %

In the $-Z$ direction, the position at 1.6 $\times$ 10$^{15}$ cm at 200 days corresponds to 1.1 $\times$ 10$^{17}$ cm at the end of the calculation. %
Therefore, the mass fractions in the regions of approximately (1.1--2.2) $\times$ 10$^{17}$ cm are further varied from those at 200 days. %
In particular, the radial distribution of O$_2$ becomes complicated; SO and NO have qualitatively similar trends. %
It is noted that the regions above correspond to the particles affected by both the heating and cooling (as seen in the middle right panel in Figure~\ref{fig:1d_n16.3_angle}). %
The complicated gas temperature (density) evolution may partly play a role in the complex distributions. %
In the regions inside 1.1 $\times$ 10$^{17}$ cm, the formation of CO, SiO, H$_2$, MgO, and FeO is recognized. %
As in the models n16.3-zp, b18.3-zp, and b18.3-zn, a dip in the mass fractions of CO, SiO, and H$_2$ can be recognized because of the same reason. %

In the $+Y$ directions, same as the model b18.3-yp, the mass fractions of molecules are barely changed from those at 200 days, since the molecule formation is already settled before 200 days. %

\section{Discussion} \label{sec:discussion}

In this section, several issues related to the results of this paper are discussed. %

\begin{figure*}
\begin{minipage}{0.5\hsize}
\begin{center}
\includegraphics[width=8.5cm,keepaspectratio,clip]{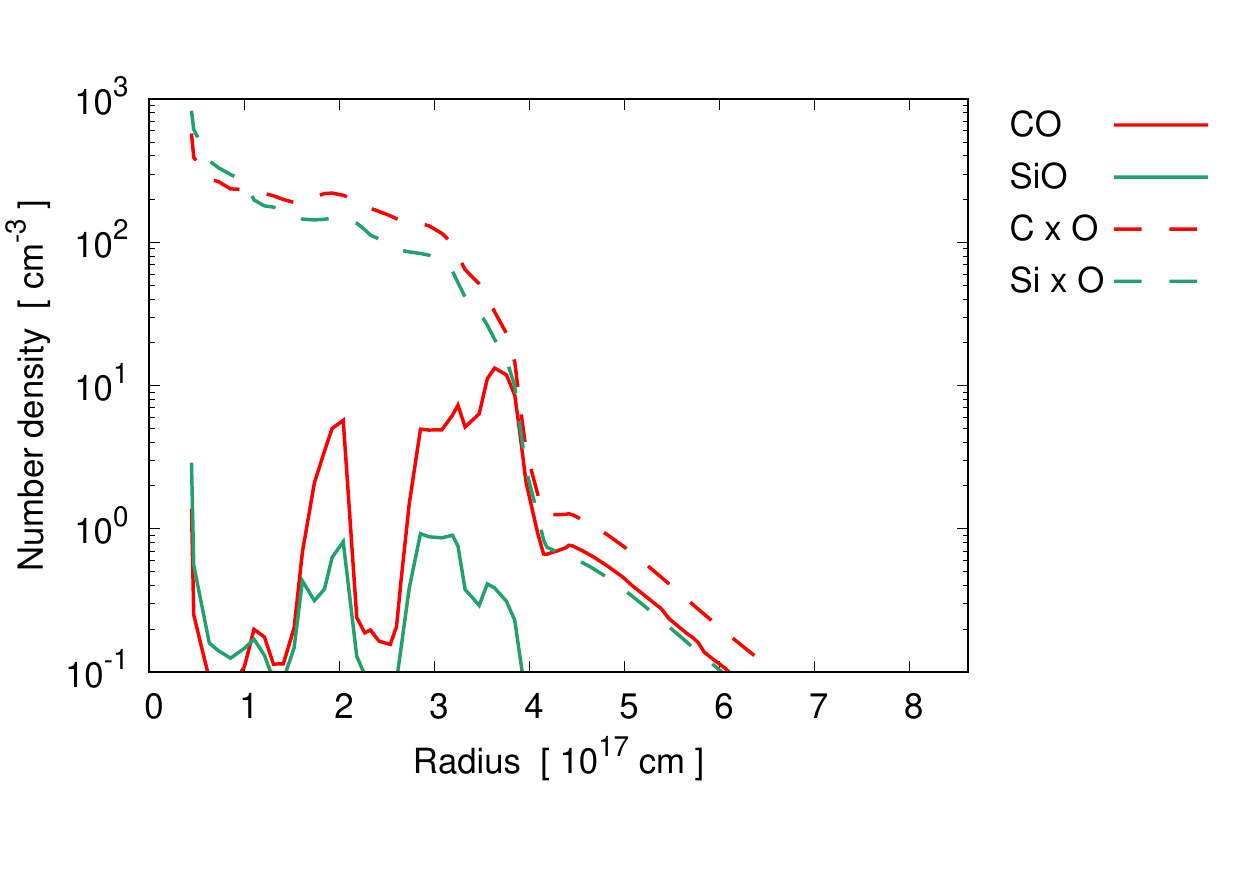}
\end{center}
\end{minipage}
\begin{minipage}{0.5\hsize}
\begin{center}
\includegraphics[width=8.5cm,keepaspectratio,clip]{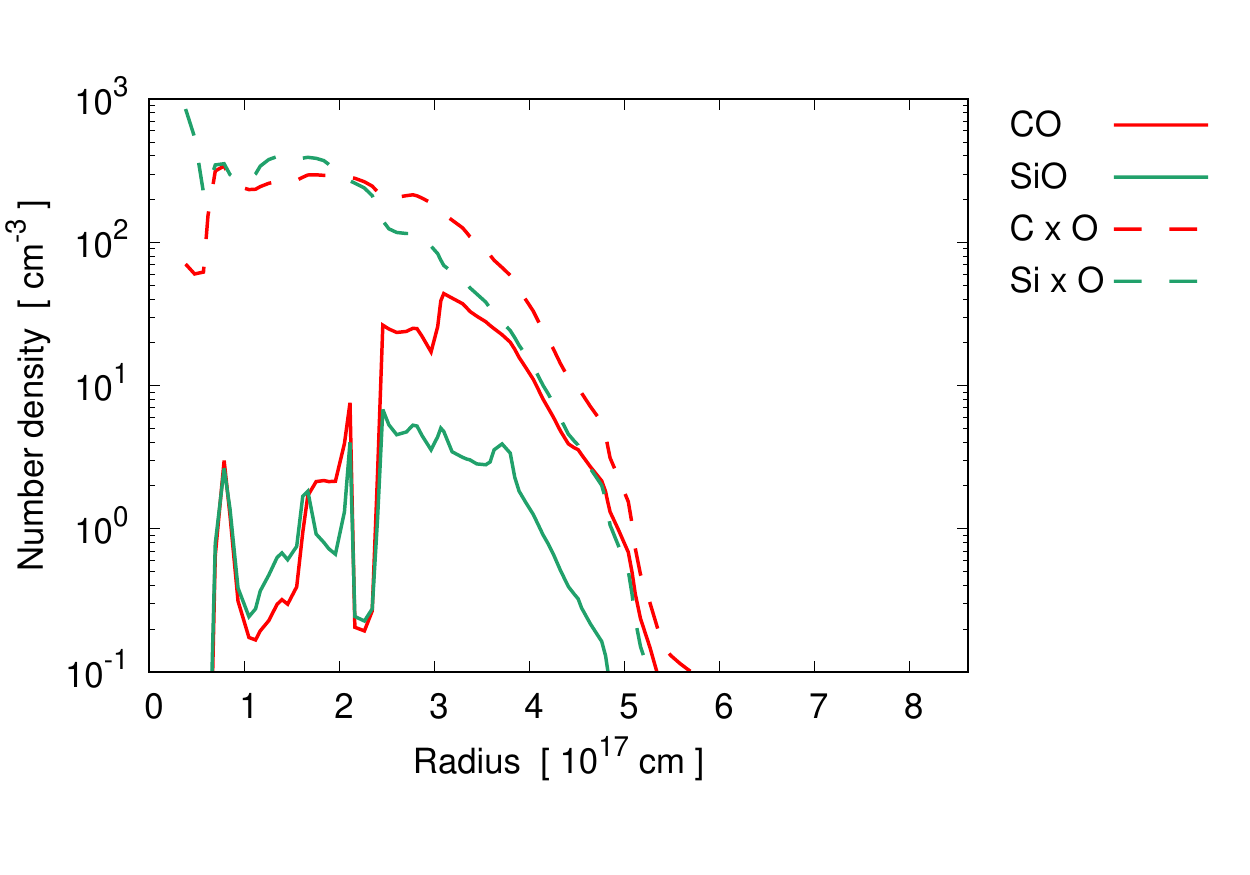}
\end{center}
\end{minipage}
\\
\begin{minipage}{0.5\hsize}
\vs{-0.7}
\begin{center}
\includegraphics[width=8.5cm,keepaspectratio,clip]{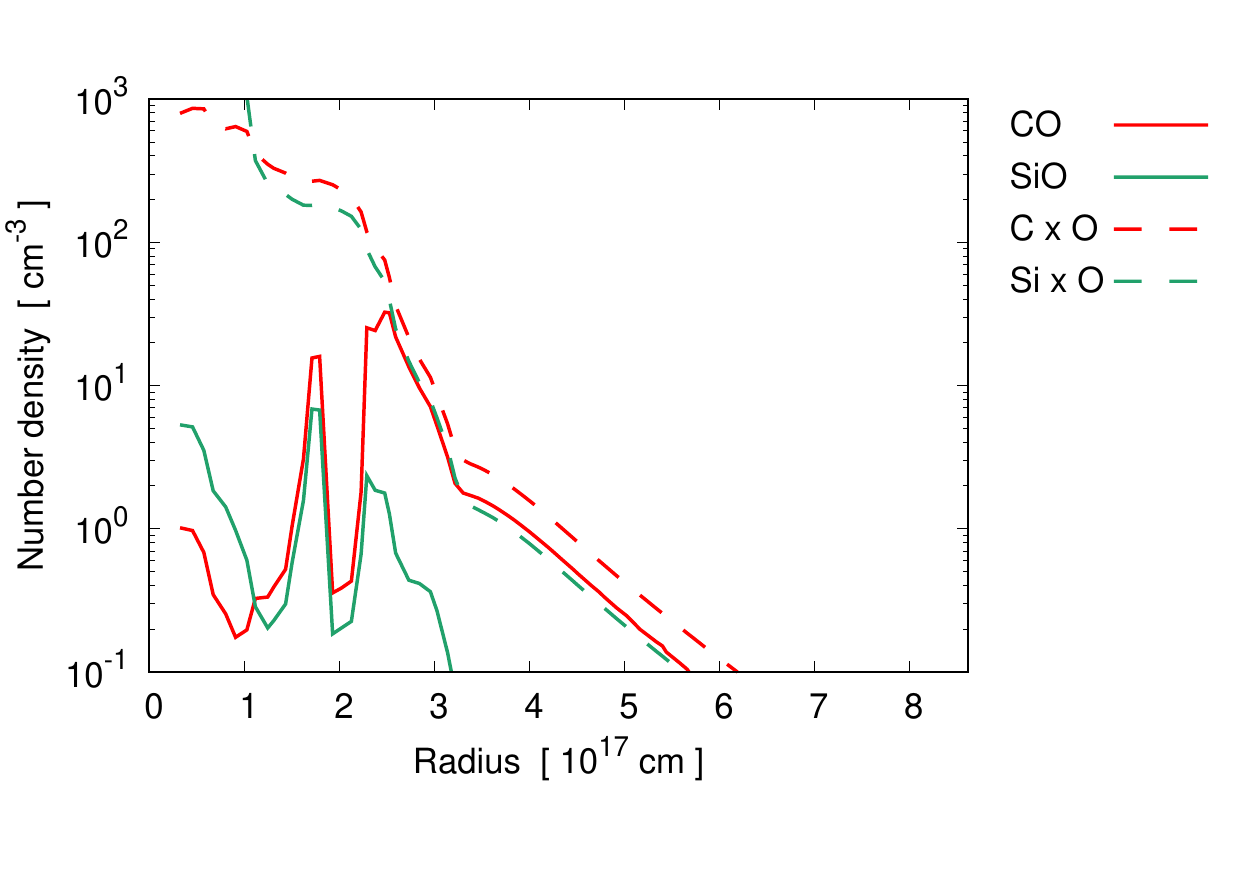}
\end{center}
\end{minipage}
\begin{minipage}{0.5\hsize}
\vs{-0.7}
\begin{center}
\includegraphics[width=8.5cm,keepaspectratio,clip]{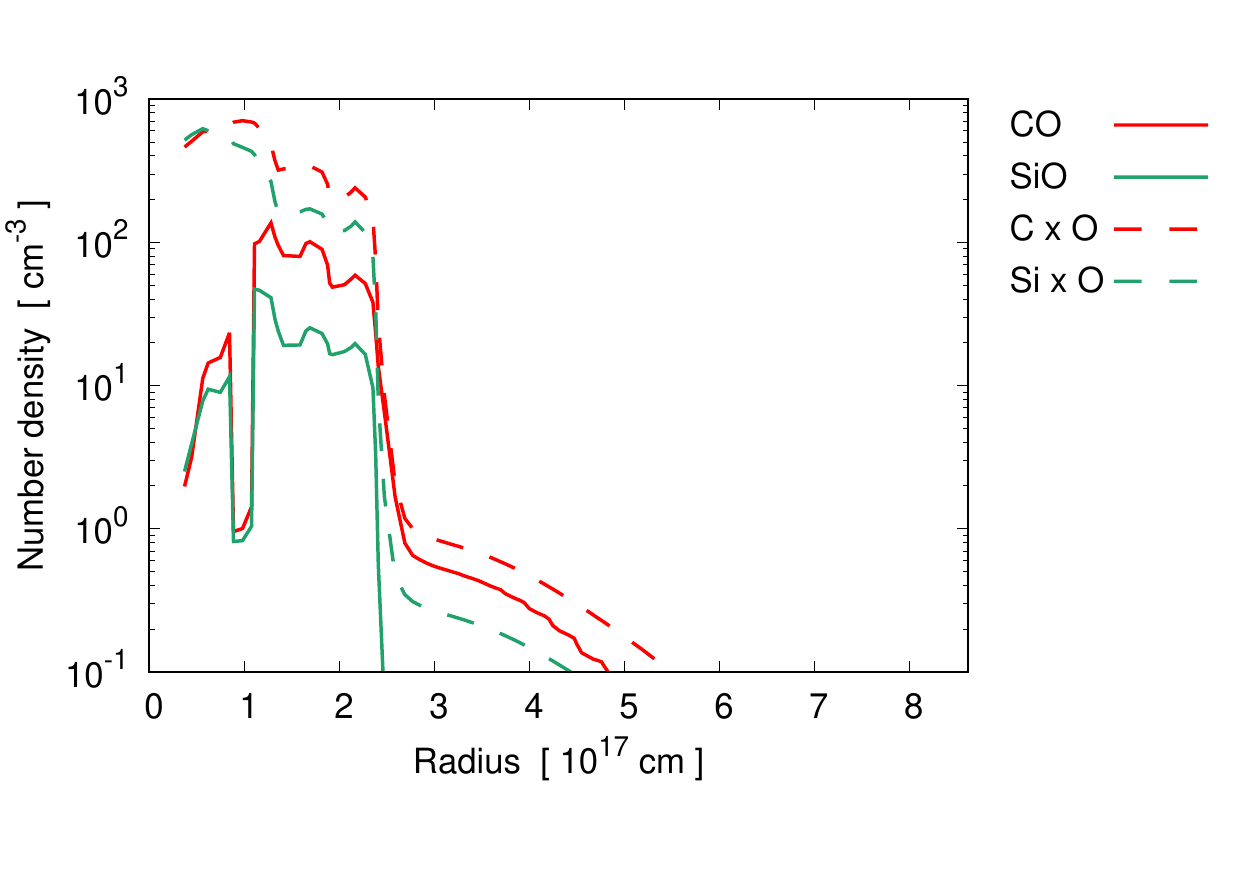}
\end{center}
\end{minipage}
\\
\begin{minipage}{0.5\hsize}
\vs{-0.7}
\begin{center}
\includegraphics[width=8.5cm,keepaspectratio,clip]{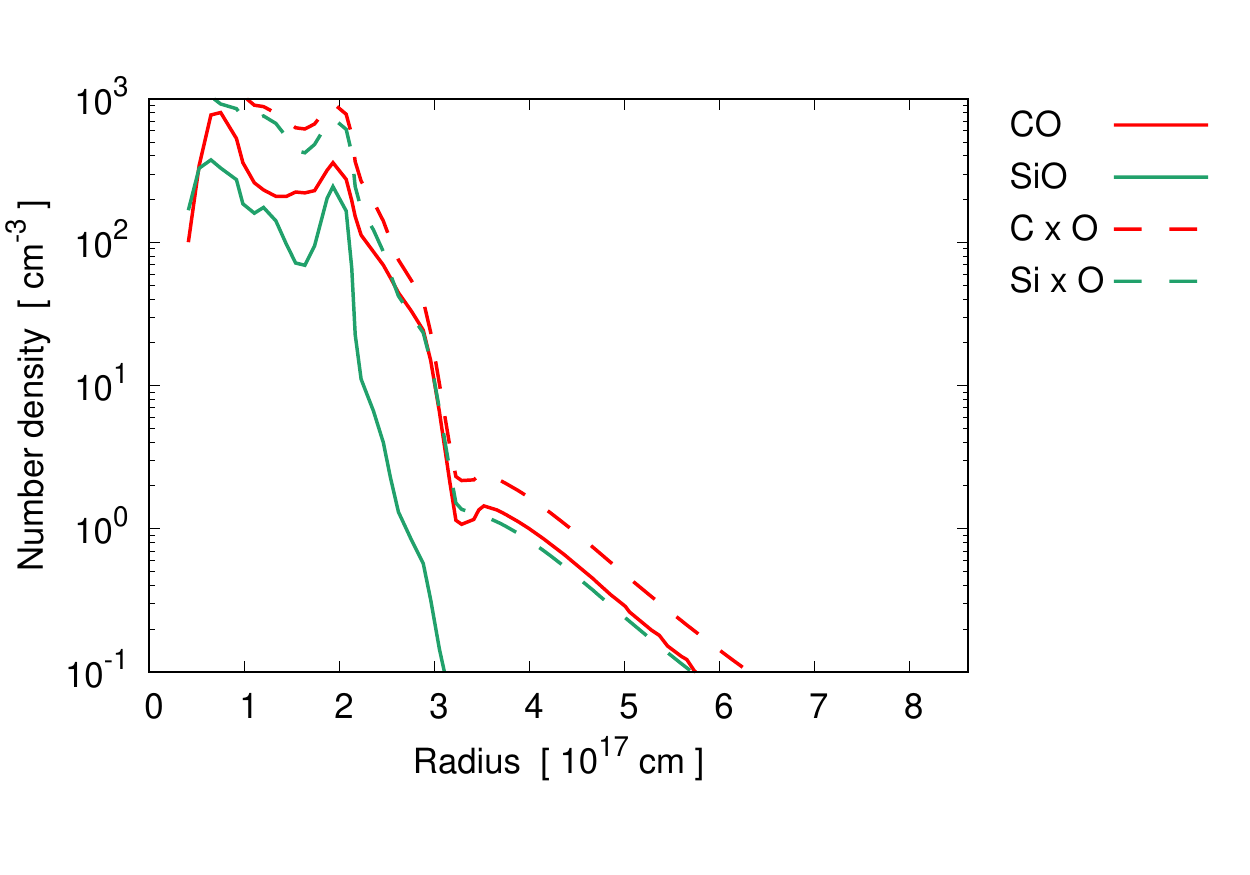}
\end{center}
\vs{-0.7}
\end{minipage}
\begin{minipage}{0.5\hsize}
\vs{-0.7}
\begin{center}
\includegraphics[width=8.5cm,keepaspectratio,clip]{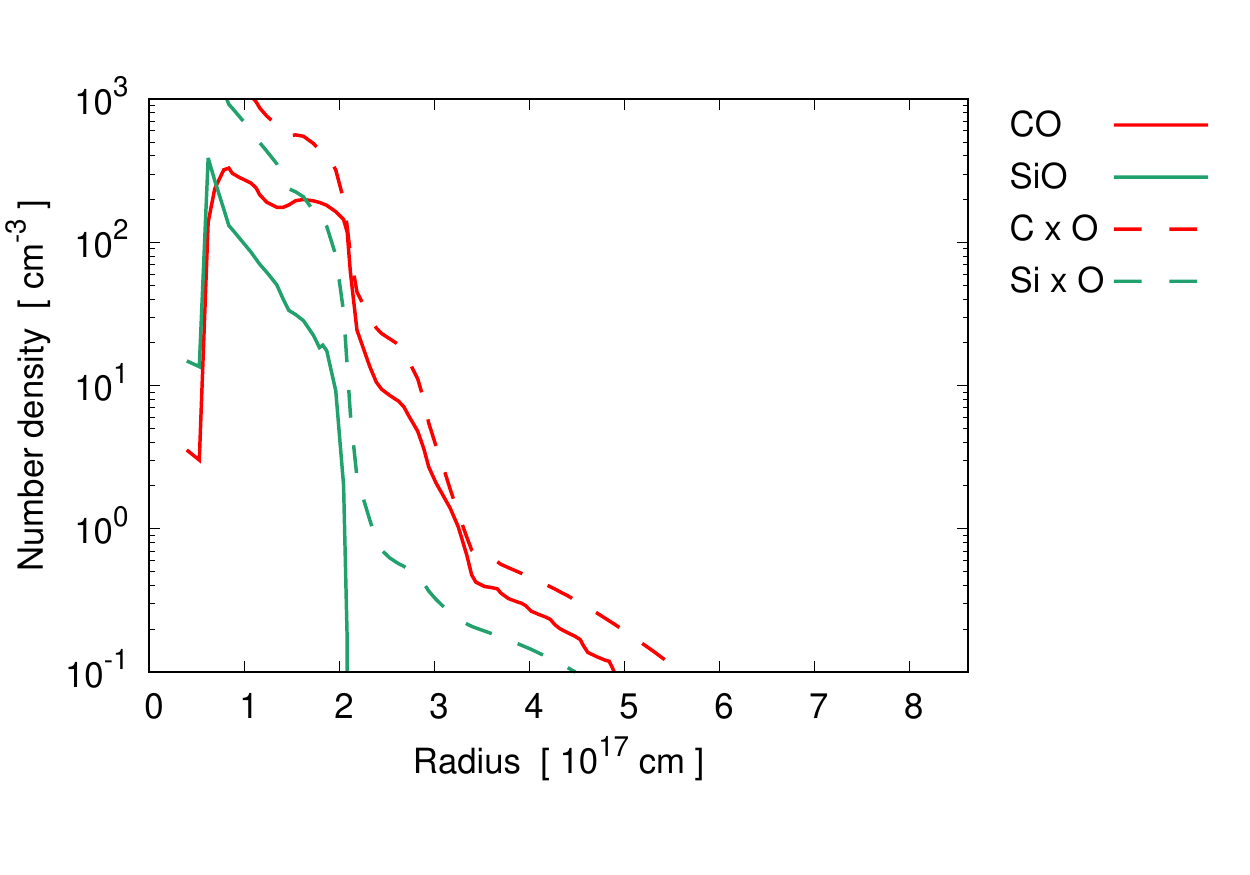}
\end{center}
\vs{-0.7}
\end{minipage}
\caption{Radial distributions of the number densities of CO and SiO (solid lines) at the end of the calculation for the models, b18.3-zp, b18.3-zn, 
b18.3-yp, n16.3-zp, n16.3-zn, and n16.3-yp. %
For reference, approximated number densities of CO and SiO (dashed lines: C $\times$ O and Si $\times$ O, respectively) are also plotted. %
Here, C $\times$ O (Si $\times$ O) denotes $\sqrt{N({\rm C}) \times N({\rm O})}$ ($\sqrt{N({\rm Si}) \times N({\rm O})}$), where $N({\rm C})$, $N({\rm O})$, and $N({\rm Si})$ are the number densities of carbon, oxygen, and silicon atoms, respectively, for the case with no chemical evolution, i.e., the case that each tracer particle keeps the initial mass fractions of the species. %
From top to bottom, the models for the directions, $+Z$, $-Z$, and $+Y$, are shown, respectively. %
The left panels (right panels) are for the models based on the binary merger progenitor model b18.3 (single-star progenitor model n16.3).} %
\label{fig:3d_dist}
\end{figure*}

\subsection{Implication to the 3D distributions of CO and SiO} \label{subsec:3d_dist}

As noted in Section~\ref{sec:intro}, ALMA observations have revealed 3D distributions of line emissions (rotational transition lines) from CO and SiO \citep{2017ApJ...842L..24A}. %
The distributions of both CO and SiO delineated by several isosurfaces seem to be rather lumpy and non-spherical. %
In particular, the distribution of CO shows a torus-like (ring-like) structure perpendicular to the equatorial ring in the nebula of SN~1987A. %
Additionally, SiO seems to be overall engulfed by CO with some exceptions (in some directions, SiO is extended further than CO) and both CO and SiO may have a hole in inner regions. %
In our previous study \citep{2020ApJ...888..111O}, the directions of the bipolar-like explosions against the equatorial ring were derived to be consistent with the observed [Fe II] lines \citep{1990ApJ...360..257H}. %
The further evolutions obtained \citep{2020A&A...636A..22O} by the 3D MHD simulations (consistent with the derived directions above) result in dipole-like iron distributions roughly parallel to the equatorial ring in the nebula. %
In \cite{2020A&A...636A..22O}, as a reference, the distribution of CO (SiO) was naively estimated from the simulation results to be the square root of the product of the (mass) densities of carbon and oxygen (silicon and oxygen). %
In the estimation, qualitatively, a torus-like structure of CO is somehow recognized \citep[see, Fig.~10 in][]{2020A&A...636A..22O} and it is also roughly perpendicular to the equatorial ring as seen in the ALMA observations. %

In this paper, the molecule formation in the representative directions (the bipolar-like explosion directions and one perpendicular to the former) is calculated. %
For an implication to the 3D distributions, here, the calculated distributions of CO and SiO in these directions are a bit more discussed. %
In Figure~\ref{fig:3d_dist}, the radial distributions of the number densities of CO and SiO (at the end of the calculation) in the $+Z$ (top panels), $-Z$ (middle panels), and $+Y$ (bottom panels) directions for both the progenitor models, i.e., the ones for the models, b18.3-zp, b18.3-zn, b18.3-yp, n16.3-zp, n16.3-zn, and n16.3-yp, are shown. %
For reference, similarly to the previous study \citep{2020A&A...636A..22O}, the distributions of the approximated number densities of CO and SiO, i.e., $\sqrt{N({\rm C}) \times N({\rm O})}$ (C $\times$ O) and $\sqrt{N({\rm Si}) \times N({\rm O})}$ (Si $\times$ O), are also plotted, where $N({\rm C})$, $N({\rm O})$, and $N({\rm Si})$ are the number densities of the seed carbon, oxygen, and silicon atoms for the case with no chemical evolution.

As can be seen, in all three directions, CO is more extended to outer regions than SiO for both the progenitor models, if the positions of the same number densities are compared. %
Compared with the approximated distributions, the distribution of CO in outer regions is roughly consistent with the approximation (C $\times$ O). %
On the other hand, the number density of SiO in the outer regions is quantitatively small compared with the approximation (Si $\times$ O) and qualitatively different from the approximation. %
The outer tails of the approximated SiO seen in some models, e.g., in the model n16.3-yp (see, $\gtrsim$ 2 $\times$ 10$^{17}$ cm in the bottom right panel), can not be seen in the calculation. %
The most distinct difference between the calculation results and the approximations is that the approximations significantly overestimate the number densities of inner regions in the $+Z$ and $-Z$ directions; %
the calculated distributions in the inner regions have complicated structures. %
The difference is at least partly attributed to the decay of $^{56}$Ni through the delay of the molecule formation by the heating, destructive processes by Compton electrons and UV photons, and the sequences induced by the decay for increasing the amounts of CO and SiO as mentioned. %

In the models based on the progenitor model b18.3 (left panels), in the $+Z$ and $-Z$ directions, at around 1 $\times$ 10$^{17}$ cm, there are some dips for both CO and SiO. %
Additionally, the position at the peak of CO is larger than that of SiO except for the innermost region. %
In the $+Y$ direction, the maximum number density is more than one order of magnitude higher than those of the two directions above and the number densities of both CO and SiO roughly decrease as the radius increases except for the innermost region. %

In the models based on the progenitor model n16.3 (right panels), in the $+Z$ and $-Z$ directions, the distributions of CO and SiO are qualitatively similar and the positions of the peaks of CO and SiO are consistent with each other. %
In the $+Y$ direction, the qualitative feature is similar to the case of b18.3. %

Since the binary merger progenitor model better explains several observational features of SN~1987A \citep{2020ApJ...888..111O,2020A&A...636A..22O}, hereafter, the models based on the progenitor model b18.3 are focused. %
According to the descriptions above, in a solid angle along the $+Z$ and $-Z$ (polar) directions, the distributions of CO and SiO may show a dipole-like (polar cap) structure and a gap (hole) in the inner regions. %
Around the perpendicular plane to the polar directions, with the gap in the polar directions and the higher number densities than those in the polar directions, the distributions may or may not look like a torus. %
For both the dipole-~and~torus-like structures, CO is overall extended to outer regions than SiO. %
Therefore, some of the observed qualitative features may be consistent with the current models. %
Such features, however, should be confirmed by the direct application to the 3D models \citep{2020ApJ...888..111O,2020A&A...636A..22O}. %
Since the distributions of CO and SiO are affected by the explosion asymmetries and the matter mixing through the distribution of $^{56}$Ni and the initial abundance rations of the seed atoms as presented, the observed 3D distribution of CO and SiO \citep{2017ApJ...842L..24A} and further observations in the future may provide insights on the explosion mechanism and/or matter mixing by comparing with such 3D theoretical models. %

\subsection{Issues of the calculation method in this study} \label{subsec:limitation}

As presented in Section~\ref{sec:results}, the evolution of the gas temperatures of the particles is an important key for the formation of molecules. %
In this study, as described in Section~\ref{sec:method}, in order to take into account the heating by the decay of $^{56}$Ni, it is assumed that some fractions of the released energies are only locally deposited depending on the optical depth for gamma rays for simplicity. %
Additionally, to control the efficiency of the gas heating, a constant parameter, $f_{\rm h}$, is introduced. %
In reality, the energy depositions, however, could occur non-locally (potential impact of non-local energy deposition is discussed in Appendix~\ref{app:non_local_edep} as mentioned; the effective non-local energy deposition model shows that it may change the amounts of molecules by at most a few tens $\%$), and the deposited energies could be consumed by not only gas heating but also ionization and excitation \citep[e.g.,][]{1995ApJ...454..472L}; %
the fraction of energy for each process (gas heating or ionization or excitation) depends on the ionization fractions \citep{1995ApJ...454..472L}. %
Therefore, the efficiency of the heating (the fraction of the deposited energies to be used for the heating), $f_{\rm h}$, is not generally constant. %
In order to obtain a more realistic temperature evolution model, the energy depositions via the decay of $^{56}$Ni should be treated by taking into account the ionization, excitation, and gas heating simultaneously with line transitions (emissions) from ionized atomic species as another cooling source, which is not taken into account in this study, as calculated e.g., in the reference \citep{1998ApJ...496..946K}. %
Recently, \cite{2023MNRAS.523..954V} reported models for spectral lines in 3D based on 3D hydrodynamical simulations incuding those effects above. %

As for the CO ro-vibrational transitions, it is difficult to reproduce the features of the observed light curves of the fluxes of the fundamental and first overtone bands \citep{1989MNRAS.238..193M,1993MNRAS.261..535M,1993A&A...273..451B,1993ApJS...88..477W}, i.e, the dome-like shapes with the peaks at around 200--250 days. %
First, the calculations in this study tend to result in fluxes higher than the observed peak fluxes before the timings of the observed peaks. %
Then, an arbitrary reduction factor, $f_{\rm red}$, to the escape probability is introduced expecting some uncertainty in the expression of the line optical depth. %
There is, however, no concrete physical basis to validate such an arbitrary factor. %
Second, in all the calculated models, the fluxes at around 200 days are apparently underestimated compared with the observations, which means that the excitations of CO vibrational levels at around the observed peaks may be insufficient. %
As mentioned, naively we would expect that supra-thermal electrons originating from the decay of $^{56}$Ni may play a role in the excitations. %
Actually, the rough coincidence between the timing of the peaks of the CO vibrational bands and the peaks of the gas temperatures of the particles heated by the decay of $^{56}$Ni may support the expectation. %
As mentioned in \cite{1995ApJ...454..472L}, the optical depth of the gamma rays from the decay of $^{56}$Co could be a key to reproducing the observed peaks. %
Additionally, in the Sobolev approximation adopted in this paper, the line optical depths (not for the gamma rays but for the CO ro-vibrational lines) are locally determined, and the trapped photons controlled by the escape probabilities, which depend on the optical depths, can only locally be absorbed. %
If the Sobolev approximation is broken, emitted photons might be trapped non-locally, and the emergence of the photons might be delayed, which may also play a role in the light curves. %
It would be necessary to take into account the energy distribution of electrons in the excitation processes of CO vibrational levels and the potential non-local effects mentioned above. %
Once a reasonable model for the CO ro-vibrational transitions is obtained, the time-dependent spectra of the CO ro-vibrational line emissions can be estimated in the future. %

The balance of the heating and cooling is important for the evolution of the gas temperatures and eventually the formation of molecules. %
Some improvements in the treatments on the energy depositions from the decay of $^{56}$Ni and the CO ro-vibrational transitions are necessary to obtain more realistic thermal histories of the ejecta particles in the future. %

It is noted that in this study, the constant parameters, $f_{\rm h}$, $f_{\rm d}$, and $t_{\rm s}$, are introduced, and the values are somehow calibrated by comparing with the observed fluxes of CO vibrational bands and the estimations in the previous studies. %
The reasonable parameter values are different between one-zone models and 1D calculations with the angle-averaged profiles based on the 3D model (b18.3-high). %
Considering the application to the 3D models without angle-averaging in the future, the selection of such parameters could matter because it is difficult to perform wide-range parameter surveys due to the numerical cost. %
The selection of the different parameter values may partly be to compensate for the different degrees of matter mixing (practically, the mixing of $^{56}$Ni) as mentioned. %
The degree of matter mixing in one-zone models would be higher than those in 1D models because of the uniform composition in one-zone models. %
Since in reality (in the 3D models), the degree of matter mixing may be milder than the angle-averaged 1D profiles, the reasonable parameter values for the direct application to the 3D models would be the extrapolation between those in the one-zone and 1D models or the boundary value (maybe for $f_{\rm d}$). %
In the 3D models, the matter mixing is probably more realistic than that in the one-zone and 1D models. %
Those parameters may be determined as more reasonable values in the 3D application models. %

However, such parameters could depend on the time and region because of the different physical conditions in reality. %
For the quantitative comparison of observations, reducing uncertainties, and disentangling degeneracies of the problem, again, more realistic treatments for the issues above, e.g., are indispensable in the future. %

\subsection{Implication to the observations by JWST} \label{subsec:jwst}

JWST\,\footnote{\texttt{https://webb.nasa.gov/}} (James Webb Space Telescope) is an infrared observatory covering wavelengths of 0.6--28.5 
$\mu$m\,\footnote{More specifically, the covering wavelengths for the four JWST instruments are as follows. NIRCam: 0.6--5 $\mu$m; NIRSpec: 0.7--5 $\mu$m; MIRI: 5--28.5 $\mu$m; FGS/NIRISS: 0.6--5 $\mu$m.} launched in December 2021. %
The fundamental ($\sim$ 4.6 $\mu$m) and first overtone ($\sim$ 2.3 $\mu$m) bands of CO vibrational transitions and the fundamental ($\sim$ 8 $\mu$m) and first overtone ($\sim$ 4 $\mu$m) bands of SiO can potentially be covered by the telescope. %
Actually, there has been an attempt for theoretical modeling \citep{2023A&A...674A.184L} of the spectra of the ro-vibrational line emissions of not only the CO and SiO but also SO and SiS for type Ib/c SNe (at a typical distance of 10 Mpc); %
it is concluded that most molecular emissions can be observed with a good signal-to-noise ratio in the 100-400 days time window. %
Unfortunately, such ro-vibrational transitions can not be expected for SN~1987A at the current age but, for general type II and type Ib/c SNe, spectral modelings of the ro-vibrational line emissions would be useful for comparison with future JWST observations, which will shed light on the formation of molecules in the ejecta of CCSNe by comparing theoretical models. %

The spatial resolution of JWST is as small as $\sim 0.1$ arcsec. %
Therefore, nearby supernova remnants, e.g., SN~1987A (distance: 51.4 kpc) and Cas A (distance: $\sim$ 3.4 kpc), can spatially be resolved. %
Actually, as mentioned in Section~\ref{sec:intro}, there has been an approved proposal for Cycle 1 JWST observations of SN~1987A \citep{2021jwst.prop.1726M}; %
the observations will spatially resolve hot dust grains and may figure out how dust is destructed by the blast wave and the reverse shock. %
There has also been an approved proposal for observations of Cas A by JWST \citep{2021jwst.prop.1947M} for investigating the properties of dust, i.e., the dust survival against the passage of the reverse shock, grain size distribution, and clump size and so on. %
It is noted that recently, the results of the first observations of SN~1987A by JWST have been reported \citep{2023ApJ...949L..27L} as mentioned in Section~\ref{sec:intro}; %
the 3D distribution of the ejecta delineated by [Fe I] shows a broken dipole-like structure, which is roughly parallel to the equatorial ring. %
The derived 3D morphology of the ejecta and the orientation seem to be in good agreement with those of our 3D models \citep{2020ApJ...888..111O,2020A&A...636A..22O}, which may further support our 3D models. %

Moreover, in the case of SN~1987A, there have also been observations of spatially resolved dust in the ejecta of SN~1987A by ALMA \citep{2019ApJ...886...51C}. %
Therefore, in light of such spatially resolved observations of SN~1987A and Cas A, it is indispensable to make models of the formation and destruction of dust based on 3D hydrodynamical models to be compared with JWST observations. %
As for SN~1987A, we plan to apply our methodology in this paper to the 3D hydrodynamical models \citep{2020ApJ...888..111O,2020A&A...636A..22O} more directly. %
Furthermore, we also plan to connect our molecule formation calculations in the ejecta to dust formation and destruction theories \citep[e.g.,][]{2010ApJ...713..356N,2013ApJ...776...24N} in the near future. %
As for Cas A, such formation and destruction models of molecules and dust can be applied to the recent 3D hydrodynamical models for Cas A \citep{2021A&A...645A..66O,2022A&A...666A...2O} in the future. %

\subsection{A caveat: possible impacts of realistic 3D explosion models} \label{subsec:caution}

In this study, the impact of matter mixing on molecule formation in CCSN ejecta was discussed based on the 3D bipolar-like explosion models for SN 1987A \citep{2020ApJ...888..111O}, in which the explosions are initiated with the rather simplified model, i.e., instantaneous asymmetric kinetic and thermal energy injections, as mentioned. %
Then, based on the obtained matter mixing, the impact of effective matter mixing was investigated by using the one-zone models, the angle-averaged 1D profiles, and the extracted radial profiles in the specific directions from the 3D models.  %
However, in reality, CCSN explosions should be driven via complicated physical processes as observed in the theories of CCSN explosion mechanisms, e.g., neutrino-driven and magnetorotationally-driven explosions (see, Section~\ref{sec:intro}). %
The 3D models adopted in this study may fail to capture some features obtained by more realistic 3D CCSN explosion models \citep[e.g.,][for neutrino-driven and magnetorotationally-driven mechanisms, respectively]{2019MNRAS.482..351V,2014ApJ...785L..29M}. %
Therefore, such more realistic 3D CCSN explosion models may result in different matter mixing, i.e., distribution of the seed atoms and $^{56}$Ni and the abundance ratios, from the adopted 3D models and potentially change the results on molecule formation presented in this paper in particular ones for the specific directions in the 3D models (Section~\ref{subsubsec:1d_angle}) and even for the one-zone and 1D models with the angle-averaged 1D profiles (Section~\ref{subsec:one_zone_results} and Sections~\ref{subsubsec:1d_param}, \ref{subsubsec:1d_single_star}, and \ref{subsubsec:1d_comp_sphel}, respectively). %
It is worth investigating the impact of more realistic 3D CCSN explosion models on the formation of molecules in the ejecta in the future.

\section{Summary} \label{sec:summary}

In this paper, the molecule formation in the ejecta of CCSNe is calculated by using a chemical reaction network based on one-zone and 1D ejecta evolution models based on the two 3D hydrodynamical models for SN~1987A \citep[b18.3-high and n16.3-high:][]{2020ApJ...888..111O}, i.e., asymmetric bipolar-like explosions with the binary merger \citep[b18.3:][]{2018MNRAS.473L.101U} and single-star \citep[n16.3:][]{1988PhR...163...13N,1990ApJ...360..242S} progenitor models. %
In Section~\ref{subsec:one_zone_results}, with one-zone models, the impact of the parameters related to the heating and cooling of gas is investigated (Section~\ref{subsubsec:one_zone_trend}). %
Then, with a reasonable set of the parameters, important chemical reactions for the formation and destruction of CO and SiO molecules are described (Section~\ref{subsubsec:chemi_reac}). %
Similar to the one-zone models, in Section~\ref{subsec:1d_results}, the impact of the parameters is investigated for 1D models and a set of parameters is derived as a fiducial case (Section~\ref{subsubsec:1d_param}). %
With the parameters of the fiducial case, by comparing the model results among the 1D models with the initial profiles obtained by the angle-averaging of the bipolar-like explosion models and ones for the spherical explosion case (the explosions are spherical but the evolution is 3D), and ones for purely spherical case, the impact of matter mixing on the molecule formation in the ejecta is effectively investigated (Section~\ref{subsubsec:1d_comp_sphel}). %
Finally, from the model results of the calculations for specified directions in the 3D hydrodynamical models ($+Z$, $-Z$, and $+Y$ directions: the positive and negative directions of the bipolar-like explosion axis and a perpendicular direction to the former two), the dependence of the molecule formation on the directions is examined (Section~\ref{subsubsec:1d_angle}). %
The main points and findings in this paper are summarized with some compliments as follows. %

\begin{enumerate}
\item 
The angle-averaged 1D profiles (Figure~\ref{fig:prof_mean}) derived from the 3D models of the asymmetric bipolar-like explosions \citep[b18.3-high and n16.3-high:][]{2020ApJ...888..111O} effectively reflect the mixing of $^{56}$Ni into outer layers and the mixing of seed atoms. %
In the angle-averaged 1D profiles, $^{56}$Ni is rather extended to outer layers compared with the spherical explosion (but the evolution is in 3D) case (Figure~\ref{fig:prof_sphel}) and purely spherical case (Figure~\ref{fig:prof_sphel_pure}). %
In the spherical explosion case, $^{56}$Ni is more concentrated in inner regions and the seed atoms are smoothly distributed. %
In the purely spherical case, $^{56}$Ni is confined only in the innermost regions and the seed atoms are rather stratified. %
The one-zone model derived from the radial averaging of the angle-averaged 1D profiles for the model b18.3-high represents the case that the mixing is further efficient compared with the angle-averaged 1D profiles because of the uniform composition. %

\item 
The decay of radioactive $^{56}$Ni and/or $^{56}$Co (hereafter, simply, the decay of $^{56}$Ni as before this section) could heat the gas and increase the gas temperature. 
If the efficiency of the gas heating by the decay of $^{56}$Ni, i.e., $f_{\rm h}$, is higher, the timing when the gas temperature goes down to $\sim$ 10$^{4}$ K, at which molecules start to form, is more delayed compared with lower $f_{\rm h}$ cases. %
Compared with lower $f_{\rm h}$ cases, molecules start to form in a low-density environment, which generally reduces the amounts of molecules, e.g., CO and SiO. %

\item 
The decay of $^{56}$Ni produces Compton electrons which could ionize and destruct molecules through the processes described in Equations~(\ref{eq:ion_ab}) and (\ref{eq:destruction}). %
The secondary UV photons due to the decay could also dissociate molecules. %
If the destruction efficiency, $f_{\rm d}$, is higher, the ionization of molecules by Compton electrons and the destruction of those by Compton electrons and/or UV photons become more effective; %
the higher $f_{\rm d}$, generally, the more effective the destruction of CO and SiO at around a few hundred days after the explosion (however, under a certain condition, ``formation" processes could locally dominate the destructive ones instead. See, the seventeenth point below). %

\item 
In the calculations in this study, fluxes of CO vibrational bands tend to become higher than the observed peak fluxes (at 200--250 days) at an early phase before 100 days. %
In order to avoid such early high fluxes, an arbitrary reduction factor, i.e., $f_{\rm red}$, had to be introduced. %
Even though, it is difficult to reproduce the observed features of the CO bands, i.e., the peaks at 200-250 days and the dome-like shapes. %
At around the observed peaks, additional excitations by supra-thermal electrons produced by the decay of $^{56}$Ni might play a role in reality. %

\item 
According to the fiducial one-zone model (see, the sixth point below), initially, CO and SiO are primarily formed by the \texttt{RA} processes in Equations~(\ref{eq:co_rad}) and (\ref{eq:sio_rad}), respectively, until a few tens days; %
after that many \texttt{NN} reactions are involved with both the formation and destruction processes. %
If the atomic ions are produced by the ionization due to Compton electrons in Equation~(\ref{eq:ion_x}), the \texttt{IN} reactions in Equations~(\ref{eq:c+_sio}), (\ref{eq:he+_co}), (\ref{eq:he+_sio_si+}), and (\ref{eq:he+_sio_o+}) (the first is for CO formation and SiO destruction, the second is for the destruction of CO, and the third and fourth are the destruction of SiO) could play an important role in the formation and/or destruction of CO and SiO. %
For contributing reactions for the formation and destruction of CO and SiO in the fiducial one-zone model, see, Figure~\ref{fig:co_reac} and Figure~\ref{fig:sio_reac}, respectively. %
The contributing reactions, however, generally depend on the thermal history and the initial composition; %
actually, some differences between the fiducial one-zone and 1D calculations can be recognized (see, the seventh point). %
The previous studies on the formation of CO and SiO in SN~1987A \citep{1990ApJ...358..262L,1992ApJ...396..679L,1994ApJ...428..769L,1995ApJ...454..472L,1996ApJ...471..480L,2018MNRAS.480.5580S,2020A&A...642A.135L} are useful for reference. %
actually, some of the chemical reactions described in this paper have also been mentioned in the studies above. %
For chemical reactions in general/other types of supernovae, e.g., the reference \citep{2009ApJ...703..642C} is useful. %

\item 
To be somehow consistent with the estimates of the amounts of CO and SiO derived from the previous studies \citep{1992ApJ...396..679L,1994ApJ...428..769L,1995ApJ...454..472L,2017MNRAS.469.3347M,2020A&A...642A.135L} and the light curves of the observed CO vibrational bands \citep{1989MNRAS.238..193M,1993MNRAS.261..535M,1993A&A...273..451B,1993ApJS...88..477W}, for the one-zone and 1D models, the parameter sets of ($f_{\rm h}$, $f_{\rm d}$, $t_{\rm s}$) = (5 $\times$ 10$^{-4}$, 10$^{-2}$, 200 days) and (5 $\times$ 10$^{-3}$, 1.0, 500 days) might be selected as the reasonable sets, respectively (for the former set, see, the middle panels in Figure~\ref{fig:one_zone_param1}; for the latter, see, e.g., the lower panels in Figure~\ref{fig:1d_param1} and the top panels in Figure~\ref{fig:1d_mass_temp_binary}). %
It is interpreted that in the one-zone models, lower values of the efficiencies of the gas heating, the ionization and destruction by Compton electrons, and the dissociation by UV photons may be preferred to compensate for the more effective matter mixing compared with the 1D models. %

\item
Contributing chemical reactions for the formation of CO and SiO in the fiducial 1D model (see, the sixth point), are not significantly different from those in the fiducial one-zone model; %
some of the primary processes, and the relative significances could be different between the two models depending on time (see, Figures~\ref{fig:co_reac}, \ref{fig:sio_reac}, \ref{fig:co_reac_b18.3-mean}, and \ref{fig:sio_reac_b18.3-mean}). %
The less effective mixing of $^{56}$Ni, hydrogen, and helium in the fiducial 1D model could decrease the contributions of the reactions involved with ionized hydrogen and helium, i.e., the \texttt{CE} reactions in Equations~(\ref{eq:c+_sio}) and (\ref{eq:h+_sio}) and the \texttt{IN} reactions in Equations~(\ref{eq:he+_co}), (\ref{eq:he+_sio_o+}), and (\ref{eq:he+_sio_si+}) compared with the fiducial one-zone model. %

\item 
The abundance ratios of the initial seed atoms are different (see, e.g., Figure~\ref{fig:prof_mean} and the top left panels in Figures~\ref{fig:1d_mass_temp_binary} and \ref{fig:1d_mass_temp_single}) between the models based on the binary merger progenitor model (b18.3) and the single-star progenitor model (n16.3); %
in the models based on n16.3, the abundance ratios of oxygen to carbon and silicon are apparently higher compared with those in the models based on b18.3; %
in the models based on n16.3, the abundance ratio of carbon to silicon is also higher than those in the models based on b18.3. %
Such different initial abundance ratios of the seed atoms could affect the consequential amounts of molecules. %
Actually, in the models based on n16.3, the amount of O$_2$ is higher than that in the models based on b18.3. %

\item
MgO and FeO are primarily formed by the \texttt{NN} reactions of magnesium and iron with O$_2$ initially and by the \texttt{3B} reactions, Mg + O + H $\lra$ MgO + H and Fe + O + H $\lra$ FeO + H, at later phases, respectively. %
In the models based on the single-star progenitor model (n16.3), the amount of MgO is higher than that of FeO. %
On the other hand, the amount of FeO dominates that of MgO at the final phase in the models based on the binary merger progenitor model (b18.3). %
This feature is attributed to the initial high magnesium abundance in the models based on n16.3. %

\item 
The formation of N$_2$ is distinctly recognized regardless of the models. Moreover, the amounts of N$_2$ in the models based on the binary merger progenitor model (b18.3) are overall higher than those in the models based on the single-star progenitor model (n16.3). %
This feature is probably attributed to the higher initial abundance of nitrogen in the models with b18.3, in particular, in the envelope (see, e.g., Figure~\ref{fig:prof_mean}); %
in the binary merger progenitor model, the CNO cycle additionally triggered during the merger process enhances the nitrogen ($^{14}$N: one of the main products of the CNO cycle) abundance in the envelope via mixing. %

\item 
The fluxes of CO vibrational bands can be affected by the mixing of $^{56}$Ni. %
In higher mixing efficiency models (the efficiencies are higher in the asymmetric explosion models, spherical explosion models, and purely spherical models, in this order; %
see the first point), the ionization by Compton electrons due to the decay of $^{56}$Ni increases the number density of thermal electrons to excite CO vibrational levels. %
Actually, it is found that the calculated fluxes of CO vibrational bands are overall higher in high mixing efficiency models compared with lower mixing efficiency (more spherical) models. %

\item 
In the purely spherical models (the models b18.3-sphel-pure and n16.3-sphel-pure), due to the rather stratified distributions of the initial seed atoms, the formation of some molecules that are not markedly formed in the other models, e.g., C$_2$, CS, SiC, SiN, are recognized. %
In particular, in both the models b18.3-sphel-pure and n16.3-sphel-pure, the amounts of C$_2$ are significantly higher than those in the other (non-spherical) models and become over 10$^{-2}$ $M_{\odot}$. %
In the model n16.3-sphel-pure, further due to the high initial oxygen abundance, O$_2$ is significantly formed and the amount is the highest among the molecules and reaches $\sim$ 1 $M_{\odot}$. %

\item 
Depending on the direction, the distributions of the seed atoms and $^{56}$Ni are different in the last snapshots of the 3D hydrodynamical models (b18.3-high and n16.3-high). %
In particular, the explosions are bipolar-like and the distributions are distinctively different between the directions along the bipolar axis ($+Z$ and $-Z$) and a direction ($+Y$) perpendicular to the axis (in the equatorial plane). %
In the bipolar axis, $^{56}$Ni is extensively distributed to outer layers (see, Figures~\ref{fig:b18.3_angle} and \ref{fig:n16.3_angle}). %
Then, in the bipolar axis, the gas temperatures of the inner particles heated by the decay of $^{56}$Ni increase making peaks at around 200 days; %
such gas heating delays the start of the formation of molecules. %
On the other hand, in the equatorial plane, $^{56}$Ni is concentrated only in the innermost region; %
the gas temperatures of most of the particles go down to $\sim$ 100 K within a few hundred days. %

\item 
In the directions along the bipolar axis ($+Z$ and $-Z$ directions) for both the progenitor models (b18.3 and n16.3), some fractions of CO and, in particular, SiO are destructed approximately from 100 days to a few hundred days. %
At this phase, The main destruction processes of CO are the ionization by Compton electrons in Equation~(\ref{eq:ion_ab}), the destruction by UV photons and Compton electrons in Equation~(\ref{eq:destruction}). %
SiO is mainly destructed by the \texttt{CE} reaction in Equation~(\ref{eq:h+_sio}) and the \texttt{NN} reactions in Equations~(\ref{eq:h_sio_sih}) and (\ref{eq:h_sio_oh}). %
On the other hand, in the equatorial plane, there is no distinct destruction by processes related to Compton electrons. %

\item 
In the $+Z$ and $-Z$ directions, FeO increases even after a few thousand days by the participation of the inner particles heated by the decay of $^{56}$Ni, which contain plenty of seed iron, to the formation of FeO by a \texttt{3B} reaction. %
Probably by reflecting the initial abundance ratios of the seed atoms, in the models based on the progenitor model n16.3, the amount of MgO dominates FeO in all three directions. %
On the other hand, in the models based on the binary merger progenitor model b18.3, in the $+Z$ and $-Z$ directions, the amount of FeO dominates MgO; %
in the $+Y$ direction, MgO dominates FeO. %

\item 
The formation of molecules depends on the radial position. Molecules start to form from the outer regions, since from these regions, the gas temperatures go down to $\sim$ 10$^4$ K. %
At 200 days, in the $+Z$ and $-Z$ directions, the formation of molecules is only recognized (from the outer regions) up to the outer part of the regions where initially, the seed atoms are abundant (see, e.g., the top and middle left panels in Figures~\ref{fig:b18.3_angle} and \ref{fig:1d_b18.3_prof}). %
On the other hand, in the $+Y$ direction, at 200 days, molecules such as CO and SiO are already remarkably formed even in the inner regions. %

\item 
The decay of $^{56}$Ni is generally effective in the ionization and destruction of molecules (in decreasing molecules) as mentioned at the third point above. %
It is, however, found that under a certain condition, if the start of the molecule formation lags behind the activation of the decay, the sequence of the formation processes of CO and SiO described below could locally dominate the destructive reactions due to the decay of $^{56}$Ni (see, the third point above); %
the ionization of atoms by Compton electrons in Equation~(\ref{eq:ion_x}) increases the number density of thermal electrons; H$^{-}$ is produced by the \texttt{REA} reaction, H + e$^-$ $\lra$ H$^-$ + $\gamma$; H$_2$ is formed by the \texttt{AD} reaction, H$^-$ + H $\lra$ H$_2$ + e$^-$; OH is formed by the \texttt{NN} reaction, O + H$_2$ $\lra$ OH + H; %
CO and SiO are eventually formed by the \texttt{NN} reactions in Equations~(\ref{eq:c_oh}) and (\ref{eq:si_oh}), respectively. %

\item 
In the $+Z$ and $-Z$ directions, even at the end of the calculations ($\sim$ 10000 days), in the inner regions, the formation of molecules is limited compared with those in the $+Y$ direction, and the molecules, e.g., CO and SiO, are not so smoothly distributed (see, e.g., the top right panels in Figure~\ref{fig:1d_b18.3_prof} and \ref{fig:1d_n16.3_prof}). %
The latter points may be partly attributed to the complicated thermal histories affected by both the heating due to the decay of $^{56}$Ni and the cooling via CO ro-vibrational transitions. %
Another factor for the feature above is that there may be a transition point (radial position) inside which the decay of $^{56}$Ni increases the amounts of CO and SiO (see, the third and seventeenth points); %
outside the point, it decreases the amounts by destructive processes related to Compton electrons. %

\item 
The survival of some of the molecules, e.g., O$_2$ and SO seem to sensitively be determined by the balance of complicated \texttt{NN} reactions depending on the thermal histories unless the seed atoms for these molecules are sufficient. %
Actually, the distributions of O$_2$ are rather complicated in the $+Z$ and $-Z$ directions as seen in the models based on the single-star progenitor model (see, the top and middle left panels in Figure~\ref{fig:1d_n16.3_prof}). %

\item 
From the results of the models based on the binary merger progenitor model b18.3, the calculated distributions of CO and SiO may partly be consistent with the observed 3D features obtained by ALMA \citep{2017ApJ...842L..24A}, e.g., a torus-like structure of CO and holes for both CO and SiO. %
Such features, however, should be confirmed by the direct application to the 3D models \citep{2020ApJ...888..111O,2020A&A...636A..22O}. %


\end{enumerate}

In this paper, we limit the application of our methodology only to the one-zone and 1D models. %
We plan to apply it to the 3D hydrodynamical models for SN~1987A \citep{2020ApJ...888..111O,2020A&A...636A..22O} in order to more directly compare the observed 3D distributions of CO and SiO \citep{2017ApJ...842L..24A} in the near future. %
We also have the plan to connect the molecule formation calculations with theoretical models of dust formation \citep[e.g.,][]{2010ApJ...713..356N,2013ApJ...776...24N} to investigate the impact of matter mixing on the formation of dust. %
The observations of not only SN~1987A but also general CCSNe and SNRs (as, for instance, Cas A) by the recently launched JWST will shed light on the molecule and dust formation in CCSNe through the comparison with theoretical models. %
It would be better to improve our current methodology in particular on the energy depositions by the decay of $^{56}$Ni and the excitation of CO ro-vibrational levels to obtain more reliable thermal histories of the ejecta and the CO line emission to be compared with observations in the future. %

As for the adopted chemical reactions, uncertainties of the rate coefficients may not be so small \cite[for a sensitivity study of chemical reaction rates to the formation of CO, see,][]{2020A&A...642A.135L} as can be partly seen in Table~\ref{table:reaction} that some of the rate coefficients have exactly the same values. %
Therefore, it would be better to use more reliable rate coefficients as much as possible, if available in the future. %
Additionally, there may be potentially important chemical reactions that are not included in this study (partly due to the limitation of the adopted species) depending on the local initial abundance. %
Related to this issue and dust formation, theoretically, there have been two approaches to the formation of clusters of dust (nucleation) in CCSNe, i.e., in one approach, dust formation is calculated with a chemical reaction network extended to clusters \citep{2013ApJ...776..107S,2015A&A...575A..95S} and in another approach, it is followed by a model of kinematic theories \citep{2010ApJ...713..356N,2013ApJ...776...24N}; %
the differences between the two approaches may affect the consequential dust formation. It would be worth seeing the impact of the differences on dust formation in the two approaches in the future. %

\acknowledgments

We acknowledge the anonymous referee for careful reading of the paper and for providing many valuable comments. %
The software used in this work was in part developed by the DOE NNSA-ASC OASCR Flash Center at the University of Chicago. The numerical computations were carried out complementarily on XC40 (YITP, Kyoto University), Cray XC50 (Center for Computational Astrophysics, National Astronomical Observatory of Japan), HOKUSAI (RIKEN). %
This work is supported by JSPS KAKENHI Grant Number 19H00693 and JP21K03545. %
MO and SN thank the support from RIKEN Interdisciplinary Theoretical and Mathematical Sciences Program. %
MO and SN also acknowledge the support from the Pioneering Program of RIKEN for Evolution of Matter in the Universe (r-EMU). %
MO and KC thank the support from the Ministry of Science and Technology, Taiwan, under grant nos. MOST 110-2112-M-001-068-MY3 and MOST 111-2112-M-001-038-MY3 and the Academia Sinica, Taiwan, under grant nos. AS-CDA-111-M04 and AS-IA-112-M04. %
SO and MM acknowledge the financial contribution from the PRIN INAF 2019 grant ``From massive stars to supernovae and supernova remnants: driving mass, energy and cosmic rays in our Galaxy". %

\appendix

\section{Chemical reaction rates} \label{app:reac}

In this appendix, the adopted chemical reactions that are expressed by the Arrhenius form as in Equation~(\ref{eq:arrhenius}), the values of the rate coefficients, and the references of the values are listed in Table~\ref{table:reaction}. %
As mentioned in the text, other than the reactions listed in Table~\ref{table:reaction}, the thermal fragmentation reactions in Equation~(\ref{eq:di}) in Section~\ref{subsubsec:thermal_frag}, the ionization and destruction by Compton electrons described in Equations~(\ref{eq:ion_x}), (\ref{eq:ion_ab}), and (\ref{eq:destruction}), and the dissociation by UV photons in Section~\ref{subsubsec:compton} are also taken into account in this paper. %

{\startlongtable
\begin{deluxetable*}{llrccrr}
\tabletypesize{\scriptsize}
\tablewidth{0pt}
\tablenum{A1}
\tablecolumns{7}
\tablecaption{Adopted rate coefficients of the included chemical reactions other than thermal fragmentation 
reactions and ones involved with Compton electrons and UV photons. The numbers in the parentheses denote the powers 
of ten. ``CD09", ``NIST", ``SL18", ``UM12", ``UM06", ``CD10" in the reference denote \cite{2009ApJ...703..642C}, 
NIST Chemical Kinetic Database (\texttt{https://kinetics.nist.gov/kinetics/index.jsp}), 
\cite{2018MNRAS.480.5580S}, UMIST database for astrochemistry 2012 \citep{2013...550A..36M}, 
UMIST database for astrochemistry 2006 \citep{2007...466.1197W}, \cite{2010ApJ...713....1C}, respectively. 
The species ``M" in the \texttt{3B} reactions denotes arbitrary species.
\label{table:reaction}}
\tablehead
{
\\ 
Type\tnm{a} \ \  & Reaction \hs{3} & $A_i$\tnm{b} & $\alpha_i$\tnm{b} & $\beta_i$\tnm{b} & Reference & Note \\
 & & (cgs) & & (K) & & 
}
\startdata
\texttt{3B} & $ {\rm H} + {\rm H} + {\rm M} \lra {\rm H}_2 + {\rm M} $ & 6.84 (-33) & -1 & 0 & CD09; NIST &  \\
\texttt{3B} & $ {\rm H} + {\rm C} + {\rm M} \lra {\rm CH} + {\rm M} $ &     1 (-33) & 0 & 0 & CD09 &  \\
\texttt{3B} & $ {\rm H} + {\rm O} + {\rm M} \lra {\rm OH} + {\rm M} $ &  4.36 (-32) & -1 & 0 & CD09; NIST &  \\
\texttt{3B} & $ {\rm Si} + {\rm H} + {\rm M} \lra {\rm SiH} + {\rm M} $ &     1 (-33) & 0 & 0 & CD09 &  \\
\texttt{3B} & $ {\rm C} + {\rm C} + {\rm M} \lra {\rm C}_2 + {\rm M} $ &  4.25 (-34) & 1 & 0 & SL18 &  \\
\texttt{3B} & $ {\rm C} + {\rm N} + {\rm M} \lra {\rm CN} + {\rm M} $ &  4.25 (-34) & 1 & 0 & SL18 &  \\
\texttt{3B} & $ {\rm C} + {\rm O} + {\rm M} \lra {\rm CO} + {\rm M} $ &  4.25 (-34) & 1 & 0 & SL18 &  \\
\texttt{3B} & $ {\rm C}^+ + {\rm O} + {\rm M} \lra {\rm CO}^+ + {\rm M} $ &  4.25 (-34) & 1 & 0 & SL18 &  \\
\enddata
\tnt{a}{Code for a reaction type listed in Table~\ref{table:types}.}
\tnt{b}{Coefficients for the Arrhenius form in Equation~(\ref{eq:arrhenius}).}
\end{deluxetable*}

\section{Potential impact of non-local energy deposition due to the radioactive decay of $^{56}$Ni} \label{app:non_local_edep}

\begin{figure*}
\begin{minipage}{0.5\hsize}
\begin{center}
\includegraphics[width=9.5cm,keepaspectratio,clip]{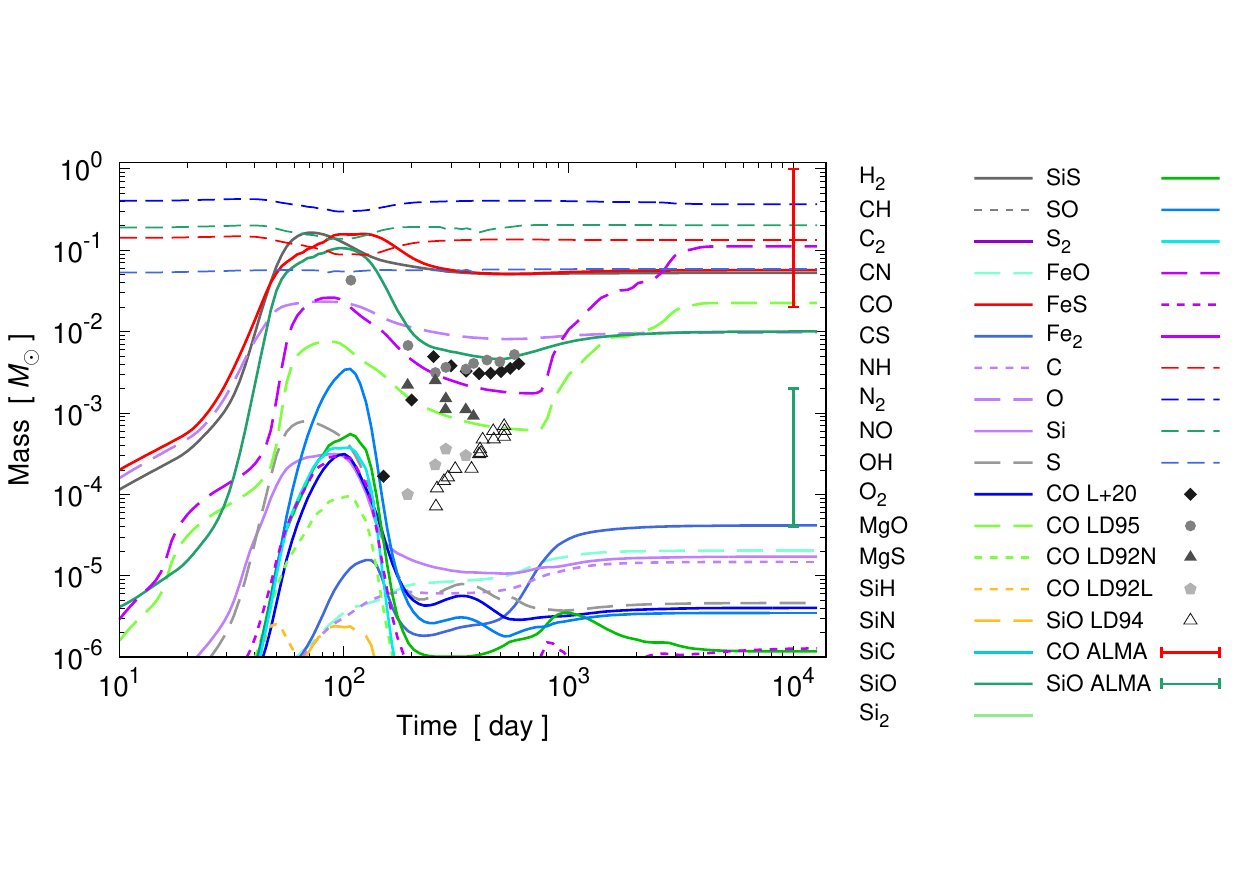}
\end{center}
\vs{-2.}
\end{minipage}
\begin{minipage}{0.5\hsize}
\begin{center}
\includegraphics[width=7.5cm,keepaspectratio,clip]{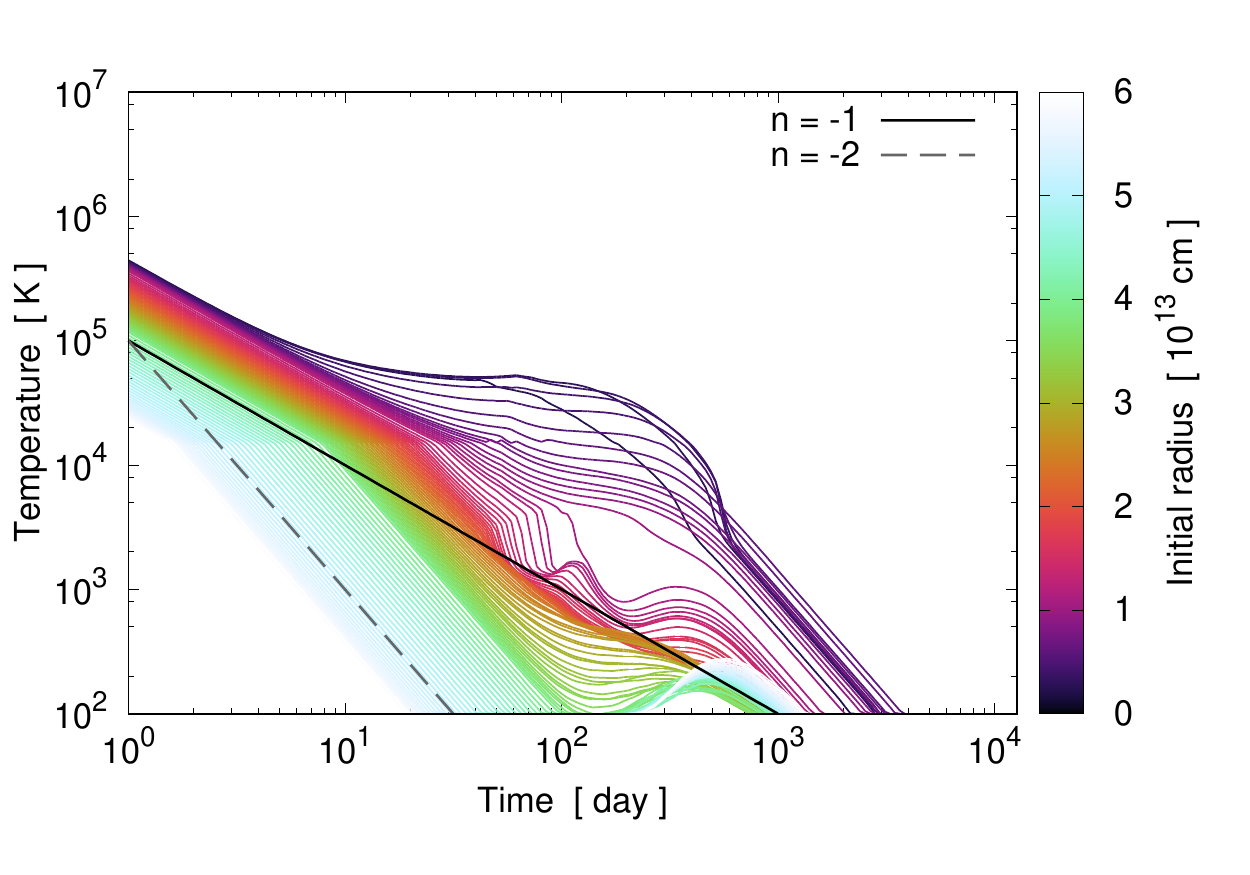}
\end{center}
\vs{-2.}
\end{minipage}
\\
\begin{minipage}{0.5\hsize}
\begin{center}
\includegraphics[width=9.5cm,keepaspectratio,clip]{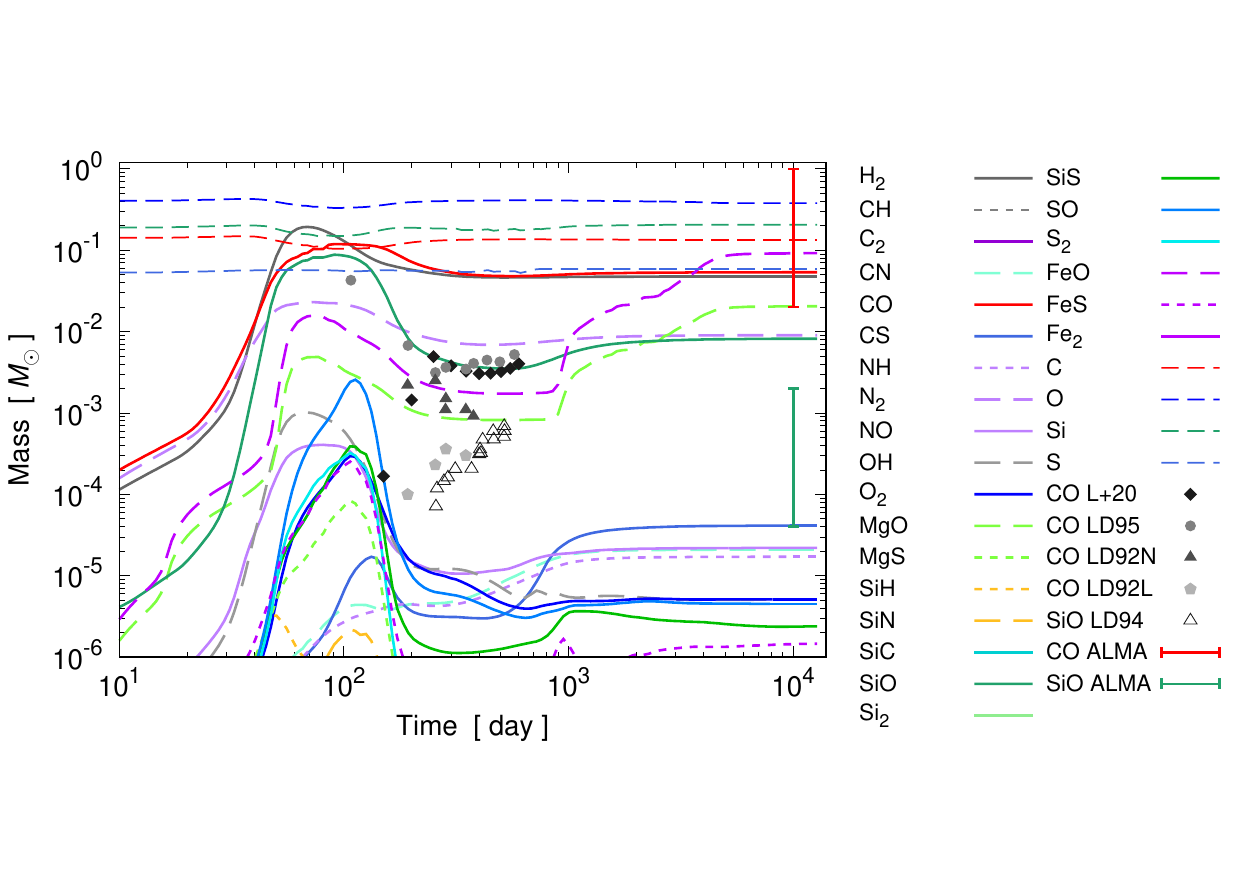}
\end{center}
\vs{-1.0}
\end{minipage}
\begin{minipage}{0.5\hsize}
\begin{center}
\includegraphics[width=7.5cm,keepaspectratio,clip]{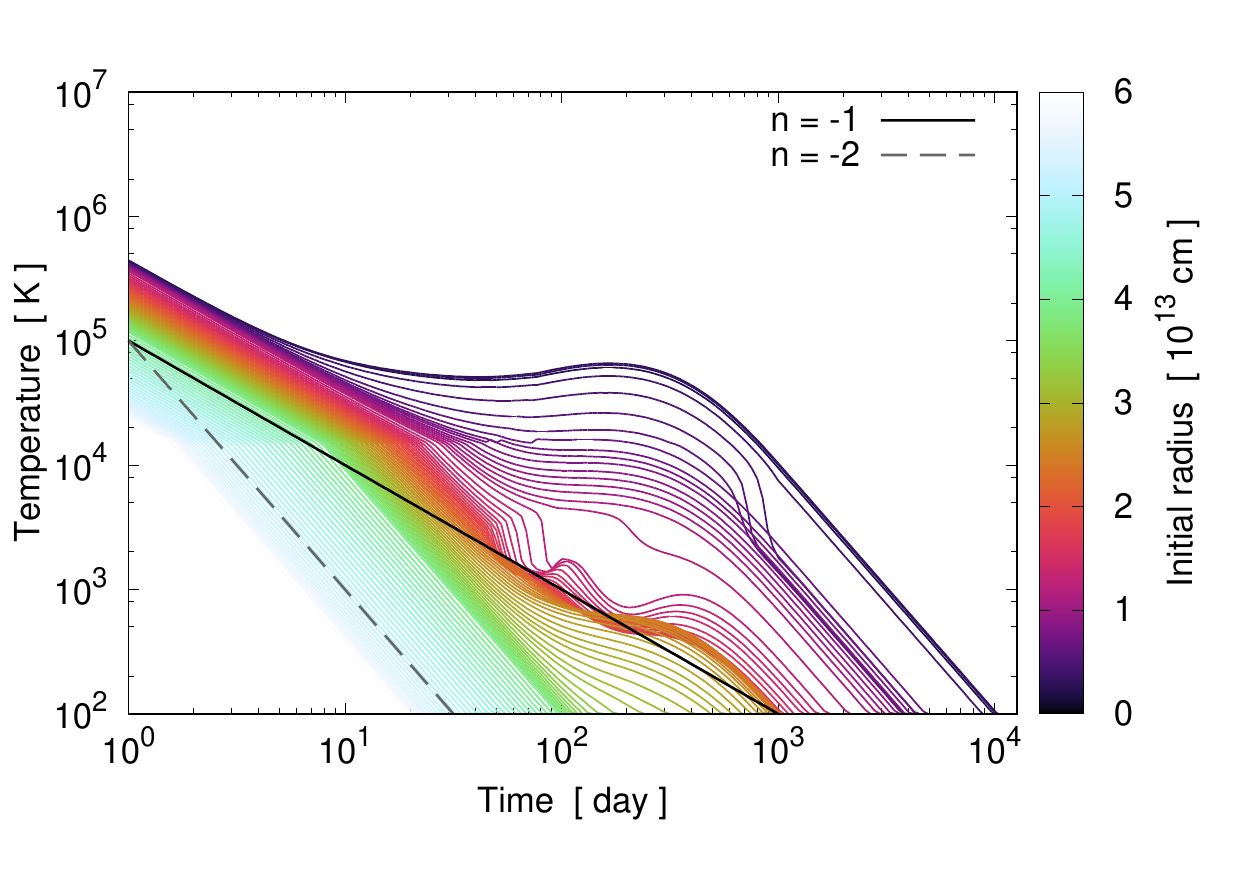}
\end{center}
\vs{-1.0}
\end{minipage}
\caption{The results of the 1D calculation with the effective energy deposition rate in Equation~(\ref{eq:qr}) with the angle-averaged profiles for the model b18.3-high \citep{2020ApJ...888..111O} with the parameters, $f_{\rm h} = 5 \times 10^{-3}$, $f_{\rm d} =$ 1.0, and $t_{\rm s} =$ 500 days (top panels: b18.3-mean-nledep) and the ones of the corresponding counterpart model without the non-local energy deposition, i.e., the model b18.3-mean (bottom panels: same as the top panels in Figure~\ref{fig:1d_mass_temp_binary}) for comparison. %
Left panels: time evolution of the total amounts of molecules and several seed atoms compared with the estimations (including theoretical calculations) for CO and SiO in previous studies: %
LD92 \citep{1992ApJ...396..679L}, LD94 \citep{1994ApJ...428..769L}, LD95 \citep{1995ApJ...454..472L}, ALMA \citep{2017MNRAS.469.3347M}, and L+20 \citep{2020A&A...642A.135L}. %
Right panels: time evolution of the gas temperatures of the tracer particles; the colors denote the initial positions of the particles. As a reference, power-law evolutions of the powers of $-1$ and $-2$ are shown on the right panel.} %
\label{fig:1d_non_local_edep}
\end{figure*}

In this section, the potential impact of ``non-local" energy deposition due to the decay of $^{56}$Ni, which is neglected in the calculations described in the main text (see, Sections~\ref{subsec:one_zone_calc} and \ref{subsec:1d_calc}), is discussed. %
As mentioned in Section~\ref{subsubsec:compton}, in reality, the emitted energies from the decay can be deposited non-locally over a region corresponding to the length scale of the effective mean free path in Equation~(\ref{eq:mean_free_gamma}). %
Here, a model of such non-local energy deposition and the results of the test calculation with the model are presented. %

Suppose the emitted energies at the radius $r'$ can be deposited at a radius $r$ ($r \geq r'$) and the energy can be deposited at $r$ from at $r'$ is proportional to $\exp\,(-(r - r')/l_{\gamma, r'})$, were $l_{\gamma, r'}$ is the mean free path in Equation~(\ref{eq:mean_free_gamma}) measured at the radius $r'$. %
It is assumed that non-local energy deposition occurs only in the outer radial direction and the time lag between the emission at $r'$ and the deposition at $r$ is neglected. Then, the effective energy generation rate per unit length (erg g$^{-1}$ s$^{-1}$ cm$^{-1}$), $q_{r'} (r)$, at $r$ contributed from at $r'$ is expressed as follows. %
\begin{equation}
q_{r'} (r) = 
\begin{cases}
\epsilon_{r'} \, N_{r'} \, \exp \left({\displaystyle -\frac{r - r'}{l_{\gamma, r'}} }\right) & {\rm if} \ r \geq r' \\
0 & {\rm if} \ r < r'
\end{cases}
,
\end{equation}
where $\epsilon_{r'} = \epsilon_{\rm Co} + \epsilon_{\rm Ni}$ is the summation of the energy generation rates in Equations~(\ref{eq:epsilon_ni}) and (\ref{eq:epsilon_co}) at $r'$. %
$N_{r'}$ is the normalization factor and the value is determined so that the equation (energy available non-locally should be the same as the energy locally generated) below is true. %
\begin{equation}
\int_{r'}^{\infty} q_{r'} (r) \,dr = \epsilon_{r'}.
\end{equation}
Therefore, $N_{r'}$ is given as %
\begin{equation}
N_{r'} = \frac{1}{\displaystyle \int_{r'}^{\infty} \exp \left( - \frac{r-r'}{l_{\gamma, r'}} \right) dr}.
\end{equation}
Finally, the effective energy deposition rate (erg g$^{-1}$ s$^{-1}$), $Q (r)$, at the radius $r$ is given by integrating the contributions from the radii $< r$ as follows. %
\begin{equation}
Q (r) = \int_{0}^{r} \frac{\rho_{r'}}{\rho (r)} \left( \frac{r'}{r} \right)^2 q_{r'} (r) \, dr',
\label{eq:qr}
\end{equation}
where $\rho_{r'}$ and $\rho (r)$ are the density at radii $r'$ and $r$, respectively. %
The factor of $(\rho_{r'}/\rho (r)) (r'/r)^2$ is introduced to conserve the energy flux since the densities and effective surface areas between the radii $r$ and $r'$ are different. %
It is noted that during the integration in Equation~(\ref{eq:qr}), $q_{r'} (r)$ is practically set to be zero if $r-r' > l_{\gamma, r'}$ to count the escape of emitted energies from the system to some extent. Hereafter, the results of the test calculation with the model are presented. %

To see the impact of the effective non-local energy deposition, the fiducial 1D model, i.e., the model b18.3-mean ($f_{\rm h} = 5 \times 10^{-3}$, $f_{\rm d} = 1.0$, $t_{\rm s} = 500$ days; %
the top panels in Figure~\ref{fig:1d_mass_temp_binary} in Section~\ref{para:1d_param2}) is selected for comparison. %
With the same values of the three parameters, $f_{\rm h}$, $f_{\rm d}$, and $t_{\rm s}$, and the same angle-averaged 1D profile based on the 3D model b18.3-high \citep{2020ApJ...888..111O} as the initial condition, the calculation is performed by replacing the effective energy deposition rate, $\epsilon$, in Equation~(\ref{eq:epsilon}) with $Q (r)$ in Equation~(\ref{eq:qr}), which may potentially affect the ionization and dissociation of molecules by Compton electrons (see, Section~\ref{subsubsec:compton}) and gas temperatures through Equation~(\ref{eq:heating_rate}). %
It is reminded that to take into account the efficiency of the heating of the gas (ionization and/or destruction of atoms and molecules), the parameter $f_{\rm h}$ ($f_{\rm d}$) is introduced. %
The test calculation model is called as b18.3-mean-nledep, hereafter. %

In Figure~\ref{fig:1d_non_local_edep}, the calculation results are shown. First, the differences in the gas temperatures (right panels) are discussed by comparing them with the fiducial model (b18.3-mean). %
Focusing on the innermost particles, it is recognized that the heating of gas in the model b18.3-mean-nledep is milder compared with the model b18.3-mean; %
the peaks at about 200 days seen in the model b18.3-mean are not so evident in the model b18.3-mean-nledep. %
The highest gas temperature at 200 days in the model b18.3-mean-nledep ($\sim$ 3 $\times$ 10$^4$ K) is less than that in the model b18.3-mean ($\sim$ 6 $\times$ 10$^{4}$ K). %
It is understood that part of locally generated energy is deposited in outer regions in the model b18.3-mean-nledep. %
As seen in the model b18.3-mean, several particles initially at around 1.5 $\times$ 10$^{13}$ cm are affected by the cooling through CO ro-vibrational transitions after 50 days. %
The innermost particles are also cooled at later phases ($\lesssim$ 1000 days). %
A distinct feature seen only in the model b18.3-mean-nledep is the heating of outer particles (initially at $\gtrsim$ 3.5 $\times$ 10$^{13}$ cm) with the peaks at around 600 days. %
As seen in the top left panel in Figure~\ref{fig:prof_mean}, initially, $^{56}$Ni is distributed inside $\sim$ 3.5 $\times$ 10$^{13}$ cm. Therefore, the heating of the outer particles is due to the effective non-local energy deposition. %

Hereafter, the the time evolution of the total amounts of molecules (left panels) is discussed focusing on the differences between the two models. %
As can be seen, the qualitative features are, actually, rather similar to each other for all the molecules, although quantitatively slight differences can be recognized. %
In the model b18.3-mean-nledep, the amounts of CO and SiO at around 100 days slightly increase by 20--30$\%$ from those in the model b18.3-mean. %
After the destruction dominant phase (after approximately 130 days), the destruction of SiO in the model b18.3-mean-nledep is slightly milder than that in the model b18.3-mean. %
The amount of SiO at 400 days in the model b18.3-mean-nledep is higher than that in the model b18.3-mean by 35$\%$. %
At the final phase, the amount of SiO in the model b18.3-mean-nledep is higher than that in the model b18.3-mean by 20$\%$. %
The amount of CO at the final phase is almost the same (the difference is about 5$\%$). %
As for other molecules, although slight qualitative and quantitative differences in the evolution of the amounts can be recognized, the quantitative differences between the two models seem to be within a few tens $\%$. %
The differences seen between the models may be attributed to the slightly milder gas heating and dissociative processes by Compton electrons (ionization, dissociation, dissociative ionization: \texttt{CM} reactions) and by UV photons (\texttt{UV} reactions) and/or destructive reactions with ionized atomic species, e.g., H$^+$ and He$^+$. %

As a summary, the impact of the effective non-local energy deposition at least on the total amounts of the molecules may not be so large compared with other uncertainties, e.g., the parameter value of $f_{\rm h}$. %
The model presented here is, however, rather simple and has not been justified by the comparison with more realistic models based on radiative transfer calculations. %
In reality, non-local energy deposition could be more significant, particularly for local quantities in 3D models. %
It would be desirable to improve the treatments in the future. %

\section{Electron impact excitation/de-excitation rates} \label{app:elec_impact} 

The rate coefficients of the electron impact excitation/de-excitation of vibrational levels in Equation~(\ref{eq:rij}), $q_{ij}$, are derived from the cross sections in \cite{2008JPCA..11212296P} (see, Figure~1 in the reference) by taking Maxwellian averaging as follows. %
\begin{equation}
\begin{aligned}
q_{ij} (T) = \frac{4}{(2 \pi \mu)^{1/2} (k_{\rm B}T)^{3/2} } \\ 
\times \int_0^{\infty} \sigma_{ij} (E) 
\,E \,\exp\left(- \frac{E}{k_{\rm B} T} \right) dE,
\end{aligned}
\end{equation}
where $\sigma_{ij} (E)$ is the cross section of the transition of the vibrational state $i$ to $j$ and $\mu$ is the reduced mass of the two body system of carbon monoxide and electron. %
The values are plotted in Figure~\ref{fig:elec_impact}. %
For the transitions involved with the vibrational level $v=6$, the cross sections can not be obtained from \cite{2008JPCA..11212296P}. %
Therefore, such rates are crudely approximated (assumed) not to violate the qualitative features, for example, among the transitions of the state $i=0$ to $j$ ($j = 1, \,\ldots, \,j_{\rm max}$), $q_{0j} > q_{0k}$ ($k>j$). %

How the rates involved with the level $v=6$ are obtained is described as follows. %
First, we assume that the rates, $q_{6i}$ ($i = 0, 1, \, \ldots, \, 5$), are proportional to the rate, $q_{50}$ (have the same temperature dependence as the rate, $q_{50}$). %
As can been seen in Figure~\ref{fig:elec_impact}, the rates, $q_{i0}$ ($i = 1, \,\ldots, \, 5$), become small as $i$ increases, i.e., the transition energy from $v=i$ to $0$ (energy difference between the levels $i$ and $0$: ${\it \Delta} E_{i0}$) increases. %
Therefore, the rate, $q_{60}$, is obtained by assuming that the ratio of the rates $q_{60}$ to $q_{50}$ is equal to the ratio of the inverse transition energies $1/{\it \Delta} E_{60}$ to $1/{\it \Delta} E_{50}$. %
The rates, $q_{6i}$ ($i = 1, \, \ldots, \, 5$) are obtained by assuming the ratios of the rates $q_{6i}$ ($i = 1, \, \ldots, \, 5$) to $q_{60}$. %
The ratio is obtained by an arithmetic averaging of the ratios of the peak rates $q_{{\rm peak}, \,ji}$ to $q_{{\rm peak}, \,j0}$ ($j=1, \, \ldots, \, 5$). %
For example, the ratio of the rates $q_{65}$ to $q_{60}$ is obtained by the arithmetic averaging of the ratios of the peak rates $q_{{\rm peak}, \,j5}$ to  $q_{{\rm peak}, \,j0}$ ($j=$ 1, 2, 3, 4). %
The rates, $q_{i6}$ ($i = 0, \, \ldots, \, 5$), are obtained by the detailed balance relation between the transitions from $v=i$ to $6$ and from $v=6$ to $i$. %
Although the assumptions here are not so physically validated, since the contributions from ro-vibrational levels higher than $v=6$ are minor, the adopted approximation above should not affect the major results in this study. %

\begin{figure*}[htbp]
\begin{center}
\includegraphics[width=16cm,keepaspectratio,clip]{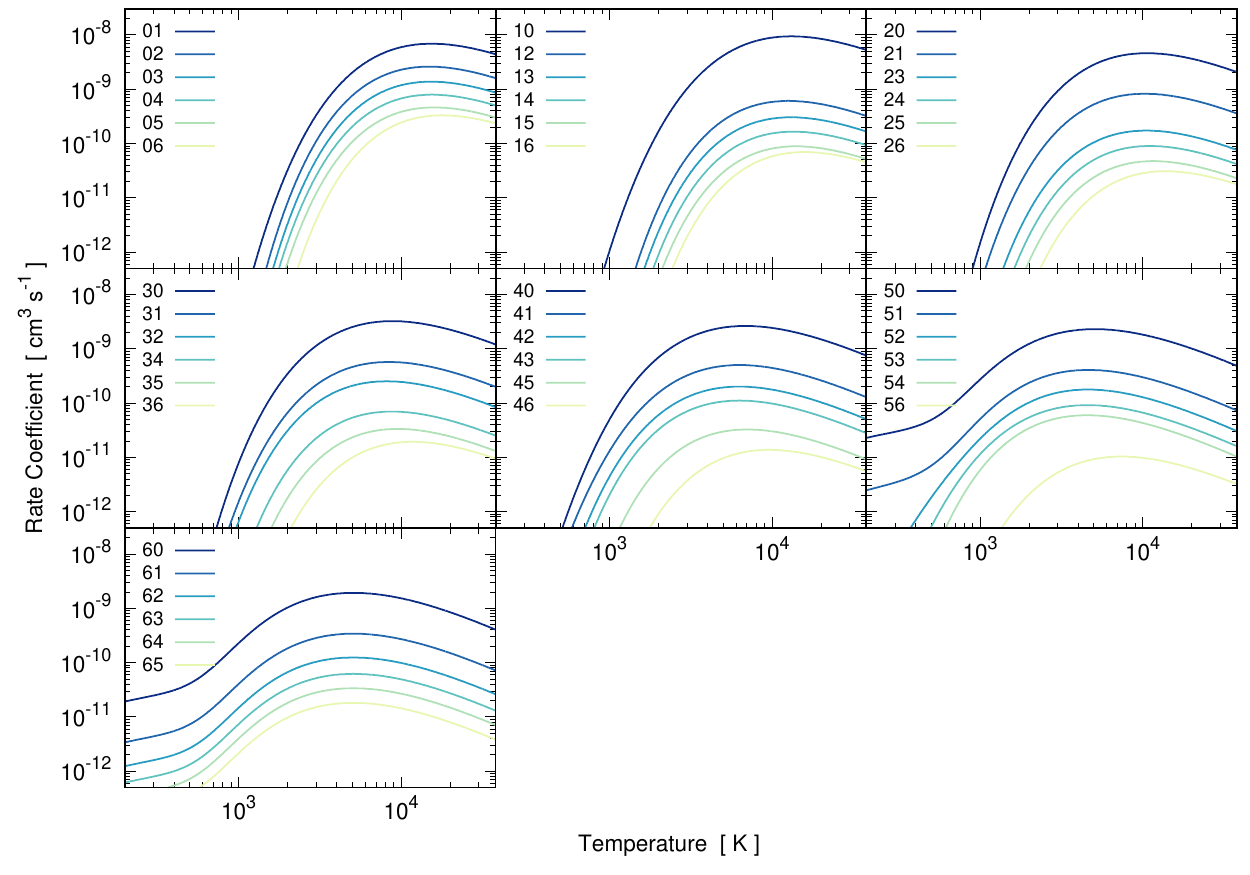}
\end{center}
\vs{-0.5}
\caption{Rate coefficients of the electron impact excitation/de-excitation of vibrational levels, $q_{ij}$, in Equation~(\ref{eq:rij}) as a function of temperature obtained based on \cite{2008JPCA..11212296P}. %
For example, the coefficient corresponding to the transition of the state $i=0$ to $j=1$ is denoted as ``01" in the legend.} %
\label{fig:elec_impact}
\end{figure*}

\section{Impact of the reduction factor $f_{\rm red}$ for the escape probabilities} \label{app:fred} 

As noted in Section~\ref{subsec:one_zone_calc}, without the reduction factor $f_{\rm red}$, at an early phase (less than about 100 days after the explosion) calculated fluxes of CO vibrational bands tend to be higher than those of the observed peak fluxes. %
In this section, how the time-dependent reduction factor $f_{\rm red}$ described in Equation~(\ref{eq:fred}) changes the CO line emissions is demonstrated by showing the results with different values of the parameter $t_{\rm s}$ with representative one-zone calculations. %

In Figure~\ref{fig:fred_app}, the reduction factor $f_{\rm red}$ is plotted as a function time for different $t_{\rm s}$ values of 200, 300, and 500 days after the explosion. %
For example, the values at 100 (300) days are approximately 1.1 $\times$ 10$^{-2}$, 5.0 $\times$ 10$^{-2}$, and 1.7 $\times$ 10$^{-1}$ (3.0 $\times$ 10$^{-2}$, 9.7 $\times$ 10$^{-2}$, and 2.5 $\times$ 10$^{-1}$) for $t_{\rm s} =$ 200, 300, and 500 days, respectively. %
In Figure~\ref{fig:fred_impact}, the one-zone calculation results with $f_{\rm h} =$ 5 $\times$ 10$^{-4}$ and $f_{\rm d} =$ 10$^{-2}$ for different $t_{\rm s}$ values are shown. %
From the values of the mean escape probabilities ${\overline \beta}$ (green dashed lines) in the left panels, it can be seen that the lower $t_{\rm s}$, the lower ${\overline \beta}$ becomes by the reduction factor $f_{\rm red}$, in particular, at early phases. %

Overall, the fluxes of CO vibrational bands (see left panels) tend to become higher than the observed peak fluxes at around 200--250 days ($\lesssim$ 10$^{-8}$ erg~s$^{-2}$~cm$^{-2}$) as early as a few tens days other than the case of $t_{\rm s} =$ 200 days (the top panels). %
Since there has been no observation before 100 days, if the fluxes before 100 days were higher than the observed peak fluxes, no detection before 100 days sounds a bit strange. %
Such early high CO line emission cases also result in rapid gas cooling and the gas temperatures (violet lines in the left panels) quickly go down to $\sim$ 100 K before 100 days except for the case of $t_{\rm s} =$ 200 days. %
It is noted that as partly mentioned in Section~\ref{para:one_zone_param1}, the previous studies \citep{1992ApJ...396..679L,1995ApJ...454..472L,2020A&A...642A.135L} have estimated the gas temperature to be $>$ 500 K before 1000 days (some of the estimations are based on the fitting of the observed CO spectra). %
Such rapid cooling is inconsistent with the previous estimations, although it may not necessarily be rejected since the adopted models and assumptions in the previous studies are at least partly different from the ones in this study. %
The formation of molecules is sensitive to the gas temperature (and density) and the fluxes of the CO line emissions depend on the amount of CO and the gas temperature again. %
Then, the degree of the gas cooling could affect nonlinearly the fluxes. %
Even though, as can be seen, the lower $t_{\rm s}$, roughly the lesser CO line emissions (gas cooling) to avoid higher fluxes before 100 days than the observed peak fluxes. %

On the other hand, the calculated fluxes with higher $t_{\rm s}$, i.e., lower reduction factor $f_{\rm red}$, better explain the observed fluxes after 200 days. %

As a summary, considering that rapid cooling seen in the calculation results with $t_{\rm s} > 200$ days is inconsistent with the previous estimations of the gas temperature and it may cause a peculiar impact on the formation of molecules at early phases (important phases for the formation of molecules), in this specific case, we prefer $t_{\rm s} =$ 200 days to avoid the overestimation compared with the observed peak fluxes and such rapid gas cooling, although the calculated fluxes after 200 days are not so consistent with the observations. %

It should be noted that the preferred parameter values could be different in other cases, in particular, in the 1D calculations as presented in Section~\ref{subsec:1d_results}. %

\begin{figure}[htbp]
\begin{center}
\includegraphics[width=8cm,keepaspectratio,clip]{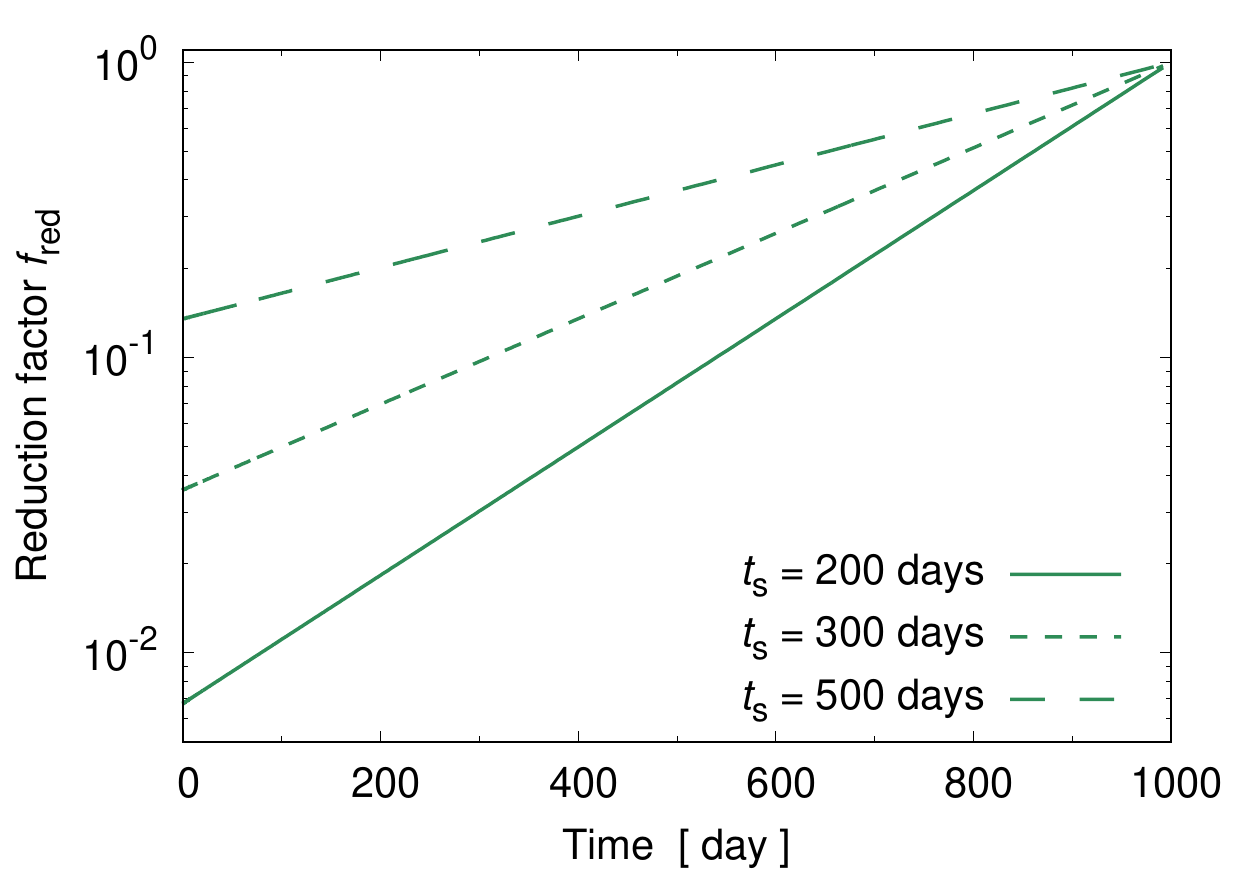}
\end{center}
\vs{-0.4}
\caption{The reduction factors $f_{\rm red}$ as a function of time for the parameter values of $t_{\rm s} =$ 200, 300, and 500 days.}
\label{fig:fred_app}
\end{figure}
%
\begin{figure*}[htbp]
\begin{minipage}{0.5\hsize}
\begin{center}
\includegraphics[width=9cm,keepaspectratio,clip]{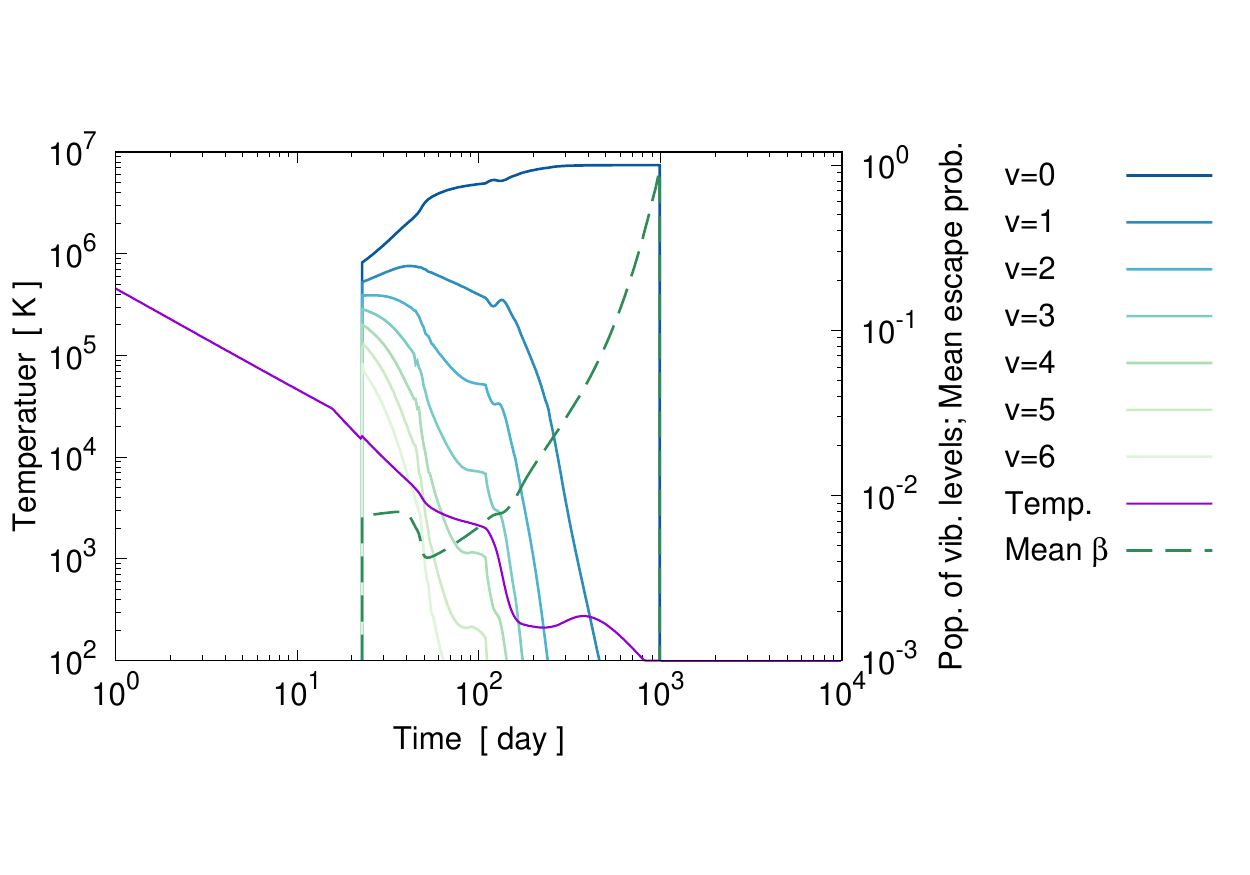}
\end{center}
\end{minipage}
\begin{minipage}{0.5\hsize}
\begin{center}
\includegraphics[width=8.5cm,keepaspectratio,clip]{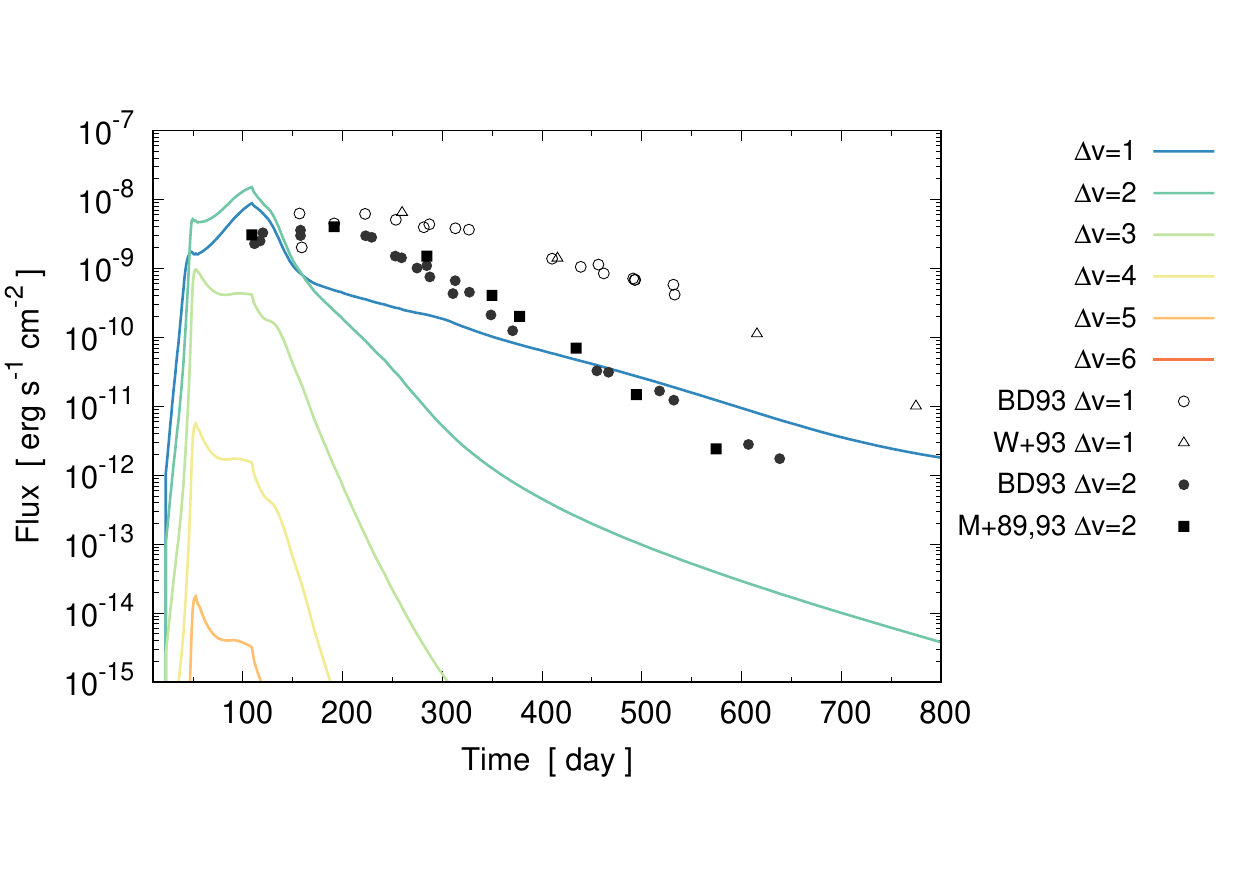}
\end{center}
\end{minipage}
\\
\begin{minipage}{0.5\hsize}
\vs{-1.6}
\begin{center}
\includegraphics[width=9cm,keepaspectratio,clip]{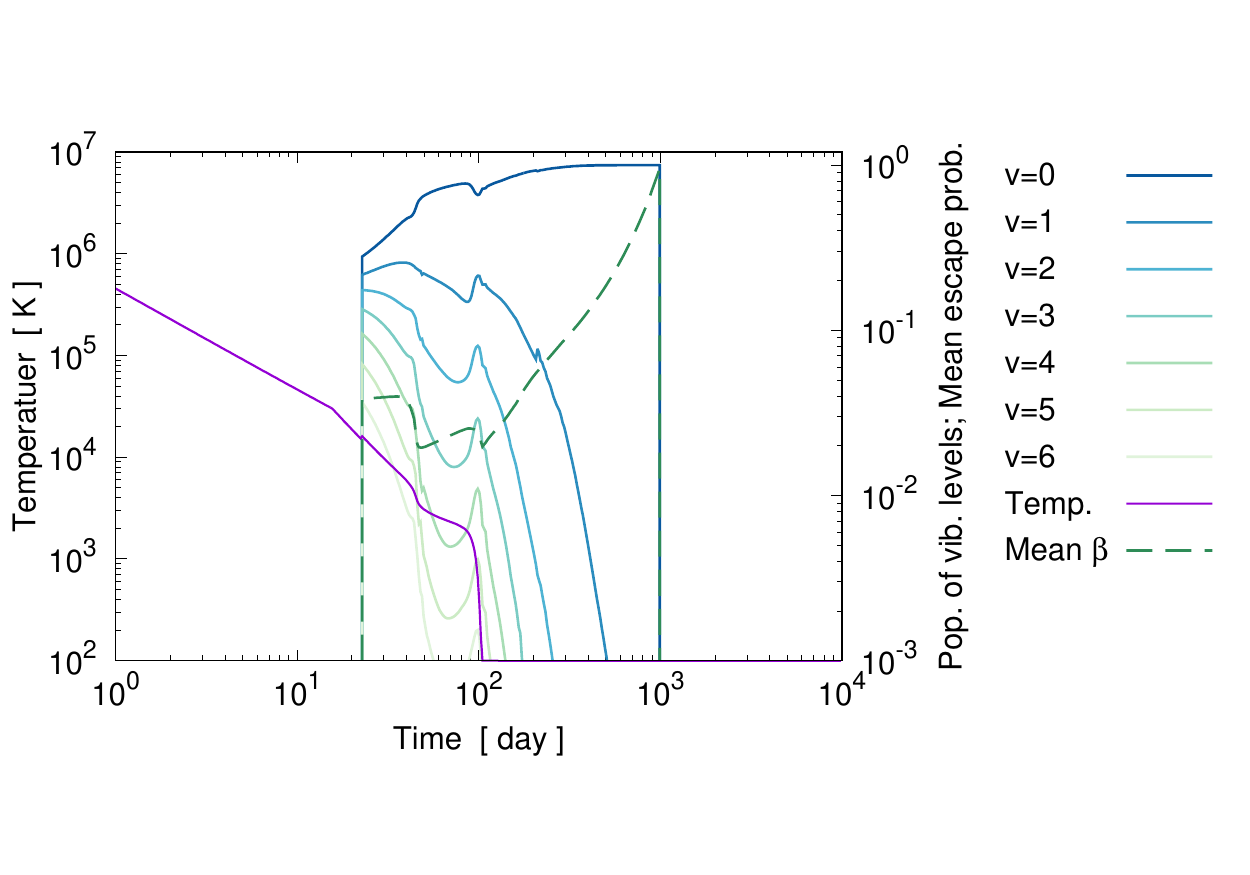}
\end{center}
\end{minipage}
\begin{minipage}{0.5\hsize}
\vs{-1.6}
\begin{center}
\includegraphics[width=8.5cm,keepaspectratio,clip]{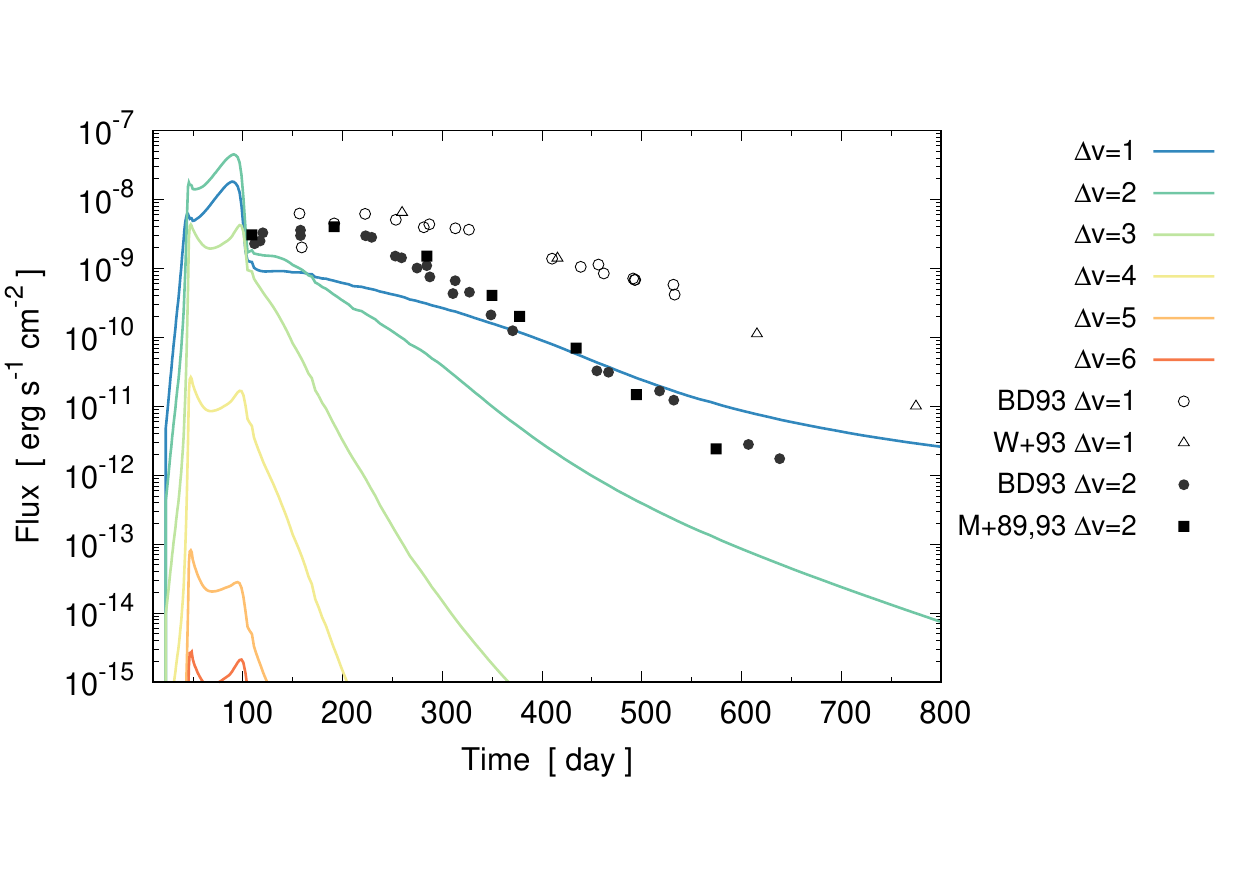}
\end{center}
\end{minipage}
\\
\begin{minipage}{0.5\hsize}
\vs{-1.6}
\begin{center}
\includegraphics[width=9cm,keepaspectratio,clip]{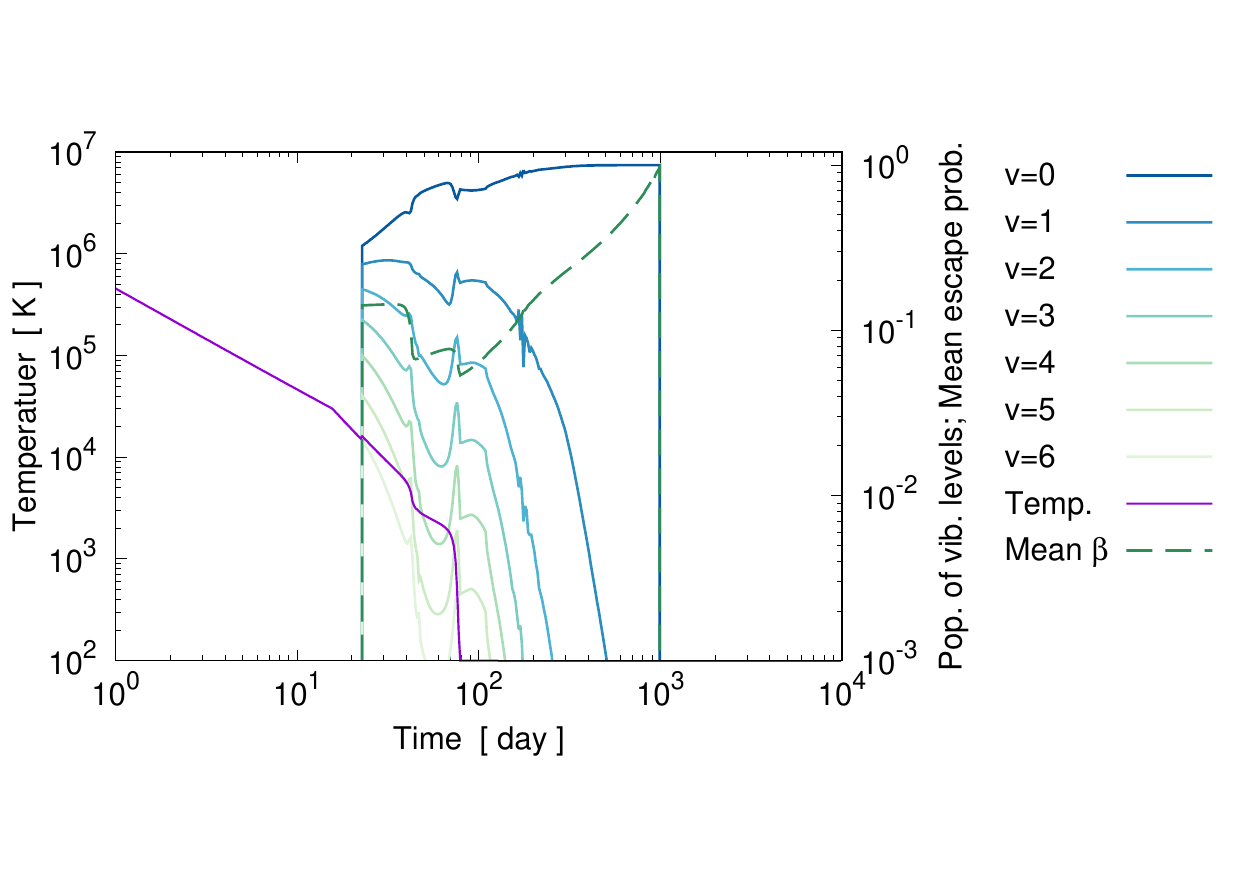}
\end{center}
\end{minipage}
\begin{minipage}{0.5\hsize}
\vs{-1.6}
\begin{center}
\includegraphics[width=8.5cm,keepaspectratio,clip]{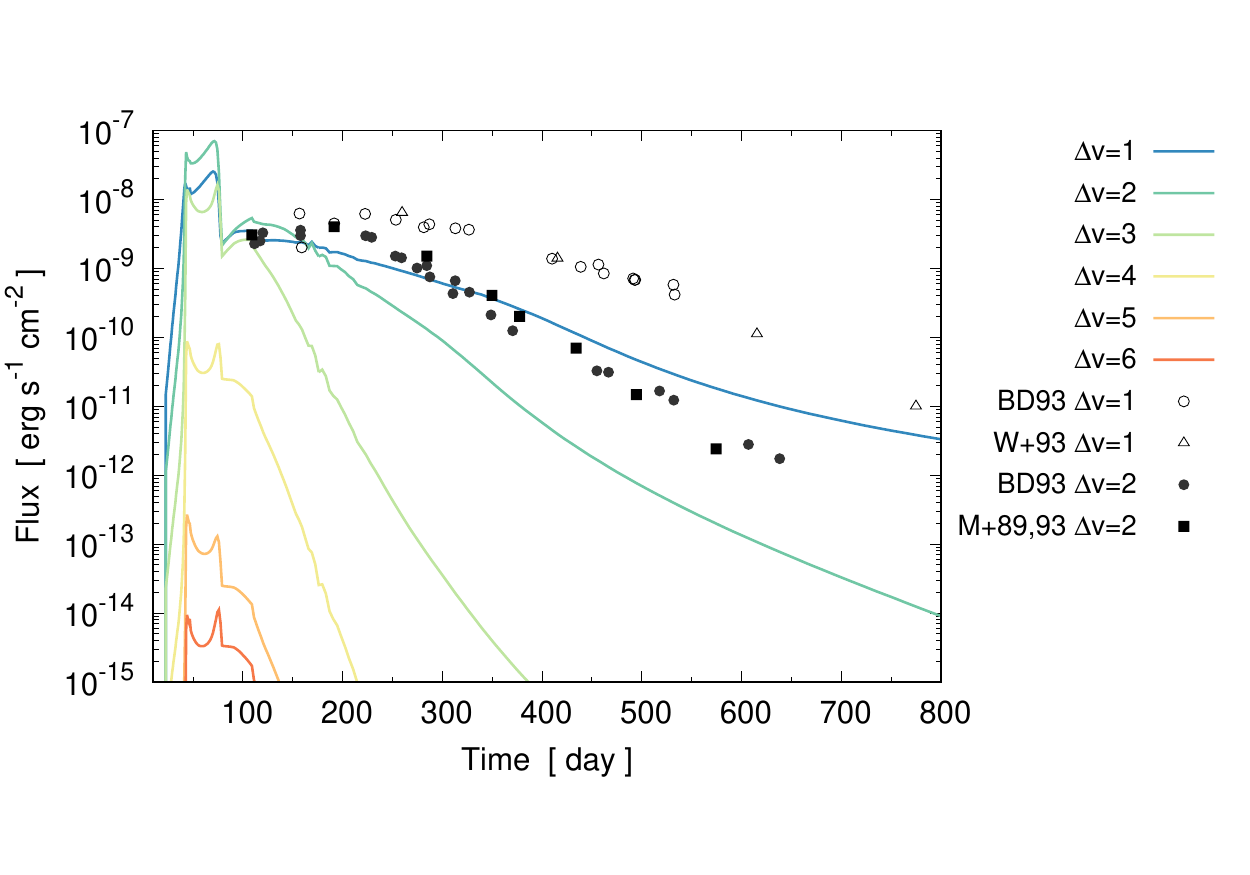}
\end{center}
\end{minipage}
\\
\begin{minipage}{0.5\hsize}
\vs{-1.6}
\begin{center}
\includegraphics[width=9cm,keepaspectratio,clip]{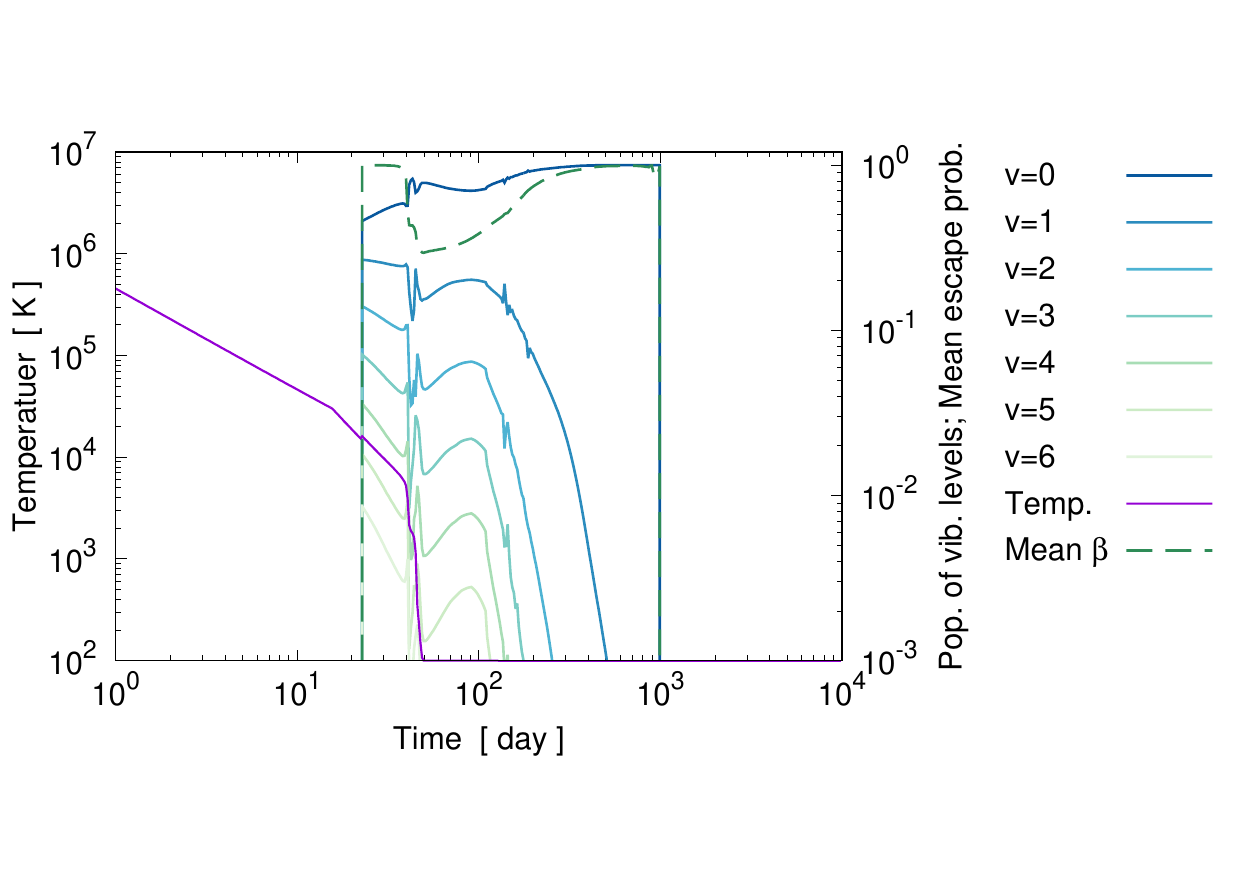}
\end{center}
\vs{-1.2}
\end{minipage}
\begin{minipage}{0.5\hsize}
\vs{-1.6}
\begin{center}
\includegraphics[width=8.5cm,keepaspectratio,clip]{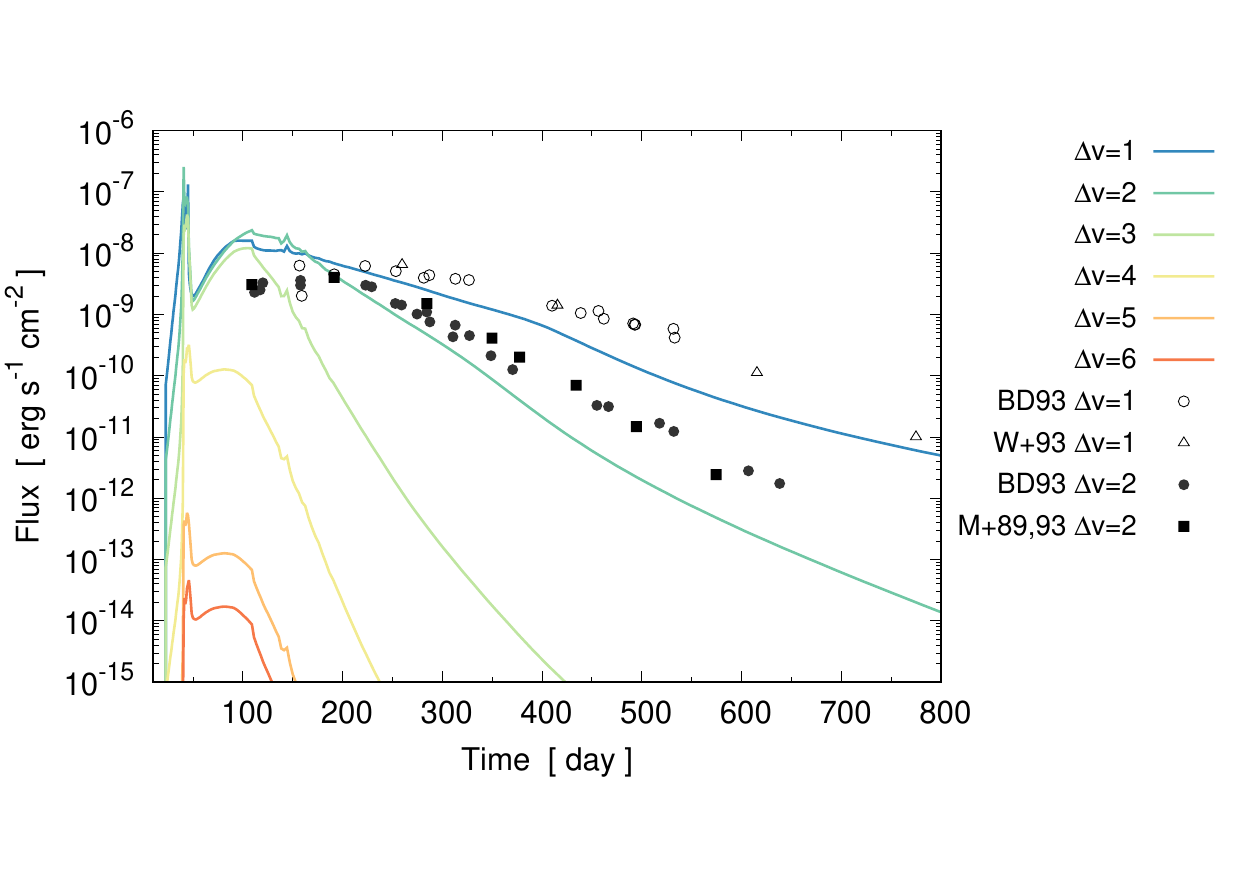}
\end{center}
\vs{-1.2}
\end{minipage}
\caption{One-zone calculation results with $f_{\rm h} =$ 5 $\times$ 10$^{-4}$ and $f_{\rm d} =$ 10$^{-2}$ for the comparison of different $t_{\rm s}$ values. %
From top to bottom, the results for $t_{\rm s} =$ 200, 300, 500, and $\infty$ days are shown, respectively. %
Left panels: time evolution of the population of vibrational levels of CO, the gas temperature, and the mean escape probability ${\overline \beta}$. %
Right panels: time evolution of the fluxes of CO vibrational bands, ${\it \Delta}v=1$ (fundamental), ${\it \Delta}v=2$ (first overtone), \ldots, ${\it \Delta}v=6$, compared with the observed light curves for ${\it \Delta}v=1$ \citep[BD93; W$+$93:][respectively]{1993A&A...273..451B,1993ApJS...88..477W} and ${\it \Delta}v=2$ \citep[M+89, 93; BD93:][respectively]{1989MNRAS.238..193M,1993MNRAS.261..535M,1993A&A...273..451B}. The top right panel is the same as the middle right panel in Figure~\ref{fig:one_zone_param1}.} %
\label{fig:fred_impact}
\end{figure*}

\bibliography{ref2,ref1}

\end{document}